\documentclass[
  aps,
  prl,
  twocolumn,
  superscriptaddress,
  preprintnumbers
]{revtex4-2}
\usepackage{orcidlink}
\usepackage{graphicx,hyperref,color,xspace}
\usepackage{amssymb,amsmath}
\usepackage{wasysym}
\usepackage{dcolumn}% Align table columns on decimal point
\usepackage{bm}% bold math
\usepackage{tikz}
\usepackage{pgfplots}
\usepackage{physics}
\usepackage{siunitx}
\usepackage[capitalise]{cleveref}
\usepackage{circuitikz}
\usepackage{xspace}
\usepackage{makecell}
\usepackage{booktabs}
\usepackage{hyperref}
\hypersetup{ 
    pdfnewwindow=true,     % links in new window
    colorlinks=true,       % false: boxed links; true: colored links
    allcolors=[RGB]{31 119 180}
} 

\pgfplotsset{compat=newest}
\usetikzlibrary{positioning,shapes,arrows,automata}
\usetikzlibrary{decorations.pathmorphing}
\usetikzlibrary{decorations.markings}
\usetikzlibrary{intersections}

\newcommand{\MyArrowIn}[5]{
    \begin{scope}[line width=1pt, color=#5,
    decoration={markings,mark=at position 0.5 with {\arrow{>}}}]
    \draw[postaction={decorate}] (#1,#2)--(#3,#4);
    \end{scope}
}
\newcommand{\MyArrowInDashed}[5]{%
    \begin{scope}[line width=1pt,dashed, color=#5,
        decoration={markings, mark=at position 0.5 with {\arrow{>}}}]
        \draw[postaction={decorate}] (#1,#2)--(#3,#4);
    \end{scope}
}
\newcommand{\MyArrowOut}[5]{
    \begin{scope}[line width=1pt, color=#5,
    decoration={markings,mark=at position 0.55 with {\arrow{>}}}]
    \draw[postaction={decorate}] (#1,#2)--(#3,#4);
    \end{scope}
}
\newcommand{\MyArrowOutDashed}[5]{%
    \begin{scope}[line width=1pt,dashed, color=#5,
        decoration={markings, mark=at position 0.55 with {\arrow{>}}}]
        \draw[postaction={decorate}] (#1,#2)--(#3,#4);
    \end{scope}
}
\newcommand{\MyReggeon}[5]{%
    \begin{scope}[line width=0.5pt, color=#5,
        decoration={snake, amplitude=1.0mm, segment length=4mm}]
        \draw[decorate] (#1,#2) -- (#3,#4);
        \draw[decorate] ($(#1,#2)+(0.6mm,0)$) -- ($(#3,#4)+(0.6mm,0)$);
    \end{scope}
}

\definecolor{jpac_blue}   {HTML}{1F77B4}
\definecolor{jpac_red}    {HTML}{D61D28}
\definecolor{jpac_green}  {HTML}{2CA02C}
\definecolor{jpac_orange} {HTML}{FF7F0E}
\definecolor{jpac_purple} {HTML}{9467BD}
\definecolor{jpac_brown}  {HTML}{8C564B}
\definecolor{jpac_pink}   {HTML}{E377C2}
\definecolor{jpac_gold}   {HTML}{BCBD22}
\definecolor{jpac_aqua}   {HTML}{17BECF}
\definecolor{jpac_grey}   {HTML}{7F7F7F}

\newcommand{\mytitle}[1]{\vspace{.5cm}{\em #1.---}}
\newcommand{\mc}{Monte Carlo\xspace}
\newcommand{\compass}{\mbox{COMPASS}\xspace}
\newcommand*{\Pom}{\ensuremath{\mathbb{P}}\xspace}
\newcommand{\ctgj}{\ensuremath{\cos{\theta_{\text{GJ}}}}\xspace}
\newcommand{\etaetaprimepi}{\ensuremath{\eta^{(\prime)}\pi^-}\xspace}
\newcommand{\etaprimepi}{\ensuremath{\eta^{\prime}\pi^-}\xspace}

\newcommand{\etaetaprime}{\ensuremath{\eta^{(\prime)}}\xspace}
\newcommand{\etapi}{\ensuremath{\eta\pi^-}\xspace}

\newcommand{\ie}{\emph{i.e.}\xspace}

\newcommand{\vs}{\emph{vs.}\xspace}

\newcommand{\aka}{\emph{a.k.a.}\xspace}

\newcommand{\mevnospace}{\ensuremath{{\mathrm{\,Me\kern -0.1em V}}}}
\newcommand{\gevnospace}{\ensuremath{{\mathrm{\,Ge\kern -0.1em V}}}}
\newcommand{\tevnospace}{\ensuremath{{\mathrm{\,Te\kern -0.1em V}}}}

\newcommand{\gev}{\gevnospace\xspace}

\newcommand*{\beamgj}{\vec{p}_\text{beam}^{\,\text{GJ}}}
\newcommand*{\targetgj}{\vec{p}_\text{target}^{\,\text{GJ}}}
\newcommand*{\etagj}{\vec{p}_{\etaetaprime}^{\,\text{GJ}}}

\newcommand*{\recoilgj}{\vec{p}_\text{recoil}^{\,\text{GJ}}}

\makeatletter
\renewcommand\@fnsymbol[1]{\@alph{#1}}
\makeatother

\begin{document}
\title{First Observation of an Exotic Reggeon}
\preprint{CERN-EP-2026–209}
\preprint{JLAB-THY-26-4814}
\preprint{MIT-CTP/6061}

% COMPASS author list -- auto-generated from AuthorsCOMPASS.tex
% by convert_compass_authors.py. Do not edit by hand; edit the script.
% Institutional funding footnotes (A-J) are relocated to Ack.tex.

% JPAC affiliations
\newcommand{\AGH}{AGH University of Krakow, Faculty of Physics and Applied Computer Science, Krak\'ow PL-30-059, Poland}
\newcommand{\barcelona}{Departament de F\'isica Qu\`antica i Astrof\'isica and Institut de Ci\`encies del Cosmos, Universitat de Barcelona, Barcelona E-08028, Spain}
\newcommand{\bochum}{Fakult\"{a}t f\"{u}r Physik und Astronomie, Ruhr-Universit\"{a}t Bochum, Bochum D-44780, Germany}
\newcommand{\catania}{INFN Sezione di Catania, Catania I-95123, Italy}
\newcommand{\ceem}{Center for  Exploration  of  Energy  and  Matter, Indiana  University, Bloomington,  Indiana  47403,  USA}
\newcommand{\gwu}{Department of Physics, The George Washington University, Washington, DC 20052, USA}
\newcommand{\hiskp}{Helmholtz-Institut f\"{u}r Strahlen- und Kernphysik (Theorie) and Bethe Center for Theoretical Physics, Universit\"{a}t Bonn, Bonn D-53115, Germany}
\newcommand{\icn}{Instituto de Ciencias Nucleares,
Universidad Nacional Aut\'onoma de M\'exico, Ciudad de M\'exico 04510, Mexico}
\newcommand{\indiana}{Department of Physics, Indiana  University, Bloomington,  Indiana 47405,  USA}
\newcommand{\jlab}{Theory Center, Thomas  Jefferson  National  Accelerator  Facility, Newport  News,  Virginia  23606,  USA}
\newcommand{\lbnl}{Nuclear Science Division, Lawrence Berkeley National Laboratory, Berkeley, California 94720, USA}
\newcommand{\messina}{Dipartimento di Scienze Matematiche e Informatiche, Scienze Fisiche e Scienze della Terra, Universit\`a degli Studi di Messina, Messina I-98166, Italy}
\newcommand{\MIT}{Center for Theoretical Physics, Massachusetts Institute of Technology, Cambridge, Massachusetts 02139, USA}
\newcommand{\odu}{Department of Physics, Old Dominion University, Norfolk, Virginia 23529, USA}
\newcommand{\ucb}{Department of Physics, University of California, Berkeley, California 94720, USA}
\newcommand{\uned}{Departamento de F\'isica Interdisciplinar, Universidad Nacional de Educaci\'on a Distancia (UNED), Madrid E-28040, Spain}
\newcommand{\wm}{Department of Physics, William \& Mary, Williamsburg, Virginia 23187, USA}
%\newcommand{\bonnhiskp}{Universit\"{a}t Bonn, Helmholtz-Institut f\"{u}r Strahlen- und Kernphysik, Bonn D-53115, Germany}
%\newcommand{\munichtu}{Technische Universit\"{a}t M\"{u}nchen, Physik Department, Garching D-85748, Germany}
% COMPASS affiliations
\newcommand{\compassauthoraanl}{A.I. Alikhanyan National Science Laboratory, 2 Alikhanyan Br. Street, 0036, Yerevan, Armenia}
\newcommand{\compassauthorbrno}{Institute of Scientific Instruments of the CAS, 61264 Brno, Czech Republic}
\newcommand{\compassauthorliberec}{Technical University in Liberec, 46117 Liberec, Czech Republic}
\newcommand{\compassauthorpraguectu}{Czech Technical University in Prague, 16636 Prague, Czech Republic}
\newcommand{\compassauthorpraguecu}{Charles University, Faculty of Mathematics and Physics, 12116 Prague, Czech Republic}
\newcommand{\compassauthorsaclay}{IRFU, CEA, Universit\'e Paris-Saclay, 91191 Gif-sur-Yvette, France}
\newcommand{\compassauthorbochum}{Universit\"at Bochum, Institut f\"ur Experimentalphysik, 44780 Bochum, Germany}
\newcommand{\compassauthorbonniskp}{Universit\"at Bonn, Helmholtz-Institut f\"ur  Strahlen- und Kernphysik, 53115 Bonn, Germany}
\newcommand{\compassauthorbonnpi}{Universit\"at Bonn, Physikalisches Institut, 53115 Bonn, Germany}
\newcommand{\compassauthorfreiburg}{Universit\"at Freiburg, Physikalisches Institut, 79104 Freiburg, Germany}
\newcommand{\compassauthormainz}{Universit\"at Mainz, Institut f\"ur Kernphysik, 55099 Mainz, Germany}
\newcommand{\compassauthormunichtu}{Technische Universit\"at M\"unchen, Physik Dept., 85748 Garching, Germany}
\newcommand{\compassauthormunichuni}{Ludwig-Maximilians-Universit\"at, 80539 M\"unchen, Germany}
\newcommand{\compassauthorcalcutta}{Matrivani Institute of Experimental Research \& Education, Calcutta-700 030, India}
\newcommand{\compassauthortelaviv}{Tel Aviv University, School of Physics and Astronomy, 69978 Tel Aviv, Israel}
\newcommand{\compassauthortriesta}{Abdus Salam ICTP, 34151 Trieste, Italy}
\newcommand{\compassauthortriesti}{Trieste Section of INFN, 34127 Trieste, Italy}
\newcommand{\compassauthortriestu}{University of Trieste, Dept.\ of Physics, 34127 Trieste, Italy}
\newcommand{\compassauthorturini}{Torino Section of INFN, 10125 Torino, Italy}
\newcommand{\compassauthorturinu}{University of Torino, Dept.\ of Physics, 10125 Torino, Italy}
\newcommand{\compassauthormiyazaki}{University of Miyazaki, Miyazaki 889-2192, Japan}
\newcommand{\compassauthoryamagata}{Yamagata University, Yamagata 990-8560, Japan}
\newcommand{\compassauthorwarsaw}{National Centre for Nuclear Research, 02-093 Warsaw, Poland}
\newcommand{\compassauthorwarsawtu}{Warsaw University of Technology, Institute of Radioelectronics, 00-665 Warsaw, Poland}
\newcommand{\compassauthorwarsawu}{University of Warsaw, Faculty of Physics, 02-093 Warsaw, Poland}
\newcommand{\compassauthoraveiro}{University of Aveiro, I3N, Dept. of Physics, 3810-193 Aveiro, Portugal}
\newcommand{\compassauthorlisbon}{LIP, 1649-003 Lisbon, Portugal}
\newcommand{\compassauthordubna}{Affiliated with an international laboratory covered by a cooperation agreement with CERN}
\newcommand{\compassauthorrussia}{Affiliated with an institute formerly covered by a cooperation agreement with CERN}
\newcommand{\compassauthorcern}{CERN, 1211 Geneva 23, Switzerland}
\newcommand{\compassauthortaipei}{Academia Sinica, Institute of Physics, Taipei 11529, Taiwan}
\newcommand{\compassauthortaipeincu}{Center for High Energy and High Field Physics and Dept.\ of Physics, National Central University, 300 Zhongda Rd., Zhongli 320317, Taiwan}
\newcommand{\compassauthorillinois}{University of Illinois at Urbana-Champaign, Dept.\ of Physics, Urbana, IL 61801-3080, USA}

% COMPASS authors
\author{G.~D.~Alexeev\,\orcidlink{0009-0007-0196-8178}}
\affiliation{\compassauthordubna}

\author{M.~G.~Alexeev\,\orcidlink{0000-0002-7306-8255}}
\affiliation{\compassauthorturinu}
\affiliation{\compassauthorturini}

\author{C.~Alice\,\orcidlink{0000-0001-6297-9857}}
\affiliation{\compassauthorturinu}
\affiliation{\compassauthorturini}

\author{A.~Amoroso\,\orcidlink{0000-0002-3095-8610}}
\affiliation{\compassauthorturinu}
\affiliation{\compassauthorturini}

\author{V.~Andrieux\,\orcidlink{0000-0001-9957-9910}}
\affiliation{\compassauthorillinois}

\author{V.~Anosov\,\orcidlink{0009-0003-3595-9561}}
\affiliation{\compassauthordubna}

\author{K.~Augsten\,\orcidlink{0000-0001-8324-0576}}
\affiliation{\compassauthorpraguectu}

\author{W.~Augustyniak}
\affiliation{\compassauthorwarsaw}

\author{C.~D.~R.~Azevedo\,\orcidlink{0000-0002-0012-9918}}
\affiliation{\compassauthoraveiro}

\author{B.~Badelek\,\orcidlink{0000-0002-4082-1466}}
\affiliation{\compassauthorwarsawu}

% ...

%People
% COMPASS collaboration
\collaboration{COMPASS Collaboration}

% JPAC collaboration
\author{C.~Fern\'andez-Ram\'irez\,\orcidlink{0000-0001-8979-5660}}\altaffiliation{Corresponding author}\email{cefera@ccia.uned.es}\affiliation{\uned}
\author{M.~Mikhasenko\,\orcidlink{0000-0002-6969-2063}}\altaffiliation{Also a COMPASS collaboration member}\affiliation{\bochum}
\author{{\L}.~Bibrzycki\,\orcidlink{0000-0002-6117-4894}}\affiliation{\AGH}
\author{G.~Foti\,\orcidlink{0009-0000-9791-3823}}\affiliation{\messina}\affiliation{\catania}
\author{N.~Hammoud\,\orcidlink{0000-0002-8395-0647}}\affiliation{\barcelona}
\author{V.~Mathieu\,\orcidlink{0000-0003-4955-3311}}\affiliation{\barcelona}
\author{G.~Monta\~na\,\orcidlink{0000-0001-8093-6682}}\affiliation{\barcelona}
\author{R.~J.~Perry\,\orcidlink{0000-0002-2954-5050}}\affiliation{\MIT}
\author{A.~Pilloni\,\orcidlink{0000-0003-4257-0928}}\affiliation{\messina}\affiliation{\catania}
\author{A.~Rodas\,\orcidlink{0000-0003-2702-5286}}\affiliation{\odu}\affiliation{\jlab}
\author{V.~Shastry\,\orcidlink{0000-0003-1296-8468}}\affiliation{\ceem} \affiliation{\indiana}
\author{W.~A.~Smith\,\orcidlink{0009-0001-3244-6889}}\affiliation{\messina}\affiliation{\lbnl}\affiliation{\ucb}\affiliation{\wm}
\author{A.~P.~Szczepaniak\,\orcidlink{0000-0002-4156-5492}}\affiliation{\jlab}\affiliation{\ceem}\affiliation{\indiana}
\author{D.~Winney\,\orcidlink{0000-0002-8076-243X}}\affiliation{\icn}
\collaboration{Joint Physics Analysis Center}
\begin{abstract}
We present new \compass high-statistics measurements of the peripheral
production of $\eta \pi^-$ and $\eta^\prime \pi^-$ pairs in the
reactions $\pi^- p \to \eta^{(\prime)}\pi^- p$. For the first time, we
perform an unbinned analysis of the high-mass region of the
$\eta\pi^-$ and $\eta^\prime\pi^-$ systems, which allows us to
disentangle the exchange mechanisms governing their production. We
report the first observation in high-energy scattering of an exotic
Reggeon with high significance exceeding $5\,\sigma$ for both
channels, which is, most likely, related to the exotic $\pi_1(1600)$.
\end{abstract}
\maketitle
\mytitle{Introduction}
Understanding the hadron spectrum in terms of quark and gluon degrees of freedom
of quantum chromodynamics (QCD) is a central goal of nonperturbative
strong-interaction physics~\cite{Shepherd:2016dni,Brambilla:2014jmp}. Many
exotic states beyond the na\"ive quark model~\cite{ParticleDataGroup:2024cfk}
have now been identified, predominantly with heavy
quarks~\cite{Esposito:2016noz,Lebed:2016hpi,Guo:2017jvc,Olsen:2017bmm,Brambilla:2019esw}
but also with light ones~\cite{Ketzer:2019wmd}. Candidates for some of the
latter are \textit{hybrid
mesons}~\cite{Barnes:1982zs,Barnes:1982tx,Horn:1977rq,Chanowitz:1982qj,Isgur:1984bm,Klempt:2007cp,Dudek:2012ag}, \ie
hadrons with an explicit gluonic excitation that provides a direct probe of
low-energy gluon
dynamics~\cite{Ketzer:2012vn,Meyer:2015eta,Gross:2022hyw}. Hybrids are
resonances and, therefore, are identified through the final states they decay
into. In particular, final states of $\eta\pi$ and $\eta^{\prime}\pi$ with odd
angular momentum are relevant because they explicitly carry exotic quantum
numbers that cannot be obtained with a quark-antiquark pair. These final states
have long served as benchmark channels in the search for hybrid mesons. Early
evidence came from two mass regions: a structure near $1.4~\gev/c^2$ in
$\eta\pi^-$, observed by E852 at BNL~\cite{Thompson:1997bs,E852:1999xev} and
Crystal Barrel at CERN~\cite{Abele:1998gn}, and a heavier one near
$1.6~\gev/c^2$ in $\eta^{\prime}\pi^-$, reported by VES~\cite{VES:1993scg} and
E852~\cite{Ivanov:2001rv}. High-statistics measurements
by \compass~\cite{Abbon:2014aex,COMPASS:2014vkj} later established both peaks
unambiguously. The puzzle of whether they correspond to one state or two was
resolved by a JPAC coupled-channel analysis of the \compass
data~\cite{JPAC:2018zyd}. The analysis showed that a single spin-exotic
resonance, the $\pi_1(1600)$, accounts for both structures. This identification
matches theoretical expectations for the lightest hybrid
meson~\cite{Lacock:1996ny,MILC:1997usn,Dudek:2013yja,Szczepaniak:2001rg,Szczepaniak:2006nx,Guo:2008yz}
and has since been reinforced by independent coupled-channel
analyses~\cite{CrystalBarrel:2019zqh,Kopf:2020yoa} and lattice QCD
calculations~\cite{Woss:2020ayi}.

\begin{figure*}[t!]
\centering
\includegraphics[width=1.9\columnwidth]{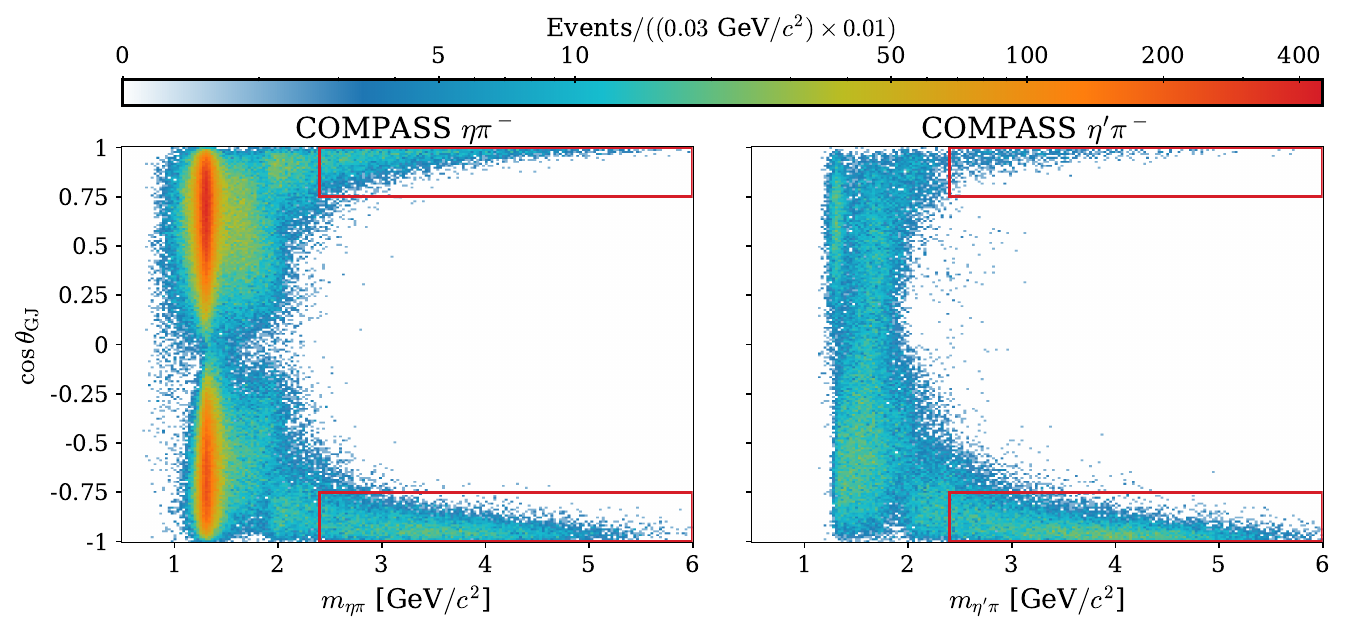}
\caption{
\etapi (left) and \etaprimepi (right) event distributions in logarithmic scale,
integrated over the four-momentum transfer to the target proton in the range
$t_p\in\left[-1,-0.1\right]\,\gev^2/c^2$ and azimuthal angle. The red boxes mark
the region fitted in this analysis. The fast-$\eta^{(')}$ region corresponds to
$\ctgj>0$. The forward-backward asymmetry $A(m_{\eta\pi})$ is obtained as the
difference between the forward and backward events for a given $m_{\eta\pi}$
bin, divided by the sum.
}
\label{fig:compass_plot}
\end{figure*}

High-energy reactions can be parametrized in terms of exchanges of towers of
increasing-spin states known collectively as Reggeons. All particles in a
Reggeon are identified by a trajectory $J=\alpha(p^2)$ that relates the spin $J$
of each particle to its invariant mass squared $p^2$, as long as $p^2 > 0$. In
the spacelike region $p^2<0$ is the momentum transferred in the reaction. In
that case, the trajectory collectively determines the energy dependence of the
cross section. Conventionally, the Reggeon takes the name of the lightest
particle lying on the trajectory~\cite{Collins:1977jy,Winney:2025tla}. In the
reactions under study, $\pi^- p \to \eta^{(\prime)}\pi^- p$, the discrete
quantum numbers of external particles are such that odd spin exchanges are
manifestly exotic. The difference between odd and even spin exchanges is
reflected in the dependence of the cross sections on the kinematic variables,
and therefore can be determined from data.

In this Letter, we report the first observation of an exotic Reggeon, with
significances of $9.9\, \sigma$ in $\eta\pi^-$ and $5.5\, \sigma$ in
$\eta^\prime\pi^-$, most plausibly identified with the exchange of the exotic
$\pi_1(1600)$. This was achieved using an event-based analysis of
recent \compass measurements of peripheral $\etaetaprimepi$ production in the
double-Regge (DR) regime that allowed us to disentangle the dominant exchanges,
and identify the presence of an exotic trajectory.

\mytitle{\compass data}
The COMPASS experiment is a forward spectrometer located at CERN
SPS~\cite{COMPASS:2007rjf}. The setup used for collecting data of this analysis
is described in Ref.~\cite{COMPASS:2014cka}. \compass collected data for two
years with a negatively charged hadron beam (97\% $\pi^-$) on a proton
target. We analyze the complete \etaetaprimepi data set, which extends up to
$m_{\eta\pi}=6\gev/c^2$; throughout this Letter, the $\eta$ subscript denotes
both $\eta$ and $\eta^\prime$ in kinematic variables. To isolate the production
of the $\eta\pi$ system, we require three charged tracks from the primary vertex
and exactly two photon clusters. Fiducial cuts on energy and kinematics are
applied to select exclusive reactions. For \etapi (\etaprimepi), we reconstruct
$\pi^0$ ($\eta$) via a kinematic fit within a tight window in the diphoton
invariant mass, then cut on \mbox{$\pi^+\pi^-\gamma\gamma$} invariant mass to
match the $\eta$ ($\eta^\prime$). The resulting samples contain $\sim270,\!000$
($\sim85,\!000$) events from threshold up to $6\gev/c^2$ with four-momentum
transfer to the target proton in the range
$t_p\in\left[-1,-0.1\right]\,\gev^2/c^2$; further details are given in
Ref.~\cite{Pekeler:2025}.

\Cref{fig:compass_plot}
shows $m_{\eta\pi}$ versus $\ctgj$, the cosine of the angle between the
$\eta^{(\prime)}$ and the beam in the Gottfried-Jackson
frame~\cite{Gottfried:1964nx,Pekeler:2025,Bibrzycki:2021rwh}. The details on the
kinematic variables can be found in the End Matter. Above $2\gev/c^2$, the
distributions peak sharply in the forward and backward regions, where the
$\eta^{(\prime)}$ or the $\pi$ carries most of the beam momentum; these are
named the fast-$\eta$ and fast-$\pi$ regions, respectively. The peaks narrow
with increasing $m_{\eta\pi}$, characteristic of diffractive processes for which
DR exchange is the expected dominant
mechanism~\cite{Bialas:1969jz,Shi:2014nea,Bibrzycki:2021rwh}.

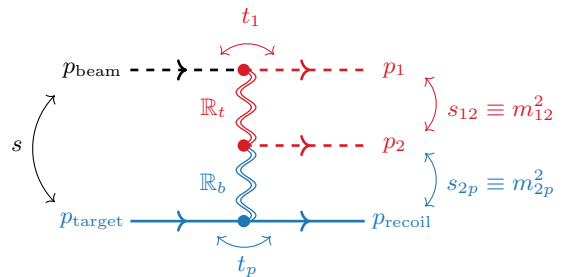
\begin{figure}
\begin{tabular}{c}
\centering
\begin{tikzpicture}[scale=1, transform shape]
\MyArrowInDashed{-1.5}{0.5}{0}{0.5}{black};
\node[align=left] at (-2,0.5) {$p_\text{beam}$};
\MyArrowOutDashed{0}{0.5}{1.6}{0.5}{jpac_red};
\node[align=left, text=jpac_red] at (2,0.5) {$p_1$};
\MyArrowOutDashed{0}{-0.5}{1.6}{-0.5}{jpac_red};
\node[align=left, text=jpac_red] at (2,-0.5) {$p_2$};
\node[align=left, text=jpac_red] at (-0.4,0) {$\mathbb{R}_t$};
\MyReggeon{0}{0.53}{0}{-0.53}{jpac_red};
\node[align=left, text=jpac_blue] at (-0.4,-1) {$\mathbb{R}_b$};
\MyReggeon{0}{-0.5}{0}{-1.55}{jpac_blue};
\MyArrowIn{-1.5}{-1.5}{0}{-1.5}{jpac_blue};
\node[align=left, text=jpac_blue] at (-2,-1.5) {$p_{\text{target}}$};
\MyArrowOut{0}{-1.5}{1.6}{-1.5}{jpac_blue};
\node[align=left, text=jpac_blue] at (2.1,-1.5) {$p_{\text{recoil}}$};
\fill[jpac_red] (0,0.5) circle (0.6ex);
\fill[jpac_red] (0,-0.5) circle (0.6ex);
\fill[jpac_blue] (0,-1.5) circle (0.6ex);
\draw[<->] (-2.5,-1.25) arc (45:-45:-1);
\node[align=left] at (-3,-0.5) {$s$};
\draw[<->,color=jpac_red] (2.4,0.4) arc (45:-45:0.5);
\node[align=left,text=jpac_red] at (3.4,0) {$s_{12}\equiv m^2_{12}$};
\draw[<->,color=jpac_blue] (2.4,-0.6) arc (45:-45:0.5);
\node[align=left,text=jpac_blue] at (3.4,-1) {$s_{2p}\equiv m^2_{2 p}$};
\draw[<->, jpac_red] (0.4,0.7) arc (45:135:0.5);
\node[align=left, text=jpac_red] at (0.1,1.2) {$t_1$};
\draw[<->, color=jpac_blue] (-0.35,-1.7) arc (45:135:-0.5);
\node[align=left, text=jpac_blue] at (0.05,-2.1) {$t_p$};    
\end{tikzpicture}
\end{tabular} 
\caption{
Diagrammatic representation of the DR production mechanism. The meson produced
with small momentum transfer $t_1$ from the beam carries the largest momentum
and is referred to as fast. Fast-$\eta$ (fast-$\pi$) amplitudes correspond to
final states $1=\eta^{(\prime)}$ ($\pi$) and $2=\pi$ ($\eta^{(\prime)}$).  The
exchanged Reggeons are $\mathbb{R}_t=a_2$, $a_2^\prime$, or $\pi_1$ ($f_2$ or
$\Pom$) at the top vertex and $\mathbb{R}_b=f_2$ or $\Pom$ at the bottom
vertex. Six amplitudes, $\mathbb{R}_t\mathbb{R}_b=a_2 f_2$, $a_2 \Pom$,
$a_2^\prime f_2$, $a_2^\prime \Pom$, $\pi_1 f_2$,  and $\pi_1 \Pom$, can
contribute to the forward region, while four, $\mathbb{R}_t\mathbb{R}_b=f_2
f_2$, $f_2 \Pom$, $\Pom f_2$, and $\Pom \Pom$, can contribute to the backward
region.}\label{fig:doubleregge-feynman}
\end{figure}

\mytitle{Production amplitude}
A $2\to 3$ process is kinematically described by three energy variables ($s$,
$s_{12}$, $s_{2p}$) and two momentum transfers ($t_1$, $t_p$). When all energies
are large and momentum transfers small, the reaction is dominated by the
exchange of a top Reggeon $\mathbb{R}_t$ between the produced mesons and a
bottom Reggeon $\mathbb{R}_b$ between the proton and the meson shown in the
middle of~\protect\cref{fig:doubleregge-feynman}---the DR
mechanism~\cite{Bialas:1969jz,Brower:1974yv,Collins:1977jy,Shi:2014nea,Bibrzycki:2021rwh}. Previous \compass
analyses were performed in the single Regge kinematics where the meson system is
produced in the resonance region. At small momentum transfer, the Regge
trajectories are approximately linear and only a few Reggeons are expected to
contribute. Reggeons with the largest trajectory are referred to as
leading. Quantum numbers of external particles restrict the possible
exchanges. For diffractive processes, the bottom exchange is dominantly Pomeron
($\Pom$), with possible subleading corrections due to the $f_2$. For the top
exchange, quantum numbers of beam and produced particles will force natural
parity, and $I^G=1^-$ ($I^G=0^+$) for the fast-$\eta$ (fast-$\pi$) amplitude,
with isospin $I$ and $G$-parity $G$. This implies that the possible top
exchanges for fast-$\pi$ are the even-spin $\Pom$ or $f_2$, with odd-spins
forbidden by Bose symmetry. For \mbox{fast-$\eta$}, the leading exchange is
expected to be the ordinary $a_2$, but as subleading Reggeons both the even-spin
(ordinary) $a_2'$ and the odd-spin (exotic) $\pi_1$ are allowed. We construct
our reaction model adopting the framework of
Refs.~\cite{Weis:1972tbu,Shimada:1978sx,Shi:2014nea,Bibrzycki:2021rwh}, with the
addition of form factors. The explicit expressions are given in the End Matter.

\begin{figure}
\begin{tabular}{c}
\includegraphics[width=0.95\linewidth]{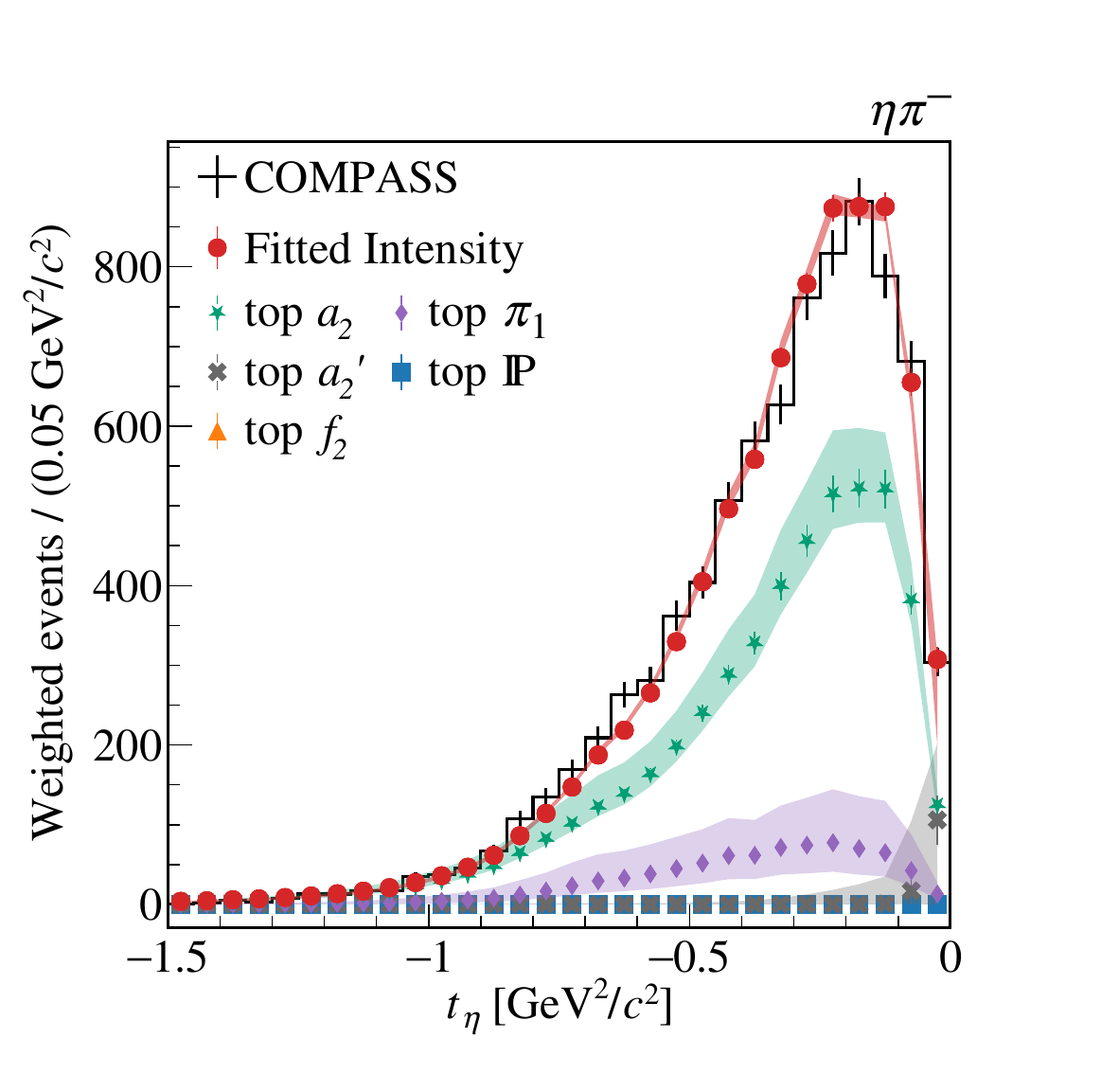} \\
\includegraphics[width=0.95\linewidth]{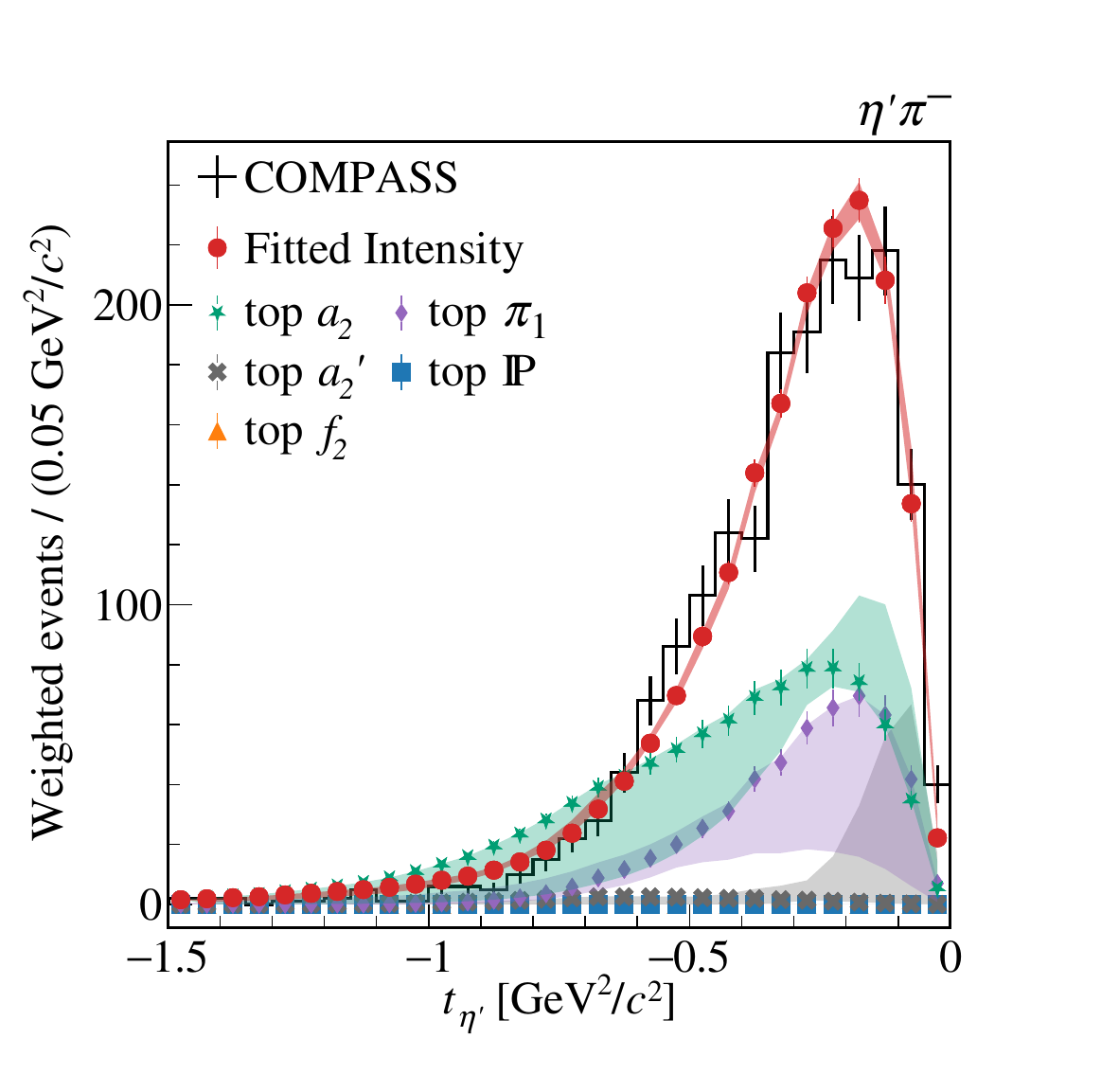} 
\end{tabular}
\caption{
Weighted intensity distributions at small $t_\eta$ for the individual
top-exchange contributions in the $\eta\pi^-$ (top) and $\eta^\prime \pi^-$
(bottom) channels. Small $t_\eta$ corresponds $\ctgj \sim 1$.  By top $\pi_1$,
$a_2$, $f_2$ and $\Pom$ we refer to the coherent sum of the amplitudes whose
exchanged $\mathbb{R}_t$ is the $\pi_1$, $a_2$, $f_2$ and $\Pom$,
respectively. Note that the top $f_2$ and $\Pom$ exchanges are negligible, as
expected. Systematic uncertainties are provided as a band. Data and fit include
the acceptance of the \compass apparatus.
}
\label{fig:weight1}
\end{figure}

\mytitle{Event fitting}
Acceptance effects are accounted for using \mc samples for the
$\eta^{(\prime)}\pi^-$ final state propagated through the \compass simulation
framework \texttt{TGEANT}~\cite{Szameitat:2017ekf}, based
on \texttt{GEANT4}~\cite{GEANT4:2002zbu}, and reconstructed and selected as for
real data. The model parameters are extracted from an unbinned event-by-event
fit to the data using the extended negative log-likelihood (ENLL)
method~\cite{Ketzer:2019wmd}. The fit and the uncertainties are computed
using \texttt{ROOT}'s \texttt{TMinuit2} package~\cite{Antcheva:2009zz}. The two
channels are fitted independently; further details are given in the End
Matter. Our event-by-event analysis is sensitive to angular correlations and
this allows us to resolve the momentum-transfer dependence.

We select the data in the range of validity of the model: forward and backward
regions, $|\ctgj| > 0.75$, and high-energies interval $2.4 <
m_{\eta\pi} \lesssim 6\gev/c^2$, that is safely above the resonance region, also
used in Ref.~\cite{Bibrzycki:2021rwh}. These regions are marked by the red boxes
in \cref{fig:compass_plot} and contain $23,\!727$ ($21,\!421$) events in the
$\eta\pi^-$ ($\eta^\prime\pi^-$) channel.

\begin{figure}
\begin{tabular}{c}
\includegraphics[width=0.95\linewidth]{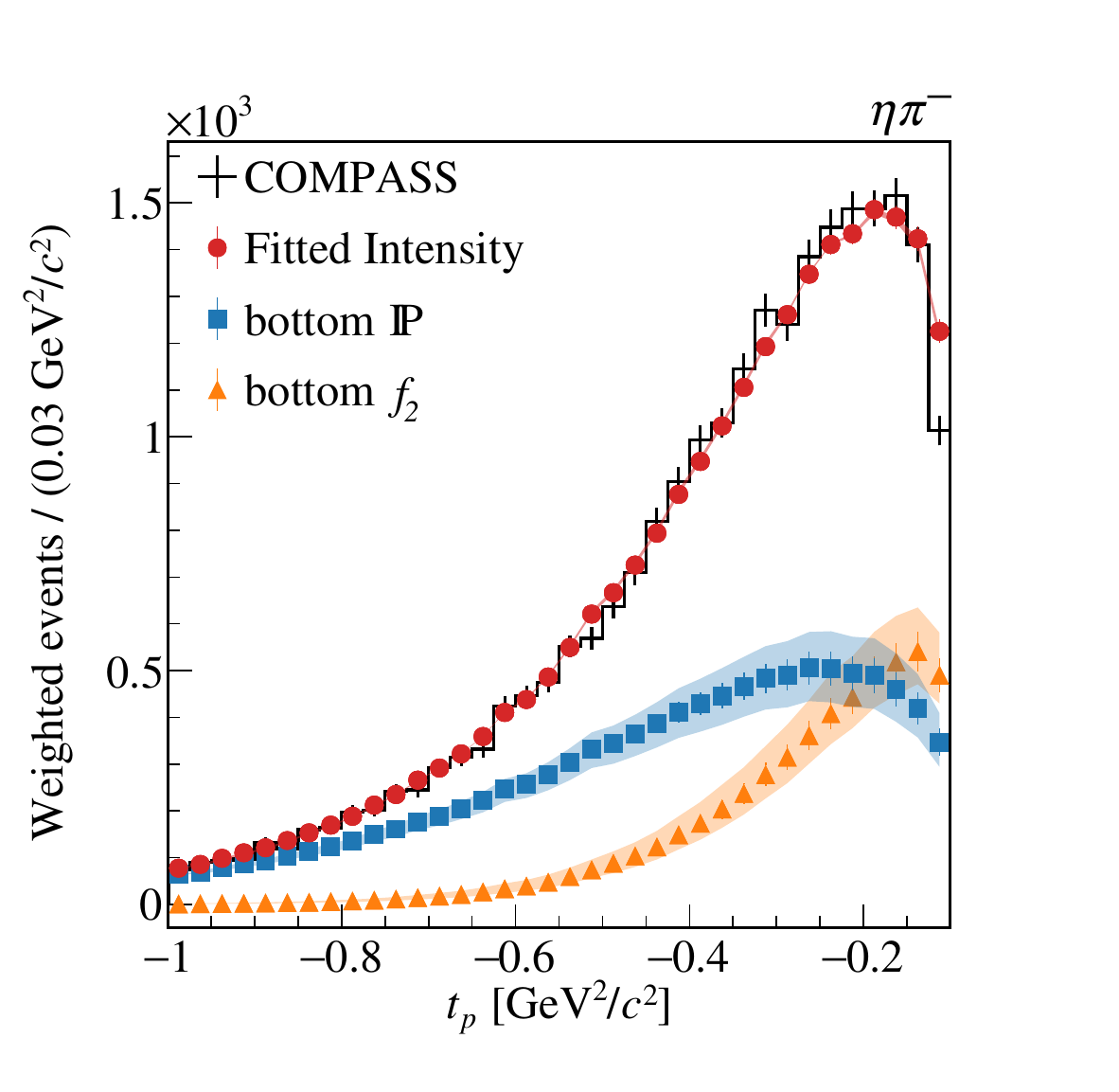} 
\end{tabular}
\caption{
Weighted intensity distributions at small $t_p$ for the individual
bottom-exchange contributions in the $\eta \pi^-$ channel. Results are similar
for the $\eta^\prime \pi^-$ channel. By bottom $\Pom$ (bottom $f_2$) we refer to
the coherent sum of the amplitudes whose exchanged $\mathbb{R}_b$ is the $\Pom$
($f_2$). Systematic uncertainties are provided as a band. Data and fit include
the acceptance of the \compass apparatus.
} 
\label{fig:weight2}
\end{figure}

\mytitle{Results}
First, we fit with only the DR amplitudes of the base model made of only the six
leading ordinary exchange amplitudes $A_i$ labeled by their top and bottom
Reggeons, namely $a_2 f_2$, $a_2 \Pom$, $f_2 f_2$, $f_2 \Pom$, $\Pom f_2$, and
$\Pom \Pom$. The model fails to describe the binned intensity distributions as
shown in the Supplemental Material~\cite{SupplementalMaterial}. A reasonable
description of the data would require raising the lower bound to
$m_{\eta\pi}=4~\gev/c^2$, which is much higher than the typical onset of the DR
region. The discrepancy originates from forward-peak events, indicating that
additional Reggeon exchanges are needed in the fast-$\eta$ amplitudes. Given
that the backward-peak events, fast-$\pi$, are well-reproduced, we do not
consider contributions by the subleading $f_2^\prime$.

We therefore include the subleading $a_2^\prime$ and the exotic $\pi_1$ as
additional top exchanges (see~\cref{fig:doubleregge-feynman}), adding the
$a_2^\prime f_2$, $a_2^\prime \Pom$, $\pi_1 f_2$, and $\pi_1 \Pom$ amplitudes to
the base model. The $\pi_1$ trajectory is distinguished by its negative
signature (odd spin) while all other trajectories considered have positive
signature (even spin); its inclusion thus tests the role of a negative-signature
exchange. Details of the trajectories are given in the End Matter. For
diagnostics, we also performed the analysis in bins of $m_{\eta\pi}$ fitted
independently, verifying that the parameters remain stable and compatible with
the global best fit. Results are presented in the Supplemental
Material~\cite{SupplementalMaterial}.

The minimal fit, which contains only the base model, is compared with fits that
add the $a_2^\prime$ exchange, the $\pi_1$ exchange, or both, and significances
are extracted from the resulting ENLLs using Wilks' theorem~\cite{Wilks:1938dza}
as explained in the End Matter. In this way, the data can select the preferred
Regge signature. In addition, we scan the parameters of the $a_2^\prime$ and
$\pi_1$ trajectories over wide, overlapping ranges. The results are summarized
in~\cref{tab:significancies}. The reported values correspond to the closest
likelihoods obtained and therefore represent minimum significances. The
additional exchanges are required at more than $11\,\sigma$ ($10\,\sigma$) in
the $\eta\pi^-$ ($\eta^\prime\pi^-$) channel. The negative-signature $\pi_1$
Reggeon in the top exchange is needed to describe the data within uncertainties,
as illustrated in~\cref{fig:weight1}: relative to a fit with the base model and
the $a_2^\prime$ exchange, adding the $\pi_1$ has a significance of
$9.9\,\sigma$ ($5.5\,\sigma$) for the $\eta\pi^-$ ($\eta^\prime\pi^-$)
channel. Best-fit parameters, their statistical and systematic uncertainties,
and a detailed comparison between model fits and data are provided in the
Supplemental Material~\cite{SupplementalMaterial}.

\begin{table}
\caption{Minimal significance studies. The base model contains only the leading ordinary exchanges.}
\label{tab:significancies}
\begin{ruledtabular}
\begin{tabular}{llrr}
\multicolumn{2}{c}{Hypotheses~~~~~~~~~~~~~~} & \multicolumn{2}{c}{ Significance} \\
$H_0$ & $H_1$&$\eta \pi^-$& $\eta^\prime \pi^-$\\
\hline
Base & Base$\,+\,a_2^\prime$ & $11.0\,\sigma$ & $10.1\,\sigma$ \\
Base & Base$\,+\,\pi_1$ & $12.8\,\sigma$& $11.7\,\sigma$ \\
Base & Base$\,+\,a_2^\prime+\pi_1$ & $15.4\,\sigma$ & $12.2\,\sigma$ \\
Base$\,+\,a_2^\prime$ & Base$\,+\,a_2^\prime+\pi_1$ & $9.9\,\sigma$ & $\phantom{0}5.5\,\sigma$ \\
Base$\,+\,\pi_1$ & Base$\,+\,a_2^\prime+\pi_1$ &$2.2\,\sigma$ & disfavored \\
\end{tabular}
\end{ruledtabular}
\end{table} 

As shown in \cref{fig:weight2}, the bottom-$\Pom$ exchange dominates over the
bottom-$f_2$, except at the smallest $t_p$ where the two are comparable,
confirming the relevance of the $f_2$ exchange suggested in
Ref.~\cite{Bibrzycki:2021rwh}. Future analyses of diffractive processes at
center-of-mass energies of tens of $\gev$ should therefore assess contributions
beyond $\Pom$. The slopes extracted for the proton-$\Pom$-proton vertex form
factor agree within uncertainties between the two channels and are consistent
with the \mbox{TOTEM-CMS} result~\cite{TOTEM:2024aso}.

\begin{figure}
\begin{tabular}{c}
\includegraphics[width=0.95\linewidth]{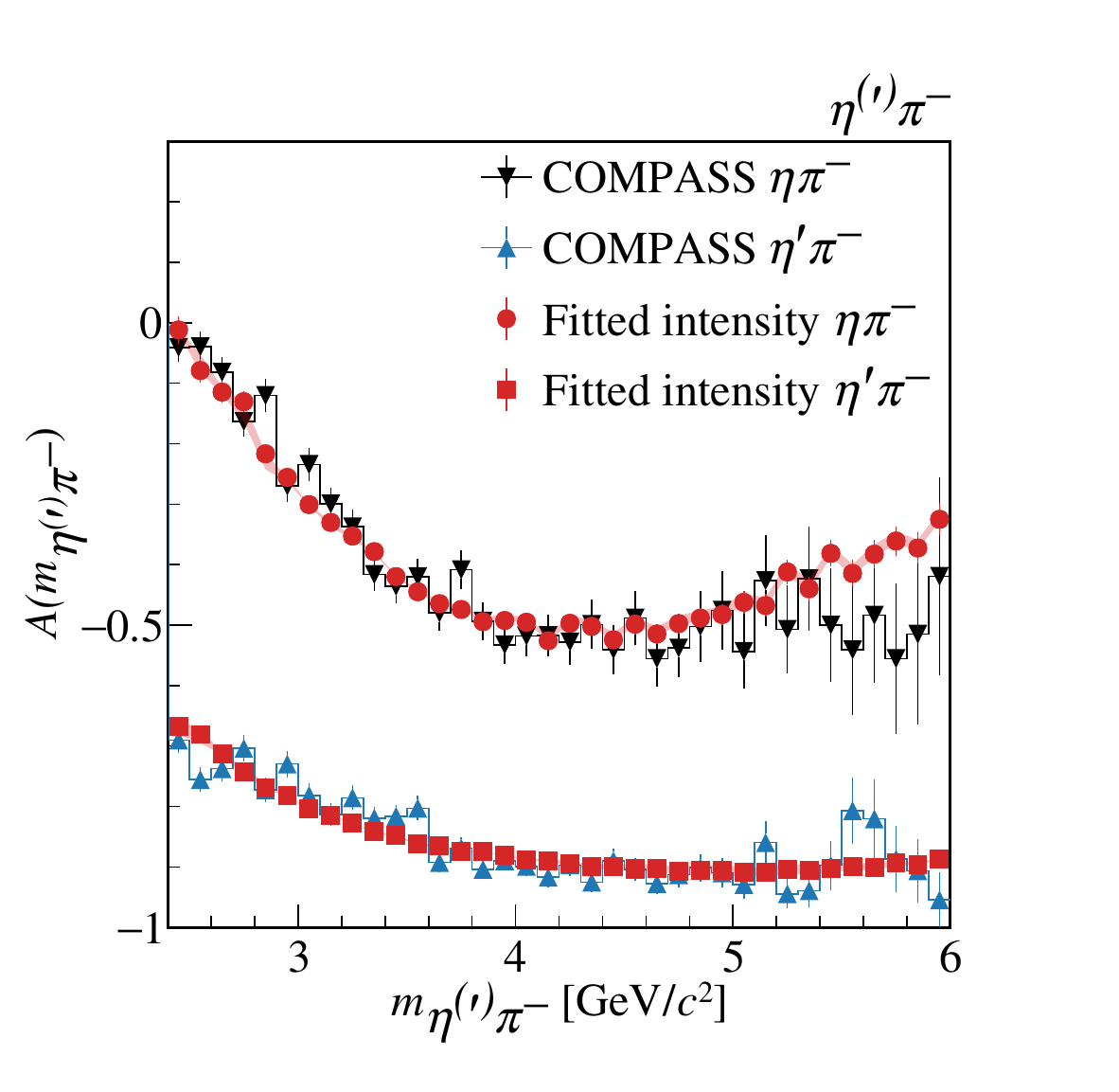}
\end{tabular}
\caption{
Forward-backward asymmetry for both channels. Note the large uncertainties at
the highest mass bins due to the small amount of forward events
(see~\cref{fig:compass_plot}.) Systematic uncertainties are provided as a
band. Data and fit include the acceptance of the \compass apparatus.
}
\label{fig:asym}
\end{figure}

Both $\eta^{(\prime)}\pi^-$ forward-backward asymmetries differ significantly
from zero, demonstrating the presence of odd exotic waves beyond the resonance
region. The asymmetry is well reproduced (\cref{fig:asym}), confirming that the
DR amplitude is the dominant production mechanism for forward and backward
events above the resonance region. A nonsignificant deviation between fit and
data appears at the highest energies, where the dynamics is dominated by the top
$\Pom$, and might require a small correction of the $\Pom$ amplitude, as shown
in the Supplemental Material~\cite{SupplementalMaterial}, were it to be
confirmed by further statistics.

\mytitle{Summary}
We report the first observation of an exotic Reggeon exchange, established at
$9.9\,\sigma$ ($5.5\,\sigma$) in the $\eta\pi^-$ ($\eta^\prime\pi^-$) channel
and most plausibly identified with the $\pi_1(1600)$.  This exotic trajectory
contributes at leading order alongside the Pomeron and is required to describe
the data. The dynamics generating exotic waves is therefore richer than
previously recognized: hybrid mesons appear here not only as produced resonances
but as exchanged trajectories, opening a complementary window on gluonic
excitations in QCD. These findings motivate analogous measurements at GlueX,
whose polarized photon beam and access to $\eta^{(\prime)}\pi^0$ final
states~\cite{GlueX:2020idb} would probe the large forward-backward asymmetry
predicted in $\eta^{(\prime)}\pi^0$ photoproduction~\cite{Montana:2025asi}, a
channel in which Pomeron exchange is forbidden and the exotic contribution can
therefore be isolated. The analysis also opens the way to incorporating
analyticity constraints in low-energy partial-wave
extractions~\cite{Mathieu:2017but}, sharpening the determination of hybrid-meson
parameters. Additionally, we find a substantial contribution of $f_2$ bottom
exchange over a wide angular range. For the \compass kinematics, this
contribution is subdominant to the favored Pomeron
exchange~{\cite{Haas:2013dbp}}, which is conventionally associated in QCD with
gluonic degrees of freedom carrying vacuum quantum
numbers~\cite{Kuraev:1977fs,Balitsky:1978ic,Donnachie:1988nj}. Quantifying such
subdominant exchanges is also essential for diffractive parton distribution
functions~\cite{H1:2007oqt}, Odderon searches~\cite{D0:2020tig}, and diffractive
production at the future Electron-Ion
Collider~\cite{AbdulKhalek:2021gbh,Anderle:2021wcy}.
%\input{Letter}
%-*-Mode: TeX; Mode: Auto-Fill; fill-column: 80; -*-
% !TEX TS-program = pdflatex
% !TEX encoding = UTF-8 Unicode
%
\begin{acknowledgments}
\mytitle{Acknowledgments}
We gratefully acknowledge the support of the CERN management and staff and the
skill and effort the technicians of our collaborating institutes. This work was
made possible by the financial support of our funding agencies: The Higher
Education and Science Committee of the Republic of Armenia (Armenia); MEYS,
Grants \mbox{LM2023040}, \mbox{CZ.02.01.01/00/22\_008/0004632} ``FORTE'' (Czech
Republic); BMBF -- Bundesministerium f\"ur Bildung und Forschung (Germany); the
B.~Sen fund (India); the Israel Academy of Sciences and Humanities (Israel);
MEXT and JSPS, Grants 18002006, 20540299, 18540281 and 26247032, the Daiko and
Yamada Foundations (Japan); NCN, Grant \mbox{2020/37/B/ST2/01547} (Poland); FCT,
Grants \mbox{DOI 10.54499/CERN/FIS-PAR/0022/2019} and \mbox{DOI
  10.54499/CERN/FIS-PAR/0016/2021} (Portugal); the Ministry of Science and
Technology (Taiwan); and the National Science Foundation, Grant
\mbox{No.~PHY-1506416} (USA), the Excellence Cluster Universe which is funded by
the Deutsche Forschungsgemeinschaft (DFG, German Research Foundation), the
Excellence Cluster ORIGINS which is funded by the DFG under Germany’s Excellence
Strategy – EXC 2094 – 390783311, the Computational Center for Particle and
Astrophysics (C2PAP), and the Leibniz Supercomputer Center (LRZ), the
U.S.~Department of Energy, Office of Science, Office of Nuclear Physics under
Contract \mbox{No.~89243126CSC00021}, U.S.~Department of Energy
contracts~\mbox{DE-FG02-87ER40365}, \mbox{DE-SC0011090}, and
\mbox{DE-SC0023598}; Simons Foundation award Simons Collaboration on Confinement
and \mbox{QCD StringsMPS-QCD-00994314} and by MIT; Projects
\mbox{CNS2022-136085}, \mbox{CEX2024-001451-M} (Unidad de Excelencia ``María de
Maeztu''), \mbox{PID2020-118758GB-I00}, and \mbox{PID2023-147112NB-C21}, all
financed by \mbox{MICIU/AEI/10.13039/501100011033/} and FEDER, UE; the Beatriu
de Pinós program by AGAUR, Grant \mbox{No. BP 2024 00189}; the COST Action
CA24159 ``Structure and Spectroscopy of Hadrons Research Project (SHARP)'',
supported by COST (European Cooperation in Science and Technology); and the
European Union’s Horizon 2020 research and innovation programme under Grant
Agreement STRONG–2020—\mbox{No.~824093}.
\end{acknowledgments}
\bibliographystyle{apsrev4-2}
\bibliography{bibliography}
\clearpage
\newpage
\appendix
\onecolumngrid
\section*{End matter}
\twocolumngrid

\begin{figure}
\begin{tabular}{c}
\includegraphics[width=\linewidth]{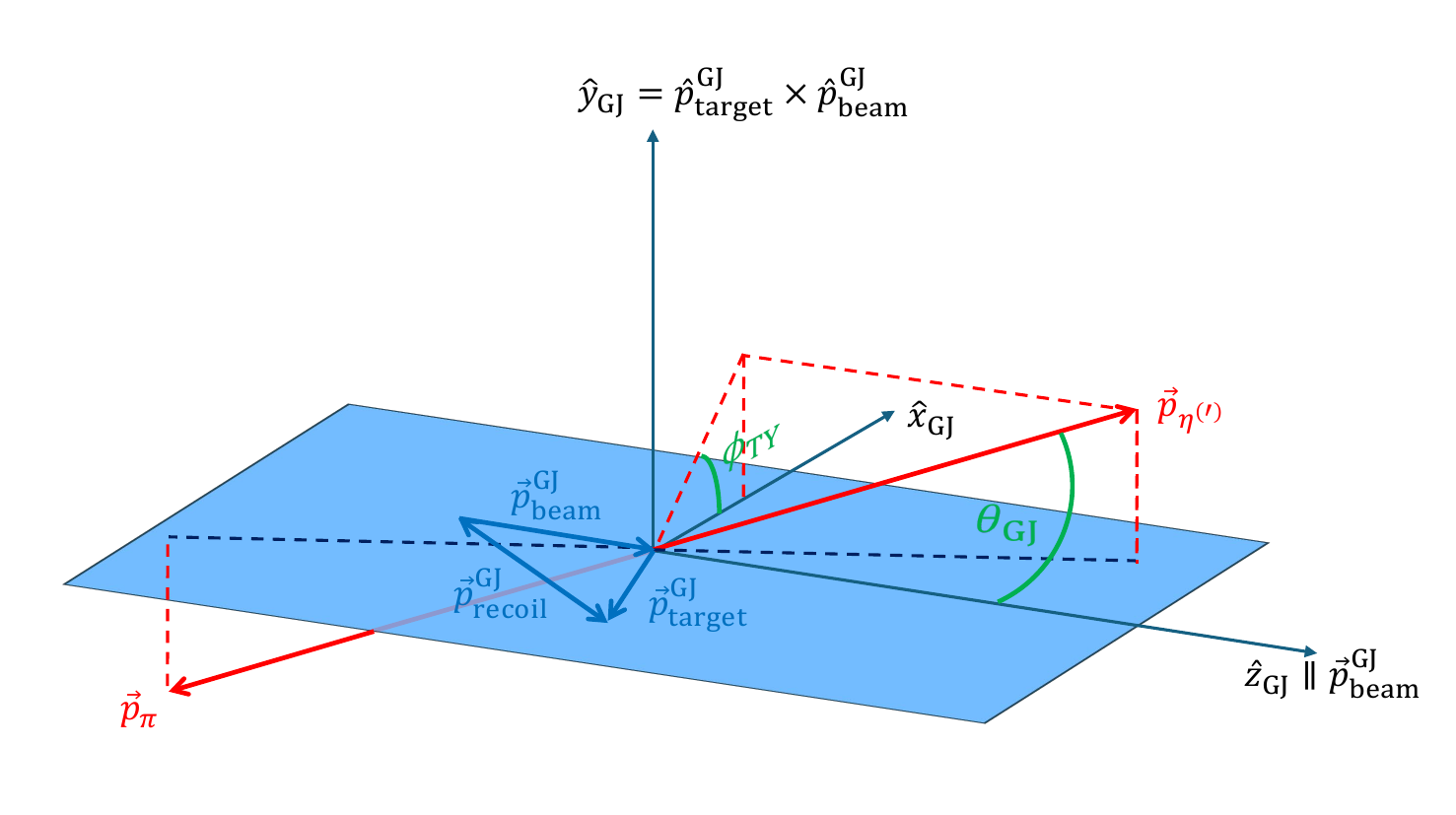}
\end{tabular}
\caption{Gottfried-Jackson reference frame. The precise definition of the vectors can be found in the text.}
\label{fig:gjframe}
\end{figure}

\mytitle{Kinematics}
The data is analyzed in the Gottfried-Jackson (GJ) reference
frame~\cite{Gottfried:1964nx}, where the $\eta^{(\prime)}\pi^-$ system is at
rest and the $\pi^-$ beam defines the $z$-axis. The production plane is
perpendicular to $\targetgj \times \beamgj$, see~\cref{fig:gjframe}. Given that
$\etagj+\vec{p}_{\pi}^{\,\text{GJ}} = \vec{0} $ the momenta are related as
follows:
\begin{align*}
\beamgj + \targetgj    & =\recoilgj,\\
\recoilgj\times\beamgj & = \left(\beamgj + \targetgj\right)\times \beamgj,\\
\targetgj\times\beamgj & =\recoilgj\times\beamgj .
\end{align*}

We define the characteristic angles in the $\eta^{(\prime)}\pi^-$ system, \ie
$\theta_{\text{GJ}}$ is the polar angle between the $\eta^{(\prime)}$ and the
\mbox{$z$-axis}, $\phi_\text{\text{TY}}$ is the azimuthal angle between the
projection of the $\vec{p}_\eta^{\,\text{GJ}}$ on the $xy$ plane and the $x$
axis, and $\epsilon$ is the angle between the recoil proton and the $\pi^-$
beam. Each event four-momenta $\zeta_k$ is characterized by the Mandelstam
variables:
\begin{align*}
s=\left(p_{\eta}+p_{\pi}+p_\text{recoil}\right)^2, \quad & 
s_{\eta \pi}=\left(p_{\eta}+p_{\pi}\right)^2,\\
s_{\eta p}=\left(p_{\eta}+p_\text{recoil}\right)^2,\quad &
s_{\pi p}=\left(p_{\pi}+p_\text{recoil}\right)^2,\\
t_{p}=\left(p_{\eta}+p_{\pi}-p_\text{beam}\right)^2, \quad &
t_{\eta}=\left(p_{\eta}-p_\text{beam}\right)^2,\\
t_{\pi}=\left(p_{\pi}-p_\text{beam}\right)^2, \quad &
\end{align*}
which can be computed in any reference frame, in particular GJ.

\mytitle{Double-Regge amplitude}
The double-Regge mechanism shown in~\cref{fig:doubleregge-feynman} dominates in
the kinematic regime given by $s, s_{12}, s_{2p} \to \infty$ with fixed $t_1$
and $t_p$, and finite $\kappa^{-1}\equiv s/( \beta s_{12}
s_{2p})$~\cite{Brower:1974yv}. The intensity distribution in terms of the
double-Regge amplitudes reads:
\begin{align*}
I(\{c, b\};\zeta_k)
= & \abs{\sum_{i} c_i \, A_i(\{b\};\zeta_k)}^2\, \\
= & \left| 
 c_{\pi_1 f_2} \, A_{\pi_1 f_2}  (\{b\};\zeta_k) +   
c_{\pi_1 \Pom} \, A_{\pi_1 \Pom} (\{b\};\zeta_k)  \right. \\
+ & c_{a_2 f_2} \, A_{a_2 f_2}  (\{b\};\zeta_k) +   
c_{a_2\Pom} \, A_{a_2 \Pom} (\{b\};\zeta_k) \\ 
+ & c_{a_2^\prime f_2} \, A_{a_2^\prime f_2}  (\{b\};\zeta_k) +   
c_{a_2^\prime\Pom} \, A_{a_2^\prime \Pom} (\{b\};\zeta_k) \\ 
+ &  \,
c_{f_2f_2}  \, A_{f_2 f_2}  (\{b\};\zeta_k) +   
c_{f_2\Pom} \, A_{f_2 \Pom} (\{b\};\zeta_k)\\
+ & \, \left.
c_{\Pom f_2}\, A_{\Pom f_2} (\{b\};\zeta_k) +
c_{\Pom\Pom} \, A_{\Pom \Pom}  (\{b\};\zeta_k)\, \right|^2 ,
\end{align*}
 where $\zeta_k$ stands for the four-momenta of each event. Each amplitude is
 given a real valued strength parameter $c_i$, that is fitted to the event
 distribution. The amplitudes include form factors $F_i(t)=\exp \left[ b_i t
   \right]$ at the top and bottom vertices which are also fitted. Hence, each
 amplitude contains two form factors, which effectively absorb any contribution
 from the middle vertex. In doing so, we end up with seven slope parameters,
 namely: $b_{\pi_1}$ associated to the exotic \mbox{$\pi \pi_1 \eta$} vertex,
 $b_{a_2}$ to the \mbox{$\pi a_2 \eta$} vertex, $b_{a_2^\prime}$ to the
 \mbox{$\pi a_2^\prime \eta$} vertex, $b_{f_2}$ to the \mbox{$\pi f_2 \pi$}
 vertex, $b_{\Pom}$ to the \mbox{$\pi \Pom \pi$} vertex, $b_f$ to the \mbox{$p
   f_2 p$} vertex, and $b_P$ to the \mbox{$p \Pom p$} vertex. The fast $\eta$
 amplitudes read:
\begin{align*}
A_{n m}(b_n,b_m;\zeta_k) & = F_n(t_\eta)\, F_m(t_p)\, T\left( \alpha_{n}(t_{\eta}),\alpha_{m}(t_p); s_{\pi p}\right),
\end{align*}
where $n= \pi_1$, $a_2$ or $a_2^\prime$ and $m=f_2$ or $\Pom$ ($m=f$ or $P$) when referring to the exchange (vertex) and the fast $\pi$ amplitudes read:
\begin{align*}
A_{nm}(b_n,b_m;\zeta_k) & = F_n(t_\pi)\, F_m(t_p)\, T\left( \alpha_{n}(t_\pi), \alpha_{m}(t_p);s_{\eta p}\right),
\end{align*}
where $n= f_2$ or $\Pom$, $m$ as in the fast-$\eta$ diagram, and $T$ is the generic amplitude:
\begin{align*}
T(\alpha_{n},\alpha_{m};s_{2p}) &=K\,\Gamma(1-\alpha_{n})\,\Gamma(1-\alpha_{m}) \\
&\times  \frac{s_0}{s} \left( \frac{s_{\eta \pi}}{ s_0}\right)^{\alpha_{n}} \left(\frac{s_{2p}}{ s_0}\right)^{\alpha_{m}} \\
&\times  \left[{\frac{\xi_{n}\xi_{mn}}{\kappa^{\alpha_{n}}}}V(\alpha_{n},\alpha_{m};\kappa)\right. \\
&+\left.{\frac{\xi_{m}\xi_{nm}}{\kappa^{\alpha_{m}}}}V(\alpha_{m},\alpha_{n};\kappa)\right],
\end{align*}
with 
\begin{align*}
K=4\, m_{\eta \pi}\abs{\Vec{p}_{\eta}^{\, \text{GJ}}}\abs{\beamgj}\abs{\recoilgj}\, \sin{\epsilon}\, \sin{\theta_\text{GJ}}\, \sin{\phi_\text{TY}},
\end{align*}
and
\begin{align*}
\xi_{n}=\frac{\tau_n+e^{-i\pi\alpha_{n}}}{2}, \quad
\xi_{nm}=\frac{\tau_n \tau_m+e^{-i\pi(\alpha_{n}-\alpha_{m})}}{2},
\end{align*}
where $\tau_n$ and $\tau_m$ are the signatures of the trajectories and
\begin{align*}
V(\alpha_{1},\alpha_{2};\kappa)=\frac{\Gamma(\alpha_{1}-\alpha_{2})}{\Gamma(1-\alpha_{2})}\,_{1}F_{1}\left(1-\alpha_{1},1-\alpha_{1}+\alpha_{2};-\kappa\right),
\end{align*}
where $_{1}F_{1}$ is the hypergeometric function of the first kind, with
$\beta=0.8\, c^4\gev^{-2}$~\cite{Bibrzycki:2021rwh}. The assumed Regge
trajectories are linear and provided in~\cref{tab:reggetrajectories}.

\begin{table}
  \caption{
    Linear Regge trajectories \mbox{$\alpha(t)=\alpha_0+\alpha^\prime\, t$} used
    in our analysis from Ketzer, Grube and Ryabchikov
    (KGR)~\cite{Ketzer:2019wmd} and the JPAC collaboration
    in~\cite{Bibrzycki:2021rwh}. $\alpha^\prime$ is given in units of
    \mbox{$c^4\gev^{-2}$}. The $\Pom$ trajectory is taken from the Review of
    Particle Physics~\cite{ParticleDataGroup:2024cfk}. The exotic $\pi_1$ and
    ordinary $a_2^\prime$ trajectory parameters are varied within the ranges
    provided.
  }
\label{tab:reggetrajectories}
\begin{ruledtabular}
\begin{tabular}{cccc}
Trajectory & Signature  & KGR & JPAC \\
\hline
$\alpha_{a_2}(t)$ & $+1$ & $ \phantom{-}0.44+0.917\, t $ & $ \phantom{-}0.53+0.90\, t $\\
$\alpha_{f_2}(t)$ & $+1$ & $ \phantom{-}0.44+0.917\, t $ & $ \phantom{-}0.47+0.89\, t $\\
$\alpha_{\, \Pom}(t)$  & $+1$ & 
\multicolumn{2}{c}{$ \phantom{-}1.08+0.25\, t\phantom{0} $}\\
$\alpha_{a_2^\prime}(t)$  & $+1$  & \multicolumn{2}{c}{$ [-0.66,-0.46] + [0.817,1.017]\, t\phantom{0} $}\\
$\alpha_{\pi_1}(t)$  & $-1$  & \multicolumn{2}{c}{$ [-0.94,-0.54] + [\phantom{0}0.48,\phantom{0}0.88]\, t\phantom{0} $}
\end{tabular}
\end{ruledtabular}
\end{table} 

\mytitle{Event fitting} The \mc accepted intensity is given by:
\begin{align*}
N(\{c,b\}\,)=\sum_{k=1}^{N_\text{acc}}I(\{c,b\};\zeta_k)=\sum_{k=1}^{N_\text{acc}}\abs{\sum_{i}c_iA_i(\{b\};\zeta_k)}^2,
\end{align*}
where $N_\text{acc}$ stands for the number of accepted events. The \mc samples
that describe the $\eta^{(\prime)}\pi^-$ final state are uniformly distributed
in phase space. Each event is propagated through the \compass \mc simulation
framework \texttt{TGEANT}~\cite{Szameitat:2017ekf}, and reconstructed and
selected in the same way as the \compass real data events. The parameters are
determined by fitting to the experimental events using the extended negative
log-likelihood (ENLL)~\cite{Ketzer:2019wmd}:
\begin{align*}
\text{ENLL}(\{c,b\}) &=N(\{c,b\})-\sum_{k=1}^{N_\text{m}}\ln{I(\{c,b\};\zeta_k)},
\end{align*}
with $N_\text{m}$ the number of measured events. The extended negative
log-likelihood takes into account that the total intensity is a fixed
quantity. The fit and the uncertainties are computed using \texttt{ROOT}'s
\texttt{TMinuit2} package~\cite{Antcheva:2009zz}. To find the best fit we
perform 20 attempts with initial parameters randomly seeded. The best fit values
together with the uncertainties from \texttt{MINUIT} are provided in the
Supplemental Material~\cite{SupplementalMaterial}.

\begin{figure}[t]
\begin{tabular}{c}
\includegraphics[width=0.8\linewidth]{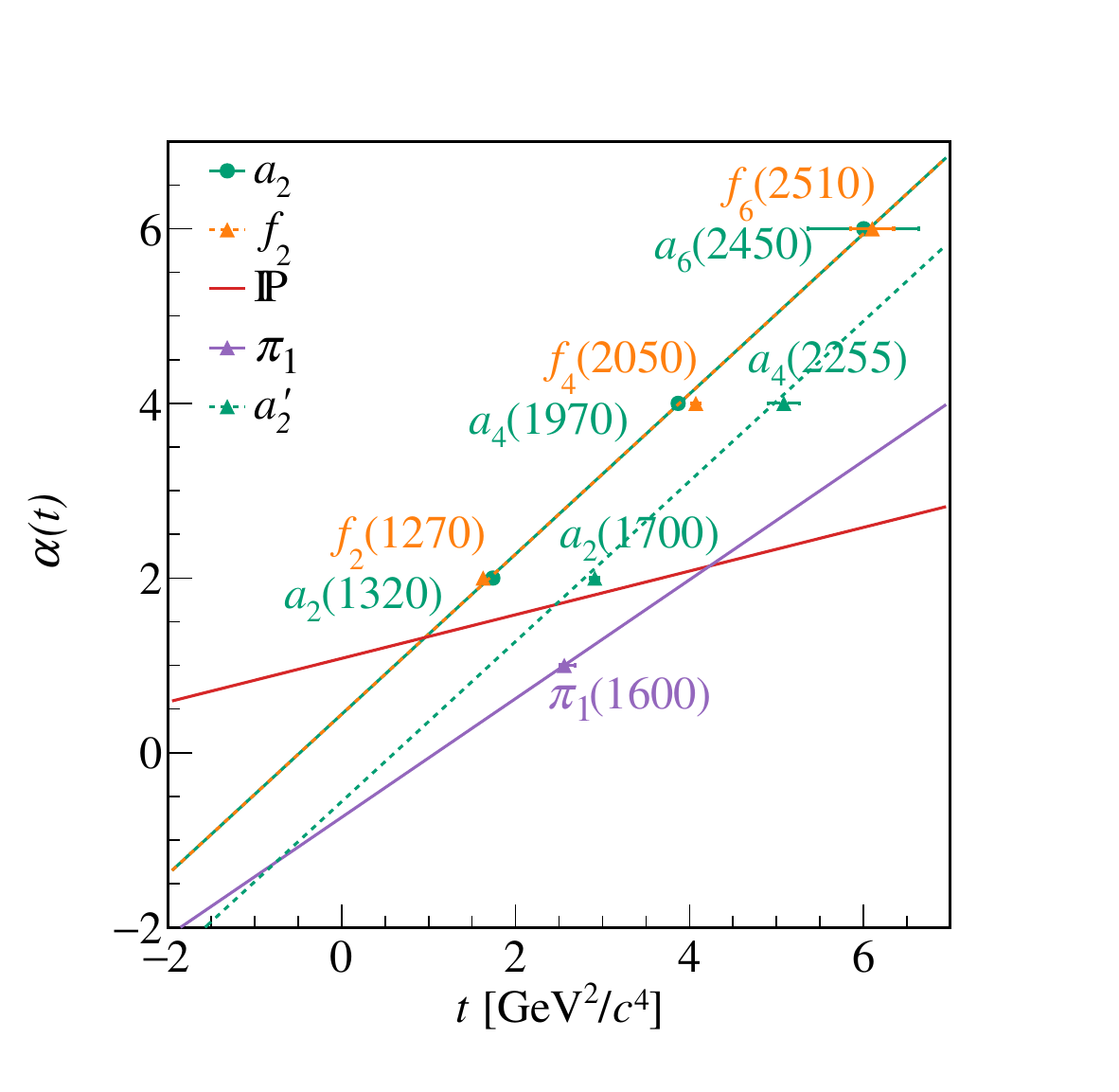} 
\end{tabular}
\caption{Regge trajectories for the KGR parameters.}
\label{fig:ChewFrautschiPlot}
\end{figure}

\mytitle{Systematics and significance}
To assess the impact of Regge‑parameter variations, we performed fits using the
KGR and JPAC parameterizations for the leading ordinary exchanges. The plots in
the Letter correspond to the KGR fits and include systematic uncertainty bands
obtained by varying the poorly established $a_2^\prime$ and $\pi_1$ trajectories
over wide ranges, see~\cref{tab:reggetrajectories}. A thorough comparison is
shown in the Supplemental Material~\cite{SupplementalMaterial}. In this way we
produce a large set of fits which sample the systematic effects associated to
the variations of the parameters of the leading, subleading and exotic
trajectories, as well as the inclusion of the $a_2^\prime$ and $\pi_1$
trajectories. We use this sample to asses the significance of including the
$\pi_1$ and the $a_2^\prime$ exchanges.

To quantify the statistical preference of extended models over the minimal
hypothesis, we employ the likelihood-ratio test in the asymptotic regime
described by Wilks’ theorem~\cite{Wilks:1938dza}. For two nested models, we take
the simpler model, with fewer parameters, as the null hypothesis $H_0$, and the
more complex model, with additional parameters, as the alternative hypothesis
$H_1$. We then define the test statistic:
\begin{align*}
q=-2 \ln \frac{\mathcal{L}_0}{\mathcal{L}_1} = 2\left(\mathrm{ENLL}_0-\mathrm{ENLL}_1\right),
\end{align*}
where $\mathcal{L}_0$ and $\mathcal{L}_1$ denote the extended likelihoods
evaluated under $H_0$ and $H_1$, respectively. Under the regularity conditions
of Wilks’ theorem (nested hypotheses), a sufficiently large sample size, and the
absence of boundary effects—the test statistic $q$ follows a $\chi^2$
distribution with $k$ degrees of freedom under the null hypothesis, where $k$ is
the difference in the number of free parameters between the two models. The
corresponding $p$-value is computed as the probability of \mbox{$\chi^2_k \ge
  q_{\mathrm{obs}}$}, \ie \mbox{$p = P(\chi^2_k \ge q_{\mathrm{obs}})$}, where
$q_{\mathrm{obs}}$ is the value of the test statistic obtained from the observed
data. This $p$-value is then translated into a Gaussian-equivalent significance,
$Z = \Phi^{-1}(1 - p)$, with $\Phi^{-1}$ denoting the inverse cumulative
distribution function of the standard normal distribution. Within this
framework, the resulting significance quantifies the degree to which the data
disfavor the minimal model relative to the extended hypotheses, thereby
providing a consistent criterion to assess whether the inclusion of additional
parameters leads to a statistically justified improvement in the description of
the data.

\clearpage\newpage
\onecolumngrid

\begin{center}
  {\huge \sc Supplemental Material}
\end{center}

\section*{Three-Dimensional Event distributions}
\begin{figure*}[h!]
\begin{tabular}{c}
\includegraphics[width=0.67\columnwidth]{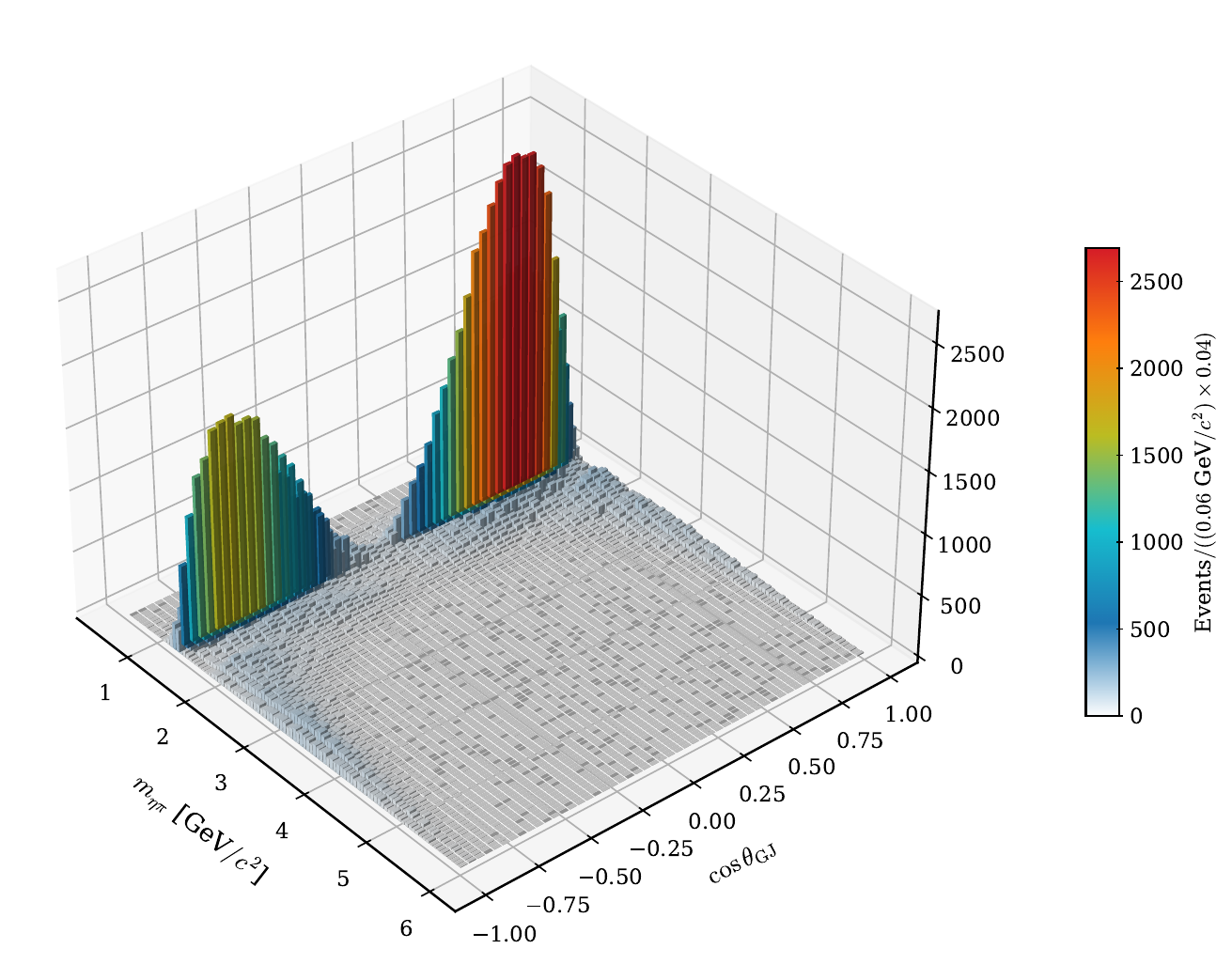} \\
\includegraphics[width=0.67\columnwidth]{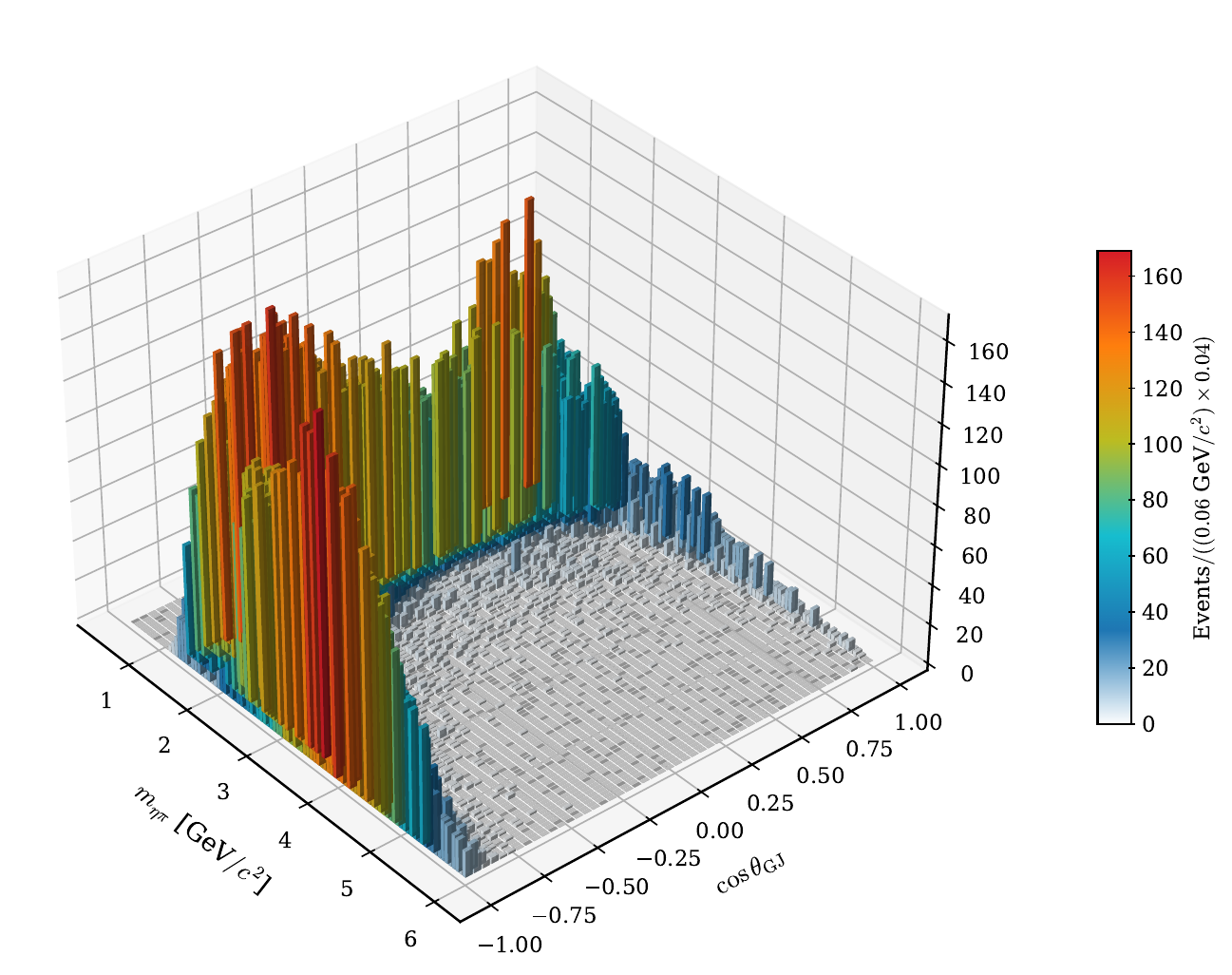}
\end{tabular}
\caption{
Three-dimensional \etapi (top) and $\eta^\prime \pi^-$ (bottom) event
distributions. The different structure of the events for both channels is very
apparent. Note that the scales are different for each histogram and that the
binning is different from the one presented in \cref{fig:compass_plot}.
}
\label{figsup:compass_plot3d_eta}
\end{figure*}

\clearpage
\newpage

\section*{Fit parameters}
In this section we provide the model parameters and the associated uncertainties
obtained by fitting the events in the double-Regge region, \ie invariant mass
range $\left[ 2.4, 6 \right]\gev$ and angular range $\left| \ctgj \right| \ge
0.75$. The following sections provide a comprehensive comparison of the fits to
the \compass data using two sets of parameters for the Regge trajectories
provided in \cref{tab:reggetrajectories}. The comparison of the fits provide a
sensible assessment of the systematics associated to the analysis.  We provide
eight fits for each channel:
\begin{itemize}
\item[--] \mbox{KGR}: Fit with the leading ordinary exchanges using the KGR trajectories; 
\item[--] \mbox{$\text{KGR}+a_2^\prime$}: Fit with the ordinary leading and subleading $a_2^\prime$ exchanges using the KGR trajectories; 
\item[--] \mbox{$\text{KGR}+\pi_1$}: Fit with the ordinary leading and exotic $\pi_1$ exchanges using the KGR trajectories;
\item[--] \mbox{$\text{KGR}+\pi_1+a_2^\prime$}: Fit with the ordinary leading and subleading $a_2^\prime$ and the exotic $\pi_1$ exchanges using the KGR trajectories;
\item[--] \mbox{JPAC}: Fit with the ordinary leading exchanges using the JPAC trajectories; 
\item[--] \mbox{$\text{JPAC}+a_2^\prime$}: Fit with the ordinary leading and subleading $a_2^\prime$ exchanges using the JPAC trajectories; 
\item[--] \mbox{$\text{JPAC}+\pi_1$}: Fit with the ordinary leading and exotic $\pi_1$ exchanges using the JPAC trajectories; and 
\item[--] \mbox{$\text{JPAC}+\pi_1+a_2^\prime$}: Fit with the ordinary leading and subleading $a_2^\prime$ and the exotic $\pi_1$ exchanges using the JPAC trajectories. 
\end{itemize}

\begin{table}[!h]
\caption{
Linear Regge trajectories \mbox{$\alpha(t)=\alpha_0+\alpha^\prime\, t$} used in
our fits from Ketzer, Grube and Ryabchikov (KGR)~\cite{Ketzer:2019wmd} and the
JPAC collaboration in~\cite{Bibrzycki:2021rwh}. \Cref{fig:ChewFrautschiPlot}
shows the trajectories for the KGR parametrization. $\alpha^\prime$ is given in
units of \mbox{$\left(c\slash \gev\right)^{2}$}. The $\Pom$ trajectory is taken
from the Review of Particle Physics~\cite{ParticleDataGroup:2024cfk}.  The
$a_2^\prime$ is the same as the $a_2$ shifted one unit.  The $\pi_1$ trajectory
is estimated from the mass of the $\pi_1(1600)$ and theoretical estimations of
other potential members of the trajectory.
}
\label{tab:reggetrajectoriessup}
\begin{ruledtabular}
\begin{tabular}{cccc}
Trajectory & Signature  & KGR & JPAC \\
\hline
$\alpha_{a_2}(t)$ & $+1$ & $ \phantom{-}0.44+0.917\, t $ & $ \phantom{-}0.53+0.90\, t $\\
$\alpha_{f_2}(t)$ & $+1$ & $ \phantom{-}0.44+0.917\, t $ & $ \phantom{-}0.47+0.89\, t $\\
$\alpha_{\, \Pom}(t)$  & $+1$ & 
\multicolumn{2}{c}{$ \phantom{-}1.08+0.25\, t\phantom{0} $}\\
$\alpha_{\pi_1}(t)$     & $-1$ &  \multicolumn{2}{c}{$-0.74 + 0.68\, t\phantom{0}$}  \\
$\alpha_{a_2^\prime}(t)$  &  $+1$ & $ -0.56+0.917\, t $ & $ -0.47+0.90\, t $ \\
\end{tabular}
\end{ruledtabular}
\end{table} 

Additionally we show fits in the $\left[ 4, 6 \right]\gev$ invariant mass and
angular $\left| \ctgj \right| \ge 0.75$ ranges using the ordinary leading
exchanges for both KGR and JPAC parameterizations of the trajectories, named
KGR-HE and JPAC-HE respectively, where HE stands for high energy. We note that
for the $\left[ 2.4, 6 \right]\gev$ range we fit $23,727$ events for the
$\eta\pi^-$ channel and $21,\!421$ for the $\eta^\prime \pi$ while for the
$\left[ 4, 6 \right]\gev$ range we fit $4,\!970$ events for the $\eta\pi^-$
channel and $6,\!388$ for the $\eta^\prime \pi$.

The fits performed using only the leading ordinary trajectories (base model)
allow us to assess whether the data can be described solely by these
contributions. We find that while the backward peak can indeed be reproduced,
the forward peak cannot. This leads to the conclusion that contributions beyond
the leading $a_2$ exchanges are required to properly account for the data. This
need is further underscored by the behavior of the asymmetry. Moreover, we show
that raising the lower-energy bound of the fitting window is insufficient: the
inclusion of additional exchanges is unavoidable. JPAC \vs KGR fits provide an
estimation of the impact of changing the parameters of the leading ordinary
trajectories

\begin{table}[!h]
\centering
\caption{
Values of the parameters for the best fit to $\eta \pi^-$ events using the KGR
trajectories. $c_i$ parameters are dimensionless and  $b_i$ parameters in units
of \mbox{ $\left(\gev/c\right)^{-2}$}.
}
\label{tabsup:bestfitkgr_eta}
\begin{ruledtabular}
\begin{tabular}{c|rrrrr}
                     & KGR-HE & KGR          &  $\text{KGR}+a_2^\prime$     & $\text{KGR}+\pi_1$ &  $\text{KGR}+\pi_1+a_2^\prime$\\ \\
Extended Log-Likelihood & 21618.3 & 84627.6 & 84544.2 & 84505 & 84479.3 \\
Strength parameters  &   & &  &  & \\
$c_{\pi_1 f_2}$ & $\text{---}$ & $\text{---}$ & $\text{---}$ & $\SI{-1.01 \pm 0.07}{}$ & $\SI{-1.02 \pm 0.06}{}$ \\
$c_{\pi_1 \Pom}$ & $\text{---}$ & $\text{---}$ & $\text{---}$ & $\SI{0.0069 \pm 0.0032}{}$ & $\SI{0.0032 \pm 0.0029}{}$ \\
$c_{a_2f_2}$ & $\SI{0.069 \pm 0.007}{}$ & $\SI{0.120 \pm 0.004}{}$ & $\SI{0.100 \pm 0.004}{}$ & $\SI{0.076 \pm 0.004}{}$ & $\SI{0.074 \pm 0.004}{}$ \\
$c_{a_2\Pom}$ & $\SI{0.0128 \pm 0.0011}{}$ & $\SI{0.0129 \pm 0.0004}{}$ & $\SI{0.0097 \pm 0.0005}{}$ & $\SI{0.0105 \pm 0.0004}{}$ & $\SI{0.0097 \pm 0.0005}{}$ \\
$c_{a_2^\prime f_2}$ & $\text{---}$ & $\text{---}$ & $\SI{-0.68 \pm 0.22}{}$ & $\text{---}$ & $\SI{-26 \pm 7}{}$ \\
$c_{a_2^\prime\Pom}$ & $\text{---}$ & $\text{---}$ & $\SI{-0.0200 \pm 0.0029}{}$ & $\text{---}$ & $\SI{0.25 \pm 0.05}{}$ \\
$c_{f_2f_2}$ & $\SI{-0.111 \pm 0.028}{}$ & $\SI{0.242 \pm 0.021}{}$ & $\SI{0.246 \pm 0.025}{}$ & $\SI{0.240 \pm 0.022}{}$ & $\SI{-0.330 \pm 0.031}{}$ \\
$c_{f_2\Pom}$ & $\SI{-0.0257 \pm 0.0028}{}$ & $\SI{0.0149 \pm 0.0010}{}$ & $\SI{0.0123 \pm 0.0011}{}$ & $\SI{0.0129 \pm 0.0009}{}$ & $\SI{-0.0063 \pm 0.0030}{}$ \\
$c_{\Pom f_2}$ & $\SI{-0.0068 \pm 0.0010}{}$ & $\SI{0.0119 \pm 0.0018}{}$ & $\SI{0.0095 \pm 0.0023}{}$ & $\SI{0.0102 \pm 0.0021}{}$ & $\SI{-0.0035 \pm 0.0021}{}$ \\
$c_{\Pom\Pom}$ & $\SI{0.00078 \pm 0.00009}{}$ & $\SI{-0.00160 \pm 0.00008}{}$ & $\SI{-0.00132 \pm 0.00009}{}$ & $\SI{-0.00131 \pm 0.00008}{}$ & $\SI{0.00136 \pm 0.00007}{}$ \\
\\
Top Vertex Form Factors  &   & &  &  & \\ 
$b_{\pi_1}$ $(\pi \pi_1 \eta)$ & $\text{---}$ & $\text{---}$ & $\text{---}$ & $\SI{-0.01 \pm 0.07}{}$ & $\SI{-0.09 \pm 0.05}{}$ \\
$b_{a_2}$ $(\pi a_2 \eta)$ & $\SI{-0.16 \pm 0.06}{}$ & $\SI{-0.136 \pm 0.021}{}$ & $\SI{-0.15 \pm 0.04}{}$ & $\SI{-0.215 \pm 0.023}{}$ & $\SI{-0.192 \pm 0.023}{}$ \\
$b_{a_2^\prime}$ $(\pi a_2^\prime \eta)$ & $\text{---}$ & $\text{---}$ & $\SI{1.68 \pm 0.35}{}$ & $\text{---}$ & $\SI{38 \pm 10}{}$ \\
$b_{f_2}$ $(\pi f_2 \pi)$ & $\SI{0.25 \pm 0.14}{}$ & $\SI{1.63 \pm 0.15}{}$ & $\SI{1.25 \pm 0.14}{}$ & $\SI{1.21 \pm 0.13}{}$ & $\SI{1.43 \pm 0.12}{}$ \\
$b_{\Pom\,}$ $(\pi \Pom \pi)$ & $\SI{0.28 \pm 0.06}{}$ & $\SI{0.732 \pm 0.031}{}$ & $\SI{0.64 \pm 0.04}{}$ & $\SI{0.64 \pm 0.04}{}$ & $\SI{0.681 \pm 0.028}{}$ \\
\\
Bottom Vertex Form Factors  &   & &  &  & \\
$b_f$ $(p f_2 p)$ & $\SI{0.42 \pm 0.19}{}$ & $\SI{1.39 \pm 0.10}{}$ & $\SI{1.04 \pm 0.10}{}$ & $\SI{1.12 \pm 0.09}{}$ & $\SI{1.05 \pm 0.08}{}$ \\
$b_P$ $(p \Pom p)$ & $\SI{1.73 \pm 0.09}{}$ & $\SI{1.654 \pm 0.034}{}$ & $\SI{1.54 \pm 0.04}{}$ & $\SI{1.54 \pm 0.04}{}$ & $\SI{1.47 \pm 0.05}{}$ \\
\\
\end{tabular}
\end{ruledtabular}
\end{table}

\begin{table}[!h]
\centering
\caption{Values of the parameters for the best fit to $\eta \pi^-$ events using the JPAC trajectories. $c_i$ parameters are dimensionless and  $b_i$ parameters in units of \mbox{ $\left(\gev/c\right)^{-2}$}.}
\label{tabsup:bestfitjpac_eta}
\begin{ruledtabular}
\begin{tabular}{c|rrrrr}
                     & JPAC-HE & JPAC          &  $\text{JPAC}+a_2^\prime$     & $\text{JPAC}+\pi_1$ &  $\text{JPAC}+\pi_1+a_2^\prime$\\ \\
Extended Log-Likelihood & 21617.3 & 84617.5 & 84550.9 & 84498.1 & 84462.6 \\
Strength parameters  &   & &  &  & \\
$c_{\pi_1 f_2}$ & $\text{---}$ & $\text{---}$ & $\text{---}$ & $\SI{-0.90 \pm 0.07}{}$ & $\SI{-0.85 \pm 0.06}{}$ \\
$c_{\pi_1 \Pom}$ & $\text{---}$ & $\text{---}$ & $\text{---}$ & $\SI{0.002 \pm 0.004}{}$ & $\SI{-0.0006 \pm 0.0033}{}$ \\
$c_{a_2f_2}$ & $\SI{0.016 \pm 0.005}{}$ & $\SI{0.0667 \pm 0.0027}{}$ & $\SI{0.0591 \pm 0.0027}{}$ & $\SI{0.0421 \pm 0.0026}{}$ & $\SI{0.0412 \pm 0.0025}{}$ \\
$c_{a_2\Pom}$ & $\SI{0.0145 \pm 0.0016}{}$ & $\SI{0.01065 \pm 0.00027}{}$ & $\SI{0.0084 \pm 0.0004}{}$ & $\SI{0.00825 \pm 0.00033}{}$ & $\SI{0.00796 \pm 0.00034}{}$ \\
$c_{a_2^\prime f_2}$ & $\text{---}$ & $\text{---}$ & $\SI{-0.35 \pm 0.14}{}$ & $\text{---}$ & $\SI{-14.2 \pm 3.4}{}$ \\
$c_{a_2^\prime\Pom}$ & $\text{---}$ & $\text{---}$ & $\SI{-0.0193 \pm 0.0033}{}$ & $\text{---}$ & $\SI{0.164 \pm 0.035}{}$ \\
$c_{f_2f_2}$ & $\SI{0.002 \pm 0.018}{}$ & $\SI{0.184 \pm 0.018}{}$ & $\SI{-0.192 \pm 0.018}{}$ & $\SI{0.192 \pm 0.019}{}$ & $\SI{-0.224 \pm 0.023}{}$ \\
$c_{f_2\Pom}$ & $\SI{-0.036 \pm 0.004}{}$ & $\SI{0.0163 \pm 0.0010}{}$ & $\SI{-0.0129 \pm 0.0011}{}$ & $\SI{0.0134 \pm 0.0009}{}$ & $\SI{-0.0110 \pm 0.0013}{}$ \\
$c_{\Pom f_2}$ & $\SI{-0.0059 \pm 0.0007}{}$ & $\SI{0.0137 \pm 0.0015}{}$ & $\SI{-0.0112 \pm 0.0017}{}$ & $\SI{0.0116 \pm 0.0017}{}$ & $\SI{-0.0091 \pm 0.0022}{}$ \\
$c_{\Pom\Pom}$ & $\SI{0.00080 \pm 0.00009}{}$ & $\SI{-0.00189 \pm 0.00009}{}$ & $\SI{0.00159 \pm 0.00010}{}$ & $\SI{-0.00152 \pm 0.00009}{}$ & $\SI{0.00155 \pm 0.00008}{}$ \\
\\
Top Vertex Form Factors  &   & &  &  & \\ 
$b_{\pi_1}$ $(\pi \pi_1 \eta)$ & $\text{---}$ & $\text{---}$ & $\text{---}$ & $\SI{0.17 \pm 0.12}{}$ & $\SI{-0.04 \pm 0.06}{}$ \\
$b_{a_2}$ $(\pi a_2 \eta)$ & $\SI{-0.18 \pm 0.06}{}$ & $\SI{-0.165 \pm 0.021}{}$ & $\SI{-0.10 \pm 0.05}{}$ & $\SI{-0.221 \pm 0.029}{}$ & $\SI{-0.202 \pm 0.026}{}$ \\
$b_{a_2^\prime}$ $(\pi a_2^\prime \eta)$ & $\text{---}$ & $\text{---}$ & $\SI{1.4 \pm 0.4}{}$ & $\text{---}$ & $\SI{31 \pm 9}{}$ \\
$b_{f_2}$ $(\pi f_2 \pi)$ & $\SI{0.11 \pm 0.10}{}$ & $\SI{1.96 \pm 0.18}{}$ & $\SI{1.54 \pm 0.15}{}$ & $\SI{1.46 \pm 0.14}{}$ & $\SI{1.59 \pm 0.14}{}$ \\
$b_{\Pom\,}$ $(\pi \Pom \pi)$ & $\SI{0.25 \pm 0.05}{}$ & $\SI{0.801 \pm 0.030}{}$ & $\SI{0.72 \pm 0.04}{}$ & $\SI{0.706 \pm 0.035}{}$ & $\SI{0.717 \pm 0.033}{}$ \\
\\
Bottom Vertex Form Factors  &   & &  &  & \\
$b_f$ $(p f_2 p)$ & $\SI{-0.70 \pm 0.30}{}$ & $\SI{1.61 \pm 0.13}{}$ & $\SI{1.27 \pm 0.12}{}$ & $\SI{1.38 \pm 0.11}{}$ & $\SI{1.32 \pm 0.10}{}$ \\
$b_P$ $(p \Pom p)$ & $\SI{2.32 \pm 0.14}{}$ & $\SI{1.770 \pm 0.033}{}$ & $\SI{1.67 \pm 0.04}{}$ & $\SI{1.61 \pm 0.04}{}$ & $\SI{1.59 \pm 0.04}{}$ \\
\\
\end{tabular}
\end{ruledtabular}
\end{table}

\begin{table}[!h]
\centering
\caption{
Values of the parameters for the best fit to $\eta^\prime \pi^-$ events using
the KGR trajectories. $c_i$ parameters are dimensionless and  $b_i$ parameters
in units of \mbox{ $\left(\gev/c\right)^{-2}$}.
}
\label{tabsup:bestfitkgr_etap}
\begin{ruledtabular}
\begin{tabular}{c|rrrrr}
                     & KGR-HE & KGR          &  $\text{KGR}+a_2^\prime$     & $\text{KGR}+\pi_1$ &  $\text{KGR}+\pi_1+a_2^\prime$\\ \\
Extended Log-Likelihood & 28211.4 & 86871.3 & 86807.1 & 86792.1 & 86763.9 \\
Strength parameters  &   & &  &  & \\
$c_{\pi_1 f_2}$ & $\text{---}$ & $\text{---}$ & $\text{---}$ & $\SI{-1.45 \pm 0.18}{}$ & $\SI{-1.26 \pm 0.10}{}$ \\
$c_{\pi_1 \Pom}$ & $\text{---}$ & $\text{---}$ & $\text{---}$ & $\SI{0.087 \pm 0.029}{}$ & $\SI{0.041 \pm 0.018}{}$ \\
$c_{a_2f_2}$ & $\SI{0.031 \pm 0.004}{}$ & $\SI{0.0511 \pm 0.0027}{}$ & $\SI{0.0329 \pm 0.0022}{}$ & $\SI{0.0172 \pm 0.0031}{}$ & $\SI{0.0144 \pm 0.0024}{}$ \\
$c_{a_2\Pom}$ & $\SI{0.0036 \pm 0.0006}{}$ & $\SI{0.00657 \pm 0.00035}{}$ & $\SI{0.00334 \pm 0.00028}{}$ & $\SI{0.0050 \pm 0.0004}{}$ & $\SI{0.00413 \pm 0.00031}{}$ \\
$c_{a_2^\prime f_2}$ & $\text{---}$ & $\text{---}$ & $\SI{-1.67 \pm 0.33}{}$ & $\text{---}$ & $\SI{0.036 \pm 0.009}{}$ \\
$c_{a_2^\prime\Pom}$ & $\text{---}$ & $\text{---}$ & $\SI{-0.041 \pm 0.005}{}$ & $\text{---}$ & $\SI{0.00280 \pm 0.00034}{}$ \\
$c_{f_2f_2}$ & $\SI{-0.148 \pm 0.015}{}$ & $\SI{-0.154 \pm 0.009}{}$ & $\SI{0.154 \pm 0.008}{}$ & $\SI{0.156 \pm 0.008}{}$ & $\SI{0.107 \pm 0.009}{}$ \\
$c_{f_2\Pom}$ & $\SI{-0.0076 \pm 0.0017}{}$ & $\SI{-0.0056 \pm 0.0005}{}$ & $\SI{0.0052 \pm 0.0005}{}$ & $\SI{0.0053 \pm 0.0005}{}$ & $\SI{0.0100 \pm 0.0006}{}$ \\
$c_{\Pom f_2}$ & $\SI{-0.0047 \pm 0.0007}{}$ & $\SI{-0.0083 \pm 0.0007}{}$ & $\SI{0.0076 \pm 0.0006}{}$ & $\SI{0.0079 \pm 0.0007}{}$ & $\SI{0.0086 \pm 0.0006}{}$ \\
$c_{\Pom\Pom}$ & $\SI{0.00045 \pm 0.00004}{}$ & $\SI{0.00079 \pm 0.00005}{}$ & $\SI{-0.00068 \pm 0.00005}{}$ & $\SI{-0.00072 \pm 0.00005}{}$ & $\SI{-0.00066 \pm 0.00005}{}$ \\
\\
Top Vertex Form Factors  &   & &  &  & \\ 
$b_{\pi_1}$ $(\pi \pi_1 \eta)$ & $\text{---}$ & $\text{---}$ & $\text{---}$ & $\SI{2.88 \pm 0.33}{}$ & $\SI{2.00 \pm 0.22}{}$ \\
$b_{a_2}$ $(\pi a_2 \eta)$ & $\SI{-0.40 \pm 0.08}{}$ & $\SI{0.32 \pm 0.05}{}$ & $\SI{-0.27 \pm 0.06}{}$ & $\SI{-0.26 \pm 0.05}{}$ & $\SI{-0.655 \pm 0.016}{}$ \\
$b_{a_2^\prime}$ $(\pi a_2^\prime \eta)$ & $\text{---}$ & $\text{---}$ & $\SI{5.2 \pm 0.4}{}$ & $\text{---}$ & $\SI{-0.674 \pm 0.013}{}$ \\
$b_{f_2}$ $(\pi f_2 \pi)$ & $\SI{0.10 \pm 0.10}{}$ & $\SI{0.81 \pm 0.09}{}$ & $\SI{0.63 \pm 0.08}{}$ & $\SI{0.70 \pm 0.09}{}$ & $\SI{0.35 \pm 0.07}{}$ \\
$b_{\Pom\,}$ $(\pi \Pom \pi)$ & $\SI{0.12 \pm 0.05}{}$ & $\SI{0.381 \pm 0.033}{}$ & $\SI{0.313 \pm 0.033}{}$ & $\SI{0.341 \pm 0.033}{}$ & $\SI{0.317 \pm 0.035}{}$ \\
\\
Bottom Vertex Form Factors  &   & &  &  & \\
$b_f$ $(p f_2 p)$ & $\SI{-0.11 \pm 0.10}{}$ & $\SI{0.19 \pm 0.09}{}$ & $\SI{0.06 \pm 0.08}{}$ & $\SI{0.15 \pm 0.08}{}$ & $\SI{-0.05 \pm 0.08}{}$ \\
$b_P$ $(p \Pom p)$ & $\SI{1.25 \pm 0.08}{}$ & $\SI{1.48 \pm 0.05}{}$ & $\SI{1.41 \pm 0.05}{}$ & $\SI{1.41 \pm 0.05}{}$ & $\SI{1.47 \pm 0.05}{}$ \\
\\
\end{tabular}
\end{ruledtabular}
\end{table}

\begin{table}[!h]
\centering
\caption{
Values of the parameters for the best fit to $\eta^\prime \pi^-$ events using
the JPAC trajectories. $c_i$ parameters are dimensionless and  $b_i$ parameters
in units of \mbox{ $\left(\gev/c\right)^{-2}$}.
}
\label{tabsup:bestfitjpac_etap}
\begin{ruledtabular}
\begin{tabular}{c|rrrrr}
                     & JPAC-HE & JPAC          &  $\text{JPAC}+a_2^\prime$     & $\text{JPAC}+\pi_1$ &  $\text{JPAC}+\pi_1+a_2^\prime$\\ \\
Extended Log-Likelihood & 28214 & 86874.3 & 86813.1 & 86792.7 & 86758.4 \\
Strength parameters  &   & &  &  & \\
$c_{\pi_1 f_2}$ & $\text{---}$ & $\text{---}$ & $\text{---}$ & $\SI{-0.38 \pm 0.04}{}$ & $\SI{-1.40 \pm 0.15}{}$ \\
$c_{\pi_1 \Pom}$ & $\text{---}$ & $\text{---}$ & $\text{---}$ & $\SI{-0.0185 \pm 0.0029}{}$ & $\SI{0.088 \pm 0.030}{}$ \\
$c_{a_2f_2}$ & $\SI{0.0169 \pm 0.0027}{}$ & $\SI{0.0261 \pm 0.0020}{}$ & $\SI{0.0189 \pm 0.0014}{}$ & $\SI{0.0265 \pm 0.0023}{}$ & $\SI{0.0061 \pm 0.0016}{}$ \\
$c_{a_2\Pom}$ & $\SI{0.0028 \pm 0.0004}{}$ & $\SI{0.00552 \pm 0.00029}{}$ & $\SI{0.00275 \pm 0.00024}{}$ & $\SI{0.00371 \pm 0.00033}{}$ & $\SI{0.00337 \pm 0.00021}{}$ \\
$c_{a_2^\prime f_2}$ & $\text{---}$ & $\text{---}$ & $\SI{-0.92 \pm 0.23}{}$ & $\text{---}$ & $\SI{0.026 \pm 0.006}{}$ \\
$c_{a_2^\prime\Pom}$ & $\text{---}$ & $\text{---}$ & $\SI{-0.048 \pm 0.005}{}$ & $\text{---}$ & $\SI{0.00241 \pm 0.00028}{}$ \\
$c_{f_2f_2}$ & $\SI{0.117 \pm 0.012}{}$ & $\SI{-0.116 \pm 0.007}{}$ & $\SI{0.121 \pm 0.007}{}$ & $\SI{-0.135 \pm 0.008}{}$ & $\SI{0.084 \pm 0.007}{}$ \\
$c_{f_2\Pom}$ & $\SI{0.0072 \pm 0.0015}{}$ & $\SI{-0.0058 \pm 0.0005}{}$ & $\SI{0.0052 \pm 0.0005}{}$ & $\SI{-0.0042 \pm 0.0006}{}$ & $\SI{0.0091 \pm 0.0005}{}$ \\
$c_{\Pom f_2}$ & $\SI{0.0044 \pm 0.0006}{}$ & $\SI{-0.0079 \pm 0.0006}{}$ & $\SI{0.0072 \pm 0.0006}{}$ & $\SI{-0.0074 \pm 0.0007}{}$ & $\SI{0.0082 \pm 0.0006}{}$ \\
$c_{\Pom\Pom}$ & $\SI{-0.00043 \pm 0.00005}{}$ & $\SI{0.00081 \pm 0.00005}{}$ & $\SI{-0.00070 \pm 0.00005}{}$ & $\SI{0.00076 \pm 0.00005}{}$ & $\SI{-0.00068 \pm 0.00005}{}$ \\
\\
Top Vertex Form Factors  &   & &  &  & \\ 
$b_{\pi_1}$ $(\pi \pi_1 \eta)$ & $\text{---}$ & $\text{---}$ & $\text{---}$ & $\SI{-0.01 \pm 0.05}{}$ & $\SI{3.07 \pm 0.32}{}$ \\
$b_{a_2}$ $(\pi a_2 \eta)$ & $\SI{-0.48 \pm 0.13}{}$ & $\SI{0.31 \pm 0.06}{}$ & $\SI{-0.27 \pm 0.06}{}$ & $\SI{0.92 \pm 0.12}{}$ & $\SI{-0.683 \pm 0.015}{}$ \\
$b_{a_2^\prime}$ $(\pi a_2^\prime \eta)$ & $\text{---}$ & $\text{---}$ & $\SI{5.2 \pm 0.4}{}$ & $\text{---}$ & $\SI{-0.700 \pm 0.012}{}$ \\
$b_{f_2}$ $(\pi f_2 \pi)$ & $\SI{0.13 \pm 0.10}{}$ & $\SI{0.81 \pm 0.09}{}$ & $\SI{0.67 \pm 0.08}{}$ & $\SI{0.81 \pm 0.10}{}$ & $\SI{0.36 \pm 0.07}{}$ \\
$b_{\Pom\,}$ $(\pi \Pom \pi)$ & $\SI{0.10 \pm 0.05}{}$ & $\SI{0.368 \pm 0.035}{}$ & $\SI{0.312 \pm 0.034}{}$ & $\SI{0.355 \pm 0.035}{}$ & $\SI{0.32 \pm 0.04}{}$ \\
\\
Bottom Vertex Form Factors  &   & &  &  & \\
$b_f$ $(p f_2 p)$ & $\SI{-0.08 \pm 0.11}{}$ & $\SI{0.13 \pm 0.09}{}$ & $\SI{0.10 \pm 0.08}{}$ & $\SI{0.24 \pm 0.08}{}$ & $\SI{0.00 \pm 0.08}{}$ \\
$b_P$ $(p \Pom p)$ & $\SI{1.29 \pm 0.09}{}$ & $\SI{1.57 \pm 0.05}{}$ & $\SI{1.47 \pm 0.05}{}$ & $\SI{1.45 \pm 0.05}{}$ & $\SI{1.49 \pm 0.05}{}$ \\
\\
\end{tabular}
\end{ruledtabular}
\end{table}

\clearpage
\newpage

\section*{Mass dependent vs. mass independent strength couplings}
\begin{figure*}[!h]
\begin{tabular}{ccc}
\includegraphics[width=0.29\linewidth]{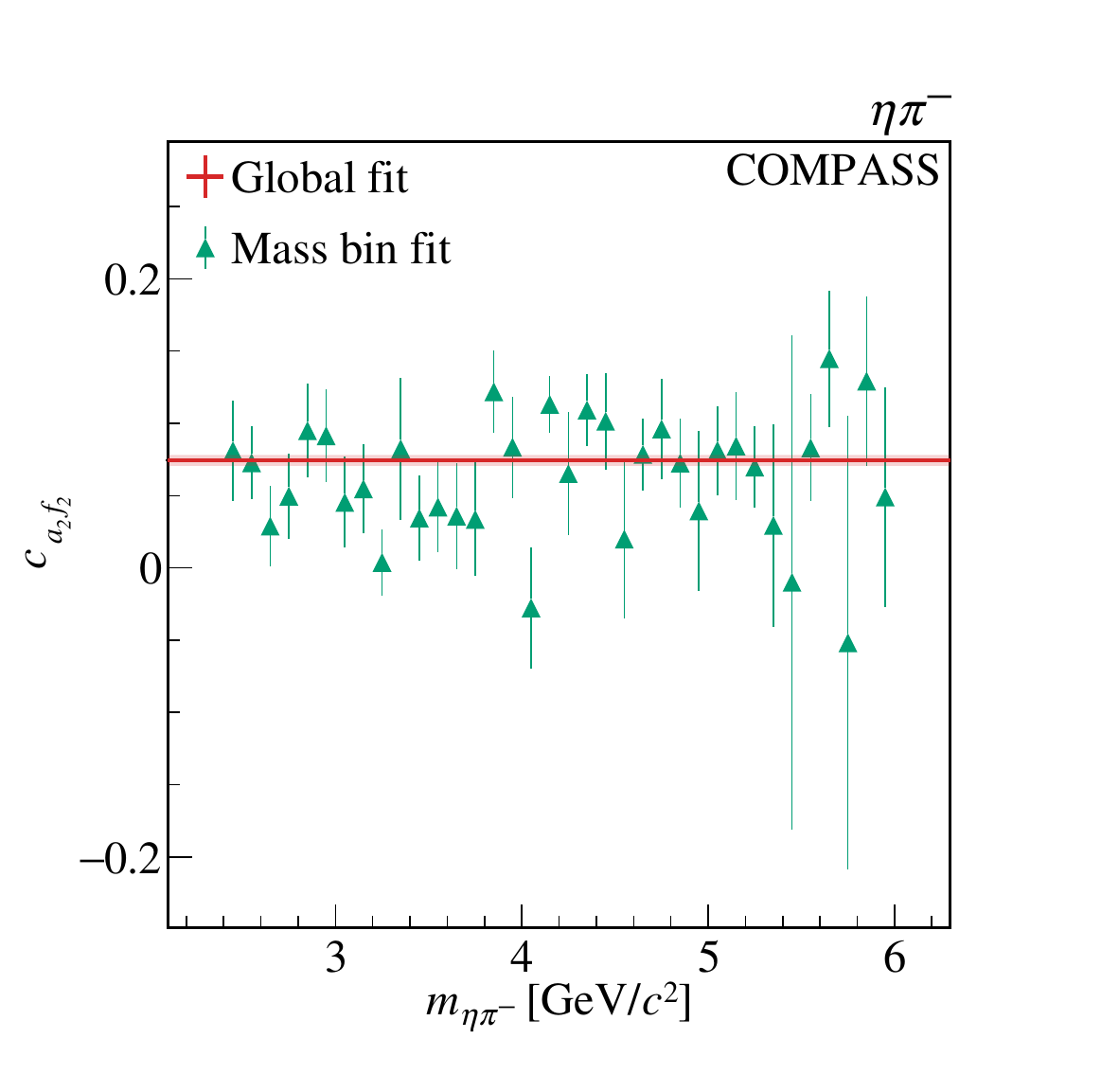} &
\includegraphics[width=0.29\linewidth]{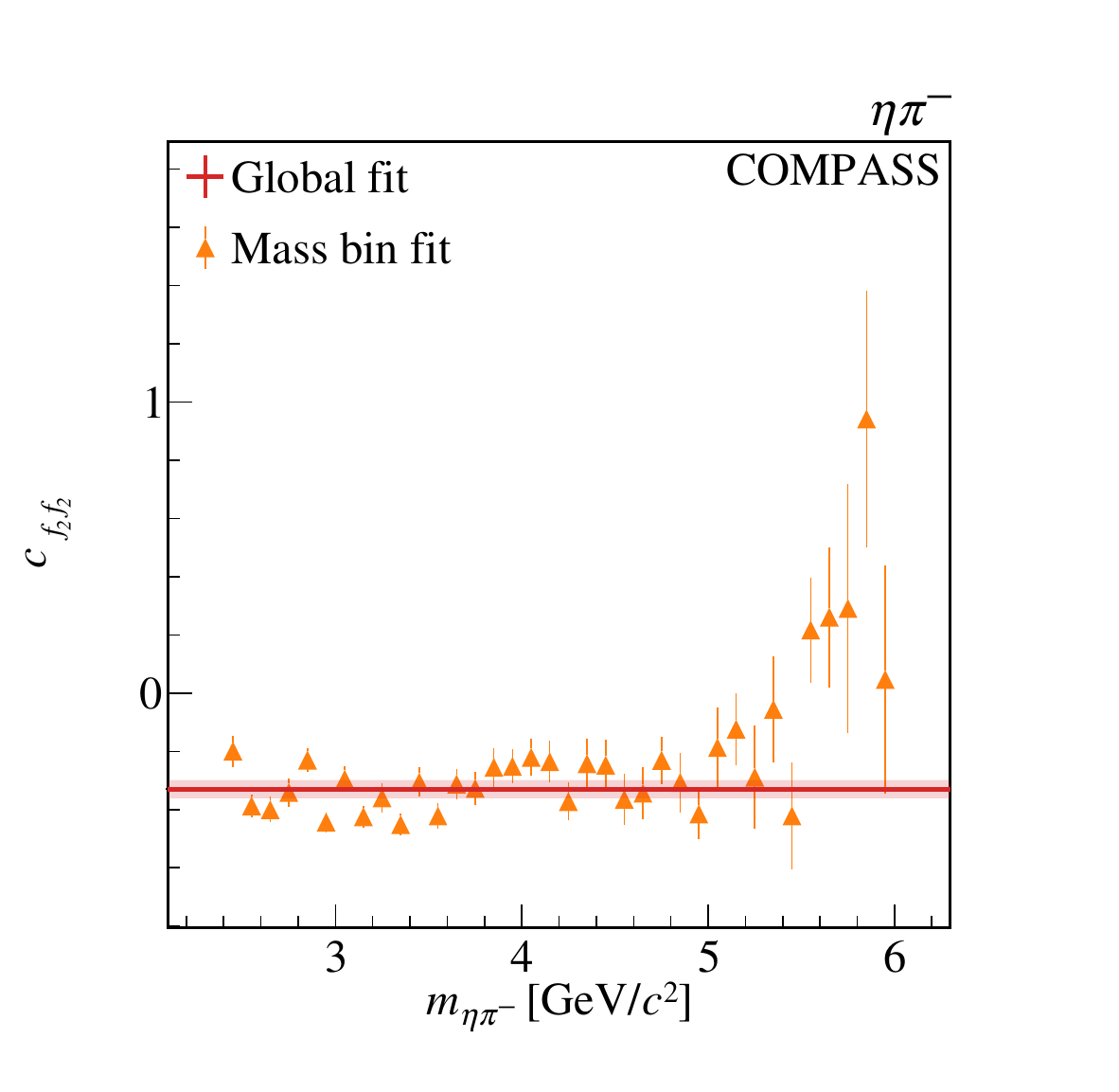} &
\includegraphics[width=0.29\linewidth]{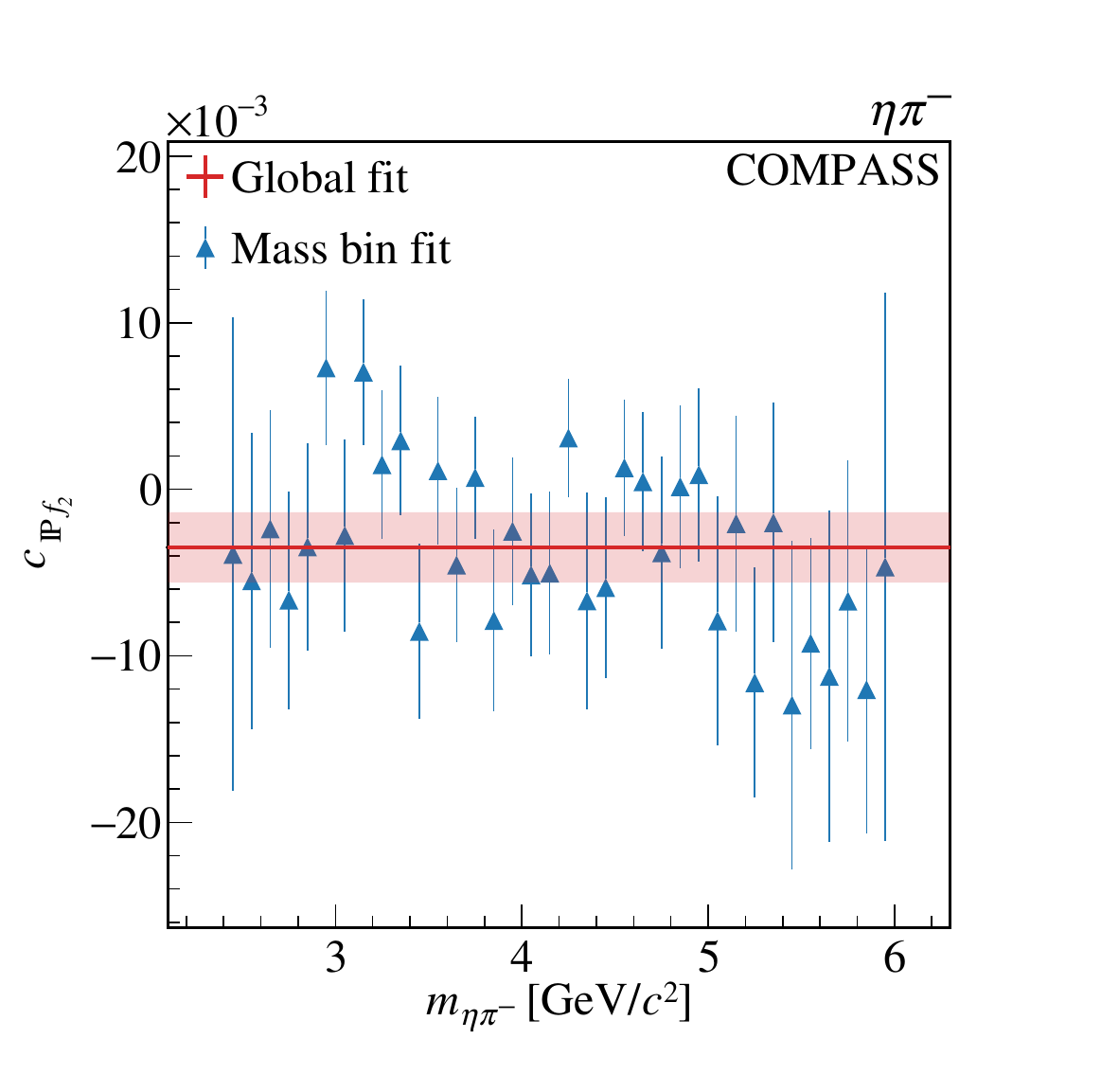} \\
\includegraphics[width=0.29\linewidth]{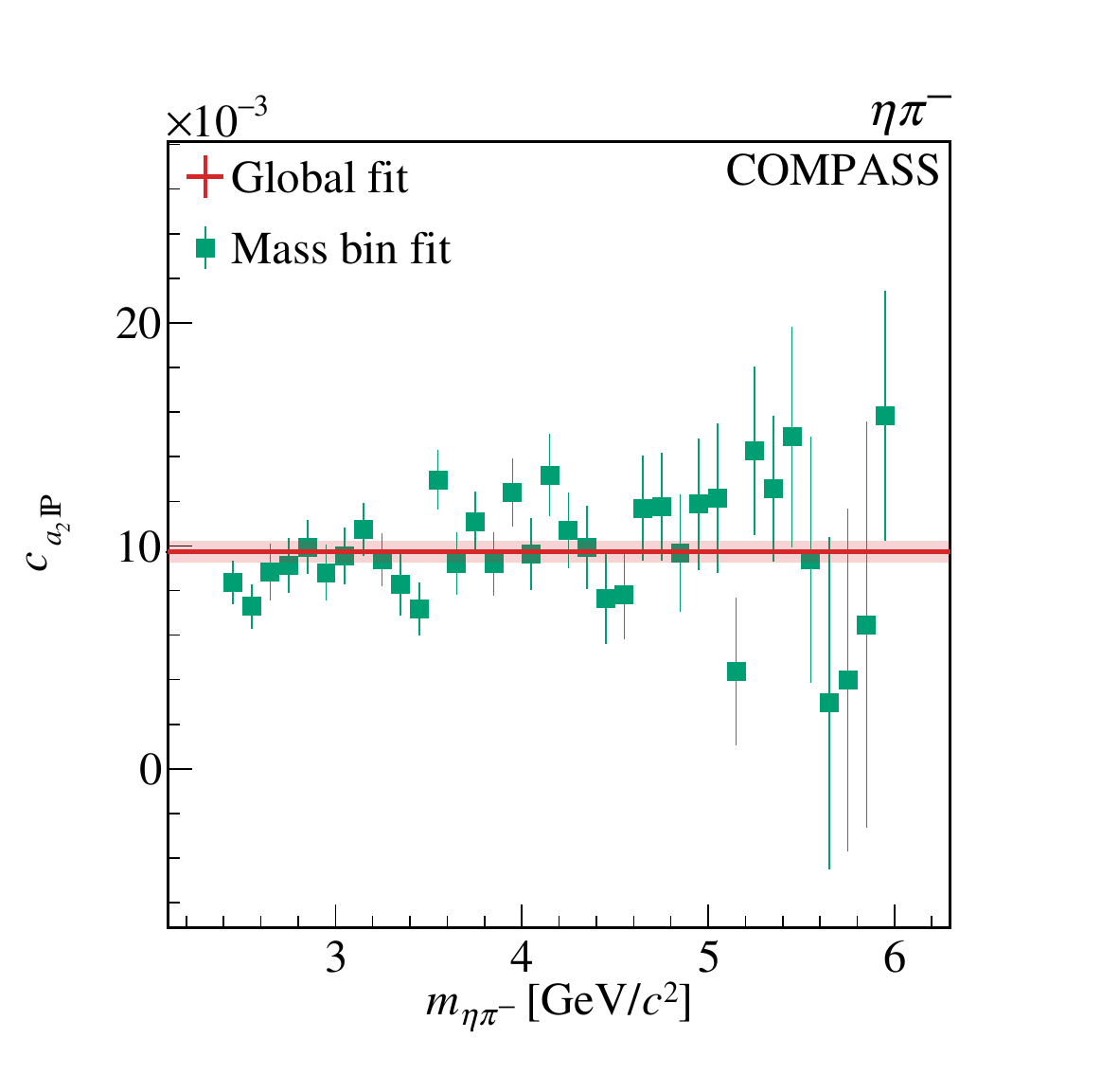} &
\includegraphics[width=0.29\linewidth]{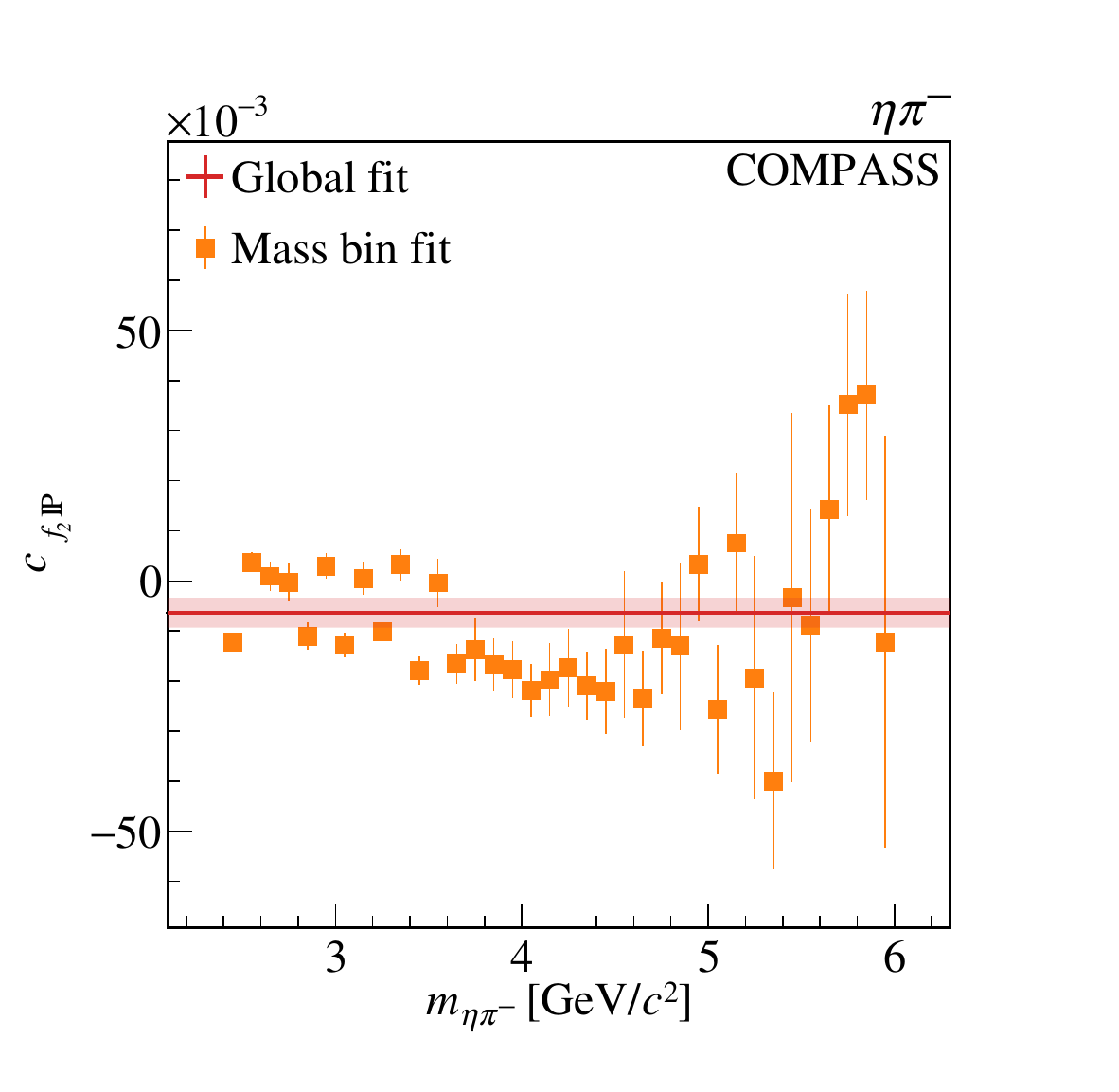} &
\includegraphics[width=0.29\linewidth]{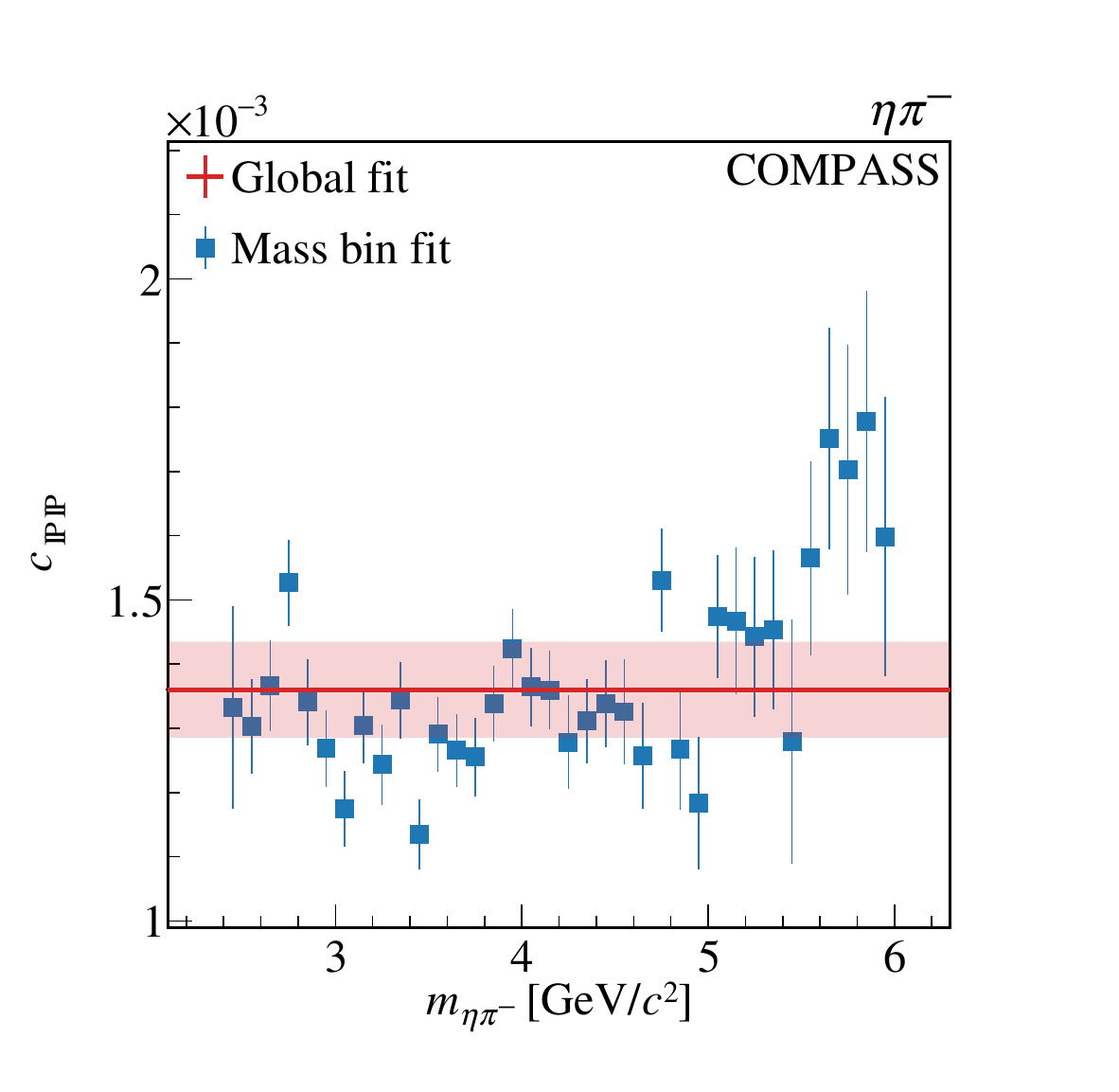}
\end{tabular}
\caption{
$\eta \pi^-$ strength parameters $\{c\}$ of
the \mbox{$\text{KGR}+a_2^\prime+\pi_1$} fit for the leading ordinary
exchanges. In red we display the best fit parameters to the data, and their
uncertainties, in the double-Regge region shown in the Letter. The data points
correspond to the event fits in a given invariant mass bin.
}
\label{figsup:etaparametersLOE}
\end{figure*}

\begin{figure*}[!h]
\begin{tabular}{cc}
\includegraphics[width=0.29\linewidth]{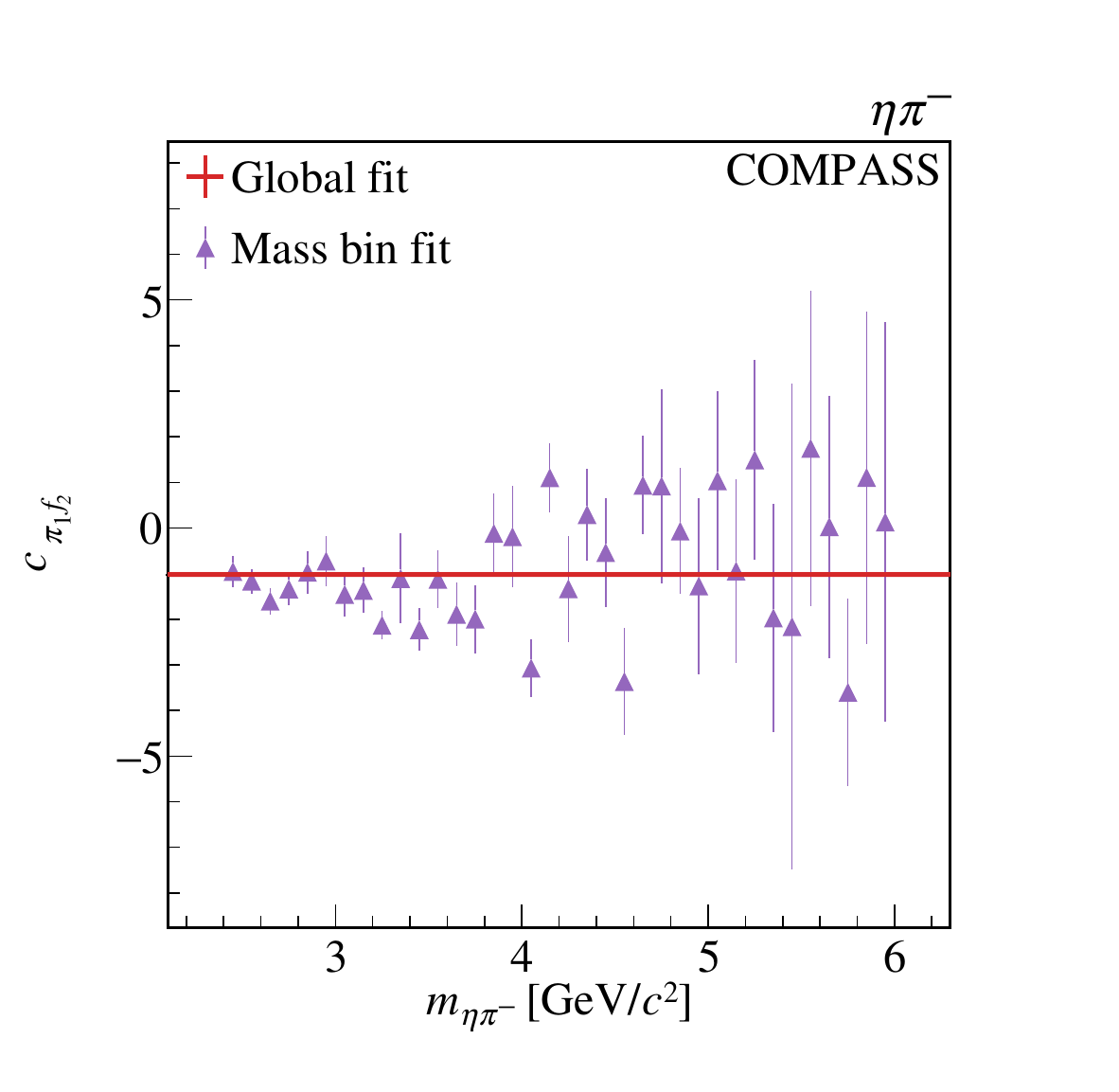} &
\includegraphics[width=0.29\linewidth]{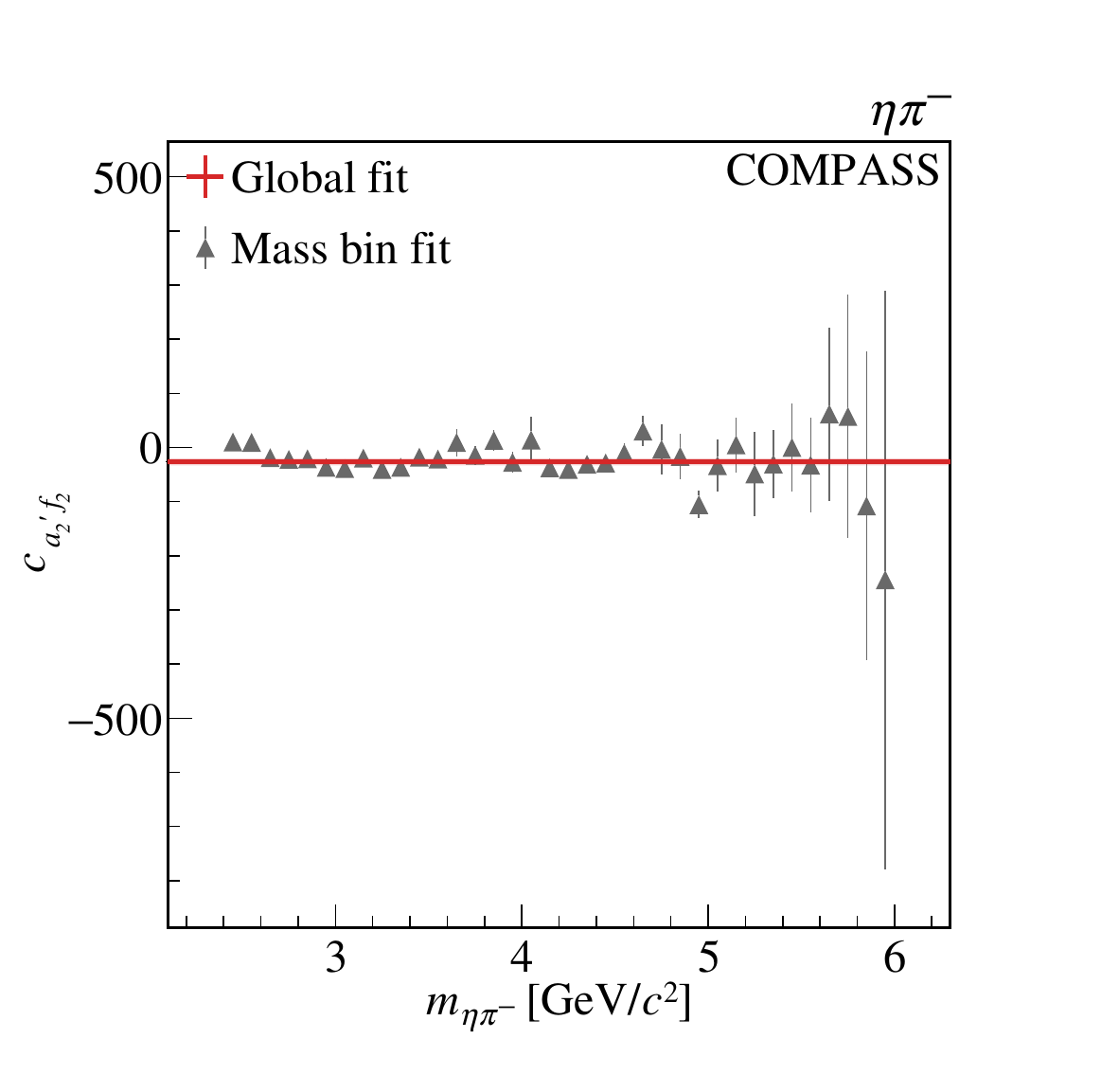} \\
\includegraphics[width=0.29\linewidth]{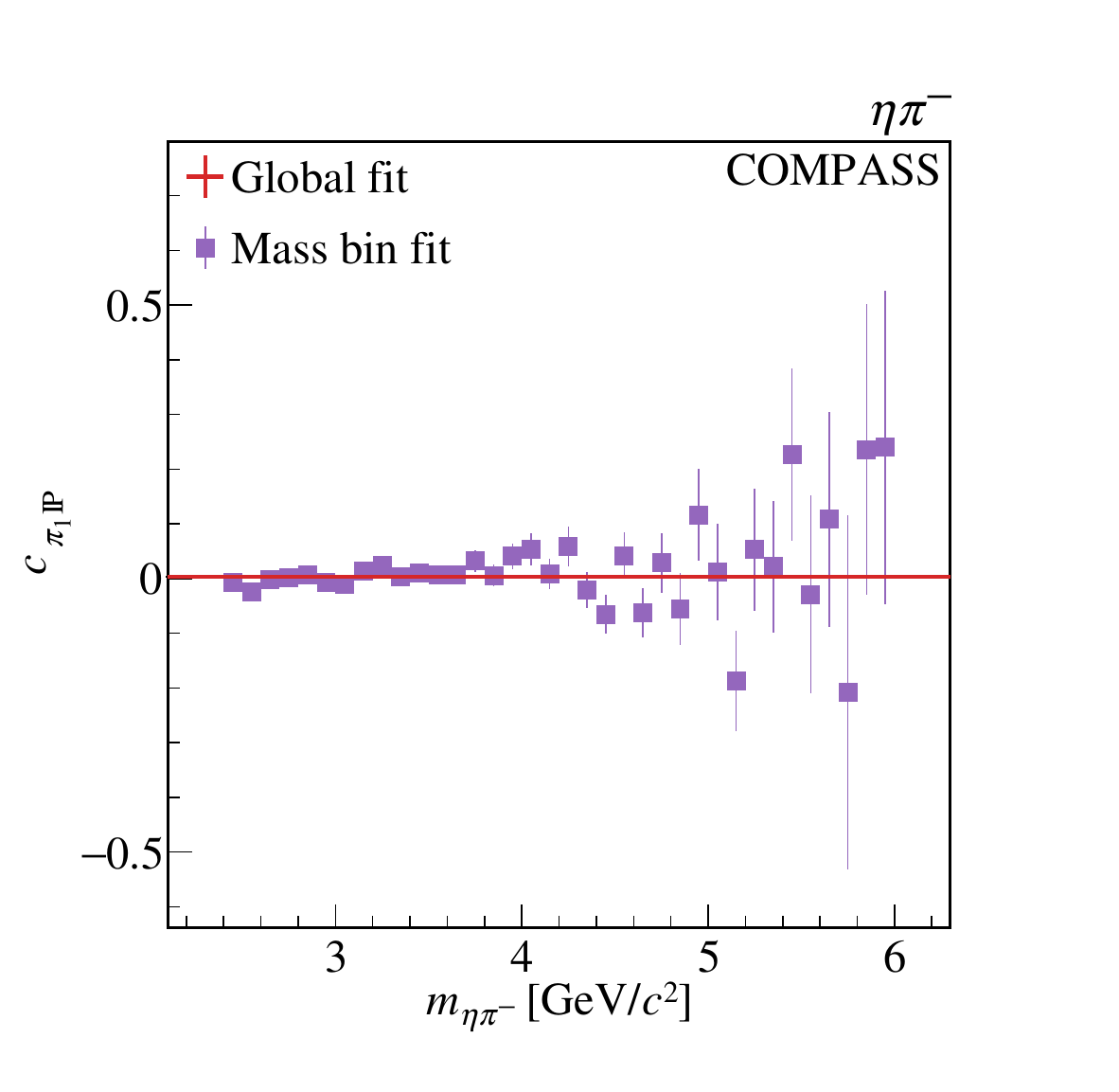} &
\includegraphics[width=0.29\linewidth]{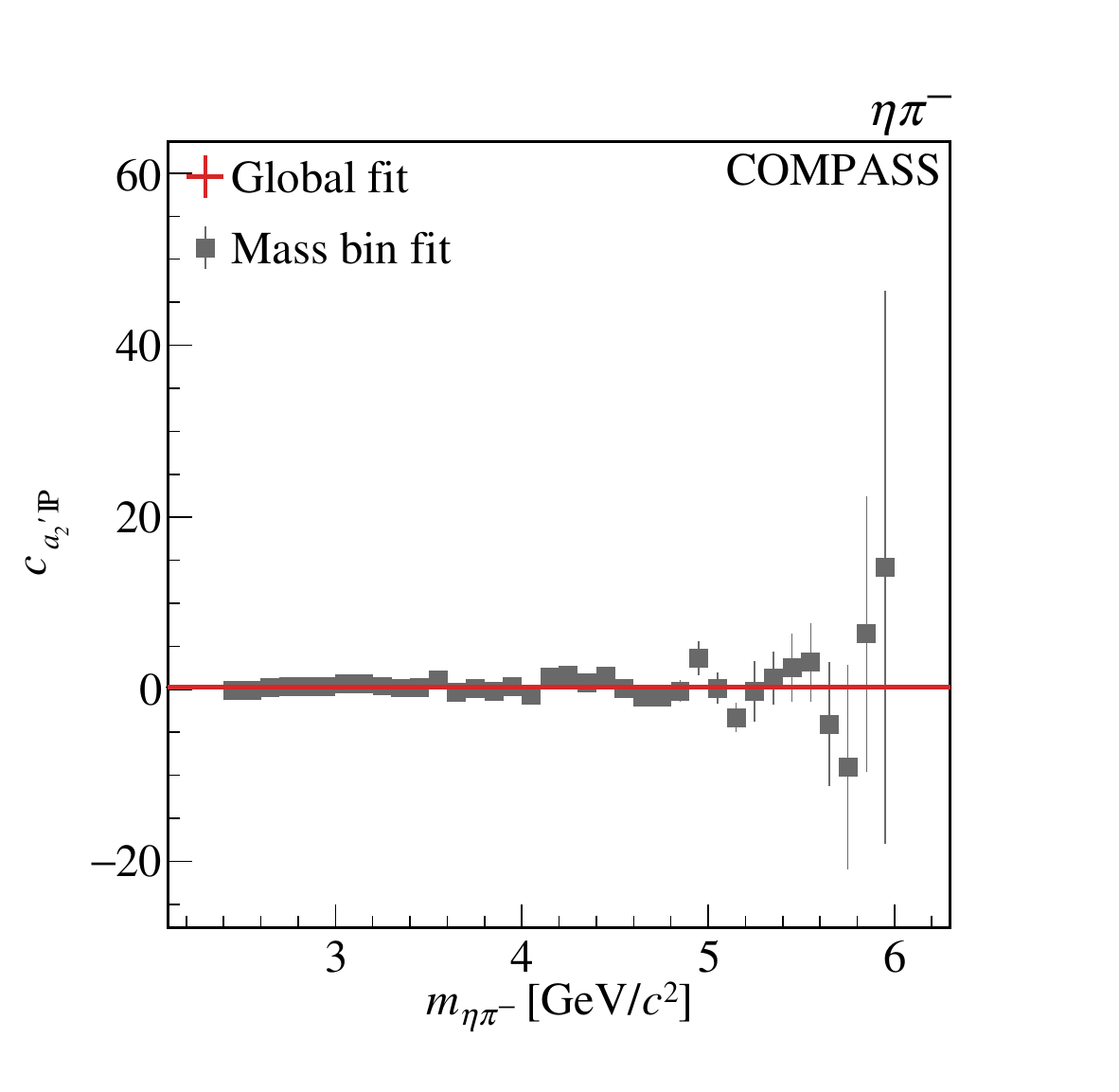} \\
\end{tabular}
\caption{
$\eta \pi^-$ strength parameters $\{c\}$ of
the \mbox{$\text{KGR}+a_2^\prime+\pi_1$} fit for the subleading $a_2^\prime$ and
exotic $\pi_1$ exchanges. Conventions as in~\cref{{figsup:etaparametersLOE}}.
}
\label{figsup:etaparametersNLOE}
\end{figure*}

\begin{figure*}[!h]
\begin{tabular}{ccc}
\includegraphics[width=0.29\linewidth]{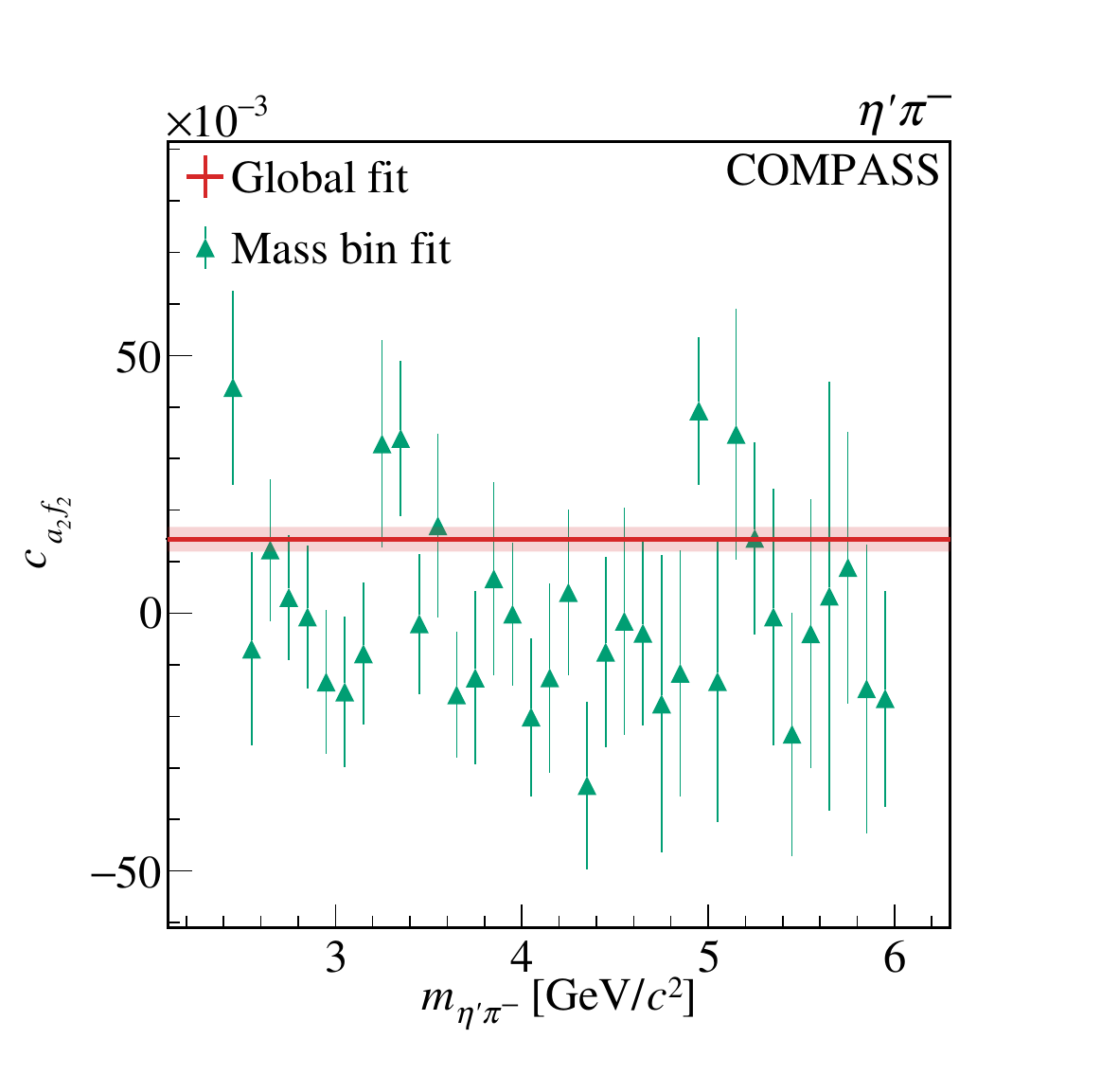} &
\includegraphics[width=0.29\linewidth]{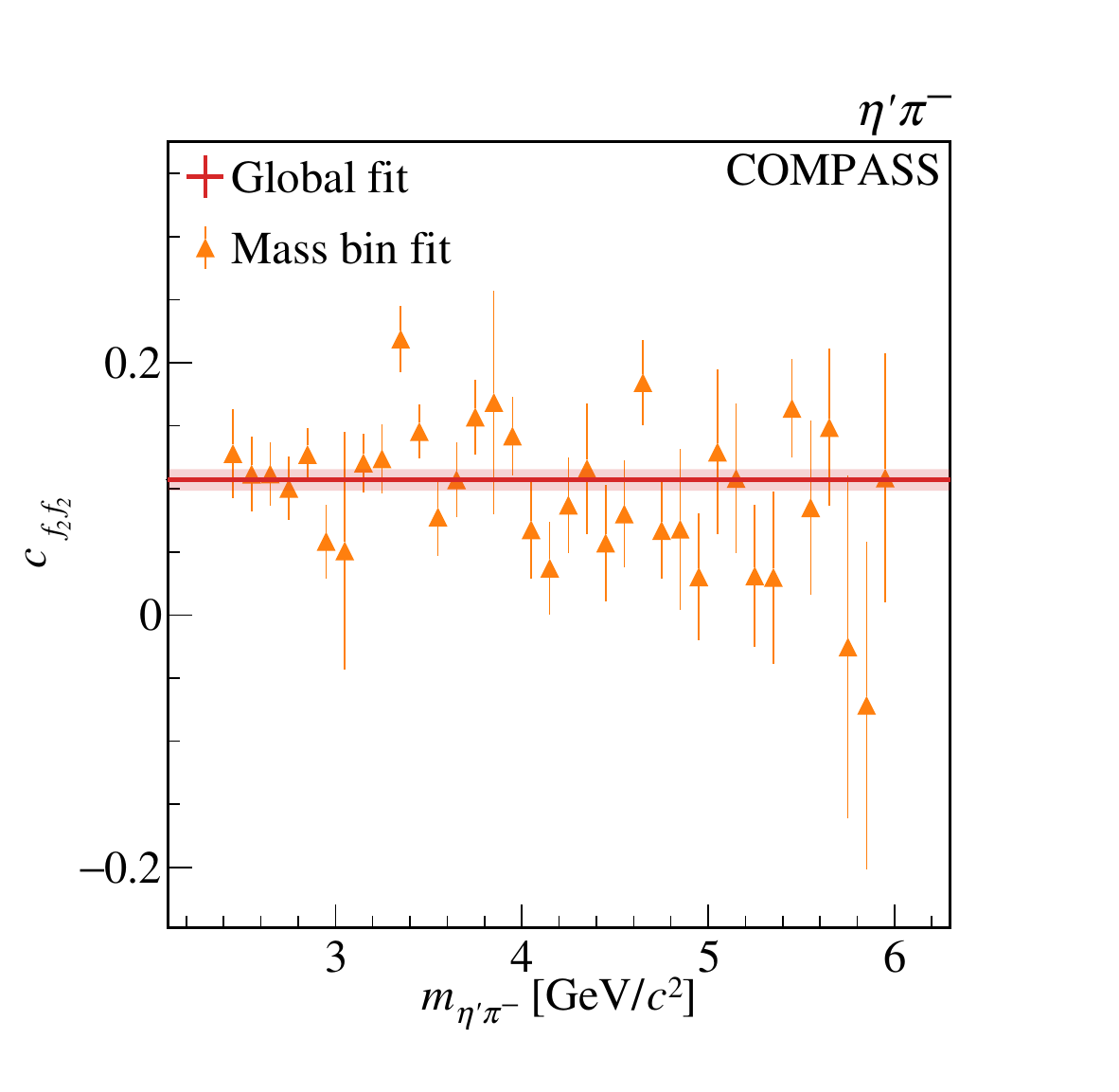} &
\includegraphics[width=0.29\linewidth]{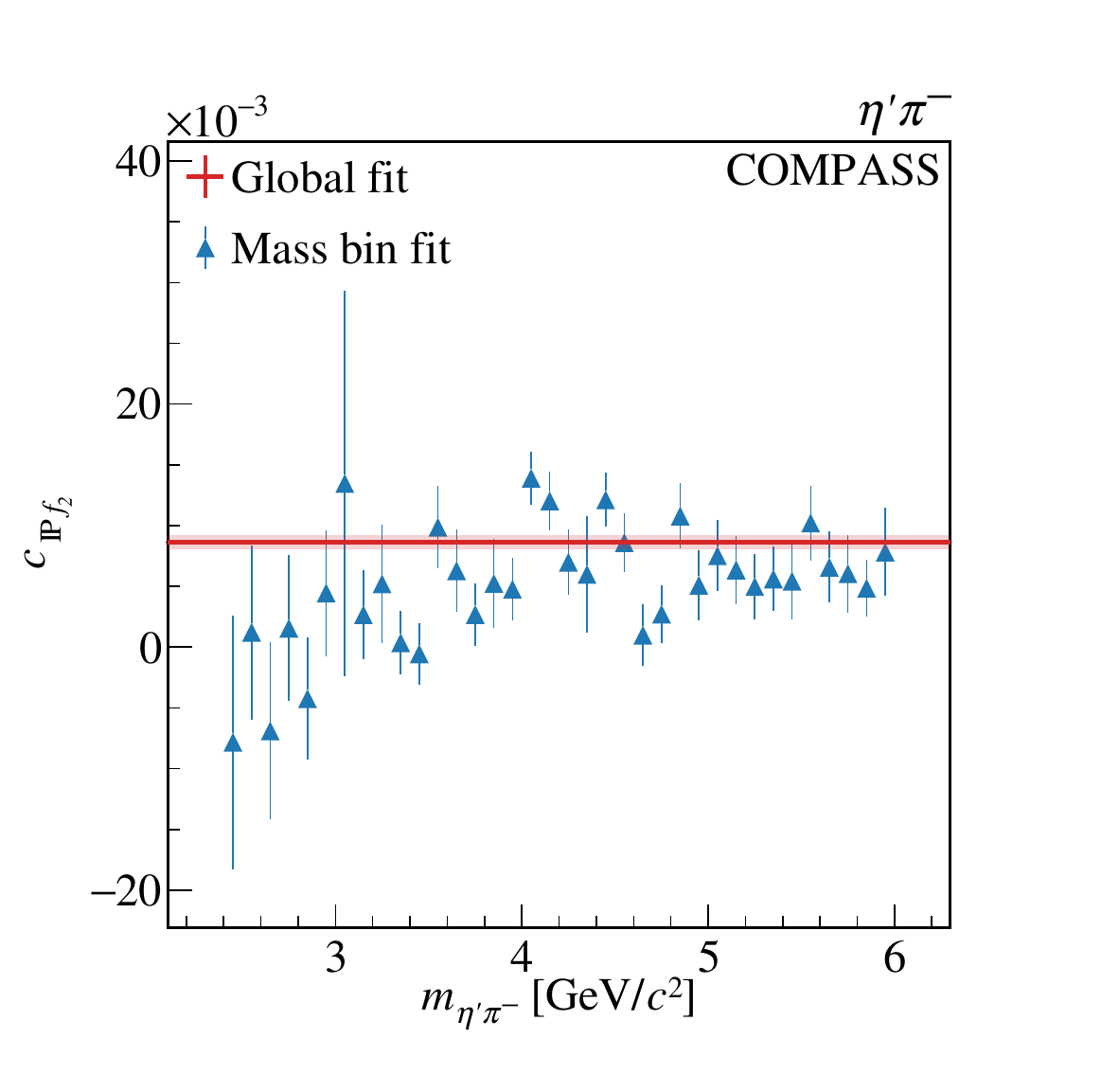} \\
\includegraphics[width=0.29\linewidth]{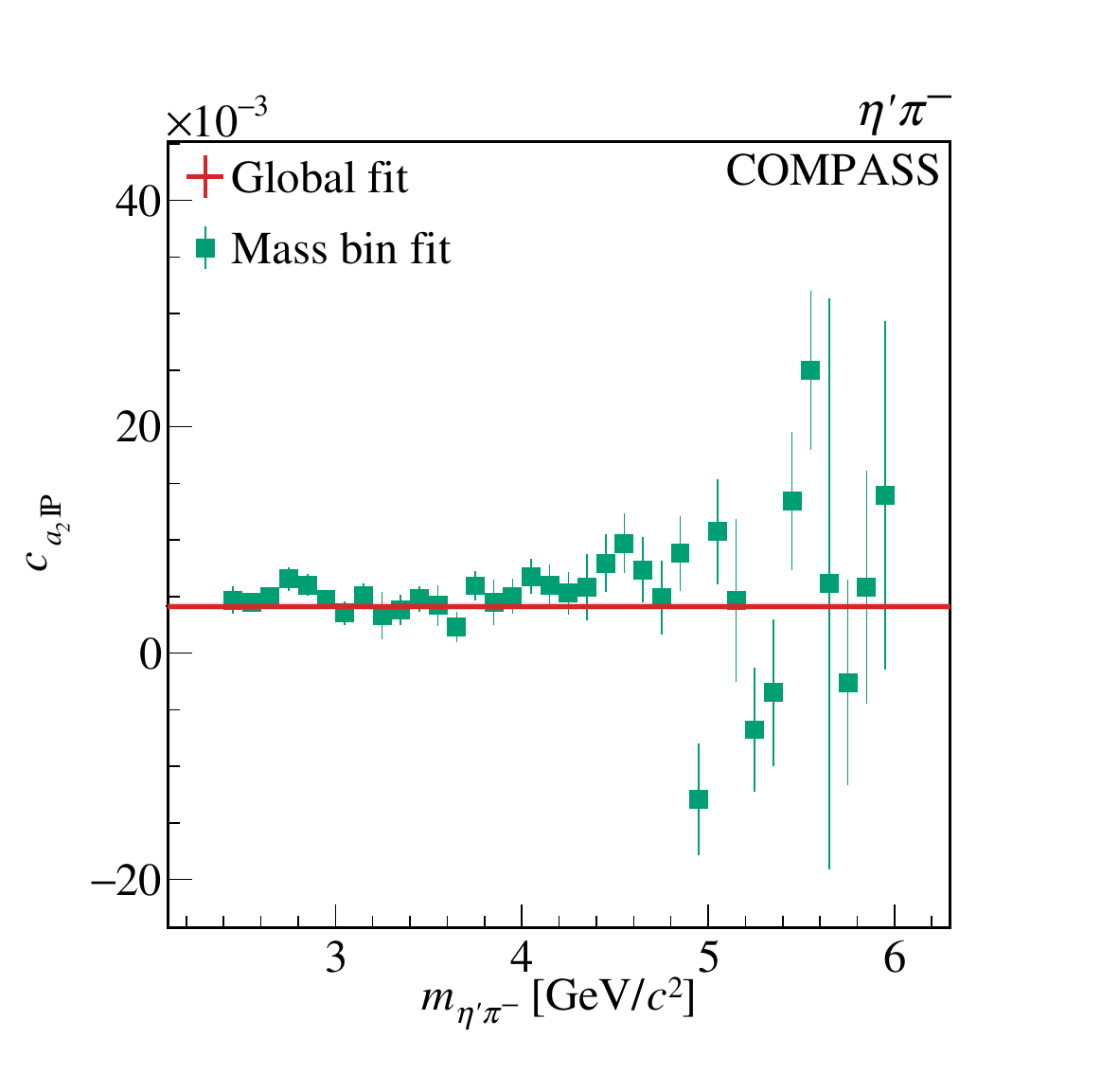} &
\includegraphics[width=0.29\linewidth]{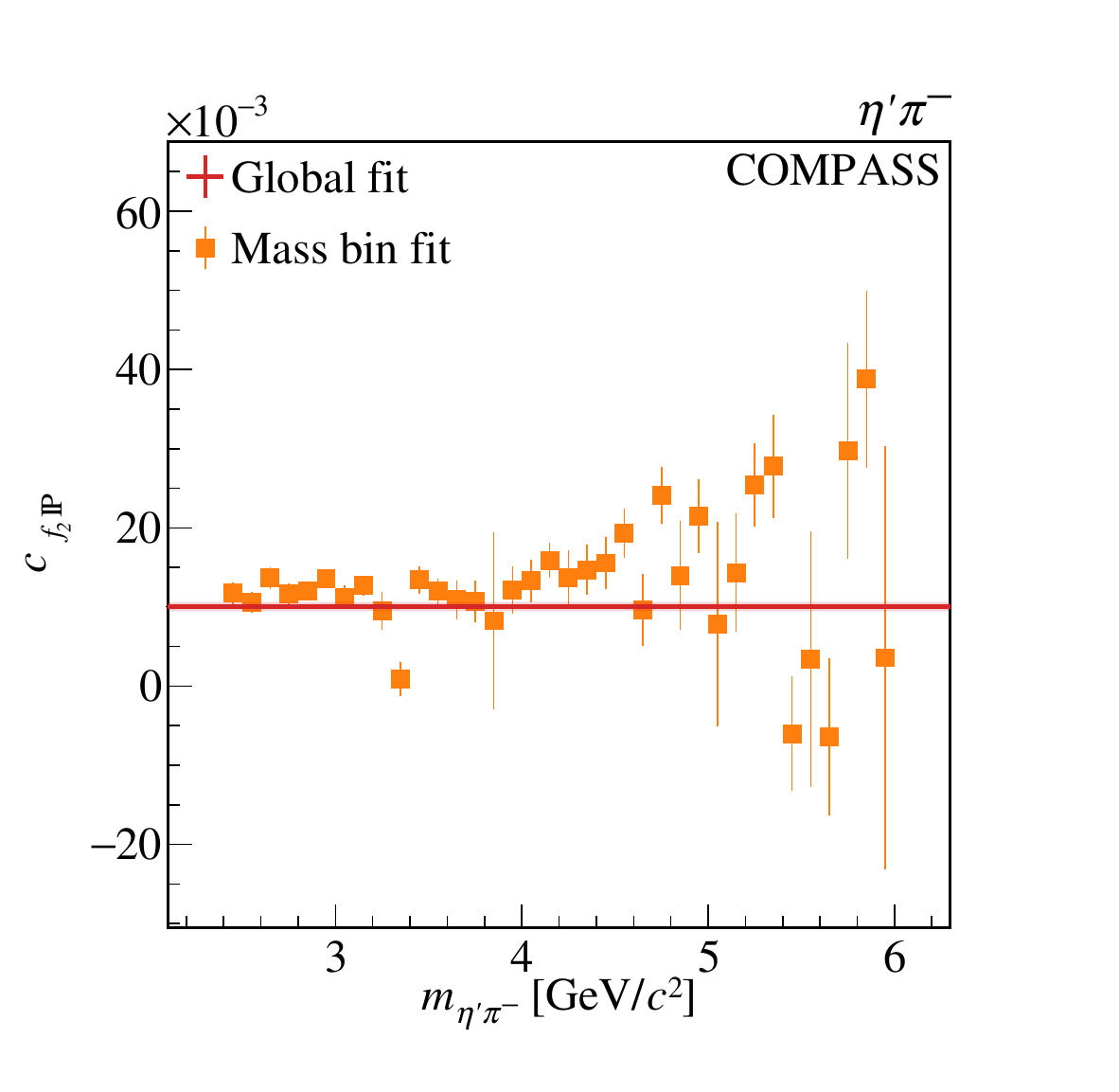} &
\includegraphics[width=0.29\linewidth]{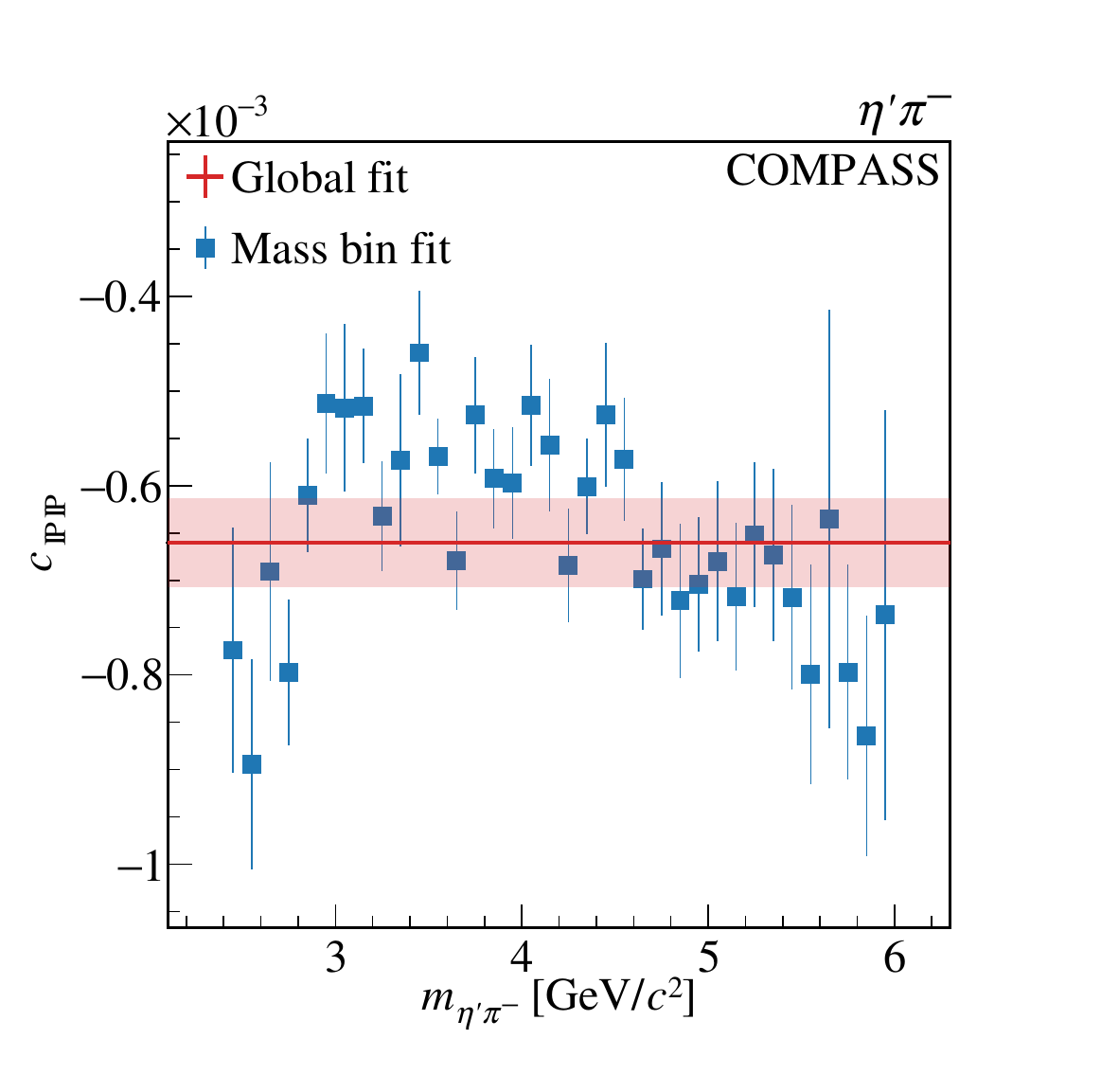}
\end{tabular}
\caption{
$\eta^\prime \pi^-$ strength parameters $\{c\}$ of
the \mbox{$\text{KGR}+a_2^\prime+\pi_1$} fit for the leading ordinary
exchanges. Conventions as in~\cref{{figsup:etaparametersLOE}}.
}
\label{figsup:etaprimeparametersLOE}
\end{figure*}

\begin{figure*}[!h]
\begin{tabular}{cc}
\includegraphics[width=0.29\linewidth]{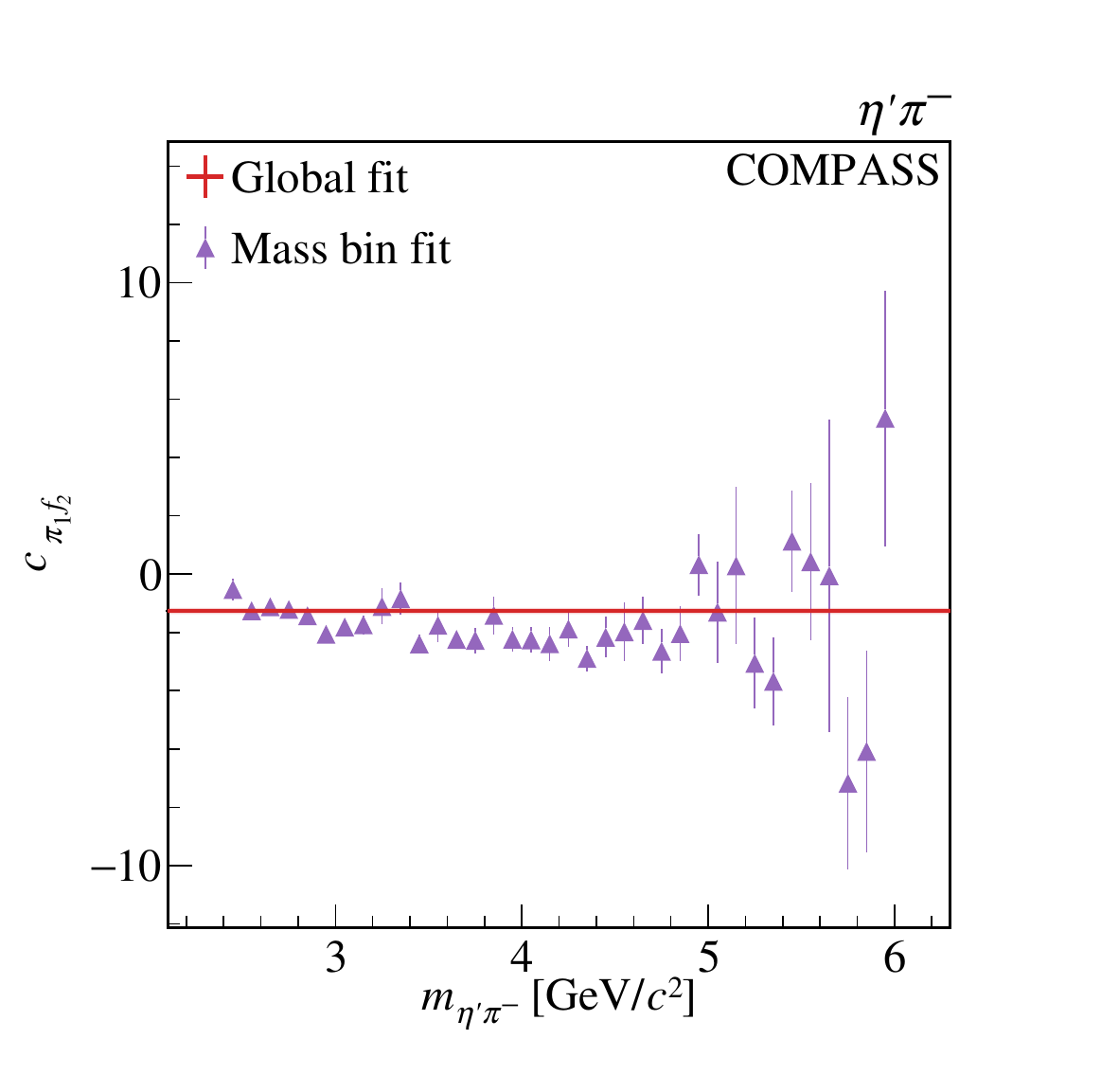} &
\includegraphics[width=0.29\linewidth]{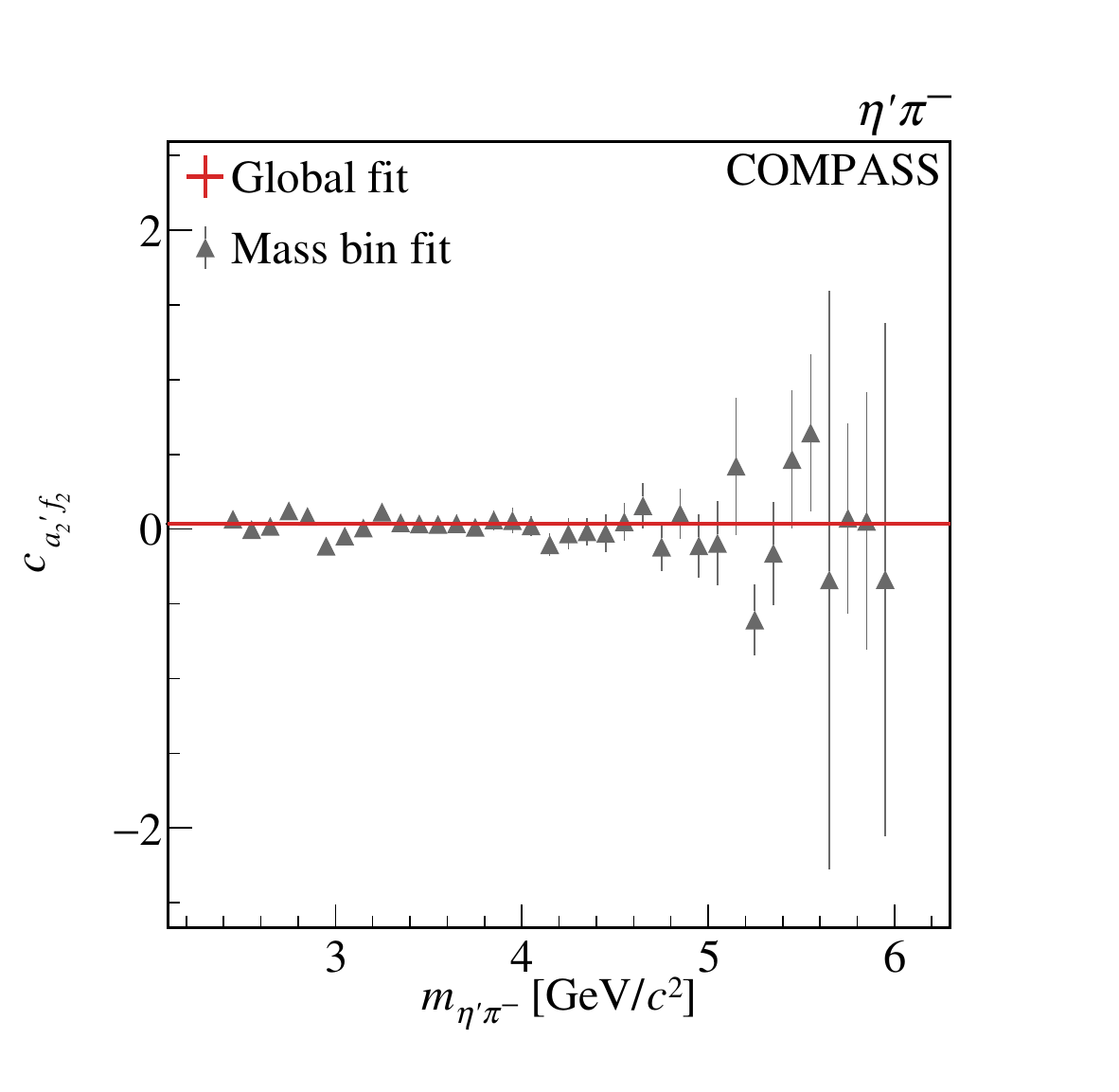} \\
\includegraphics[width=0.29\linewidth]{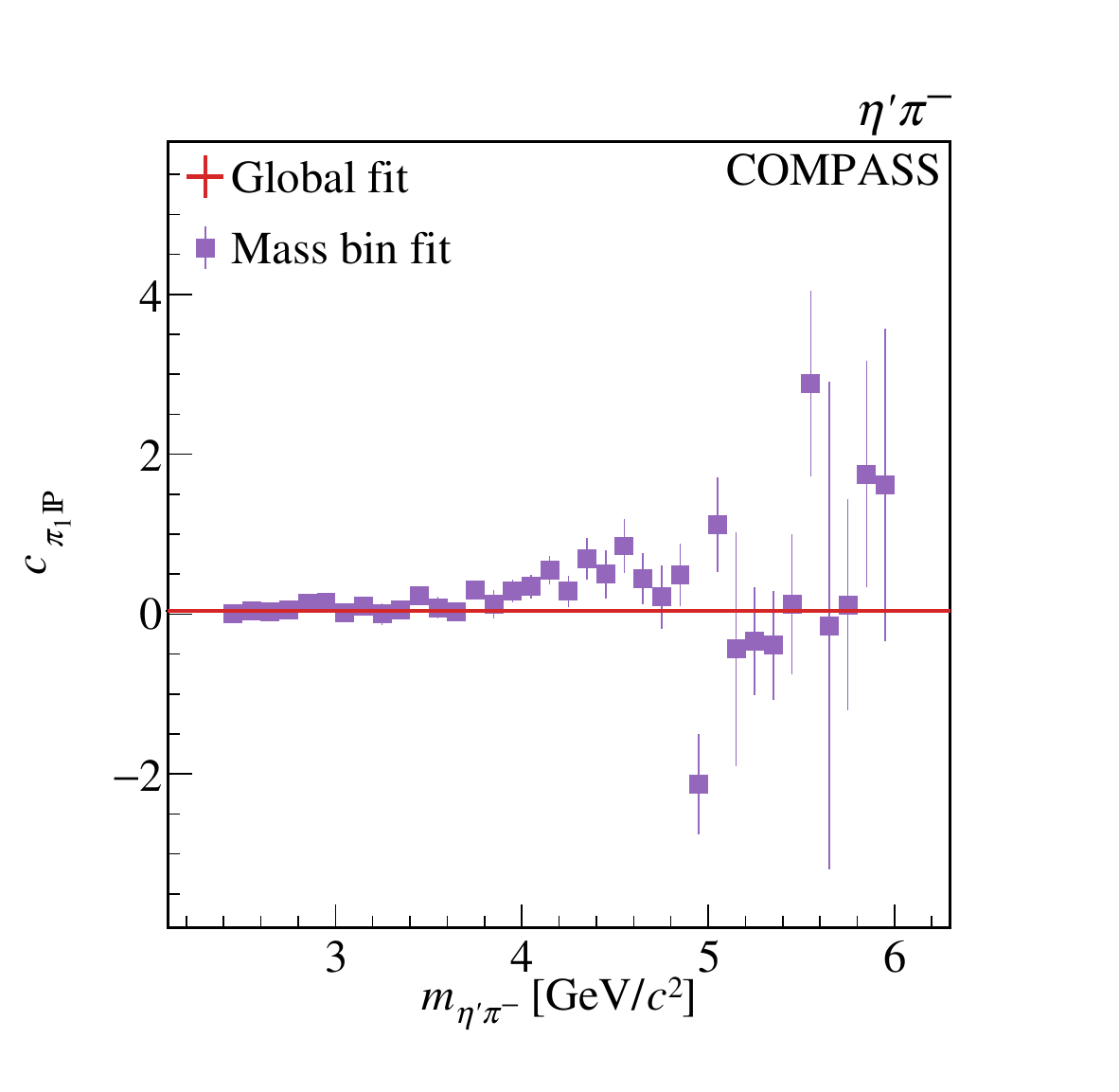} &
\includegraphics[width=0.29\linewidth]{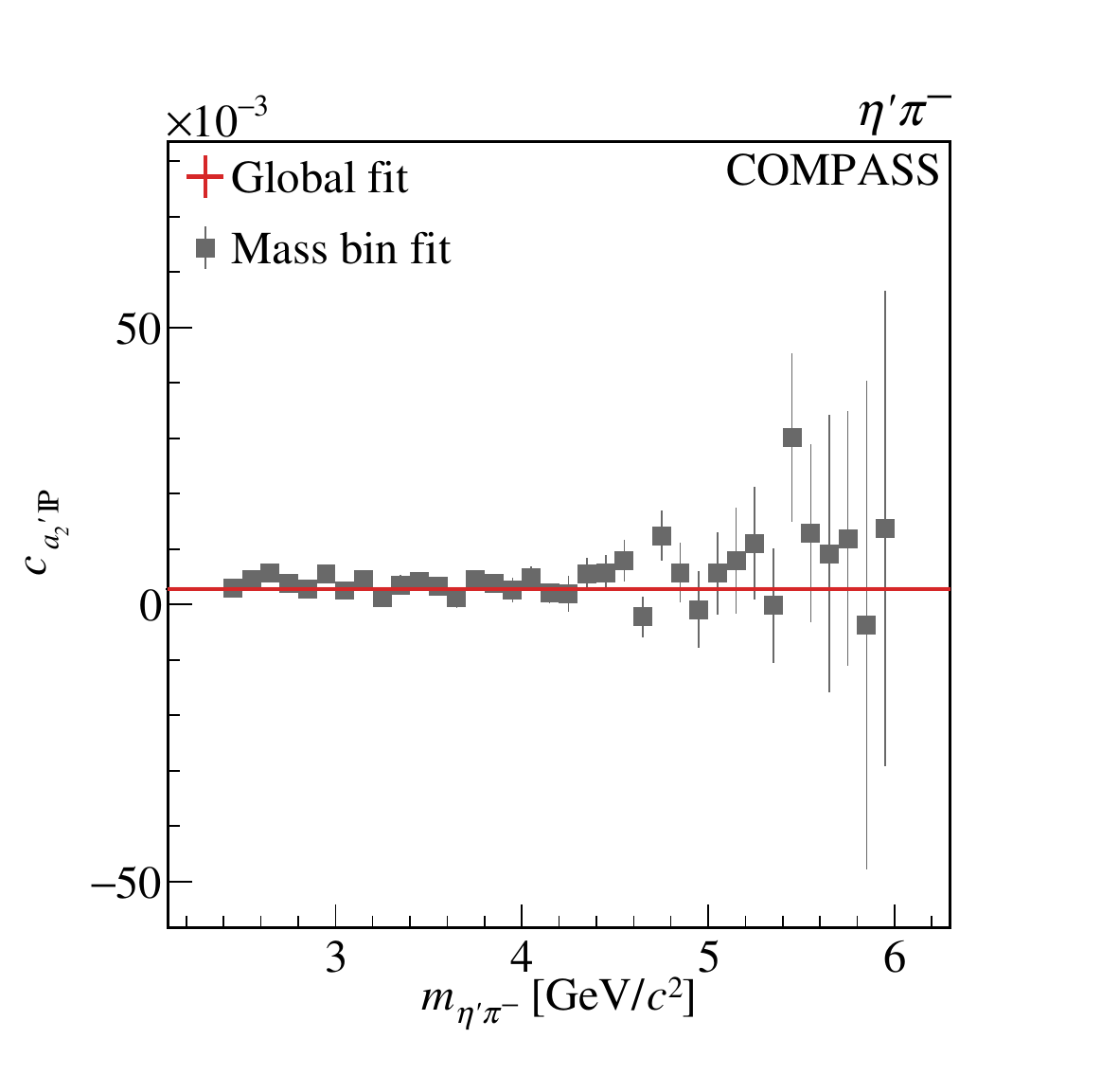} \\
\end{tabular}
\caption{
$\eta^\prime \pi^-$ strength parameters $\{c\}$ of
the \mbox{$\text{KGR}+a_2^\prime+\pi_1$} fit for the subleading $a_2^\prime$ and
exotic $\pi_1$ exchanges. Conventions as in~\cref{{figsup:etaparametersLOE}}.
}
\label{figsup:etaprimeparametersNLOE}
\end{figure*}

We conducted several exploratory fits to assess the energy range of
applicability of the double-Regge amplitude. In particular, we performed
individual fits to the events in narrow invariant mass bins (\aka mass dependent
fits) obtaining the strength couplings $\{ c\}$ for each bin. The form factors
were fixed to the best mass-independent event fit. Regge theory constrains the
energy dependence of the amplitude, hence, if the double-Regge amplitude is the
correct description of the data, the mass dependent strength couplings $\{ c\}$
should remain approximately constant, within uncertainties, throughout the
energy range. In~\cref{figsup:etaparametersLOE,figsup:etaprimeparametersLOE}
--$\eta \pi^-$ and $\eta^\prime \pi^-$, respectively--, we show the strength
couplings $\{ c\}$ of the leading ordinary exchanges for each bin compared to
the mass-independent event fit in the whole invariant mass range
$m_{\eta \pi}\in[2.4,6]\gev/c^2$ using the \mbox{$\text{KGR}+a_2^\prime+\pi_1$}
parametrization. \Cref{figsup:etaparametersNLOE,figsup:etaprimeparametersNLOE}
display the same for the nonleading exchanges. We find that in that invariant
mass range the mass-independence assumption holds and the double-Regge amplitude
correctly describes the energy dependence of the data. We note that the relative
strength of the parameters do not correspond to the relative strength of the
contributions as these are affected by the vertex form factors and the inherent
strength of the specific diagram.

\clearpage
\newpage
\section*{$\eta \pi^-$ distributions}
\begin{figure*}[!h]
\begin{tabular}{cccc}
\includegraphics[width=0.25\linewidth]{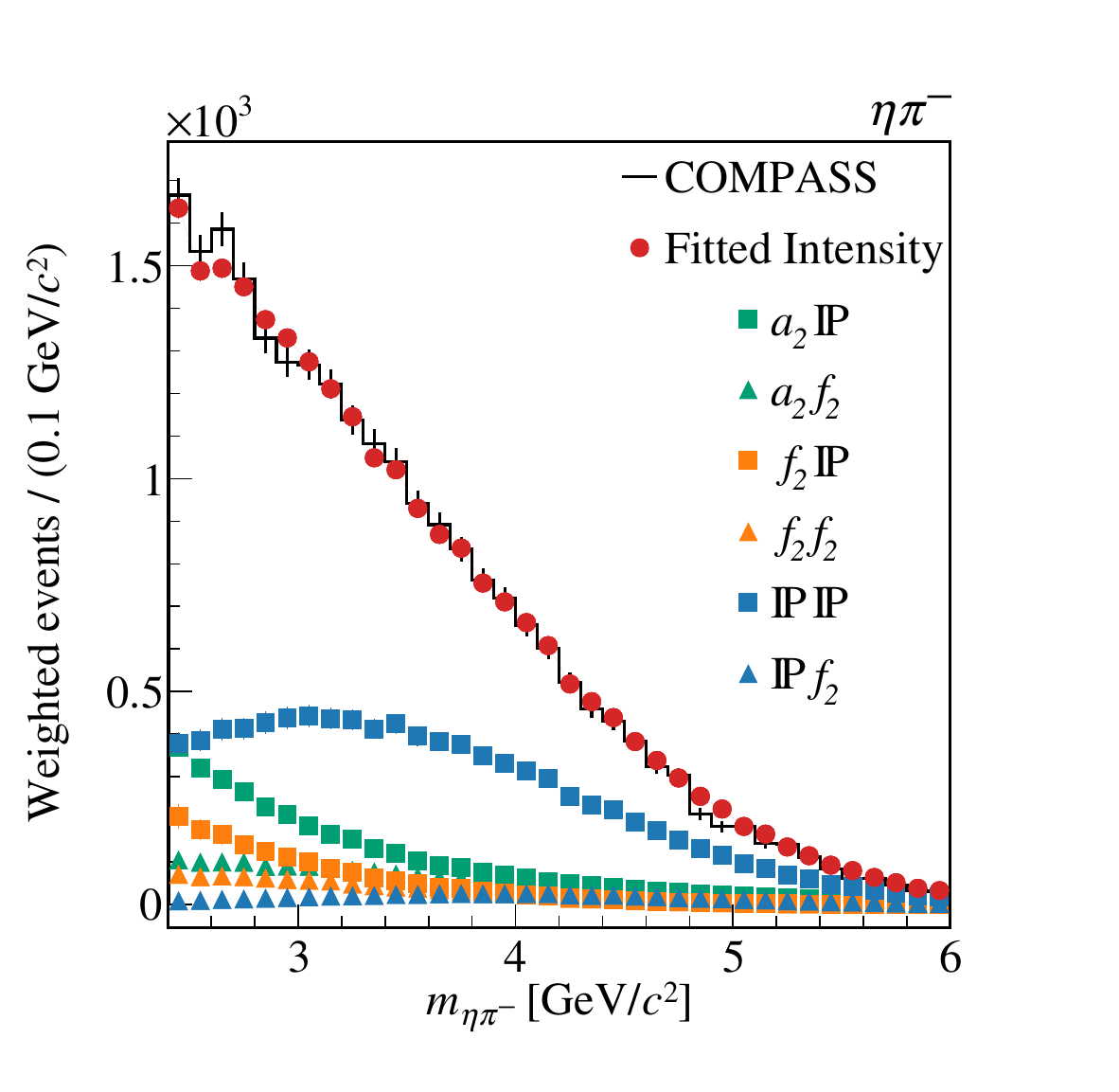} & 
\includegraphics[width=0.25\linewidth]{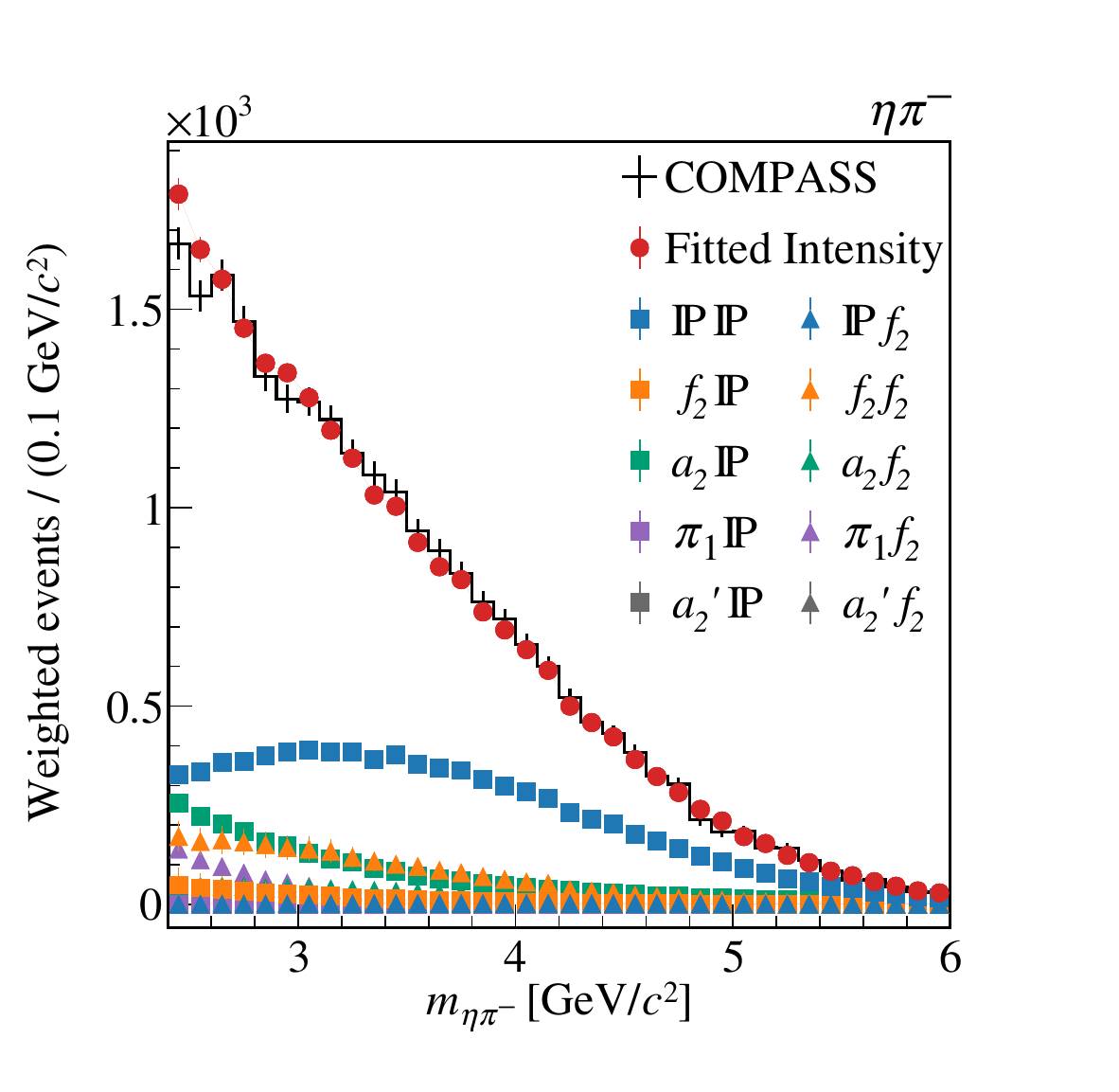} & 
\includegraphics[width=0.25\linewidth]{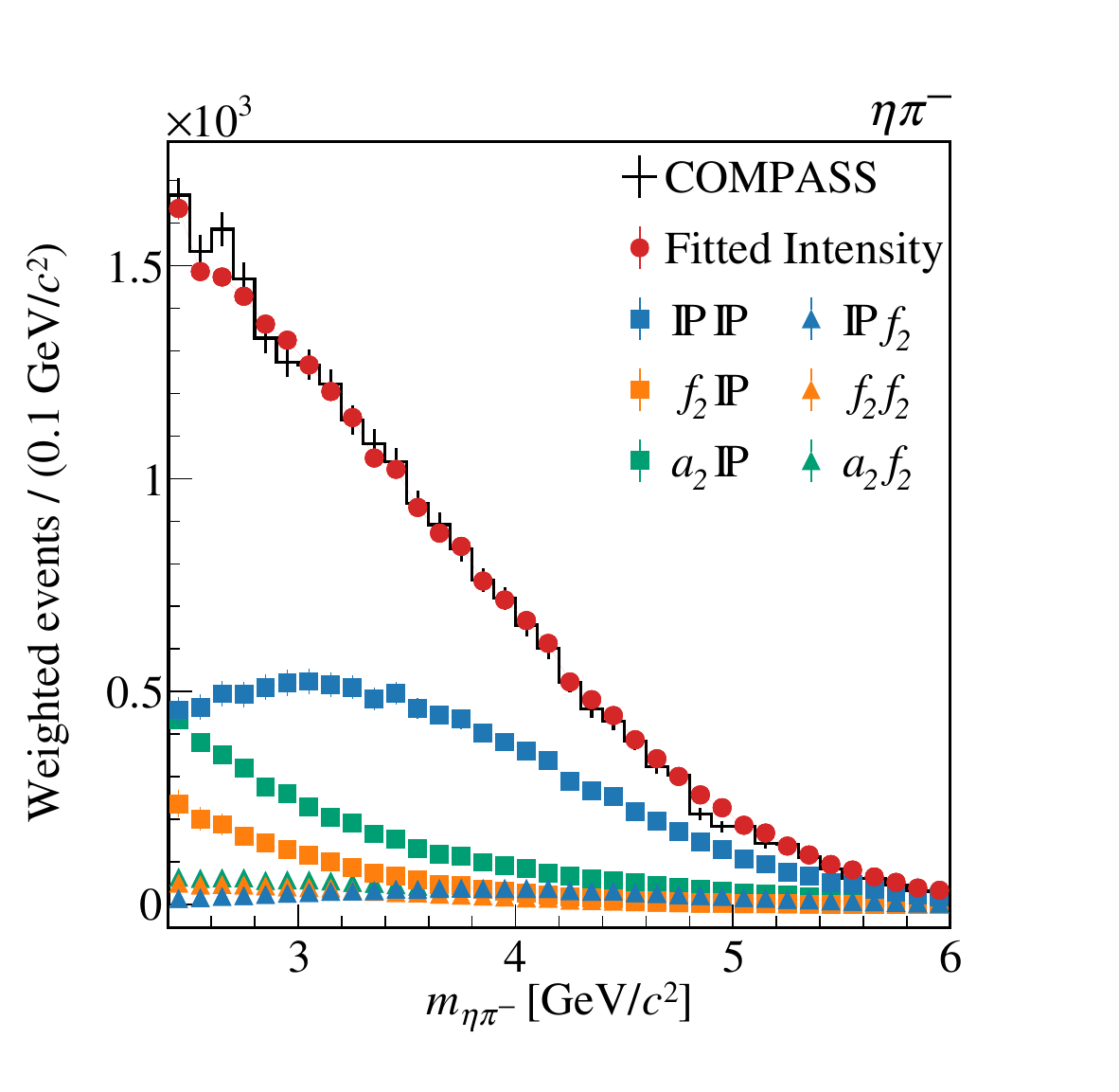} & 
\includegraphics[width=0.25\linewidth]{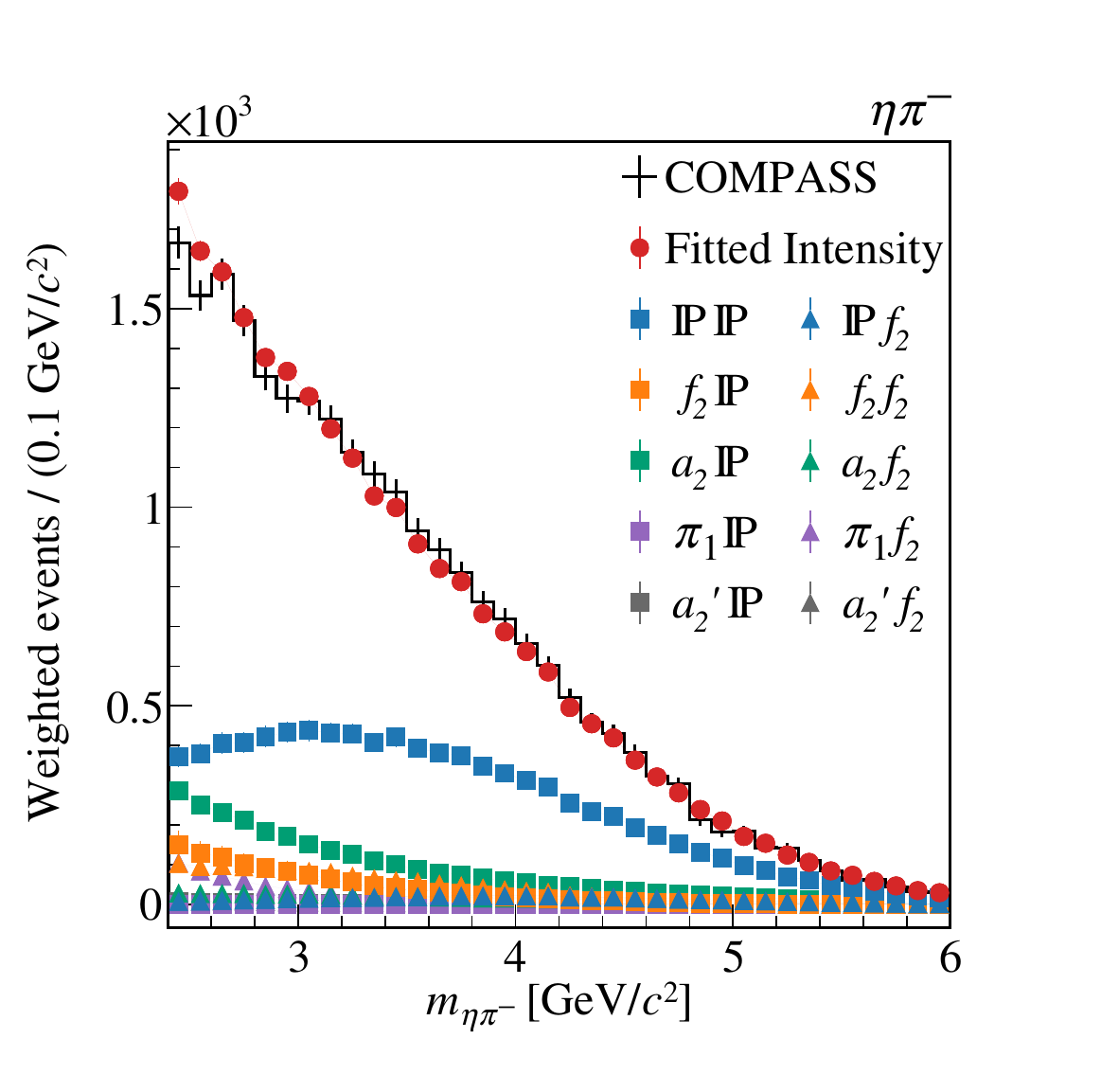} 
\end{tabular}
\caption{
$\eta \pi^-$ weighted intensity distributions dependence on
$m_{\eta \pi}$. Left: \mbox{KGR} fit;
Center-left: \mbox{$\text{KGR}+\pi_1+a_2^\prime$} fit; Center-right: \mbox{JPAC}
fit; Right: \mbox{$\text{JPAC}+\pi_1+a_2^\prime$} fit. Binned \compass data is
drawn in black. Diagrams with bottom $\Pom$ are marked with squares and those
with bottom $f_2$ with triangles.
}
\label{figsup:etaweight1}
\end{figure*}

\begin{figure*}[!h]
\begin{tabular}{cccc}
\includegraphics[width=0.25\linewidth]{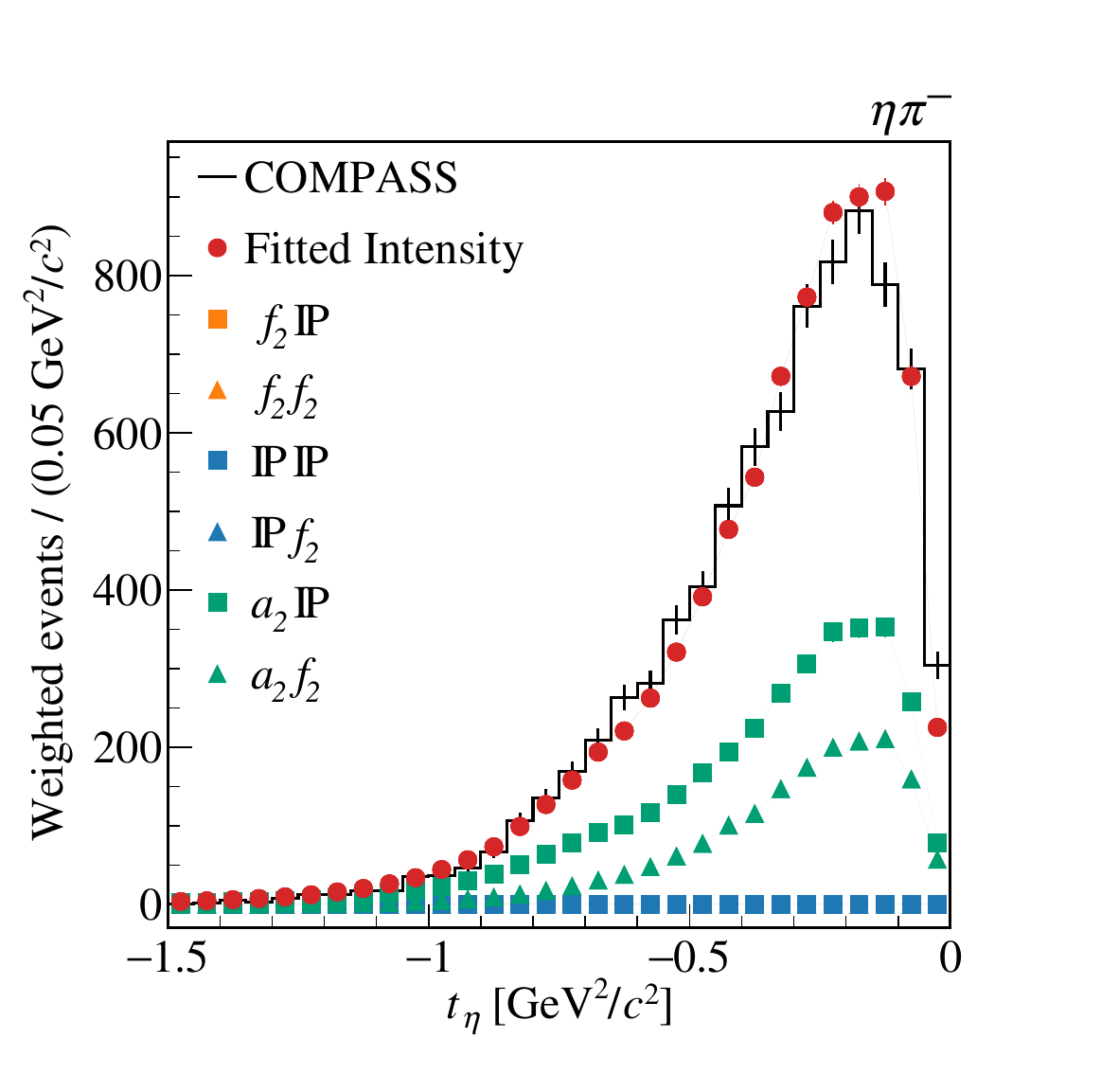} & 
\includegraphics[width=0.25\linewidth]{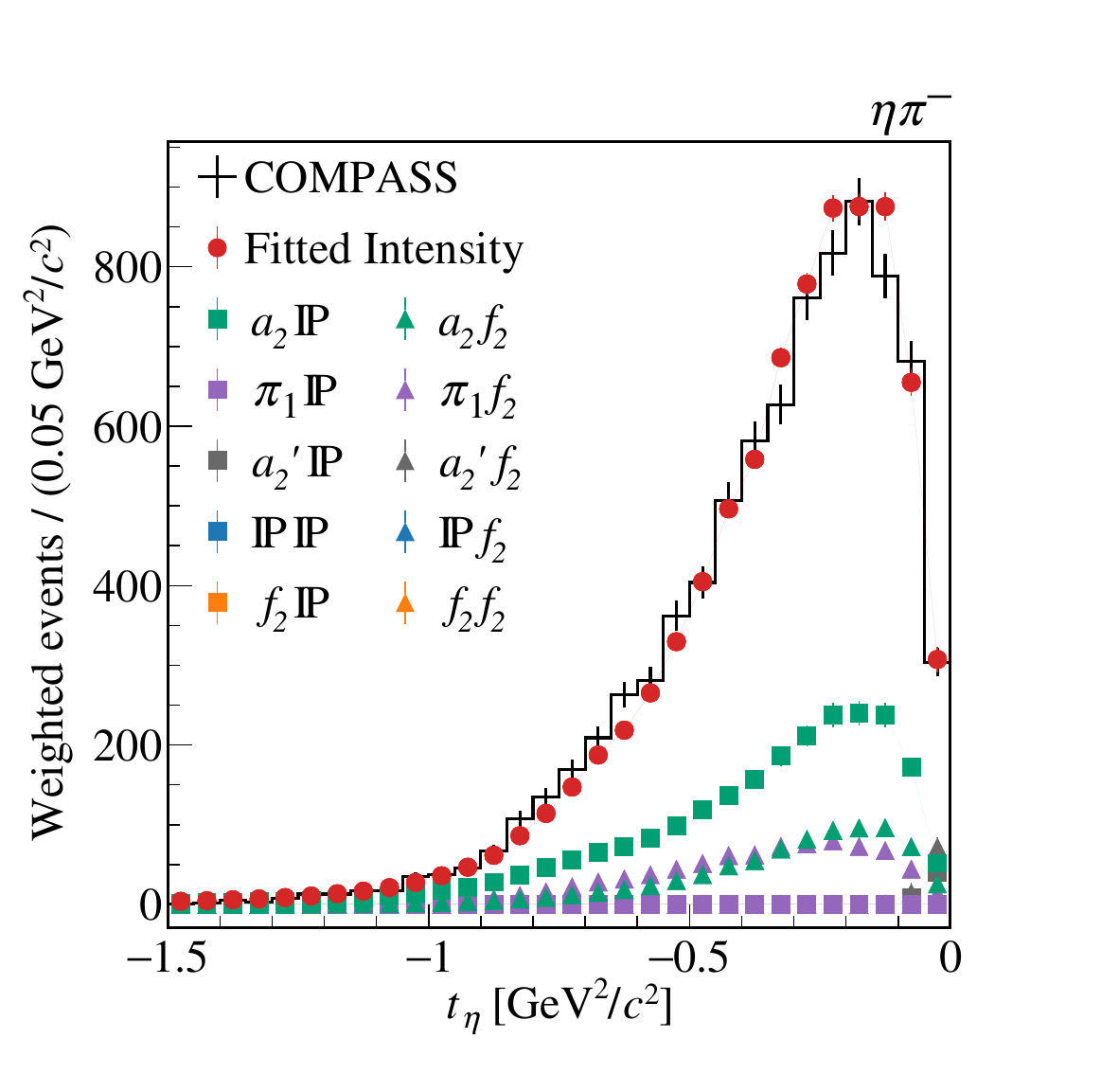} & 
\includegraphics[width=0.25\linewidth]{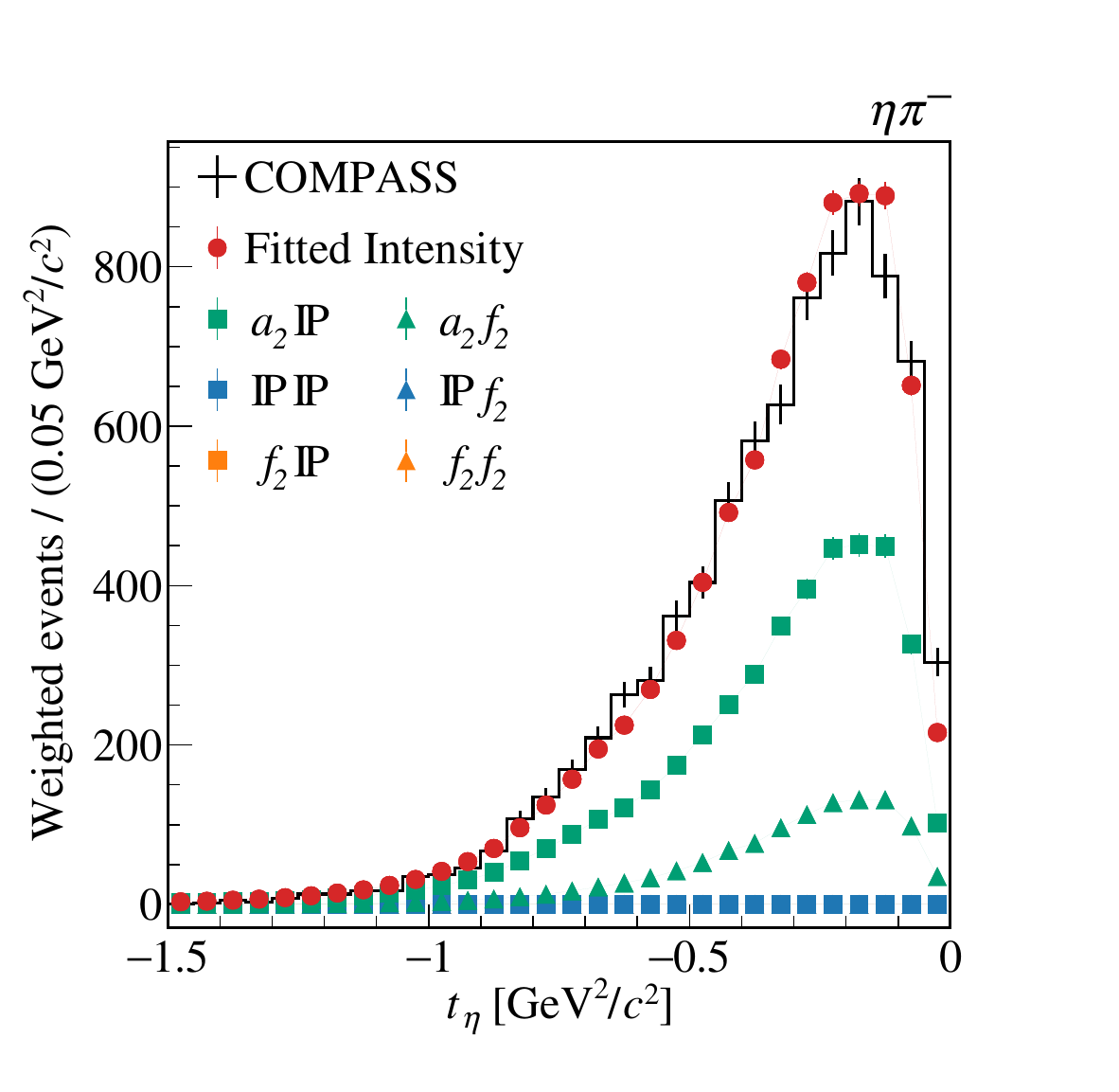} & 
\includegraphics[width=0.25\linewidth]{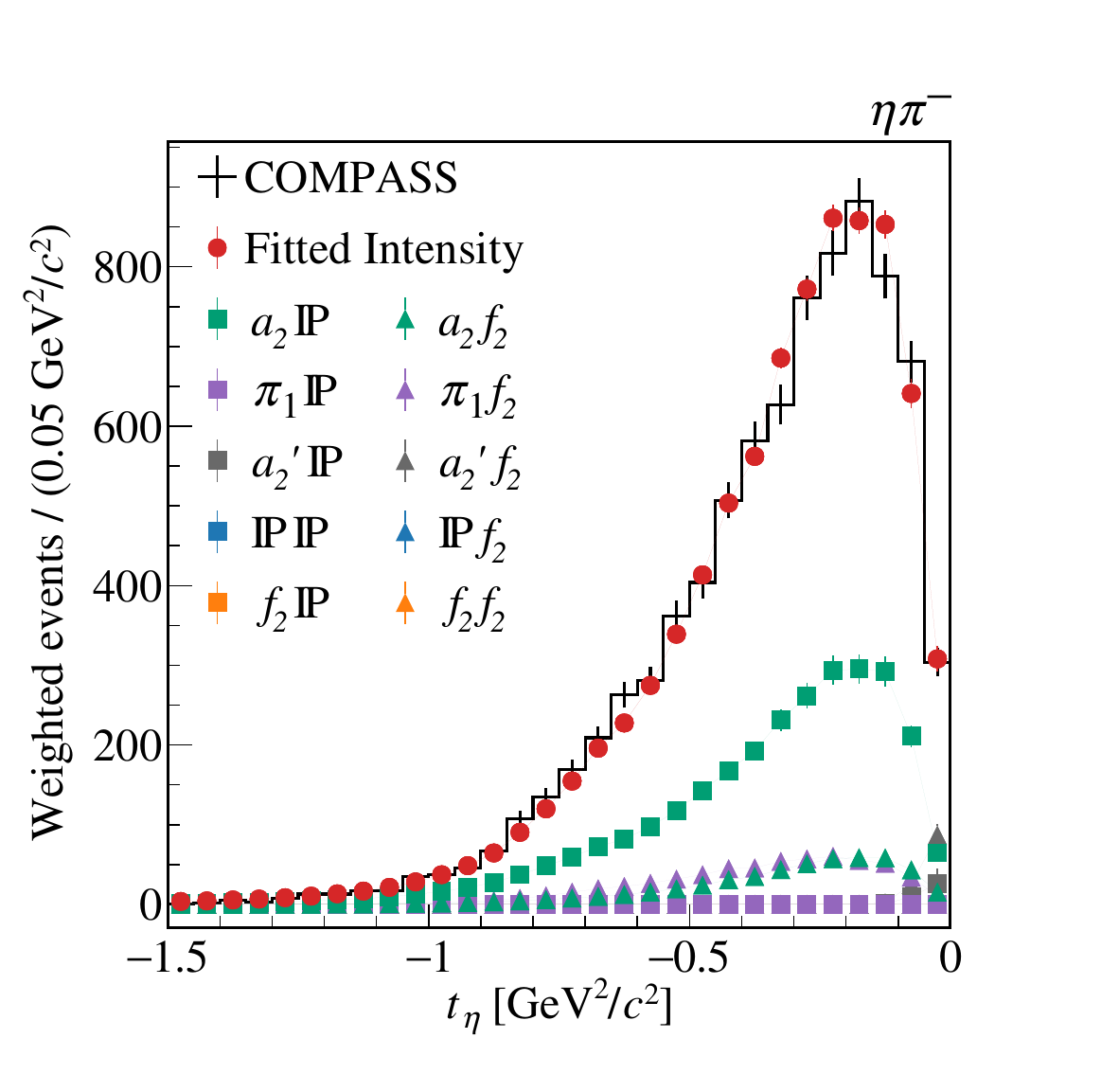} 
\end{tabular}
\caption{
$\eta \pi^-$ weighted intensity distributions dependence on $t_\eta$. Individual
contributions. Left: \mbox{KGR} fit;
Center-left: \mbox{$\text{KGR}+\pi_1+a_2^\prime$} fit; Center-right: \mbox{JPAC}
fit; Right: \mbox{$\text{JPAC}+\pi_1+a_2^\prime$} fit. Conventions as
in~\cref{figsup:etaweight1}.
}
\label{figsup:etaweight2_1}
\end{figure*}

\begin{figure*}[!h]
\begin{tabular}{cccc}
\includegraphics[width=0.25\linewidth]{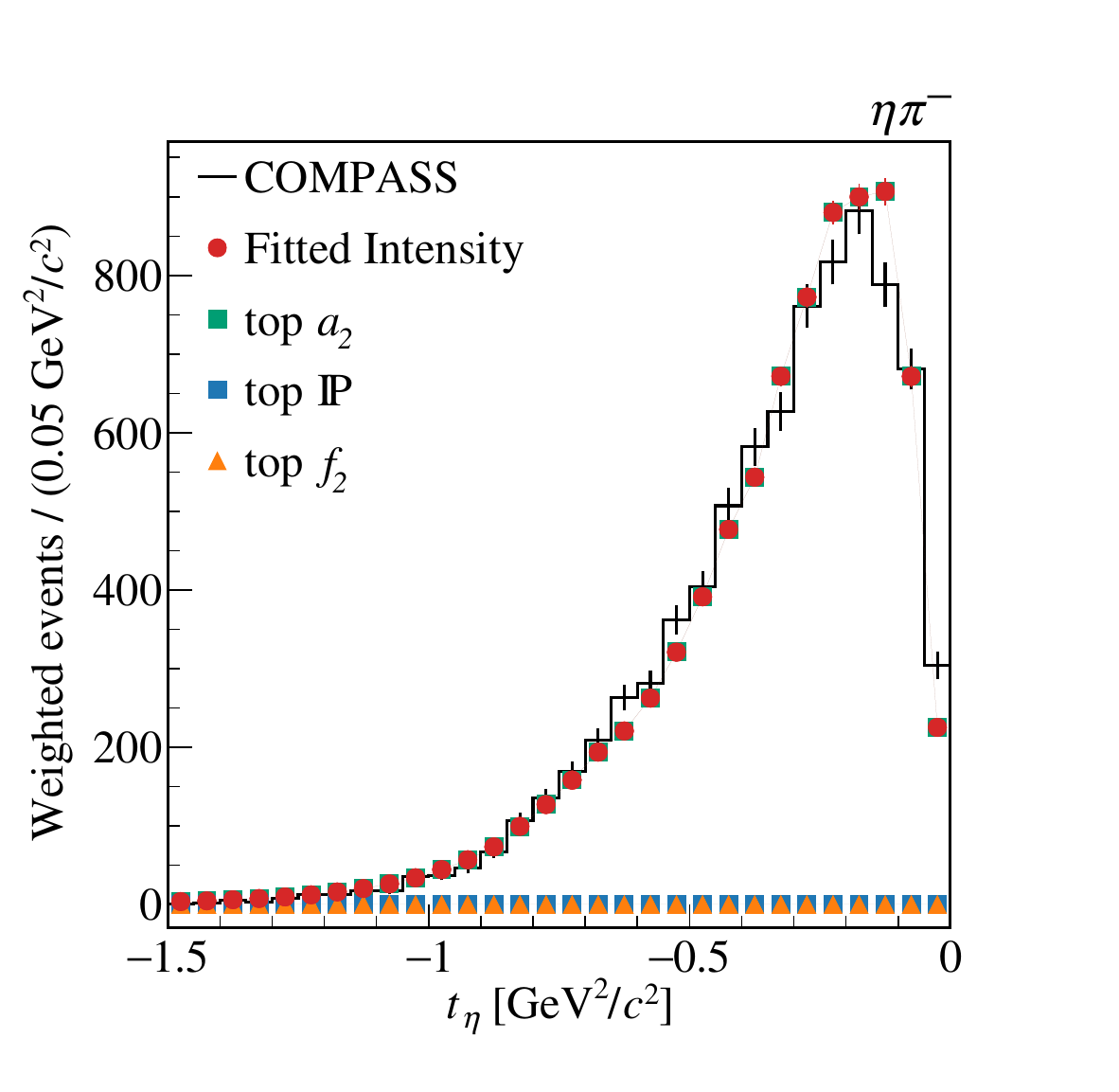} & 
\includegraphics[width=0.25\linewidth]{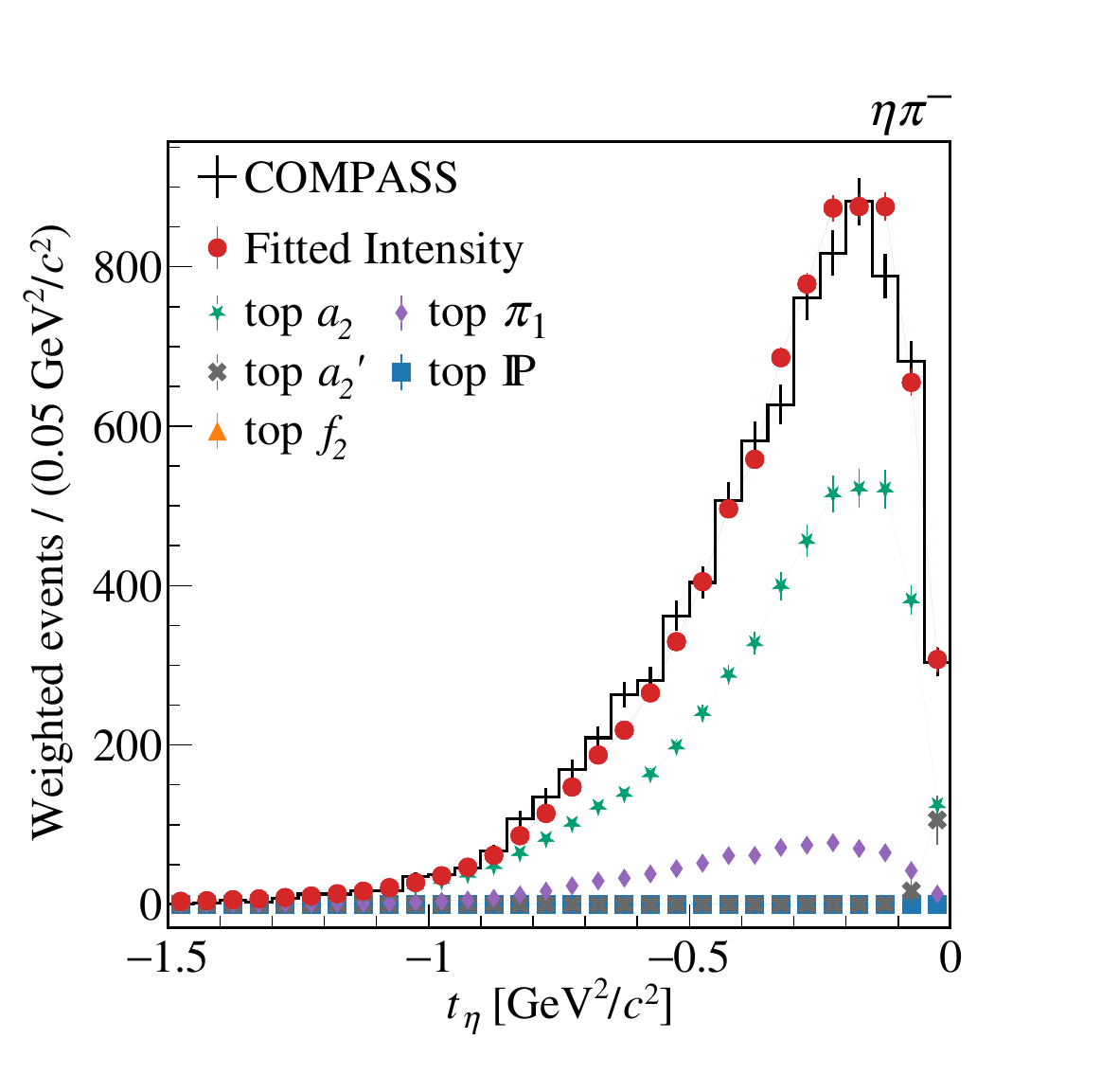} & 
\includegraphics[width=0.25\linewidth]{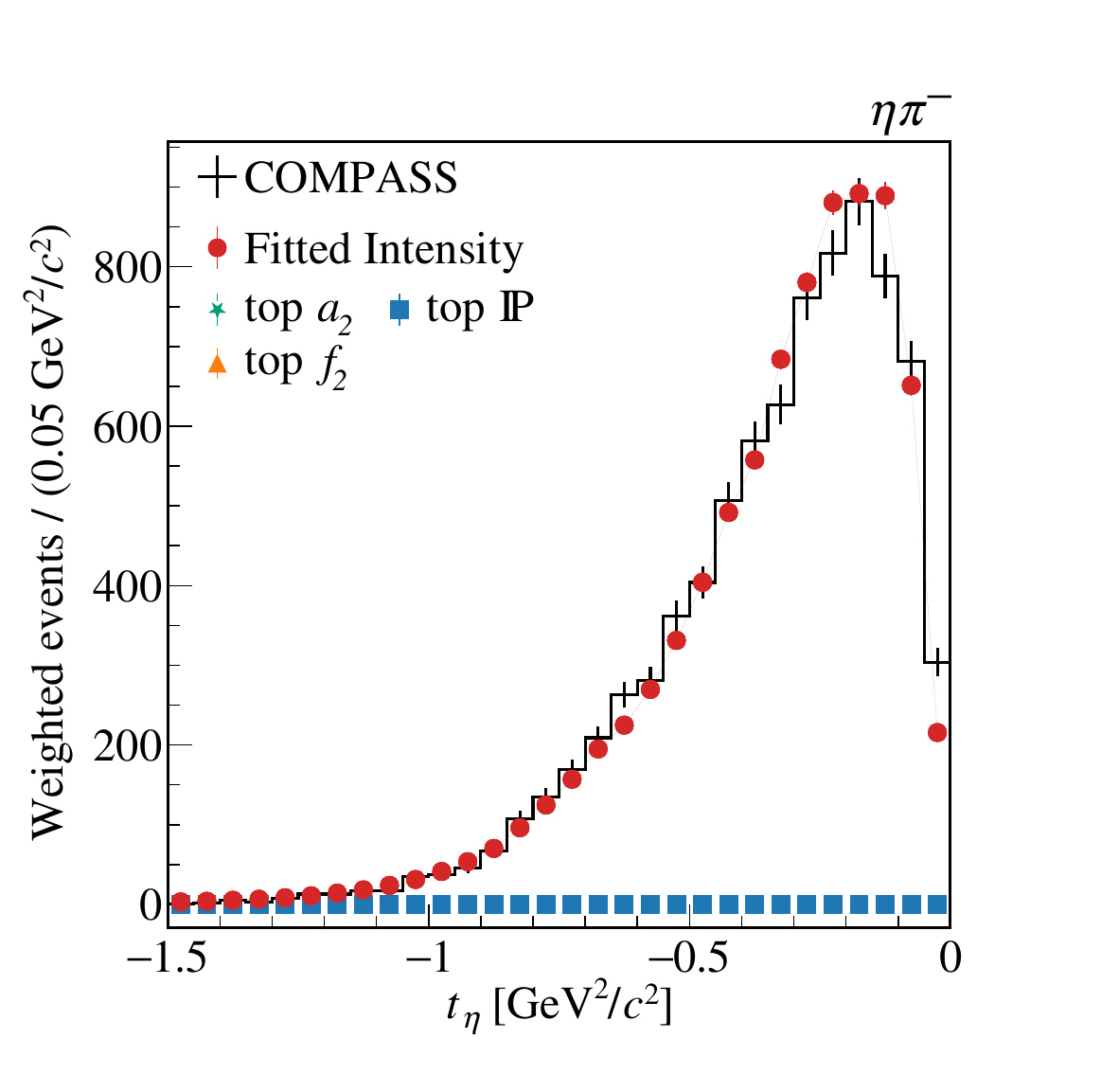} & 
\includegraphics[width=0.25\linewidth]{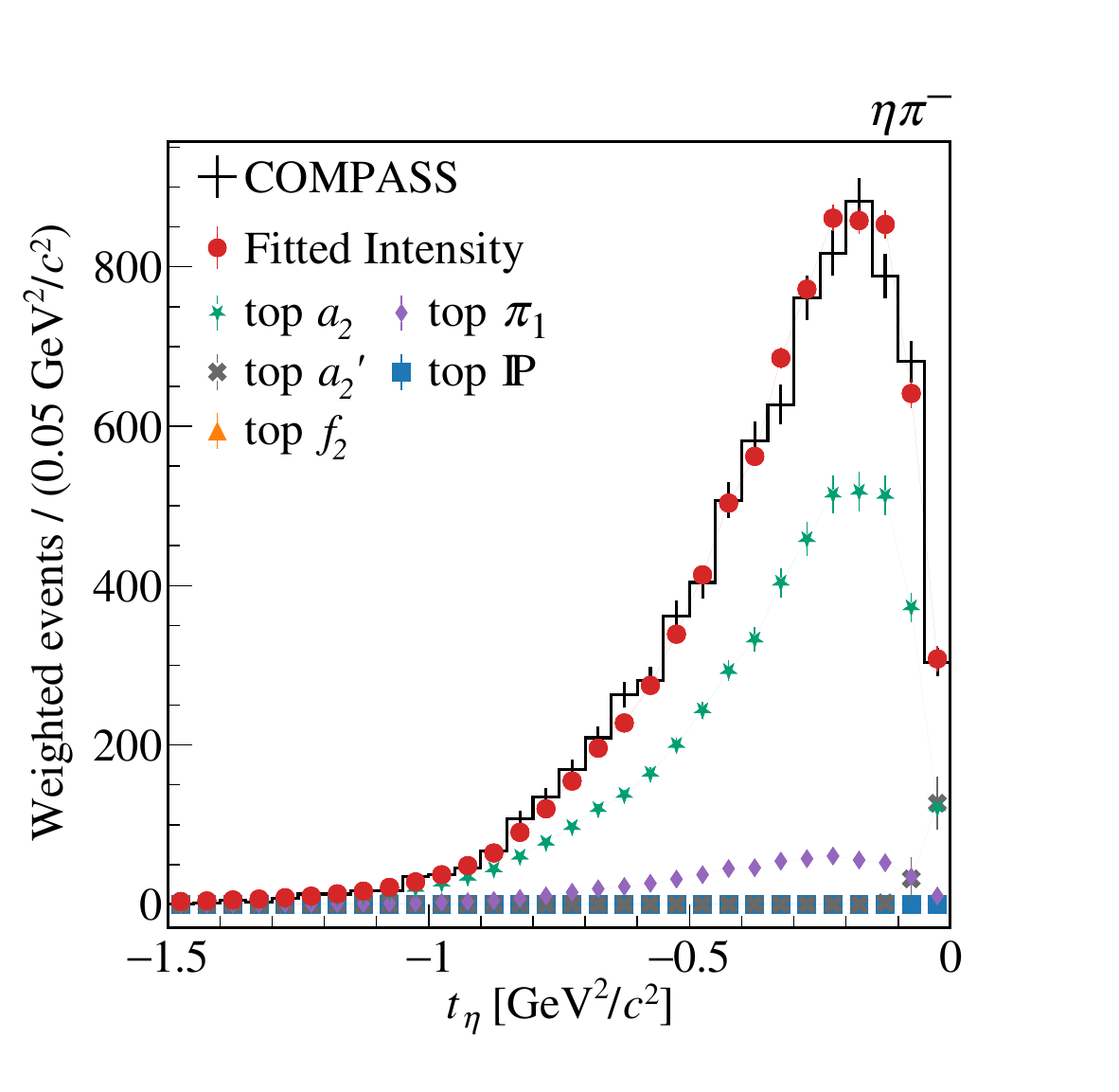} 
\end{tabular}
\caption{
$\eta \pi^-$ weighted intensity distributions dependence on $t_\eta$. Top
Reggeon exchange contributions coherently summed. By top $\Pom$, top $f_2$, top
$a_2$, top $a_2^\prime$ and top $\pi_1$ we refer to the coherent sum of the
amplitudes exchanged $\mathbb{R}_t$ is the $\Pom$, $f_2$, $a_2$, $a_2^\prime$
and , $\pi_1$, respectively. Left: \mbox{KGR} fit;
Center-left: \mbox{$\text{KGR}+\pi_1+a_2^\prime$} fit; Center-right: \mbox{JPAC}
fit;Right: \mbox{$\text{JPAC}+\pi_1+a_2^\prime$} fit.
}
\label{figsup:etaweight2_2}
\end{figure*}

\begin{figure*}[!h]
\begin{tabular}{cccc}
\includegraphics[width=0.25\linewidth]{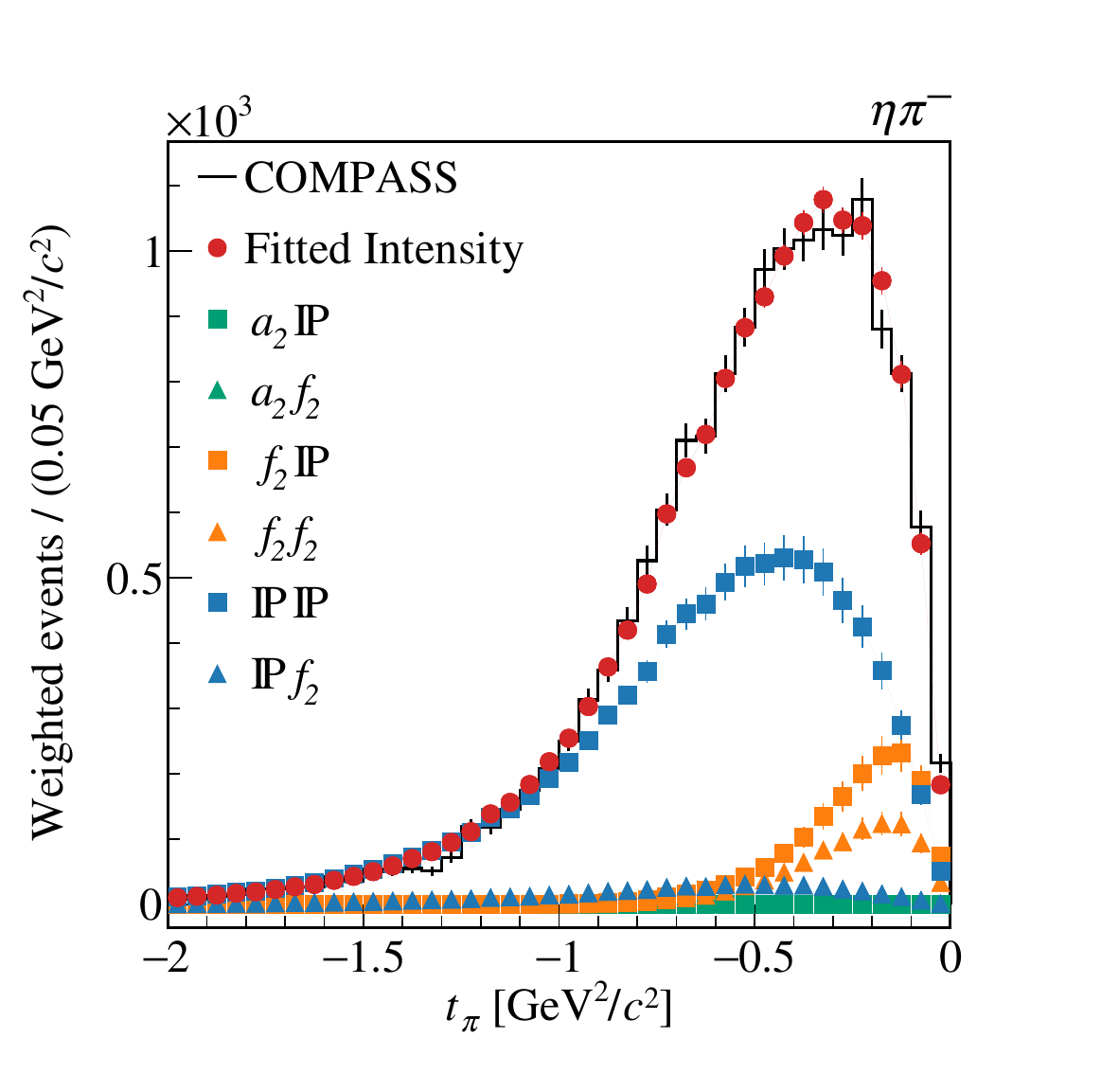} & 
\includegraphics[width=0.25\linewidth]{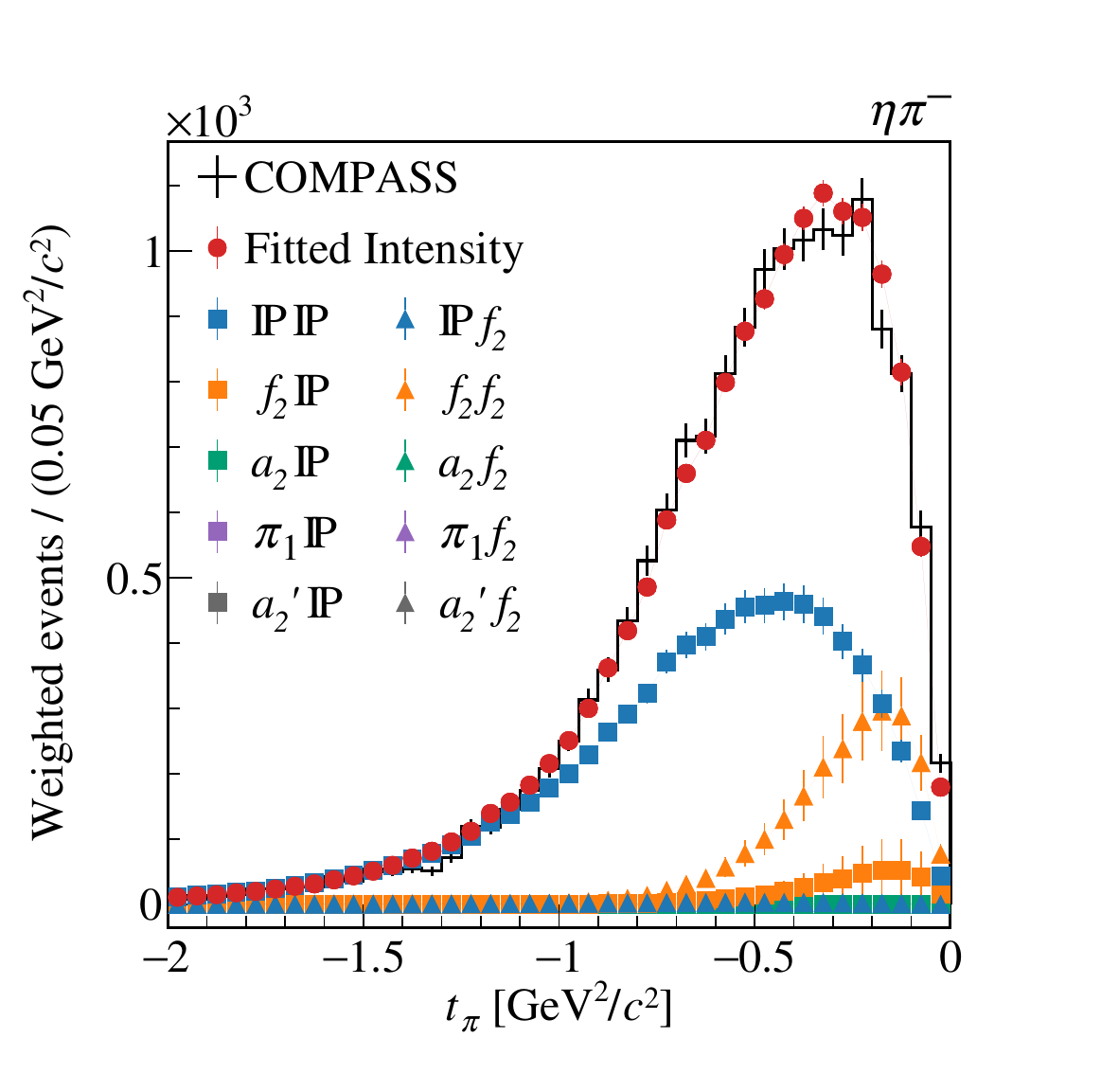} & 
\includegraphics[width=0.25\linewidth]{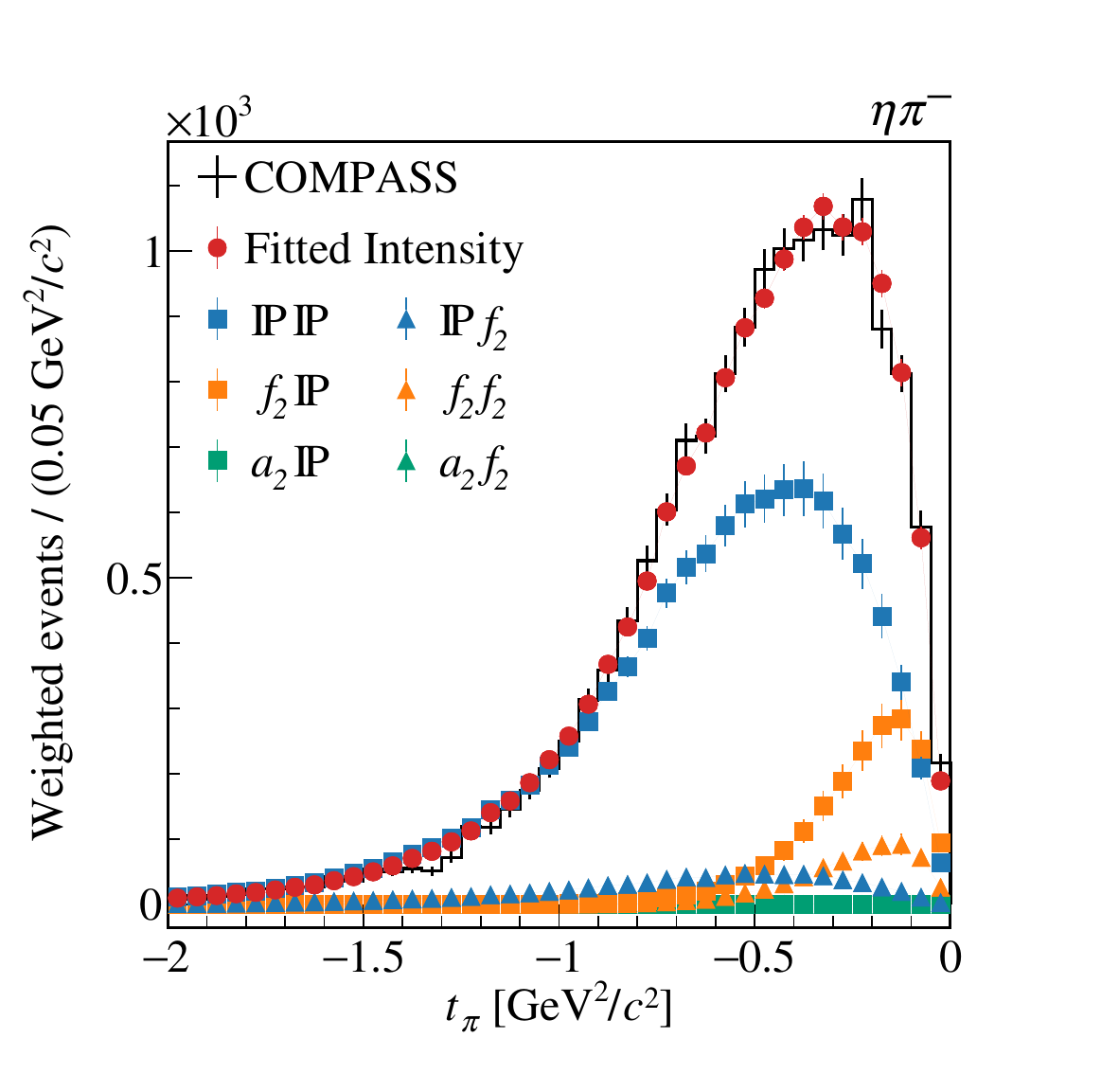} & 
\includegraphics[width=0.25\linewidth]{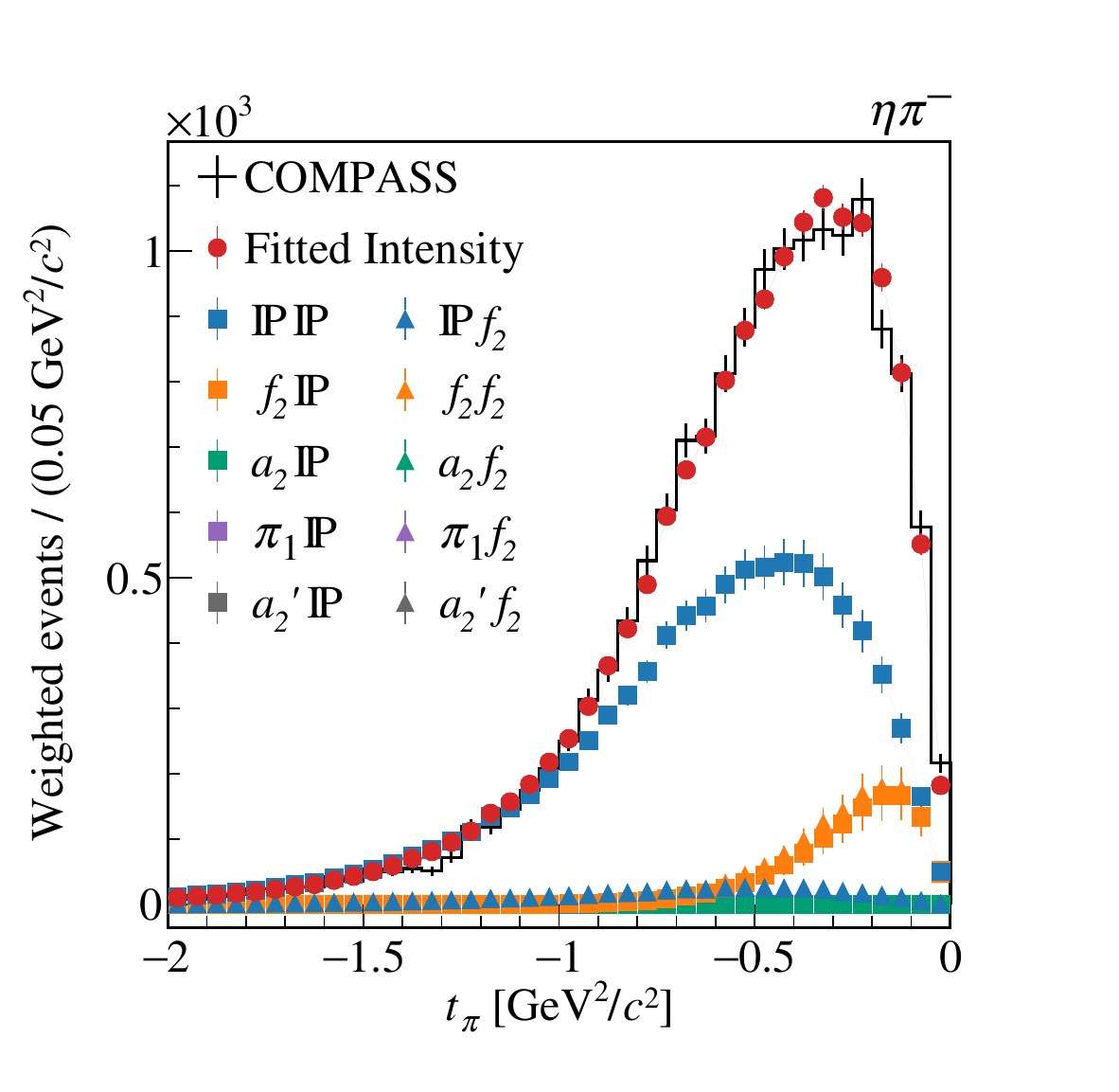} 
\end{tabular}
\caption{
$\eta \pi^-$ weighted intensity distributions dependence on $t_\pi$. Individual
contributions. Left: \mbox{KGR} fit;
Center-left: \mbox{$\text{KGR}+\pi_1+a_2^\prime$} fit; Center-right: \mbox{JPAC}
fit; Right: \mbox{$\text{JPAC}+\pi_1+a_2^\prime$} fit. Conventions as
in~\cref{figsup:etaweight1}.
}
\label{figsup:etaweight3_1}
\end{figure*}

\begin{figure*}[!h]
\begin{tabular}{cccc}
\includegraphics[width=0.25\linewidth]{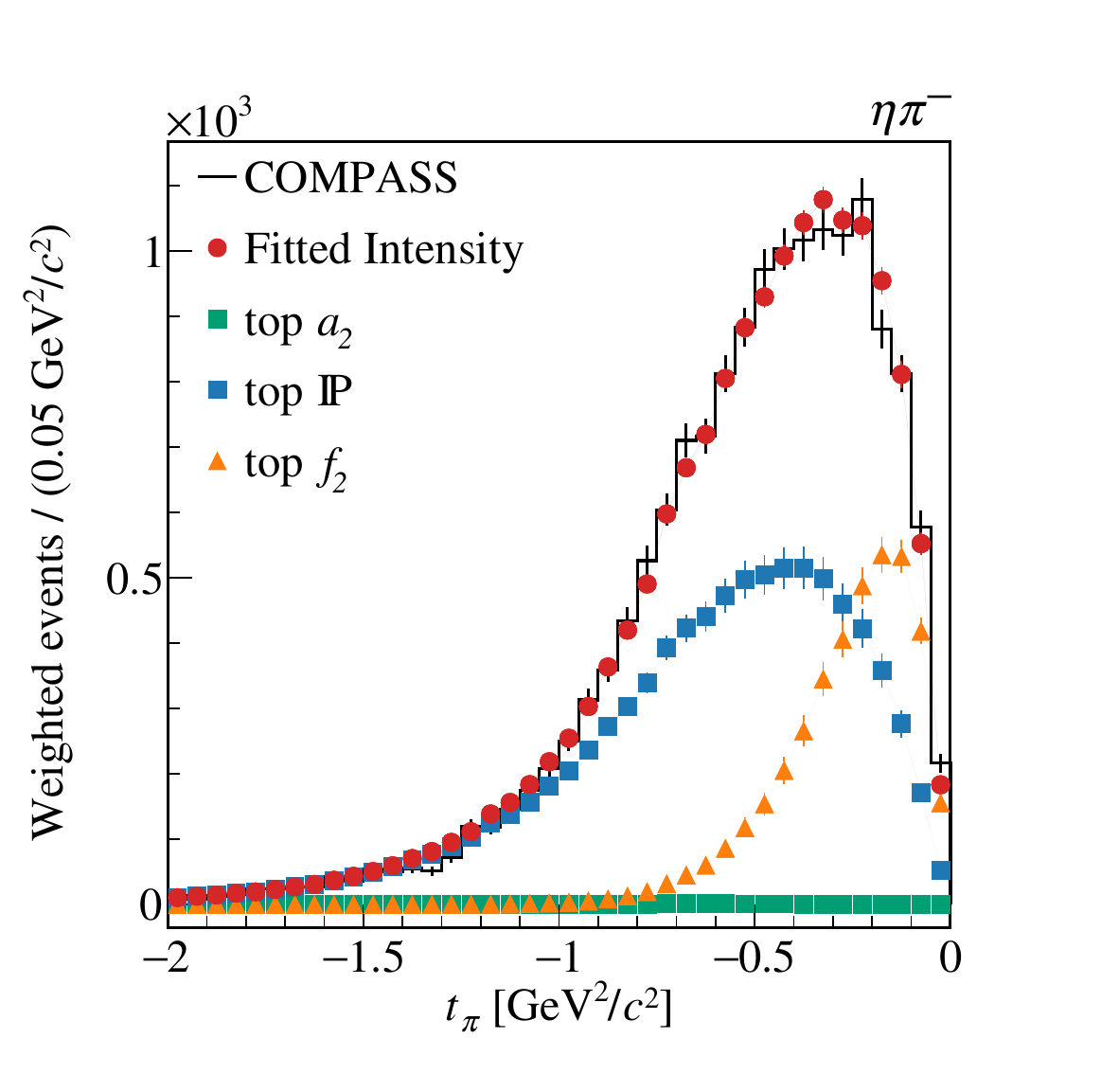} & 
\includegraphics[width=0.25\linewidth]{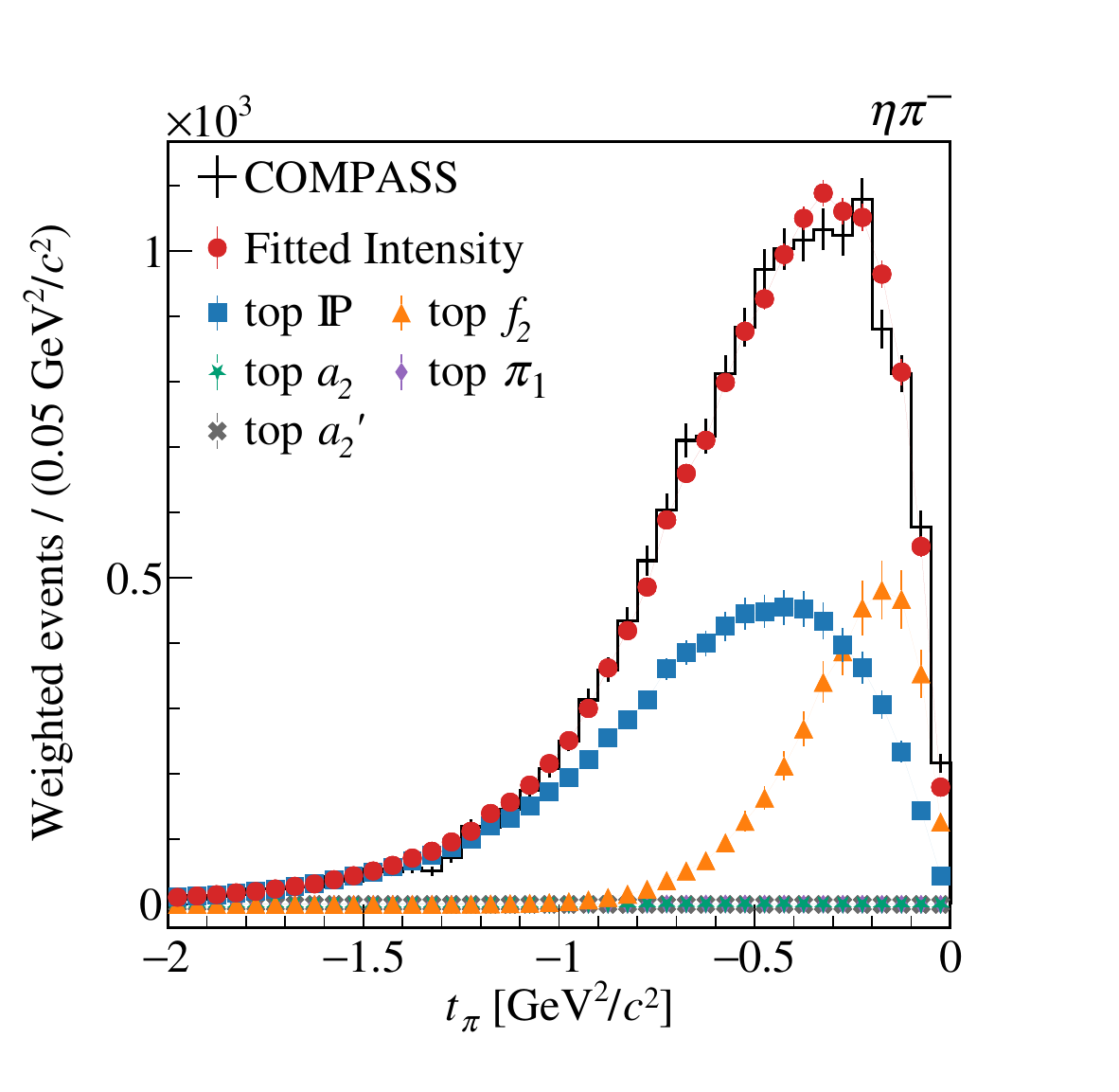} & 
\includegraphics[width=0.25\linewidth]{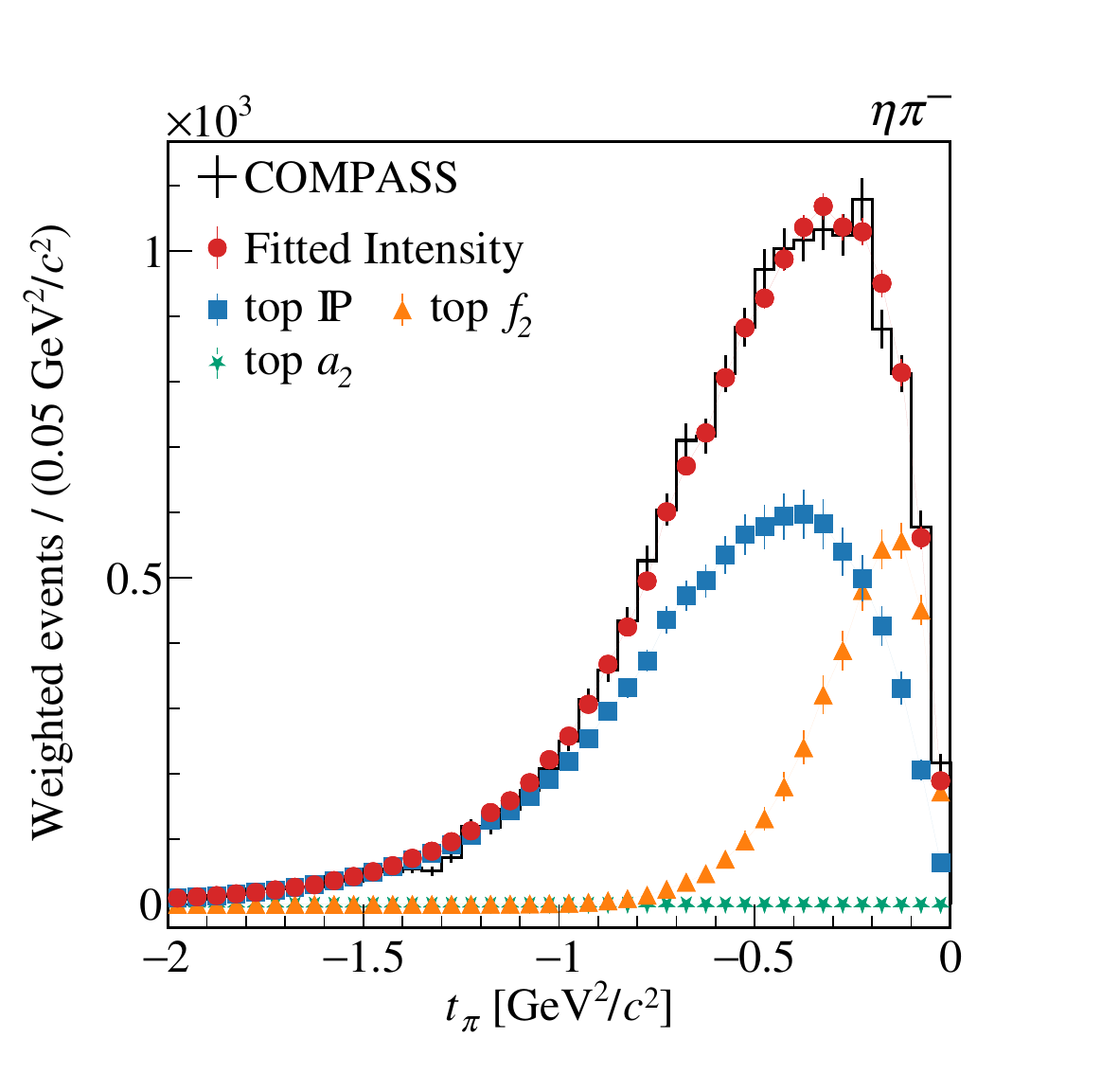} & 
\includegraphics[width=0.25\linewidth]{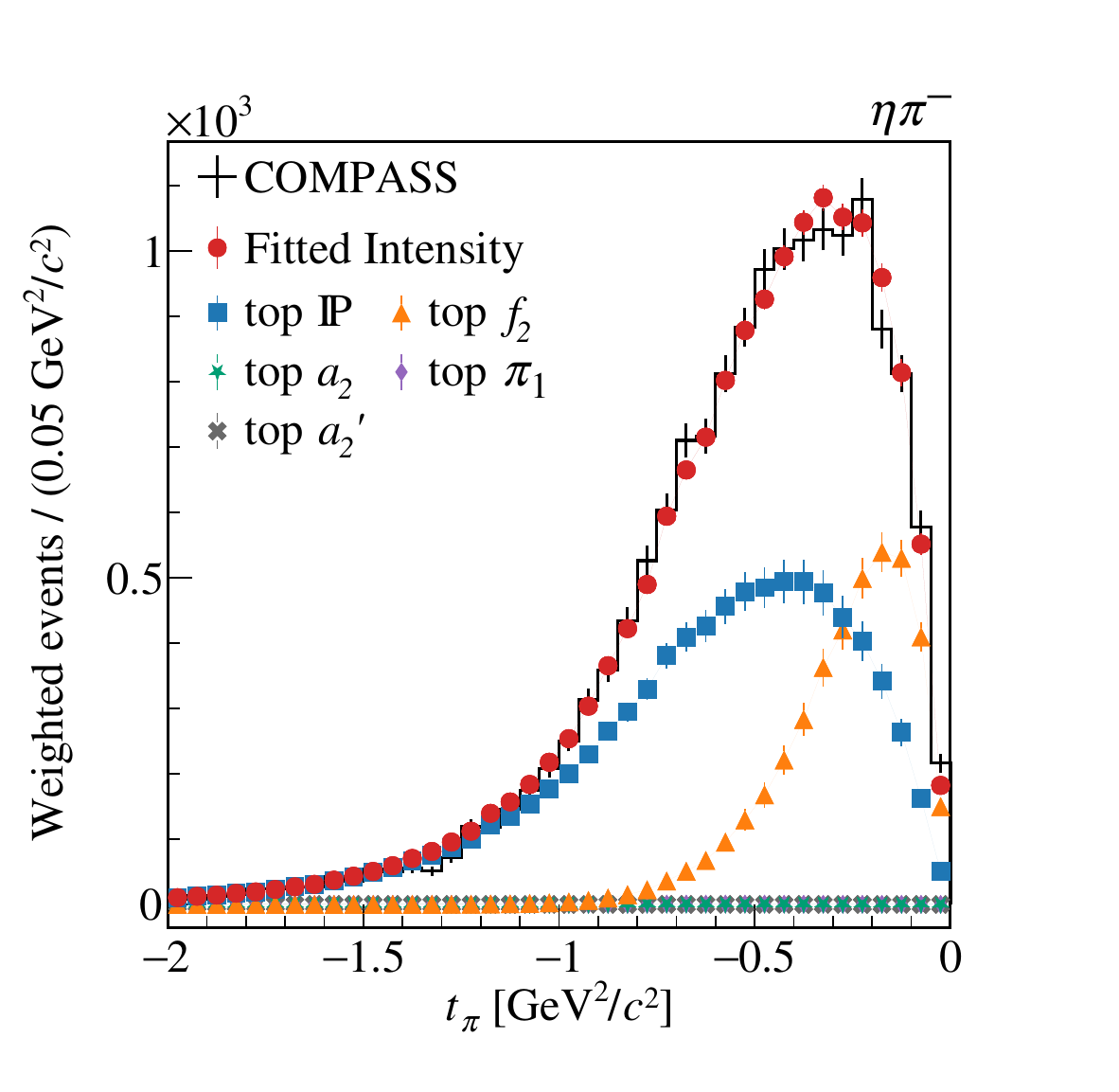} 
\end{tabular}
\caption{
$\eta \pi^-$ weighted intensity distributions dependence on $t_\pi$. Top Reggeon
exchange contributions coherently summed. Left: \mbox{KGR} fit;
Center-left: \mbox{$\text{KGR}+\pi_1+a_2^\prime$} fit; Center-right: \mbox{JPAC}
fit; Right: \mbox{$\text{JPAC}+\pi_1+a_2^\prime$} fit. Conventions as
in~\cref{figsup:etaweight2_2}.
}
\label{figsup:etaweight3_2}
\end{figure*}

\begin{figure*}[!h]
\begin{tabular}{cccc}
\includegraphics[width=0.25\linewidth]{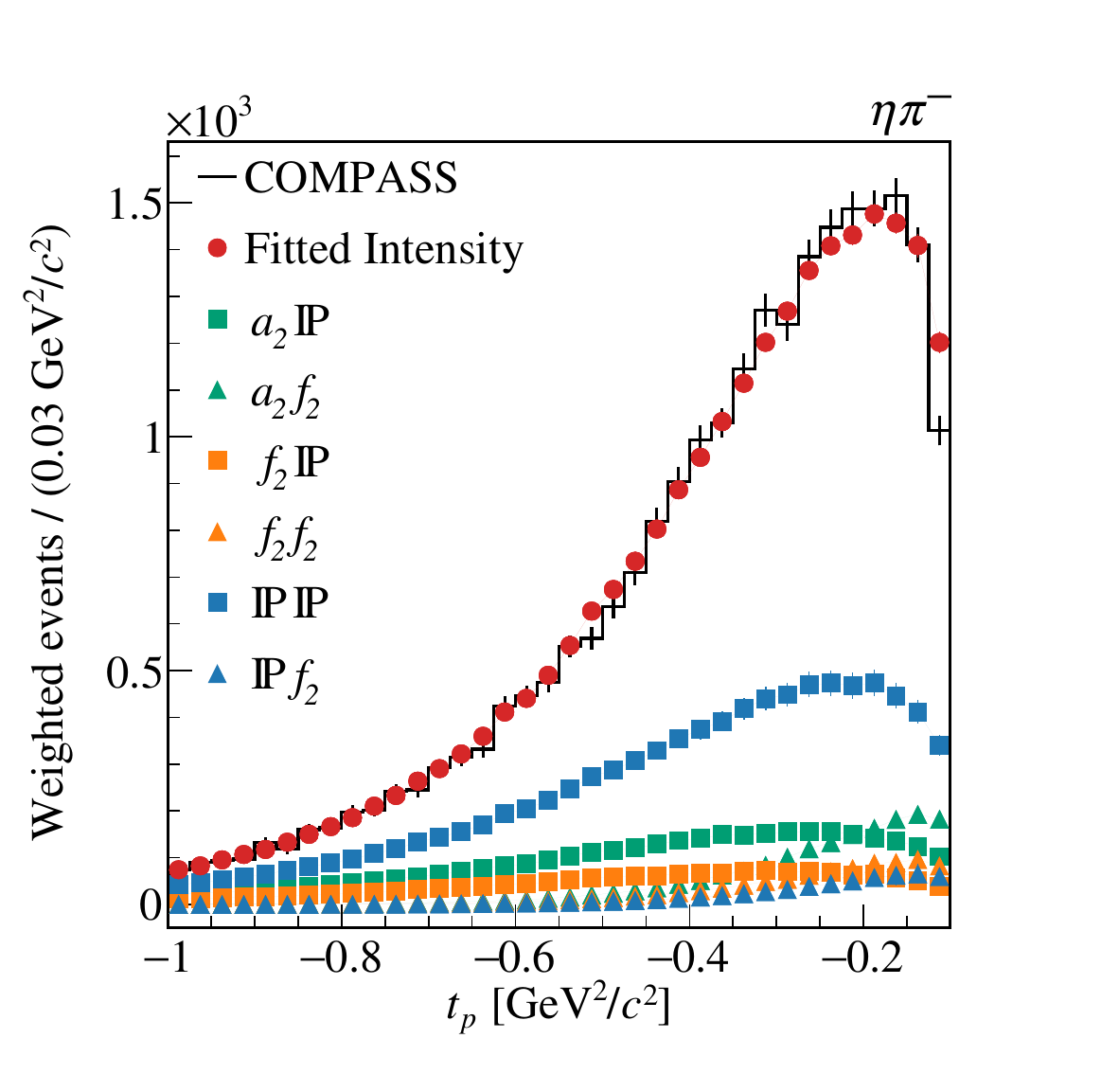} & 
\includegraphics[width=0.25\linewidth]{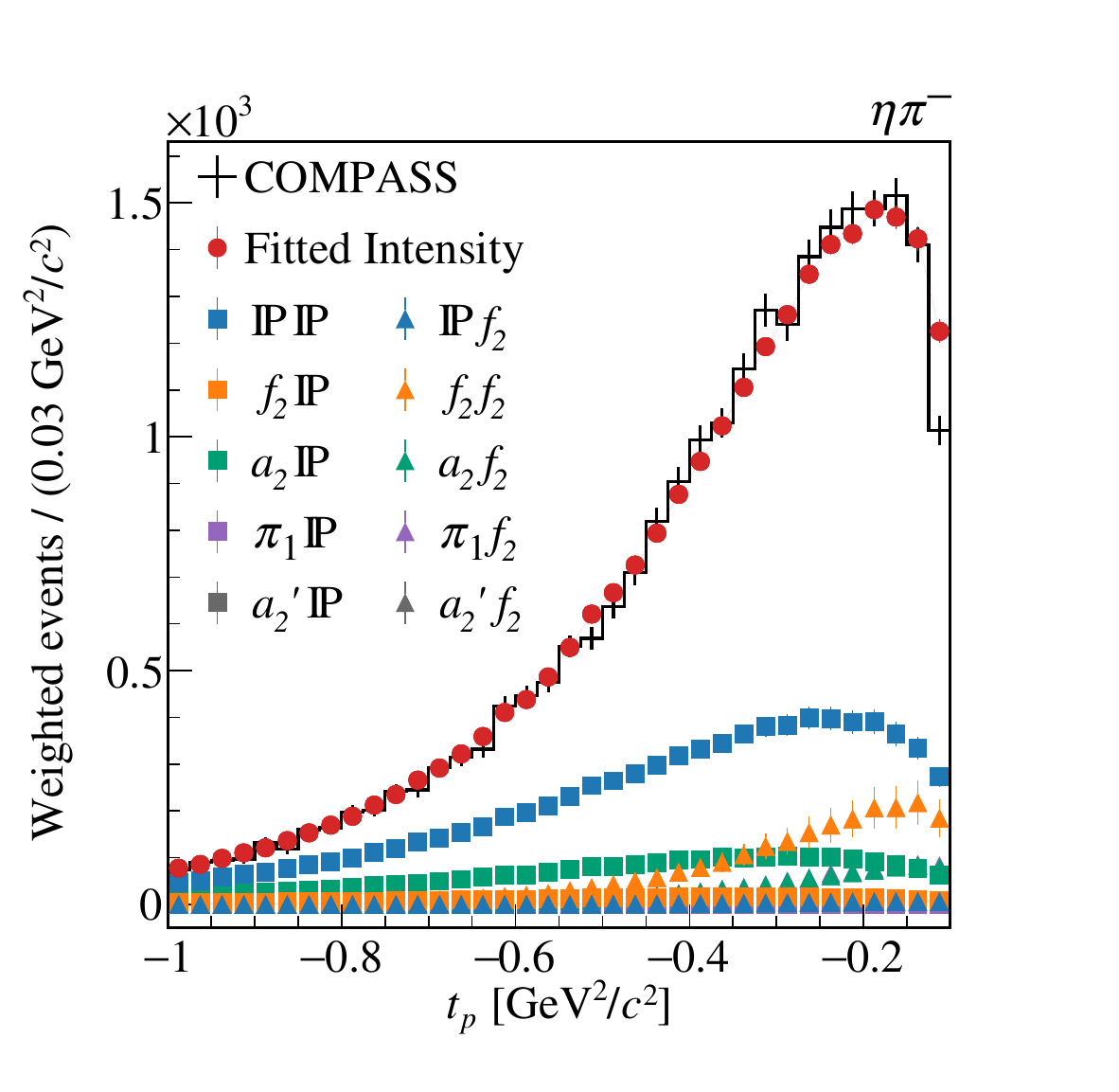} & 
\includegraphics[width=0.25\linewidth]{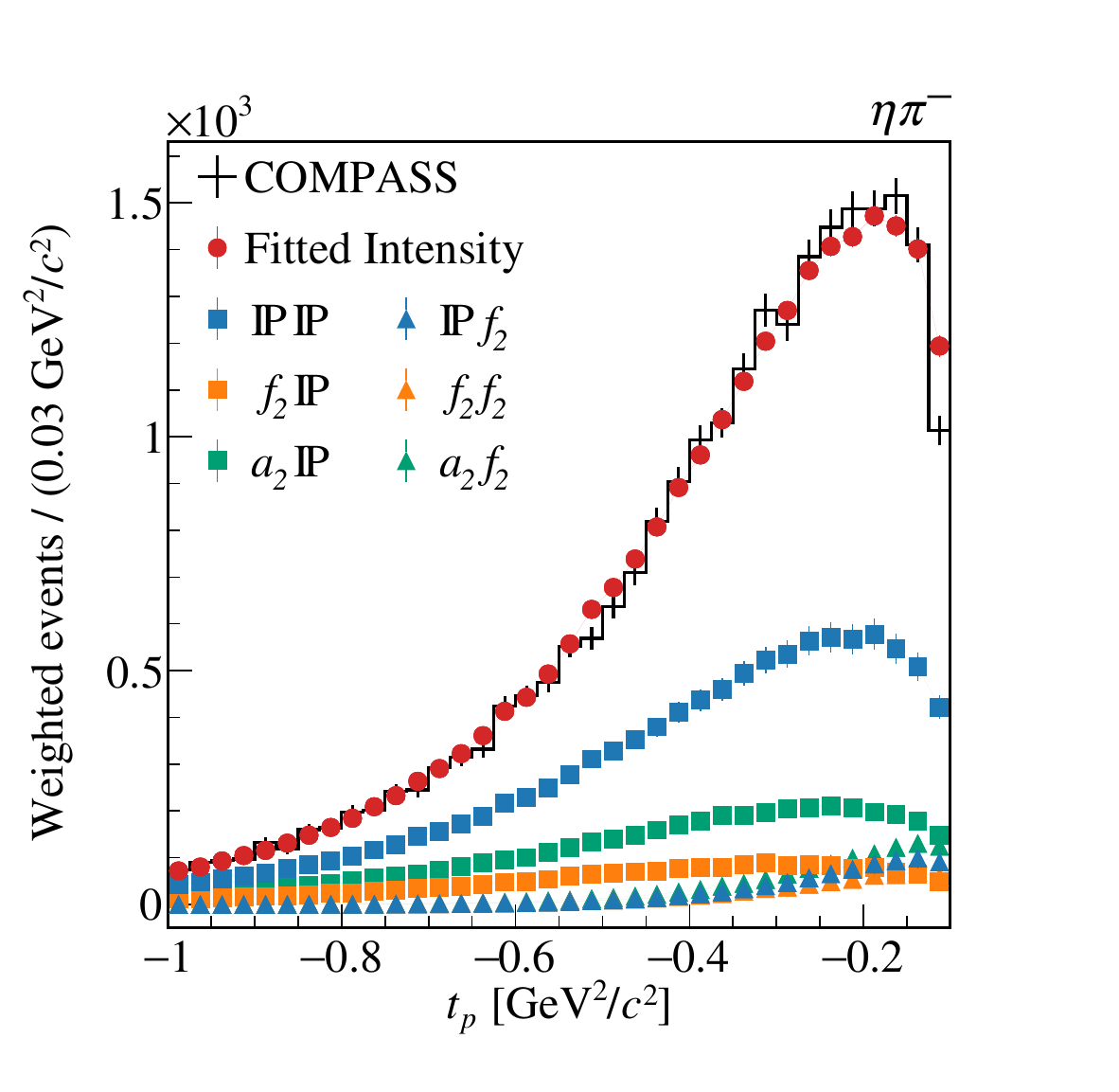} & 
\includegraphics[width=0.25\linewidth]{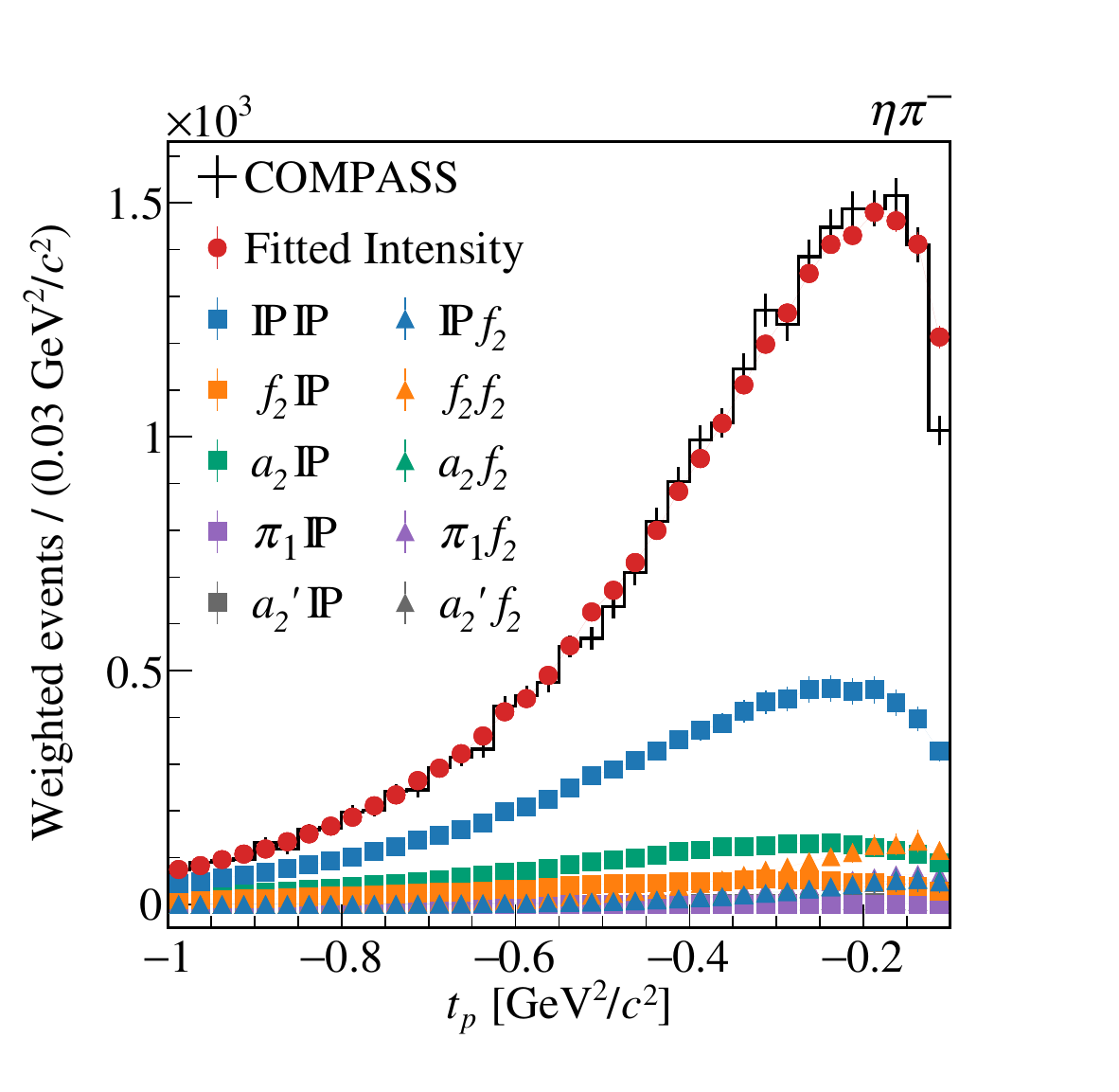} 
\end{tabular}
\caption{
$\eta \pi^-$ weighted intensity distributions dependence on $t_p$. Individual
contributions. Left: \mbox{KGR} fit;
Center-left: \mbox{$\text{KGR}+\pi_1+a_2^\prime$} fit; Center-right: \mbox{JPAC}
fit; Right: \mbox{$\text{JPAC}+\pi_1+a_2^\prime$} fit. Conventions as
in~\cref{figsup:etaweight1}.
}
\label{figsup:etaweight4_1}
\end{figure*}

\begin{figure*}[!h]
\begin{tabular}{cccc}
\includegraphics[width=0.25\linewidth]{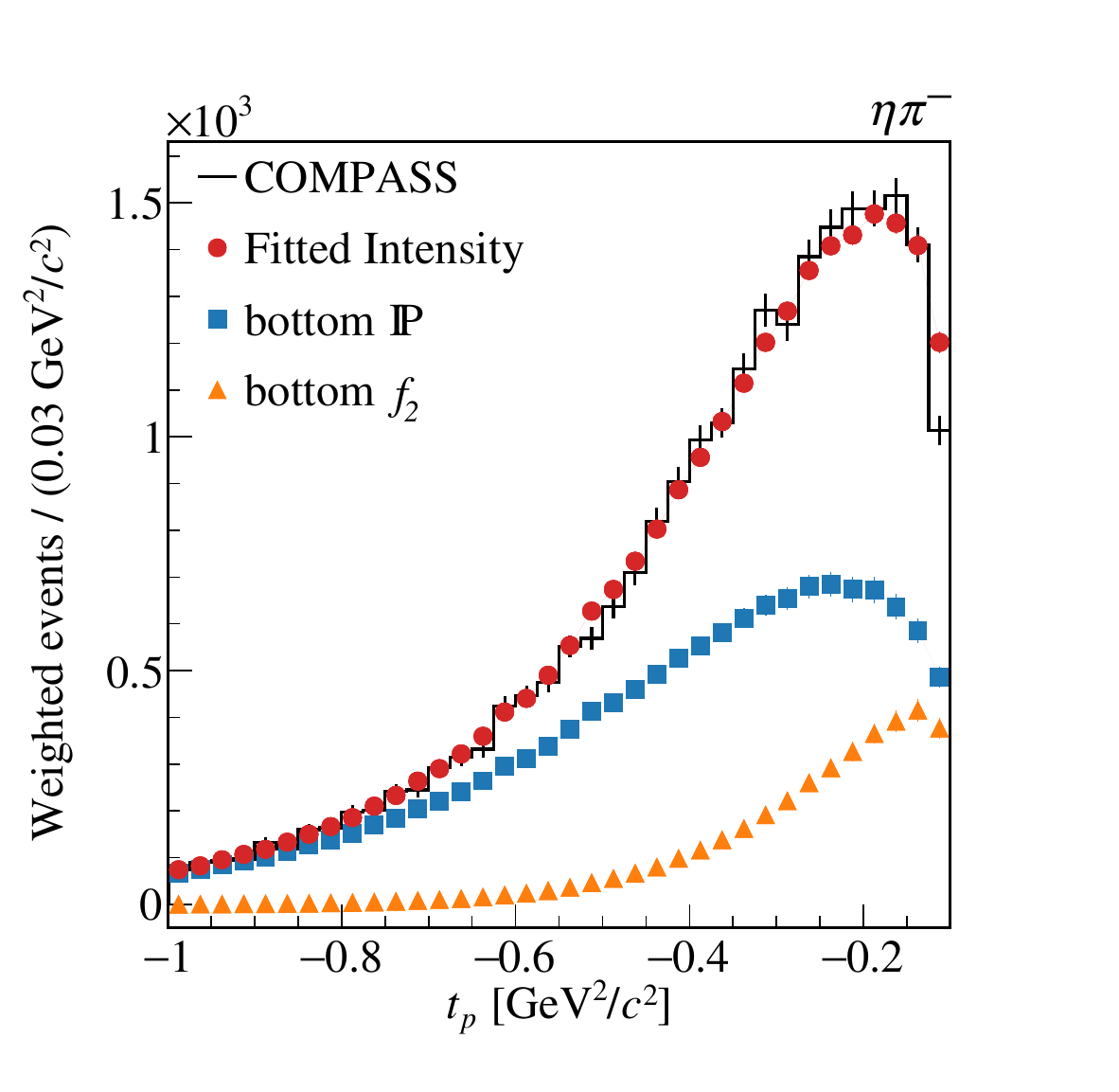} & 
\includegraphics[width=0.25\linewidth]{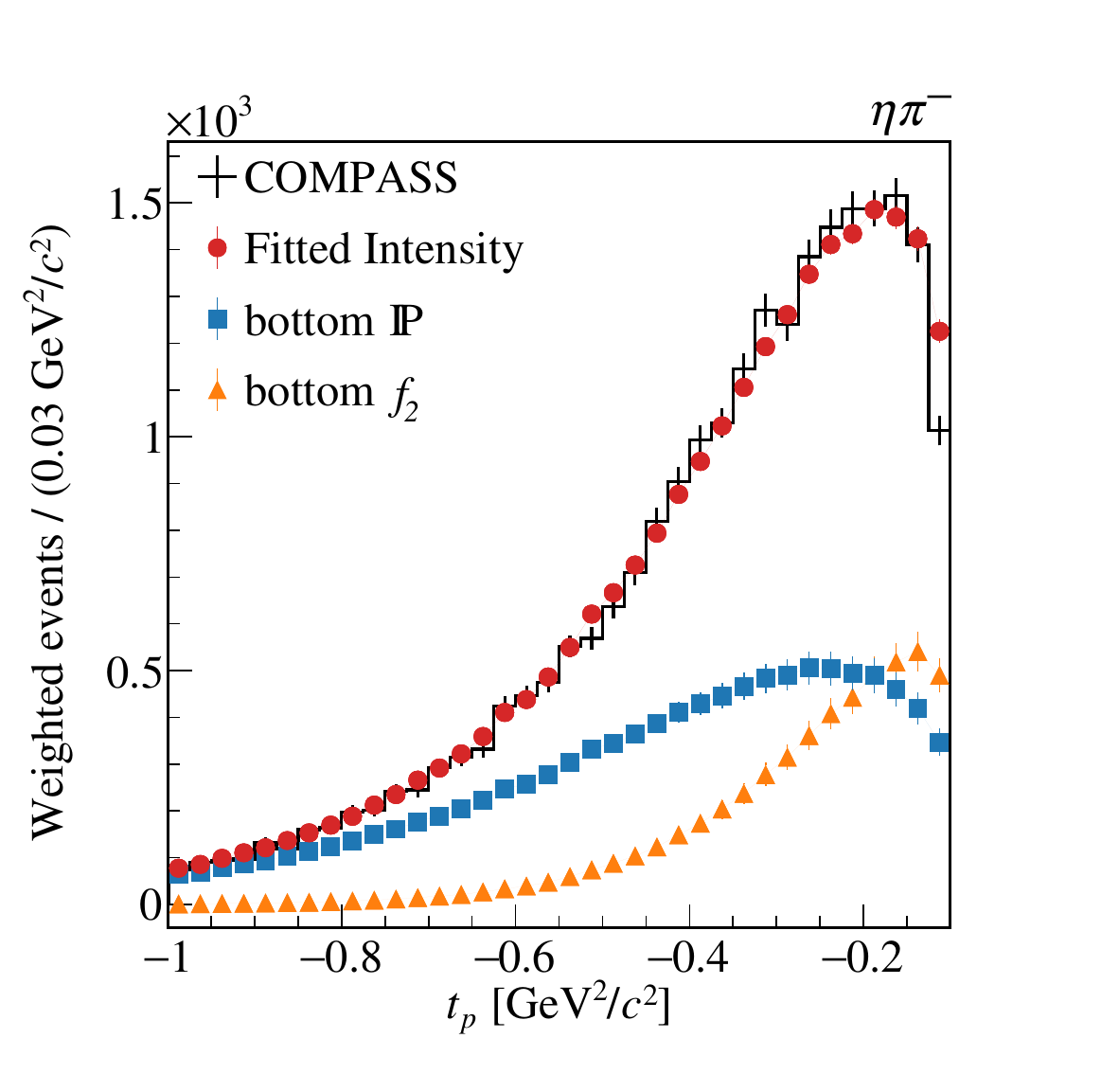} & 
\includegraphics[width=0.25\linewidth]{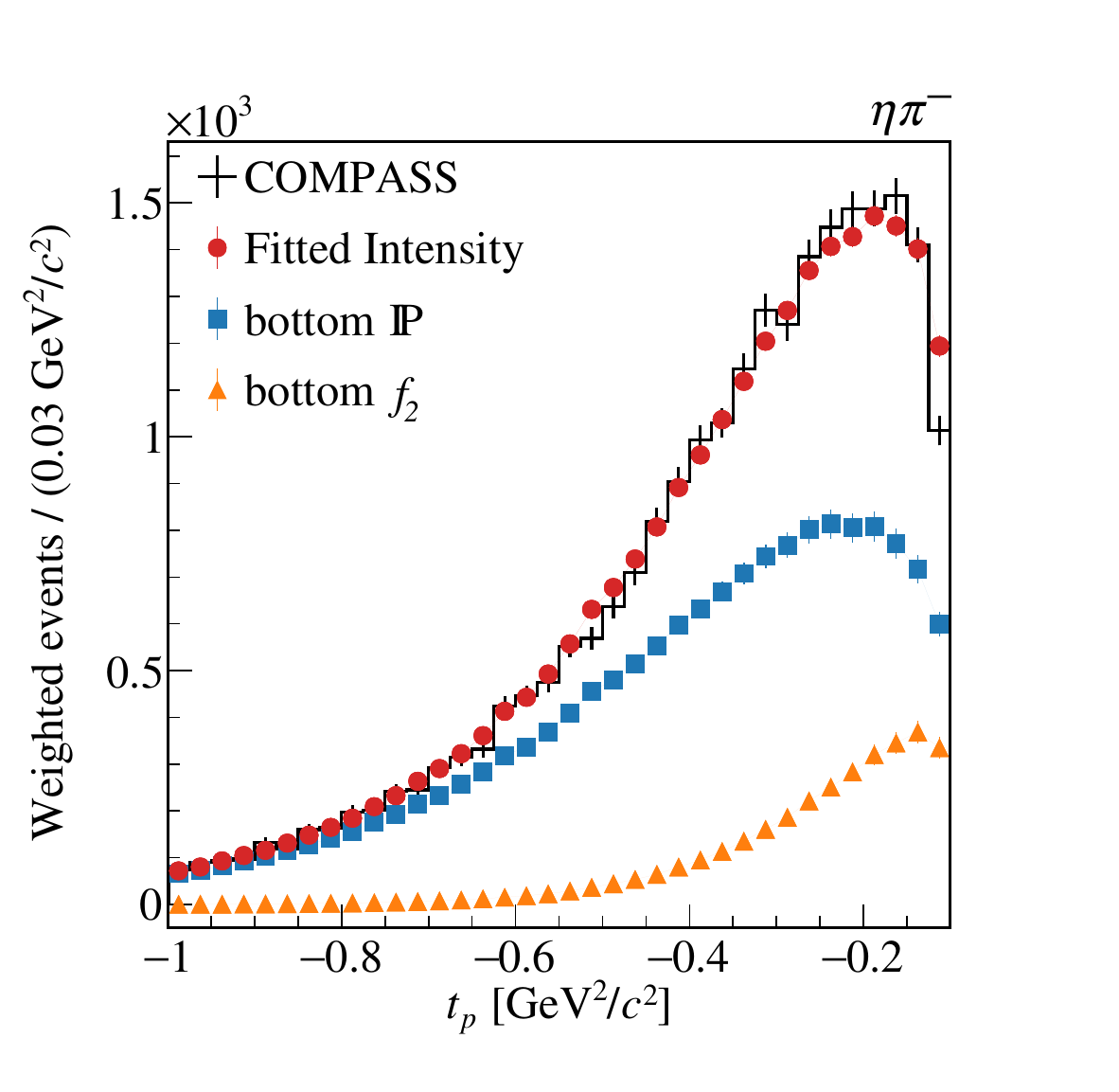} & 
\includegraphics[width=0.25\linewidth]{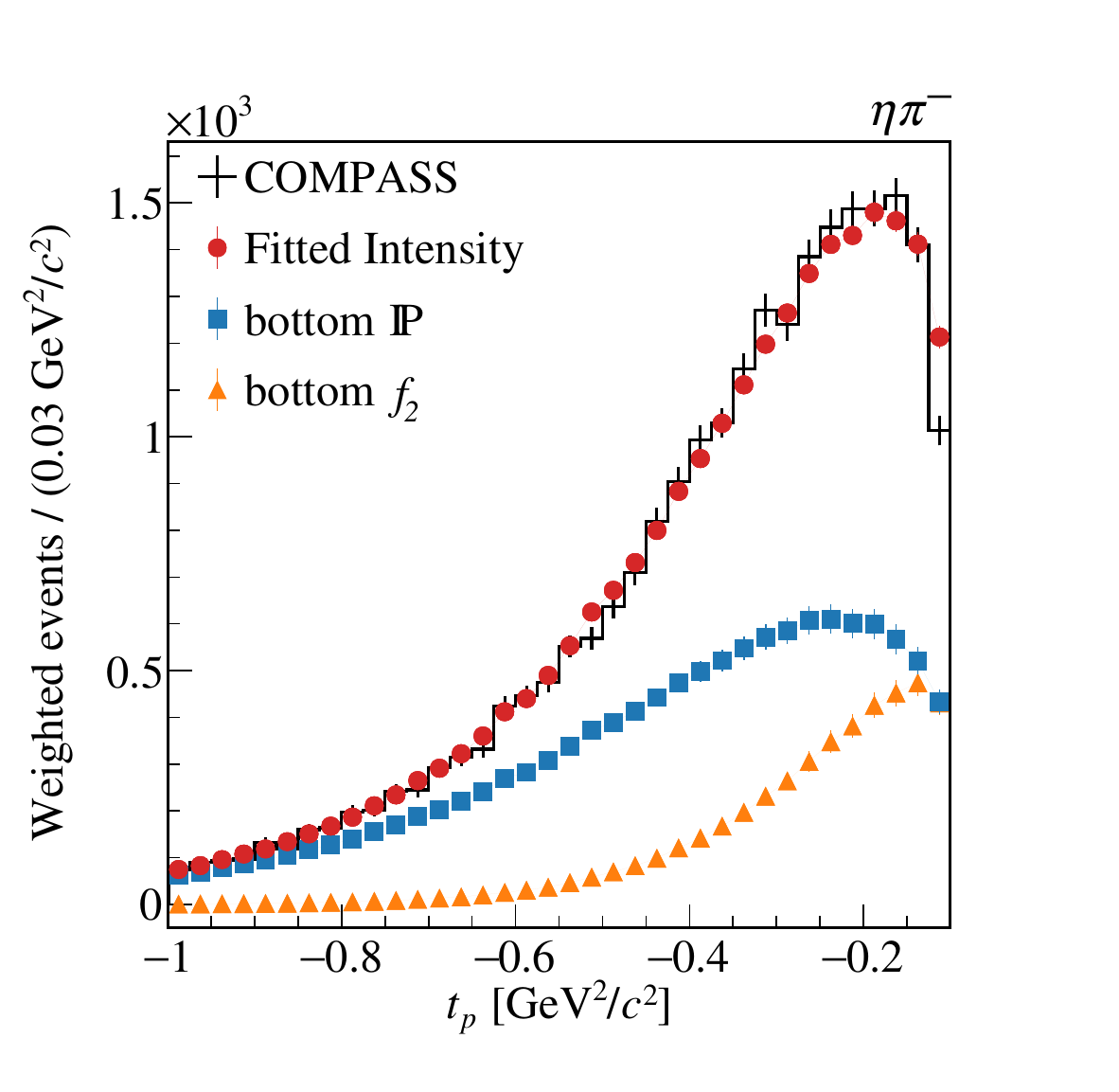} 
\end{tabular}
\caption{
$\eta \pi^-$ weighted intensity distributions dependence on $t_p$. Bottom
Reggeon exchange contributions coherently summed. By bottom $\Pom$ (bottom
$f_2$) we refer to the coherent sum of the amplitudes whose exchanged
$\mathbb{R}_b$ is the $\Pom$ ($f_2$). Left: \mbox{KGR} fit;
Center-left: \mbox{$\text{KGR}+\pi_1+a_2^\prime$} fit; Center-right: \mbox{JPAC}
fit; Right: \mbox{$\text{JPAC}+\pi_1+a_2^\prime$} fit.
}
\label{figsup:etaweight4_2}
\end{figure*}

\begin{figure*}[!h]
\begin{tabular}{cccc}
\includegraphics[width=0.25\linewidth]{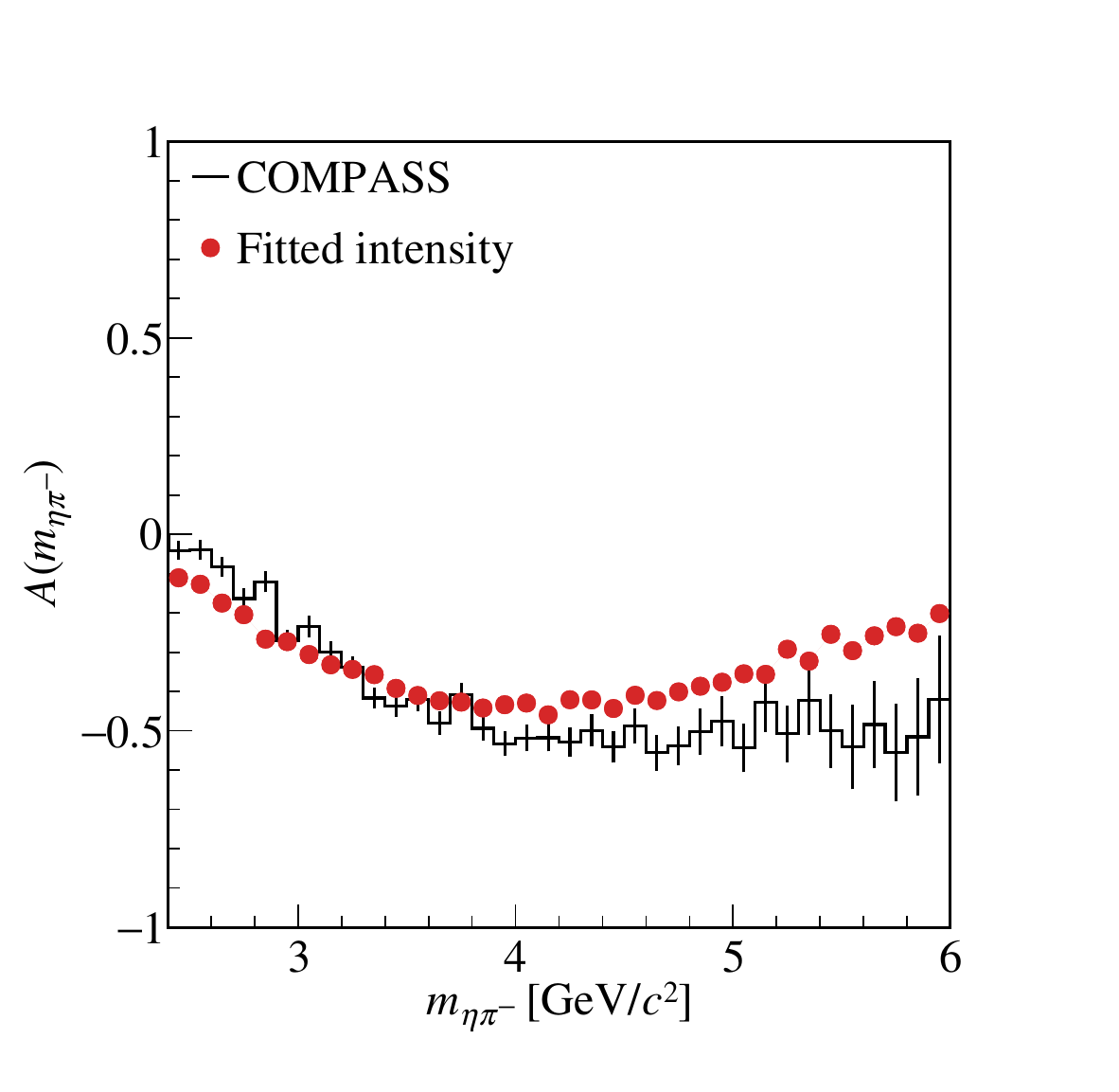} & 
\includegraphics[width=0.25\linewidth]{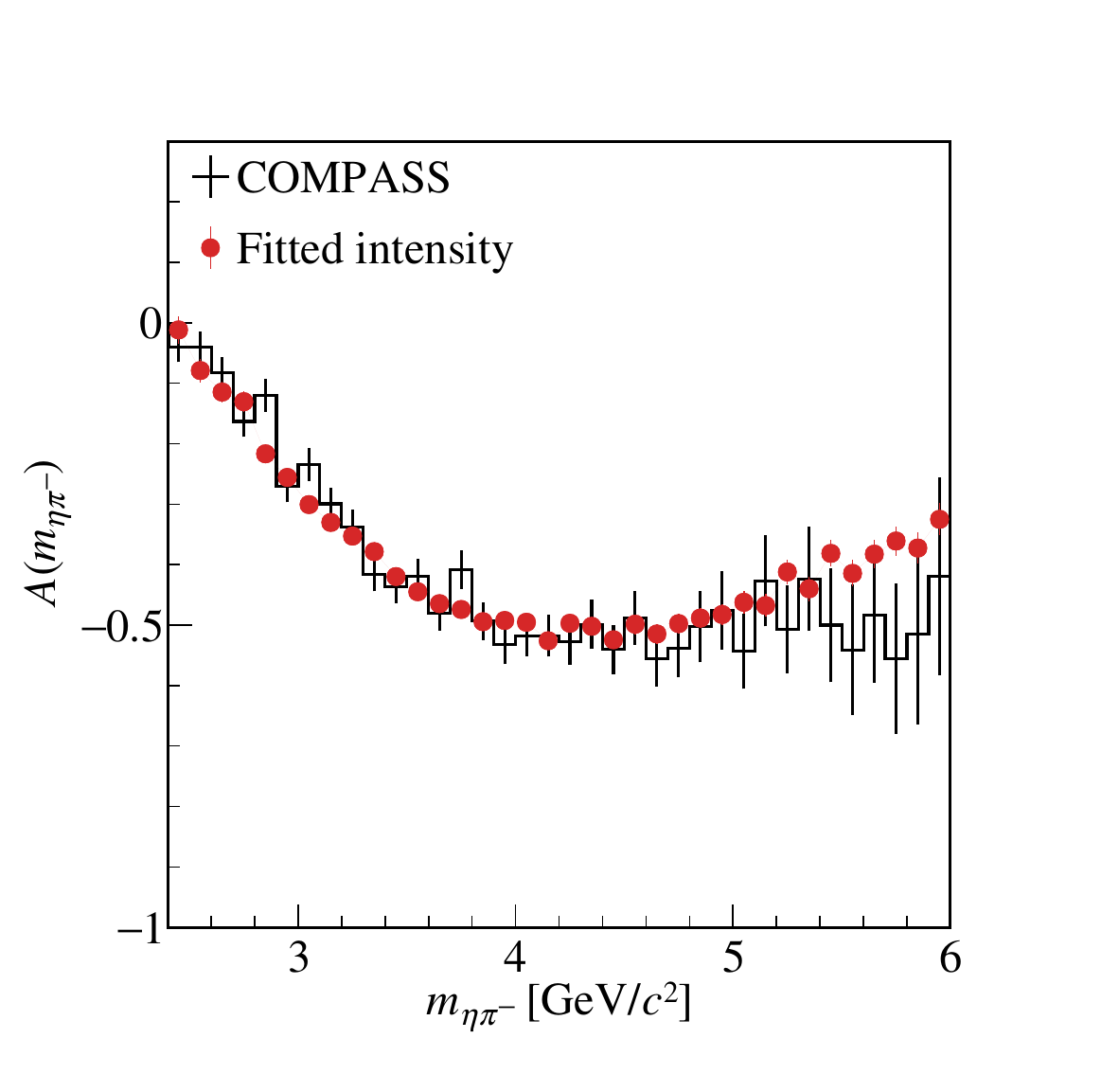} & 
\includegraphics[width=0.25\linewidth]{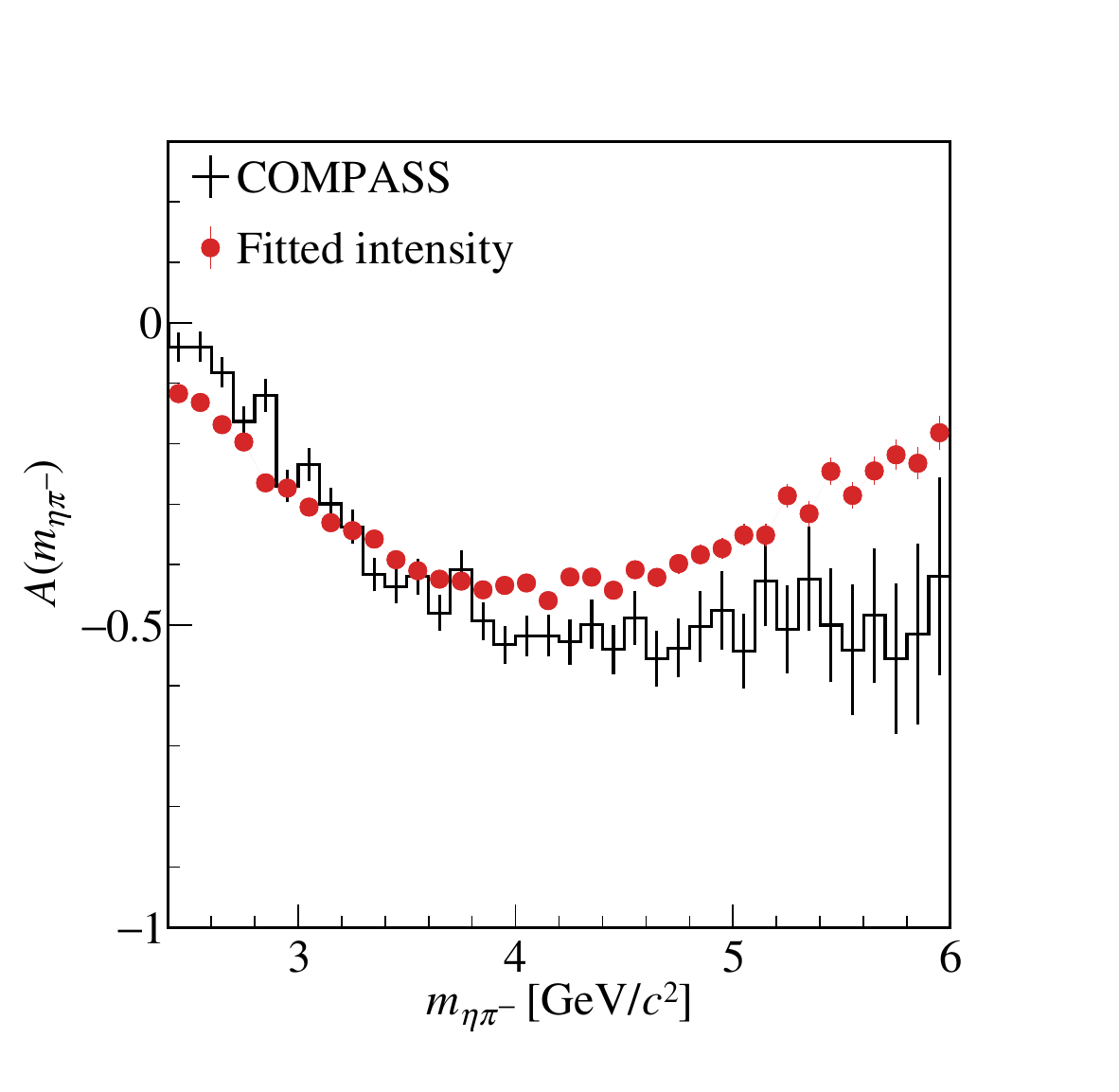} & 
\includegraphics[width=0.25\linewidth]{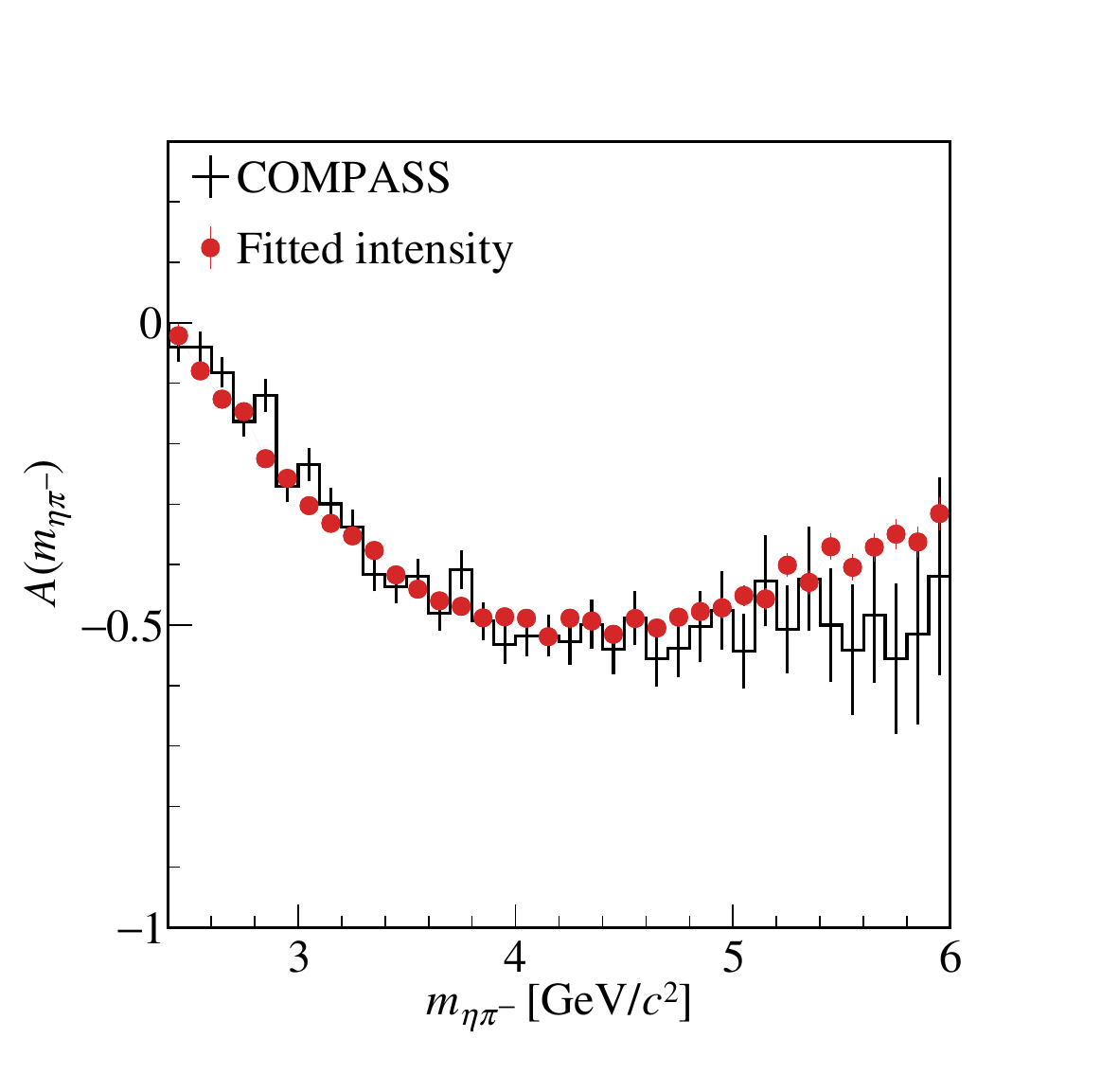} 
\end{tabular}
\caption{
$\eta \pi^-$ weighted forward-backward asymmetry distribution dependence on
$m_{\eta \pi}$. Left: \mbox{KGR} fit;
Center-left: \mbox{$\text{KGR}+\pi_1+a_2^\prime$} fit; Center-right: \mbox{JPAC}
fit; Right: \mbox{$\text{JPAC}+\pi_1+a_2^\prime$} fit. Binned \compass data is
drawn in black.
}
\label{figsup:etaweight5}
\end{figure*}

In this section we compare the $\eta \pi^-$ weighted intensity distributions of
the fits using the \mbox{KGR} (left), \mbox{$\text{KGR}+\pi_1+a_2^\prime$}
(center-left), \mbox{JPAC} (center-right)
and \mbox{$\text{JPAC}+\pi_1+a_2^\prime$} (right) models to the binned \compass
data. The difference between \mbox{KGR} and \mbox{JPAC} fits provide an
estimation of the systematic uncertainty associated to the leading ordinary
exchanges. We present results depending on the invariant mass distribution
$m_{\eta \pi}$ (\cref{figsup:etaweight1}), $t_\eta$
(\cref{figsup:etaweight2_1,figsup:etaweight2_2}), $t_\pi$
(\cref{figsup:etaweight3_1,figsup:etaweight3_2}), and $t_p$
(\cref{figsup:etaweight4_1,figsup:etaweight4_2}.) We provide the individual
contributions of the exchanges as well as top $\pi_1$, top $a_2^\prime$, top
$a_2$, top $f_2$ and top $\Pom$ aggregates for $t_\eta$ and $t_\pi$
distributions (see~\cref{figsup:etaweight2_2,figsup:etaweight3_2}), and bottom
$f_2$ and bottom $\Pom$ aggregates for $t_p$ distribution
(see \cref{figsup:etaweight4_2}.) Finally, we show the results for the
forward-backward asymmetry in~\cref{figsup:etaweight5}.

\clearpage
\newpage
\section*{$\eta^\prime \pi^-$ distributions}
\begin{figure*}[!h]
\begin{tabular}{cccc}
\includegraphics[width=0.25\linewidth]{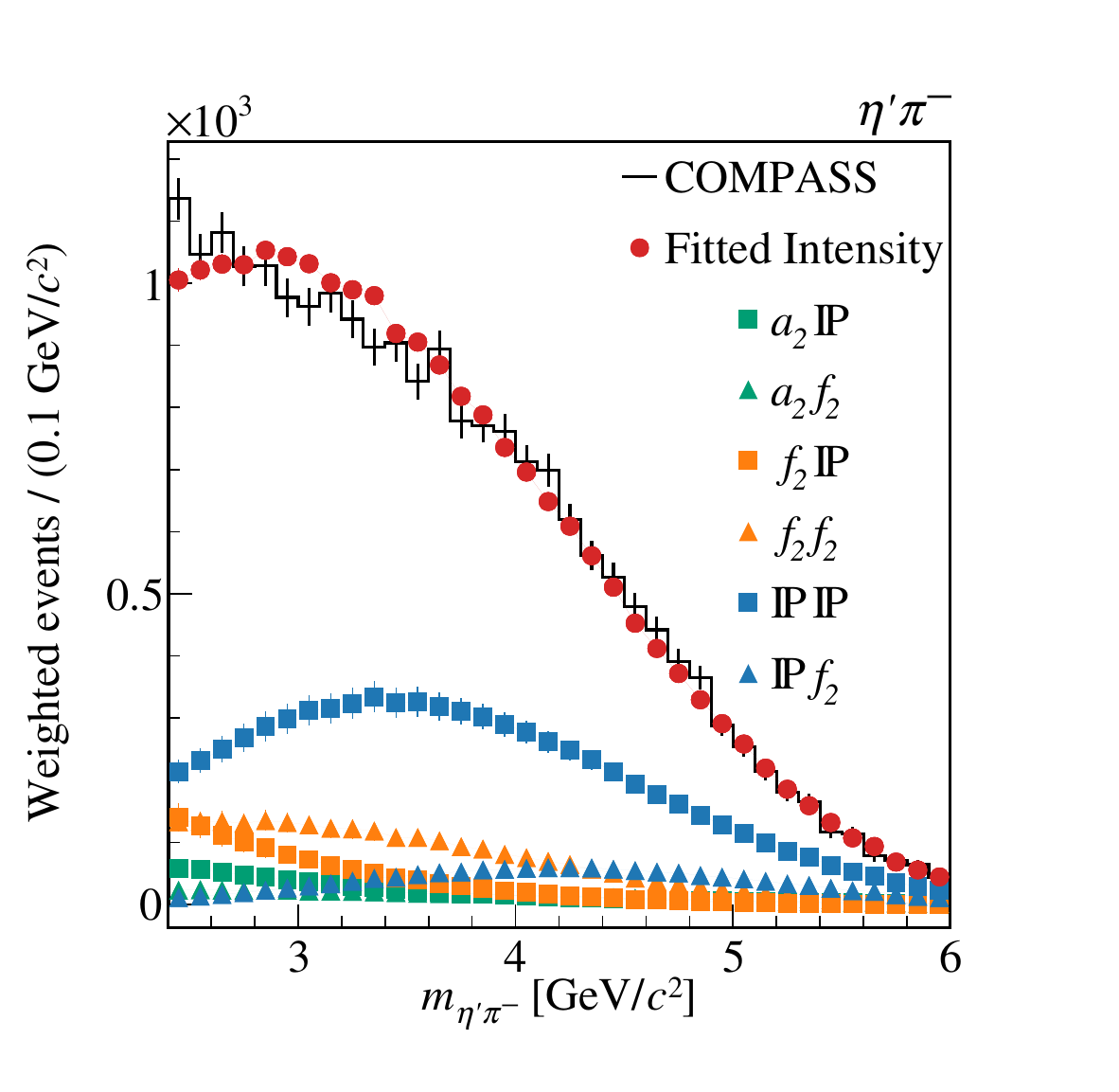} & 
\includegraphics[width=0.25\linewidth]{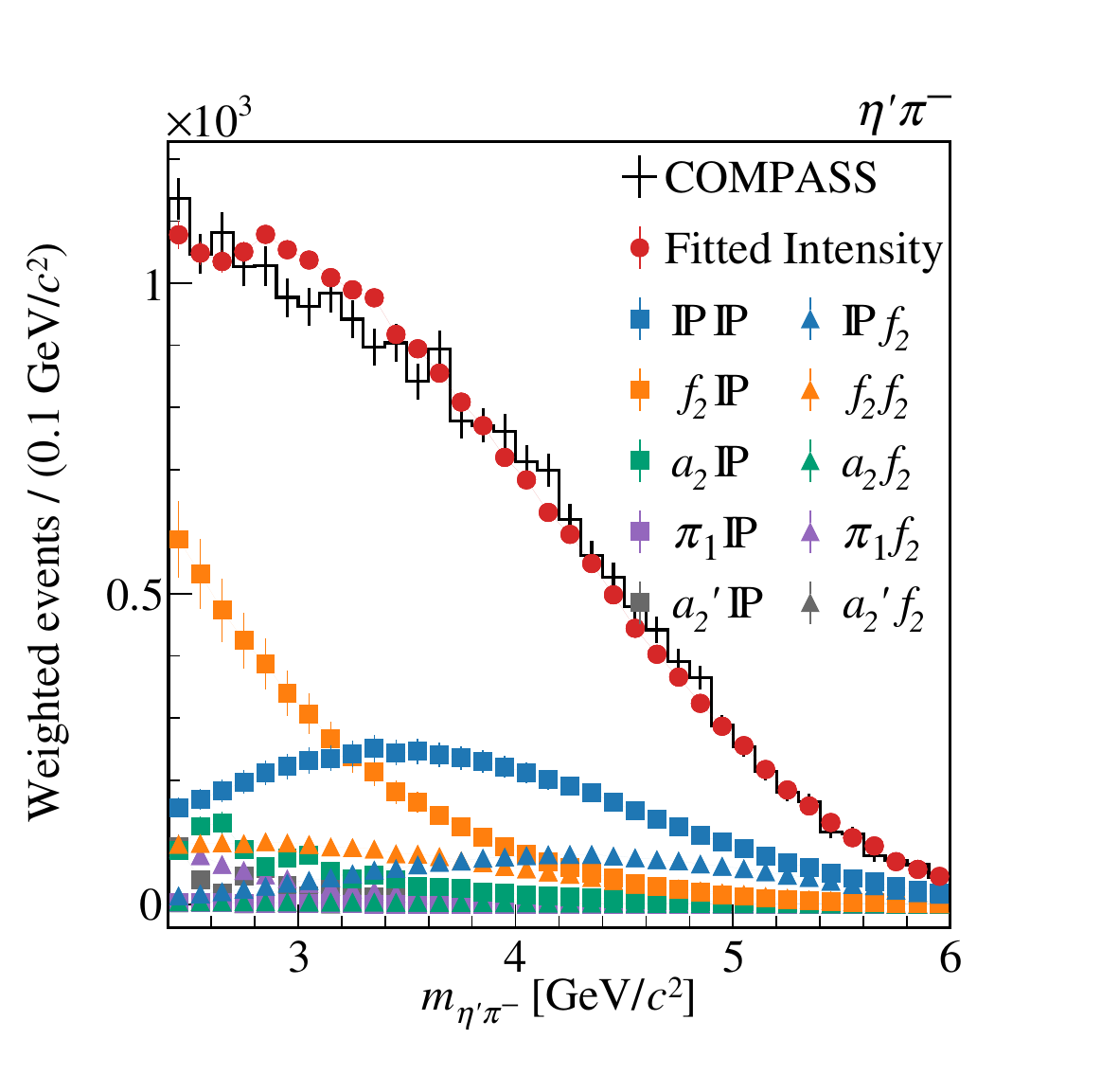} & 
\includegraphics[width=0.25\linewidth]{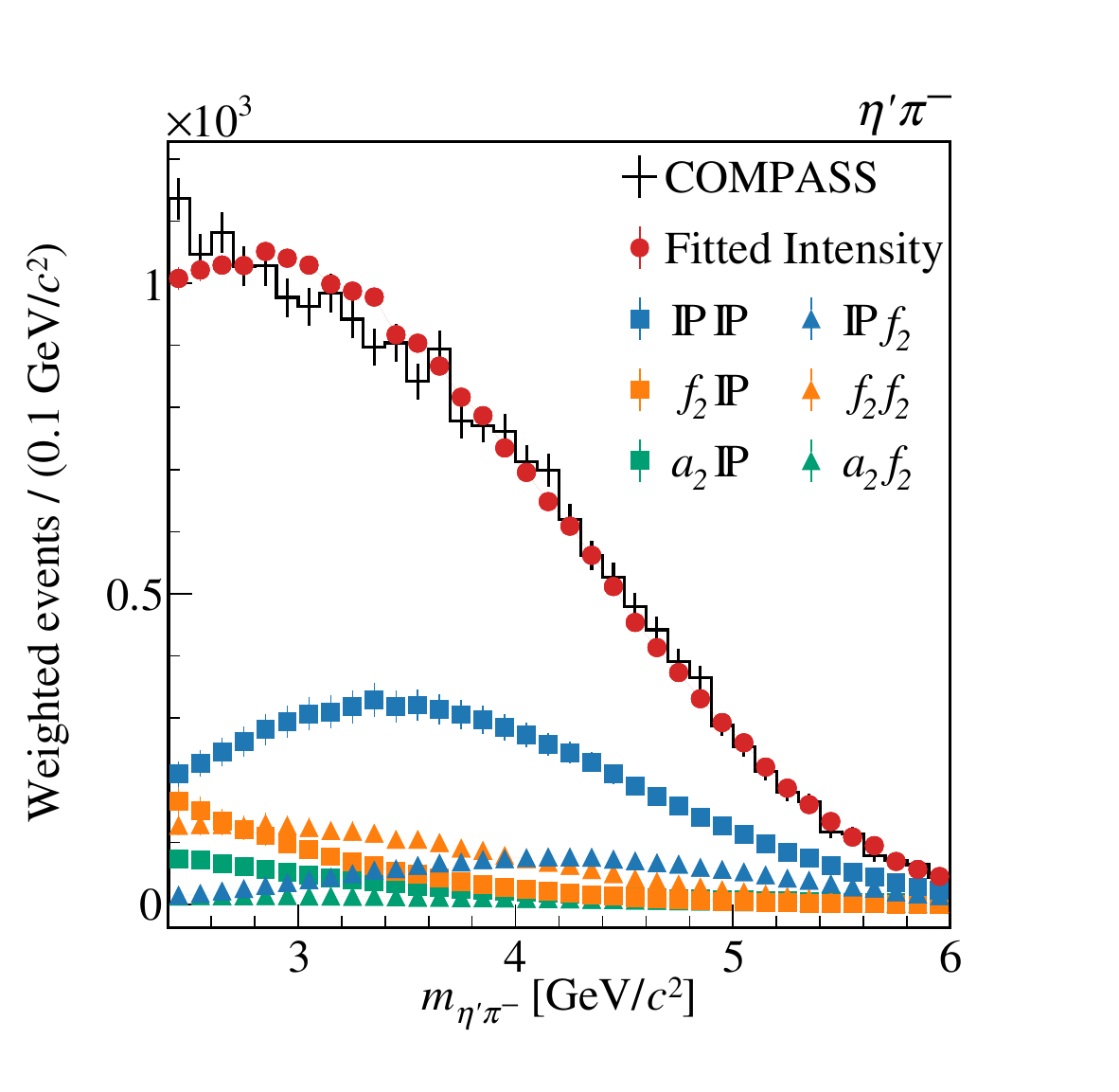} & 
\includegraphics[width=0.25\linewidth]{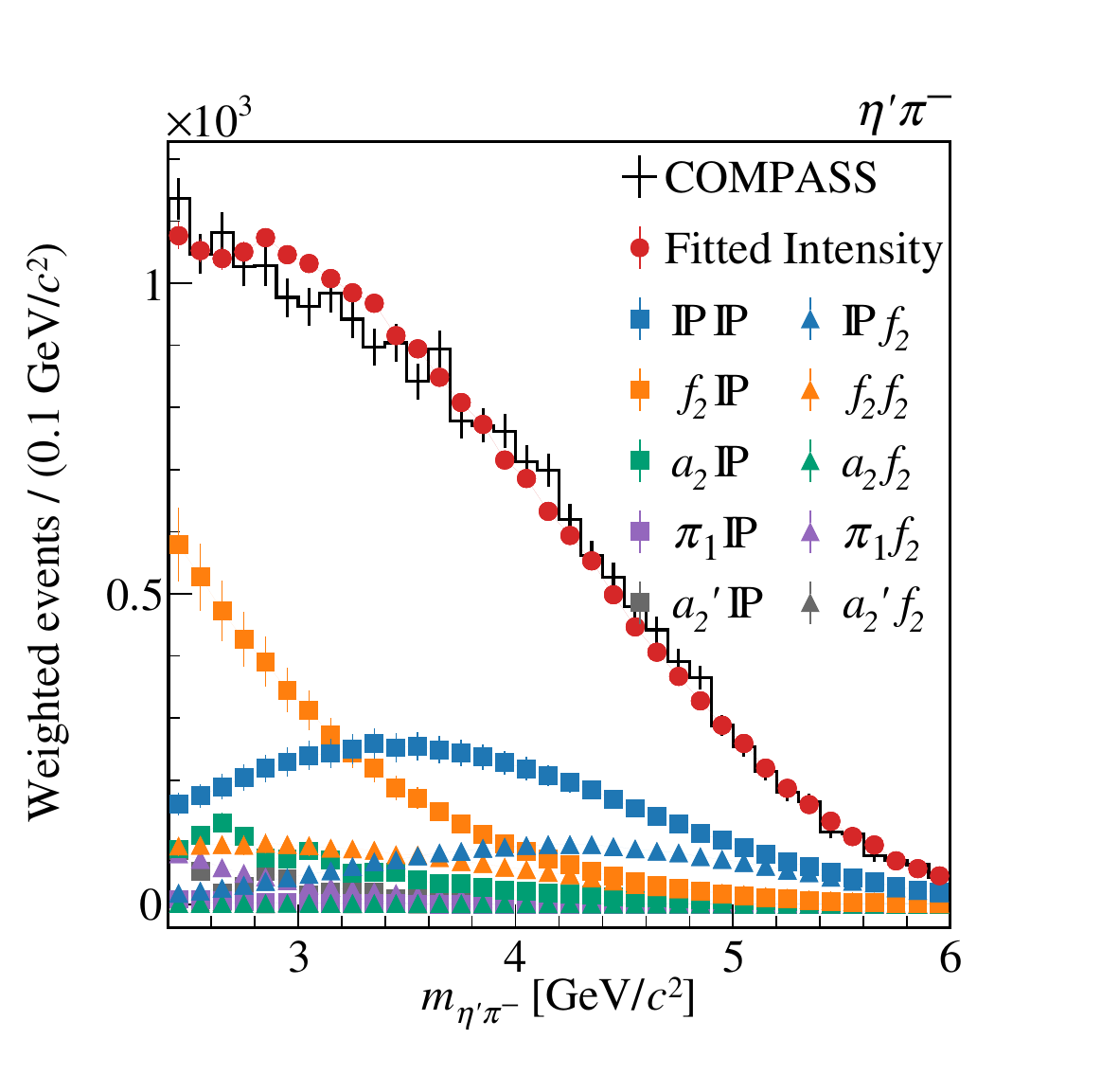} 
\end{tabular}
\caption{
$\eta^{\prime} \pi^-$ weighted intensity distributions dependence on
$m_{\eta \pi}$. Left: \mbox{KGR} fit;
Center-left: \mbox{$\text{KGR}+\pi_1+a_2^\prime$} fit; Center-right: \mbox{JPAC}
fit; Right: \mbox{$\text{JPAC}+\pi_1+a_2^\prime$} fit. Conventions as
in~\cref{figsup:etaweight1}.
} 
\label{figsup:etaprimeweight1}
\end{figure*}

\begin{figure*}[!h]
\begin{tabular}{cccc}
\includegraphics[width=0.25\linewidth]{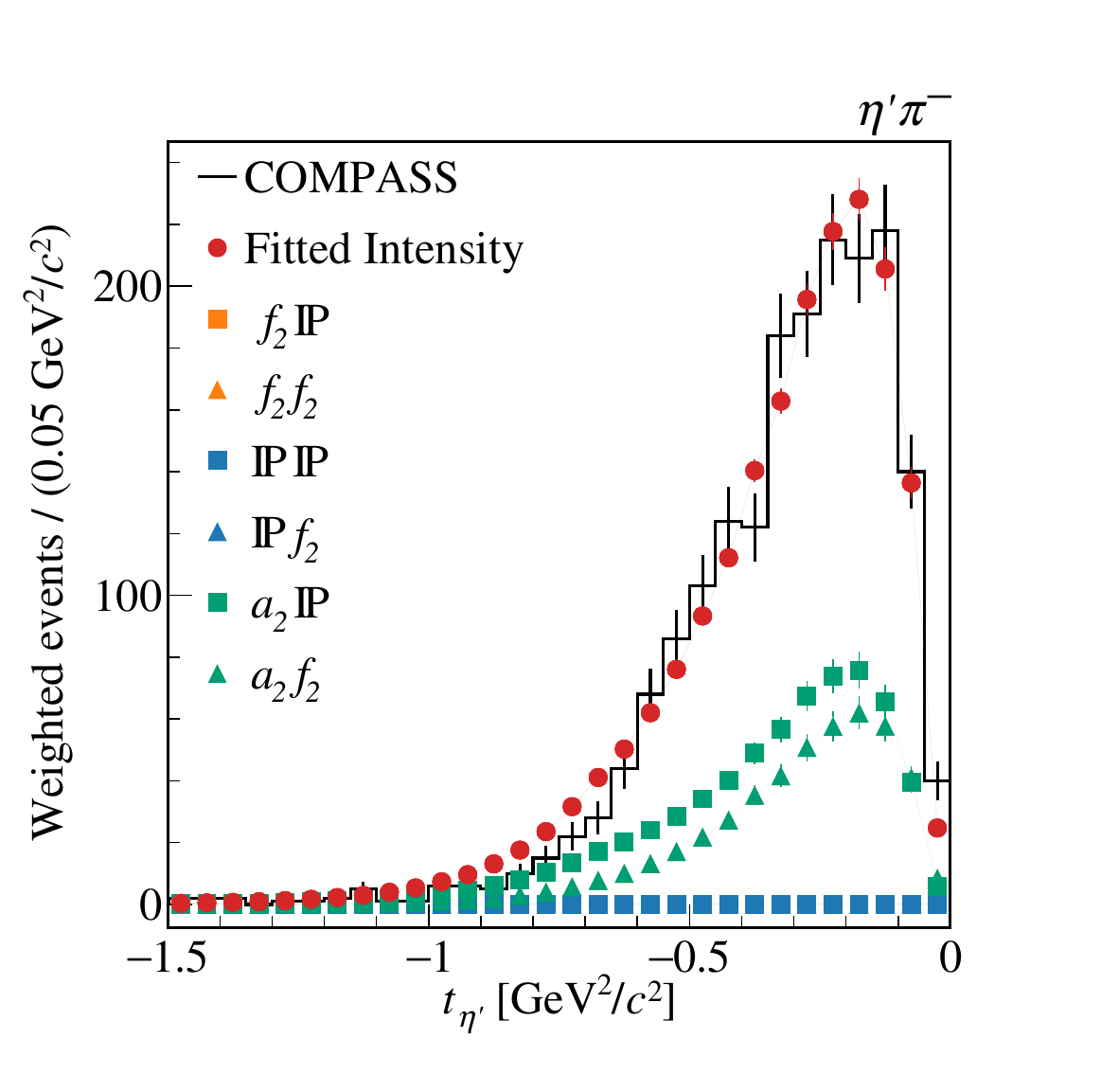} & 
\includegraphics[width=0.25\linewidth]{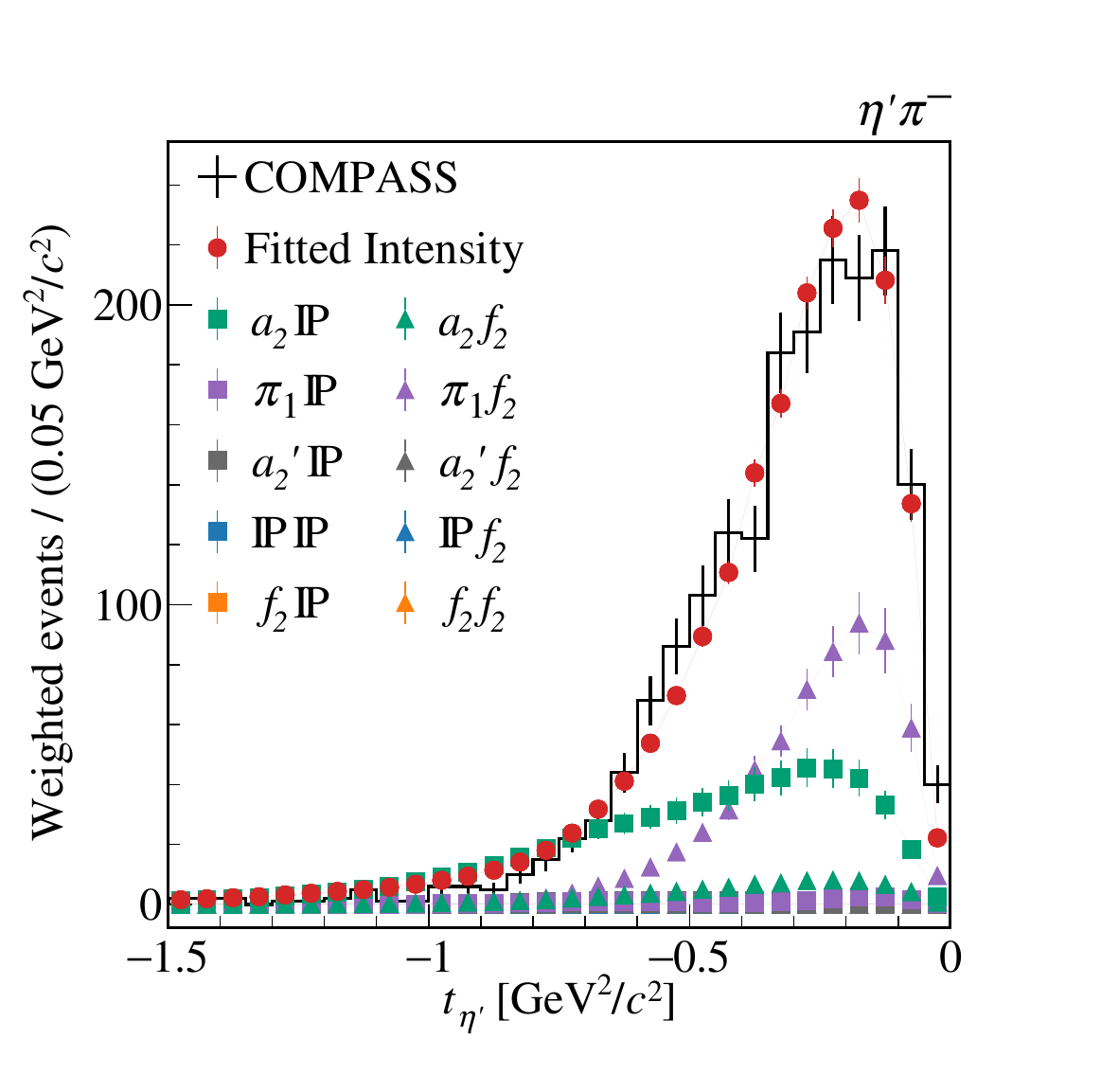} & 
\includegraphics[width=0.25\linewidth]{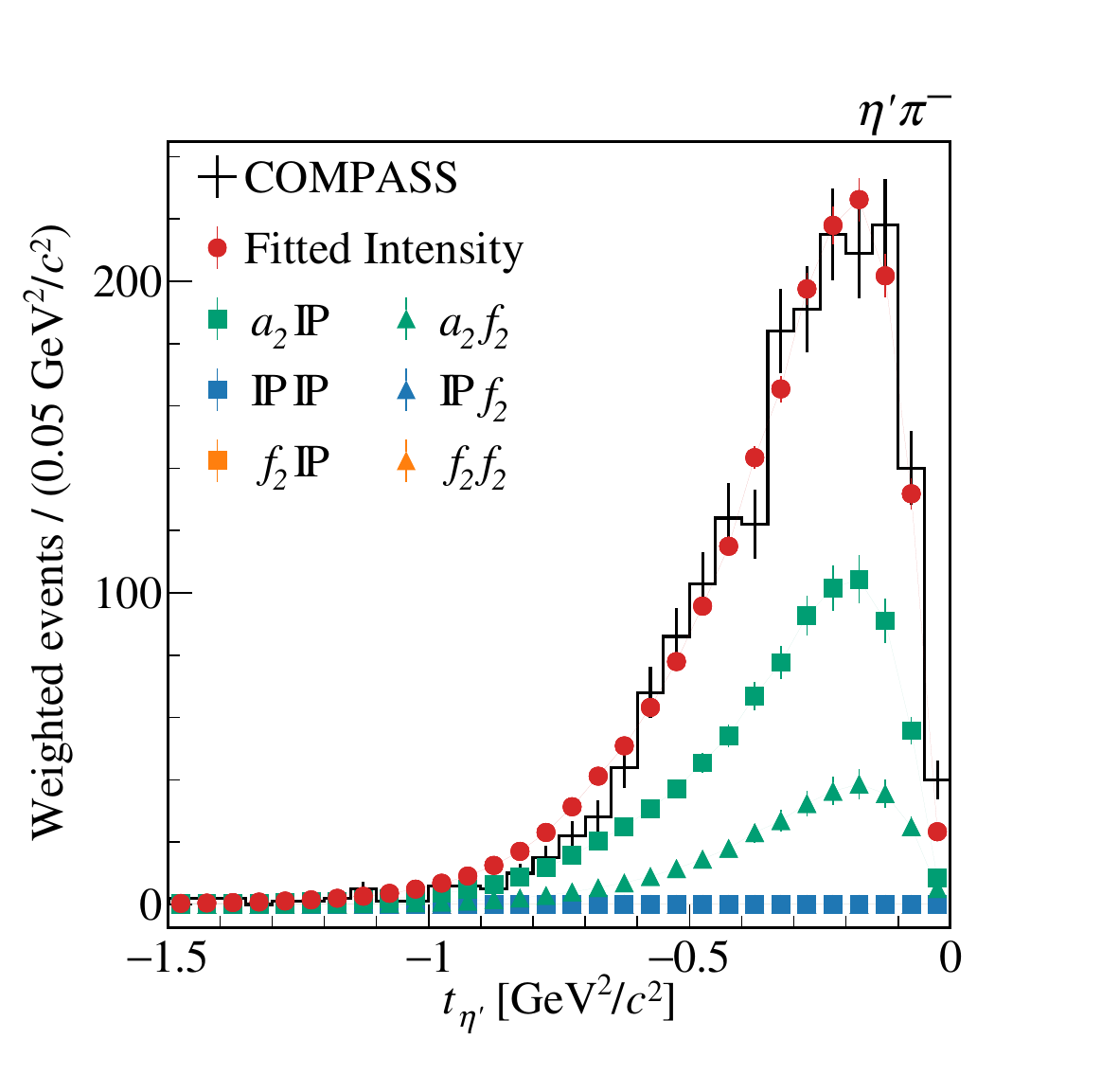} & 
\includegraphics[width=0.25\linewidth]{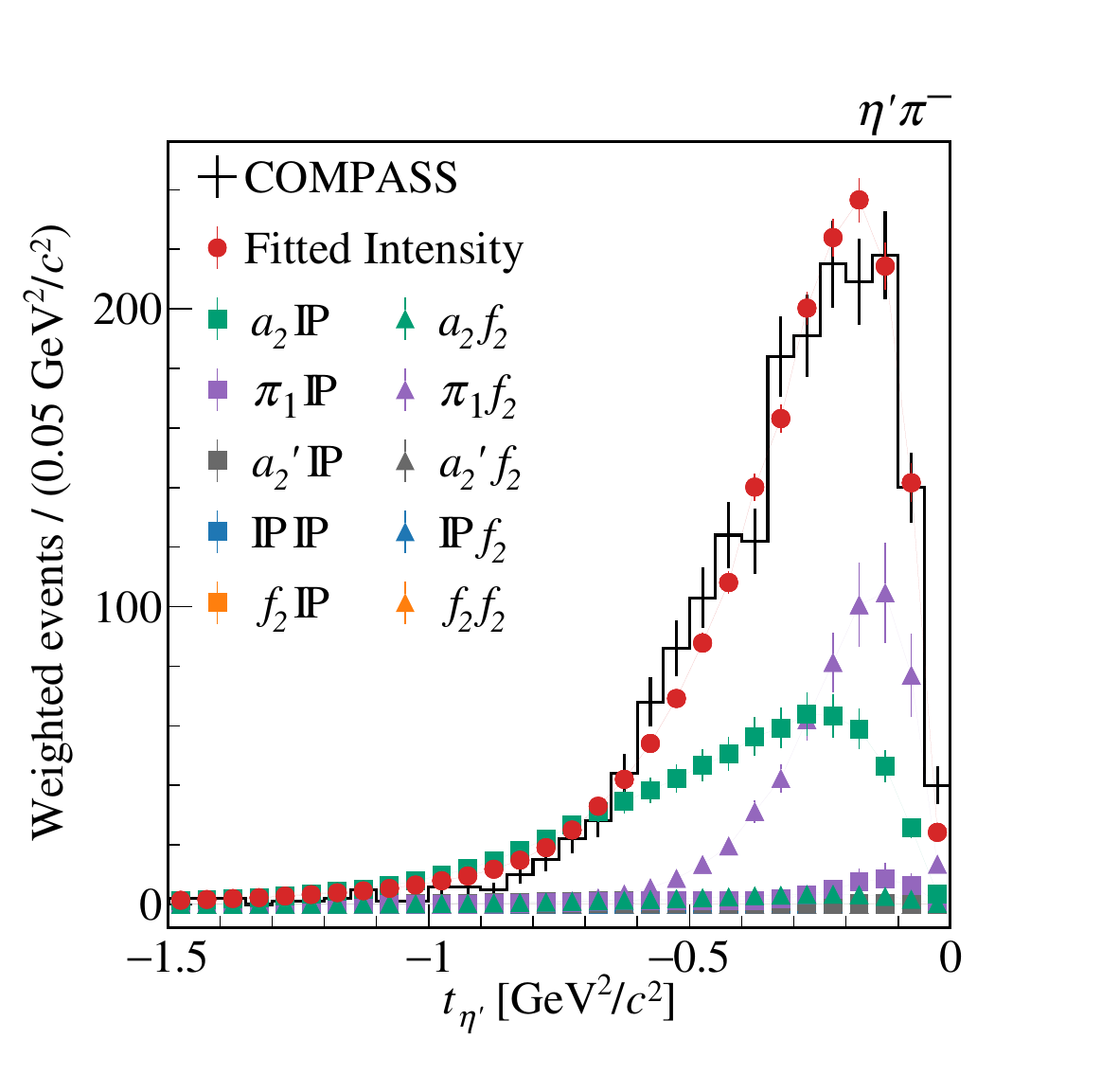} 
\end{tabular}
\caption{
$\eta^{\prime} \pi^-$ weighted intensity distributions dependence on
$t_\eta$. Individual contributions. Left: \mbox{KGR} fit;
Center-left: \mbox{$\text{KGR}+\pi_1+a_2^\prime$} fit; Center-right: \mbox{JPAC}
fit; Right: \mbox{$\text{JPAC}+\pi_1+a_2^\prime$} fit. Conventions as
in~\cref{figsup:etaprimeweight1}.
} 
\label{figsup:etaprimeweight2_1}
\end{figure*}

\begin{figure*}[!h]
\begin{tabular}{cccc}
\includegraphics[width=0.25\linewidth]{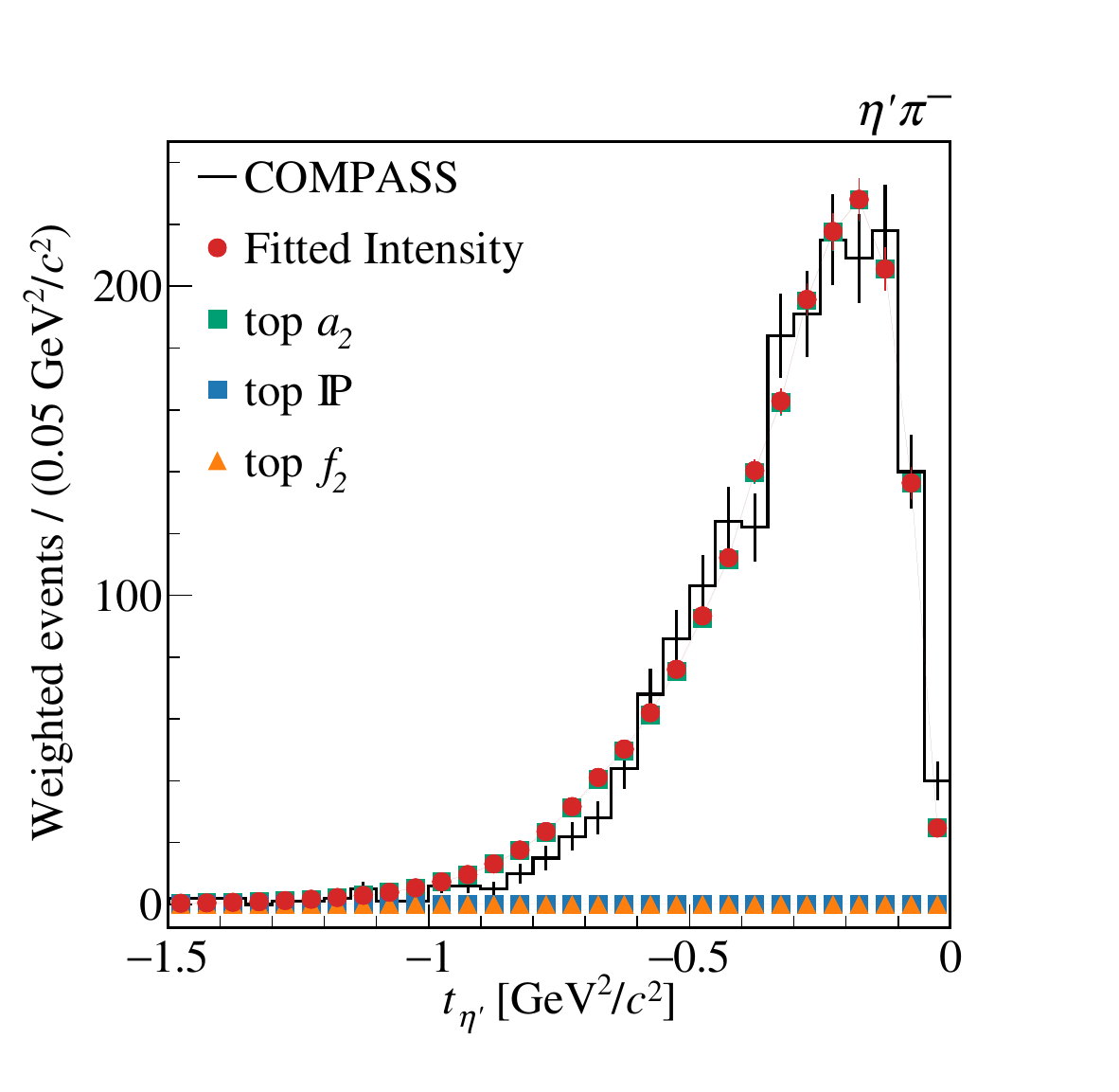} & 
\includegraphics[width=0.25\linewidth]{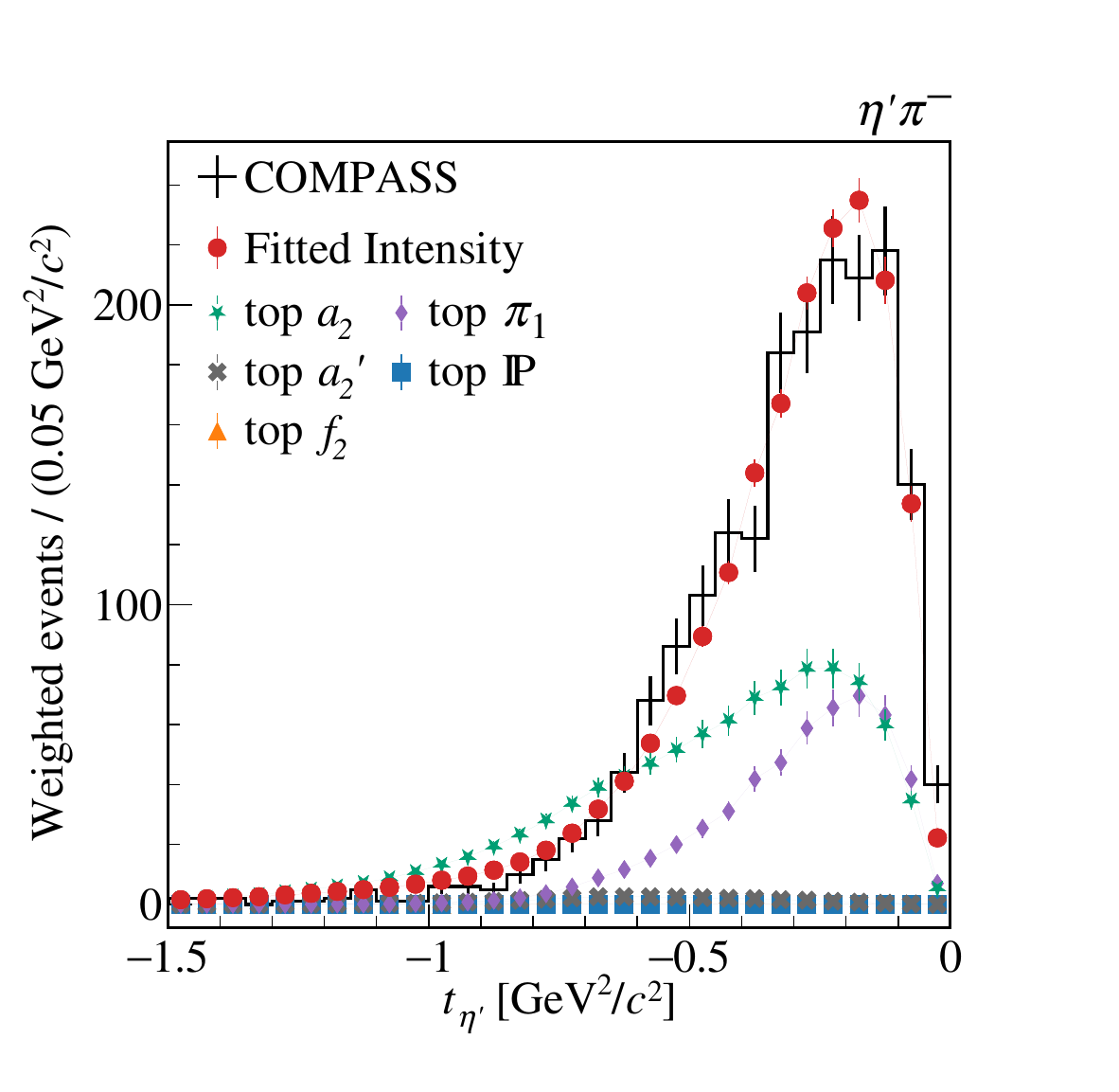} & 
\includegraphics[width=0.25\linewidth]{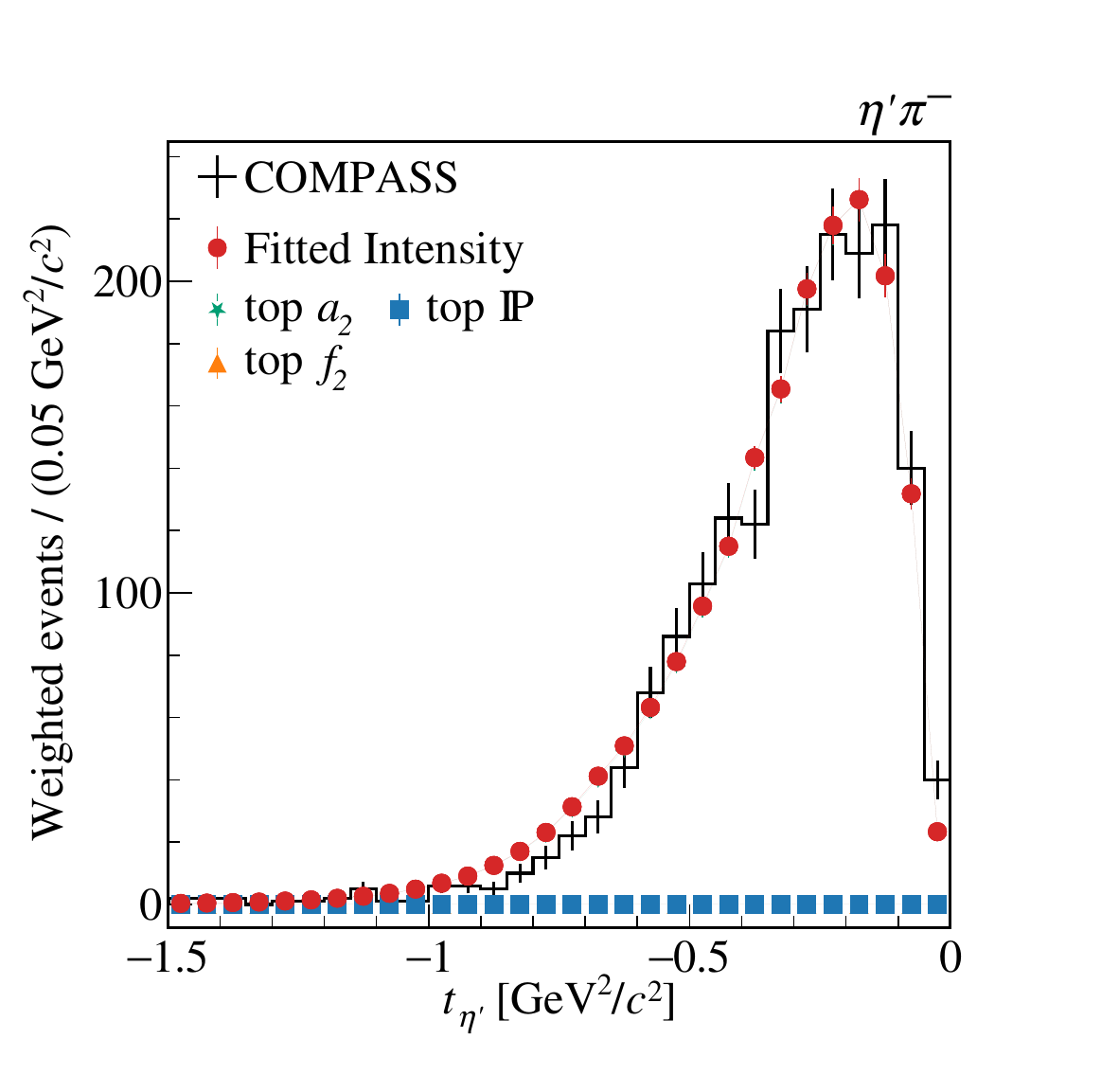} & 
\includegraphics[width=0.25\linewidth]{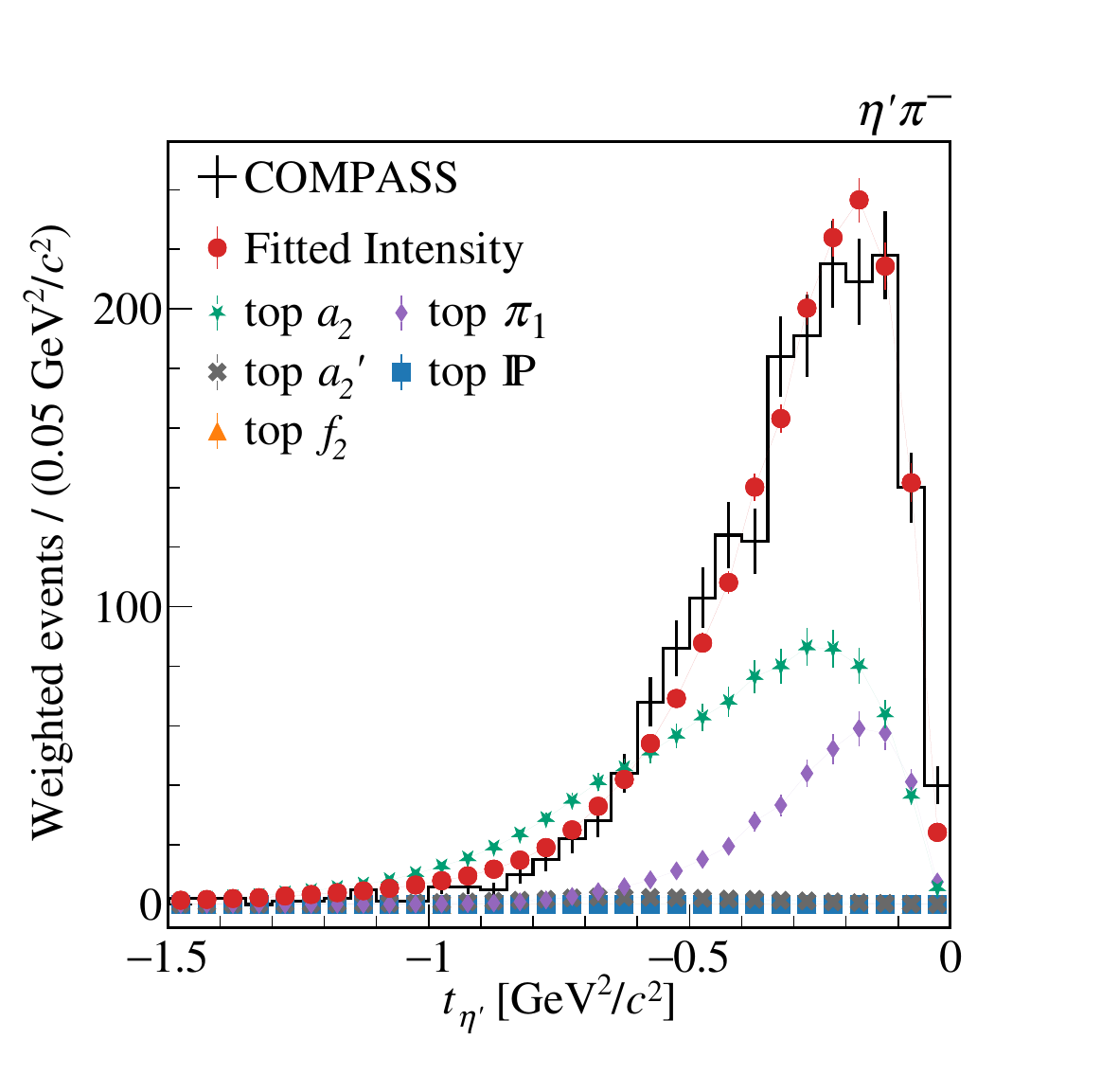} 
\end{tabular}
\caption{
$\eta^{\prime} \pi^-$ weighted intensity distributions dependence on
$t_\eta$. Top Reggeon exchange contributions coherently summed. Left: \mbox{KGR}
fit; Center-left: \mbox{$\text{KGR}+\pi_1+a_2^\prime$} fit;
Center-right: \mbox{JPAC} fit; Right: \mbox{$\text{JPAC}+\pi_1+a_2^\prime$}
fit. Conventions as in~\cref{figsup:etaprimeweight1}.
}
\label{figsup:etaprimeweight2_2}
\end{figure*}

\begin{figure*}[!h]
\begin{tabular}{cccc}
\includegraphics[width=0.25\linewidth]{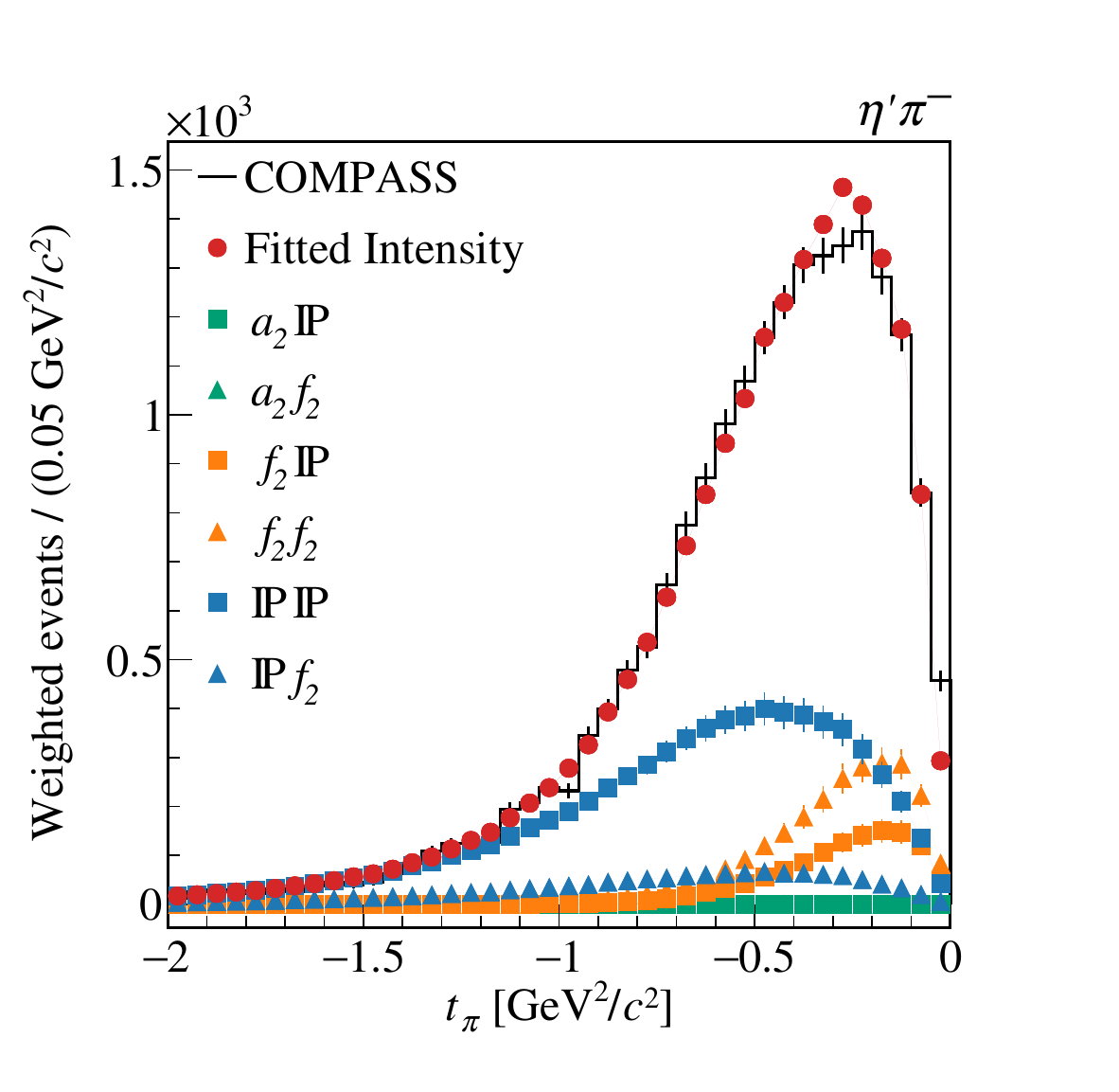} & 
\includegraphics[width=0.25\linewidth]{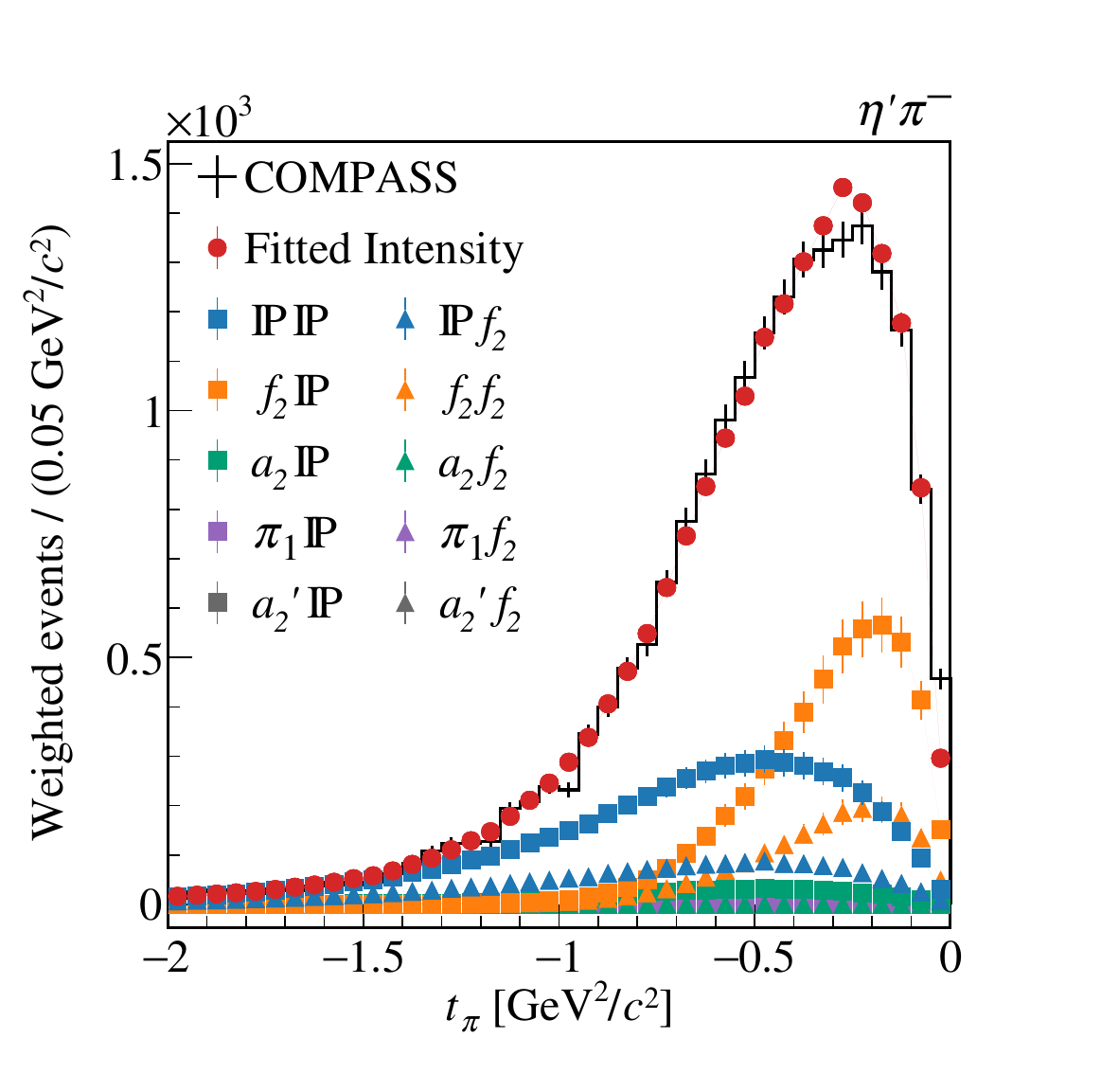} & 
\includegraphics[width=0.25\linewidth]{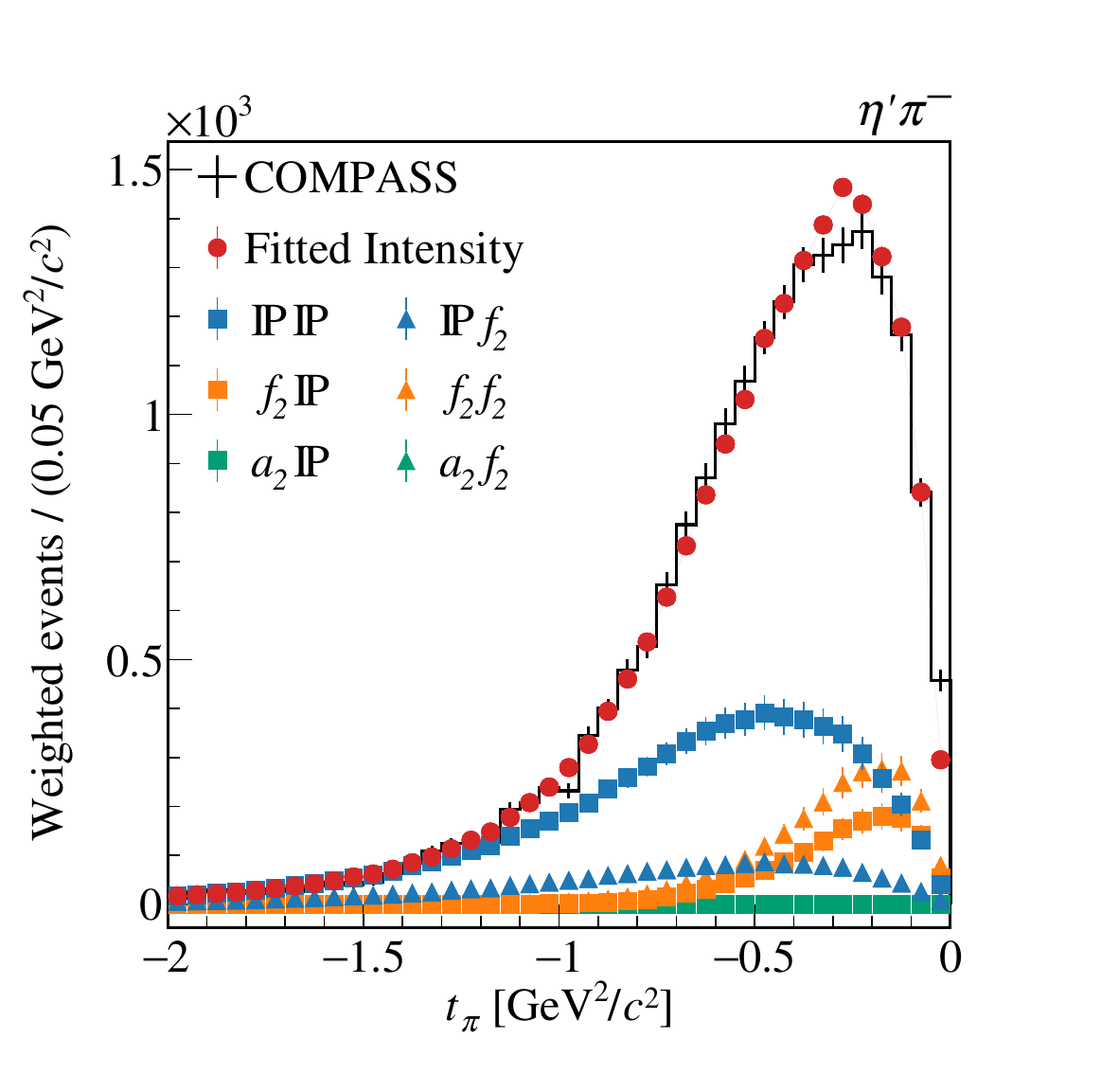} & 
\includegraphics[width=0.25\linewidth]{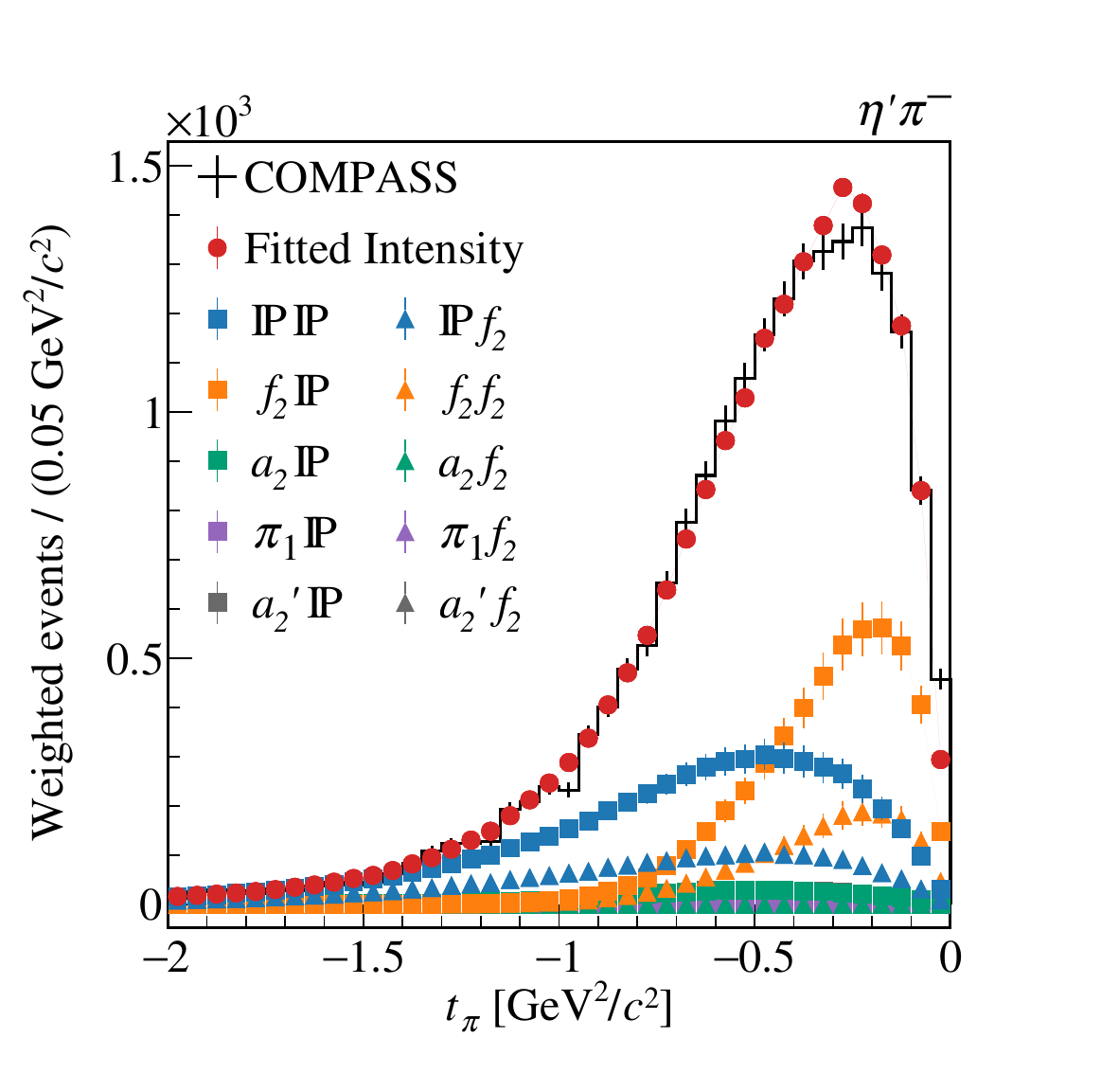} 
\end{tabular}
\caption{
$\eta^{\prime} \pi^-$ weighted intensity distributions dependence on
$t_\pi$. Individual contributions. Left: \mbox{KGR} fit;
Center-left: \mbox{$\text{KGR}+\pi_1+a_2^\prime$} fit; Center-right: \mbox{JPAC}
fit; Right: \mbox{$\text{JPAC}+\pi_1+a_2^\prime$} fit. Conventions as
in~\cref{figsup:etaprimeweight1}.
}
\label{figsup:etaprimeweight3_1}
\end{figure*}

\begin{figure*}[!h]
\begin{tabular}{cccc}
\includegraphics[width=0.25\linewidth]{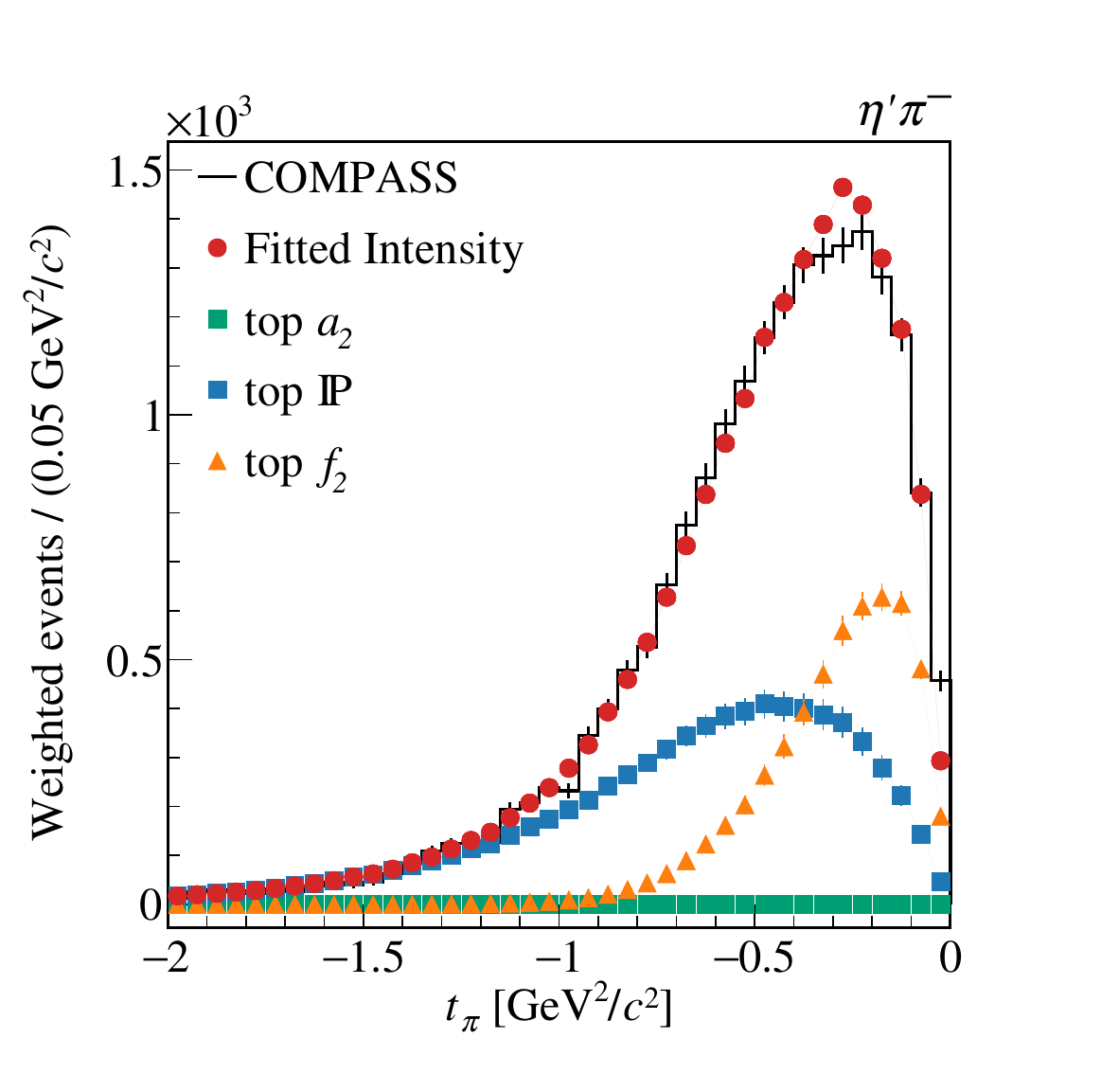} & 
\includegraphics[width=0.25\linewidth]{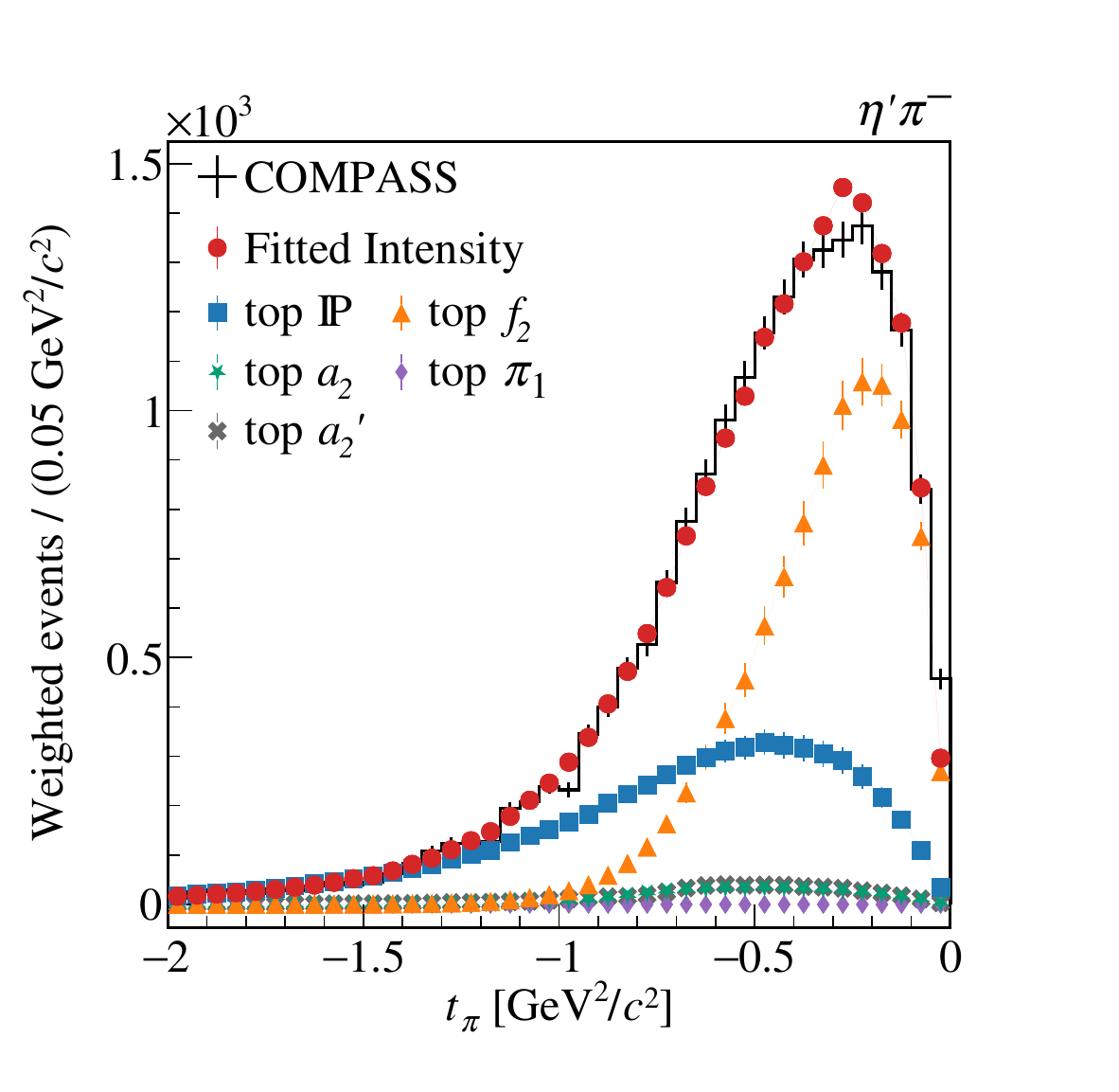} & 
\includegraphics[width=0.25\linewidth]{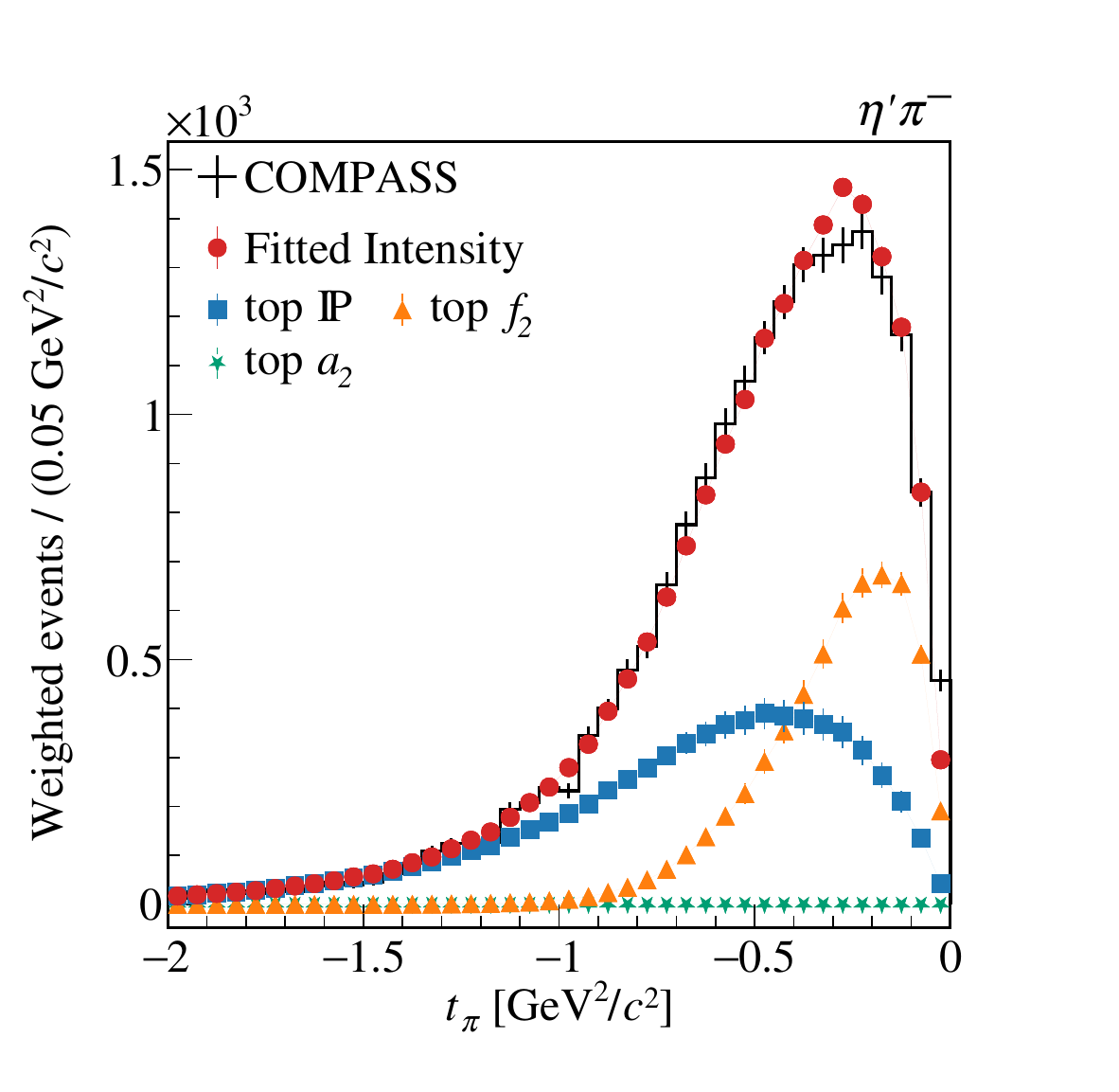} & 
\includegraphics[width=0.25\linewidth]{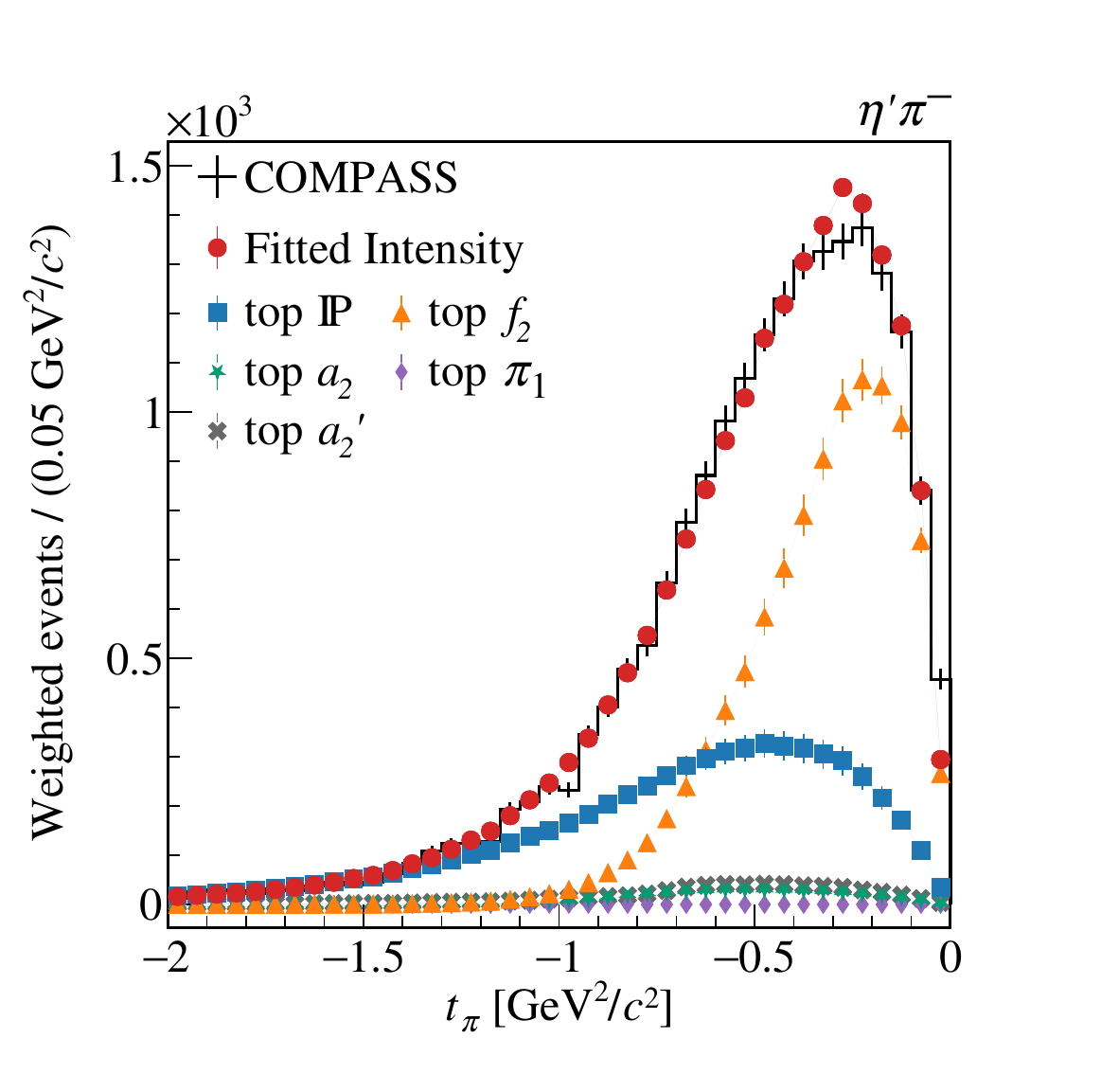} 
\end{tabular}
\caption{
$\eta^{\prime} \pi^-$ weighted intensity distributions dependence on
$t_\pi$. Top Reggeon exchange contributions coherently summed. Left: \mbox{KGR}
fit; Center-left: \mbox{$\text{KGR}+\pi_1+a_2^\prime$} fit;
Center-right: \mbox{JPAC} fit; Right: \mbox{$\text{JPAC}+\pi_1+a_2^\prime$}
fit. Conventions as in~\cref{figsup:etaprimeweight1}.
}
\label{figsup:etaprimeweight3_2}
\end{figure*}

\begin{figure*}[!h]
\begin{tabular}{cccc}
\includegraphics[width=0.25\linewidth]{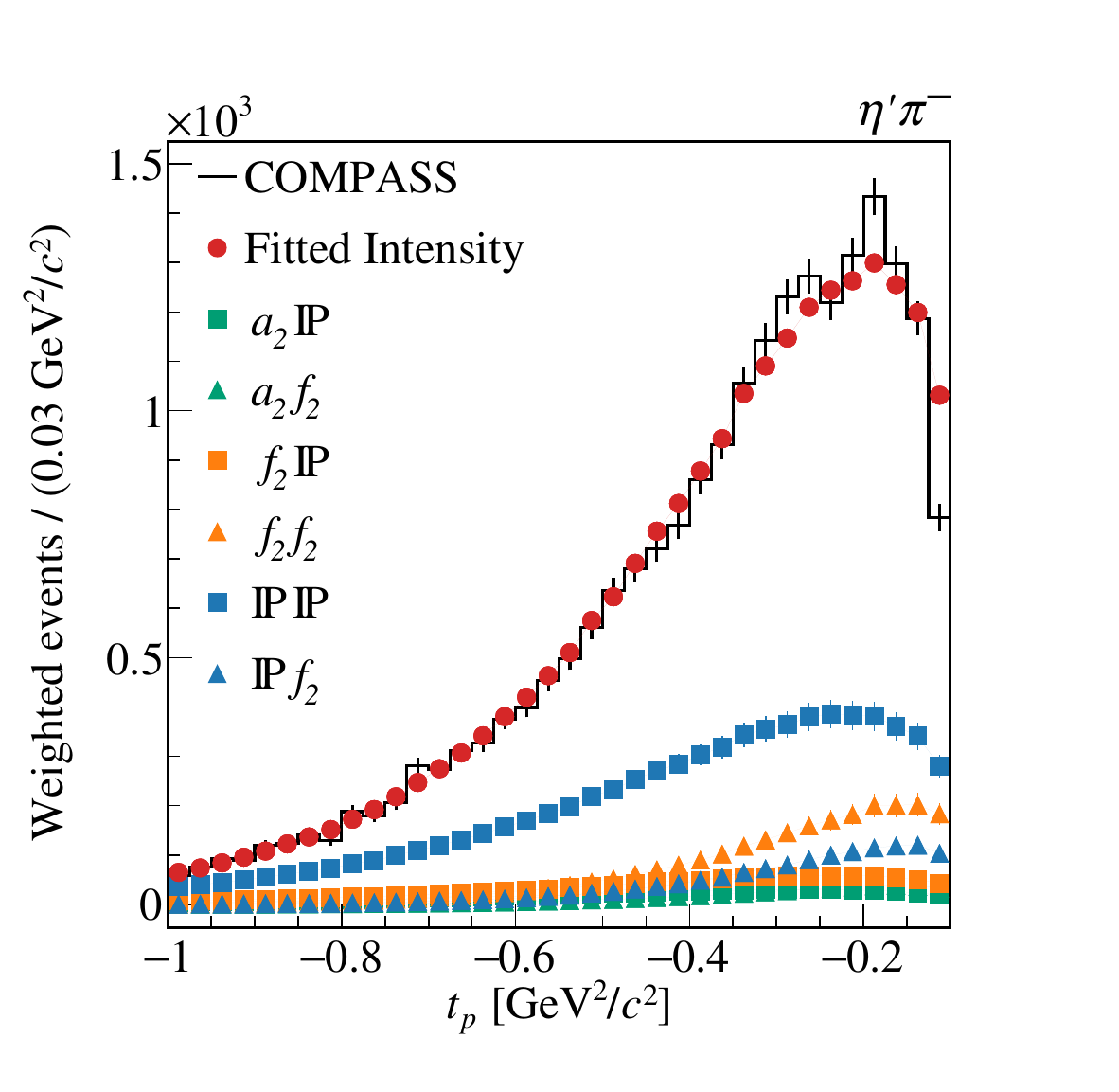} & 
\includegraphics[width=0.25\linewidth]{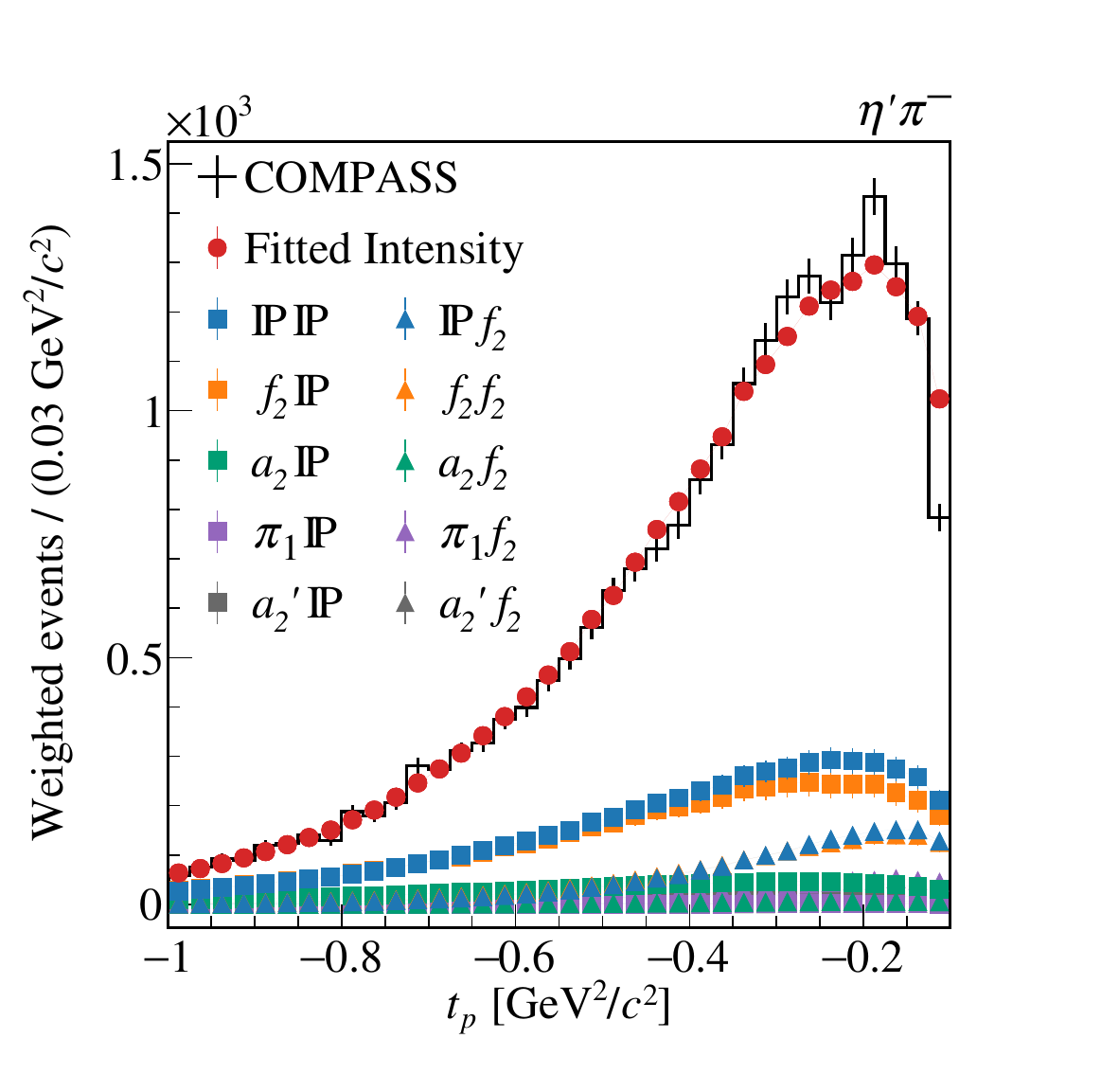} & 
\includegraphics[width=0.25\linewidth]{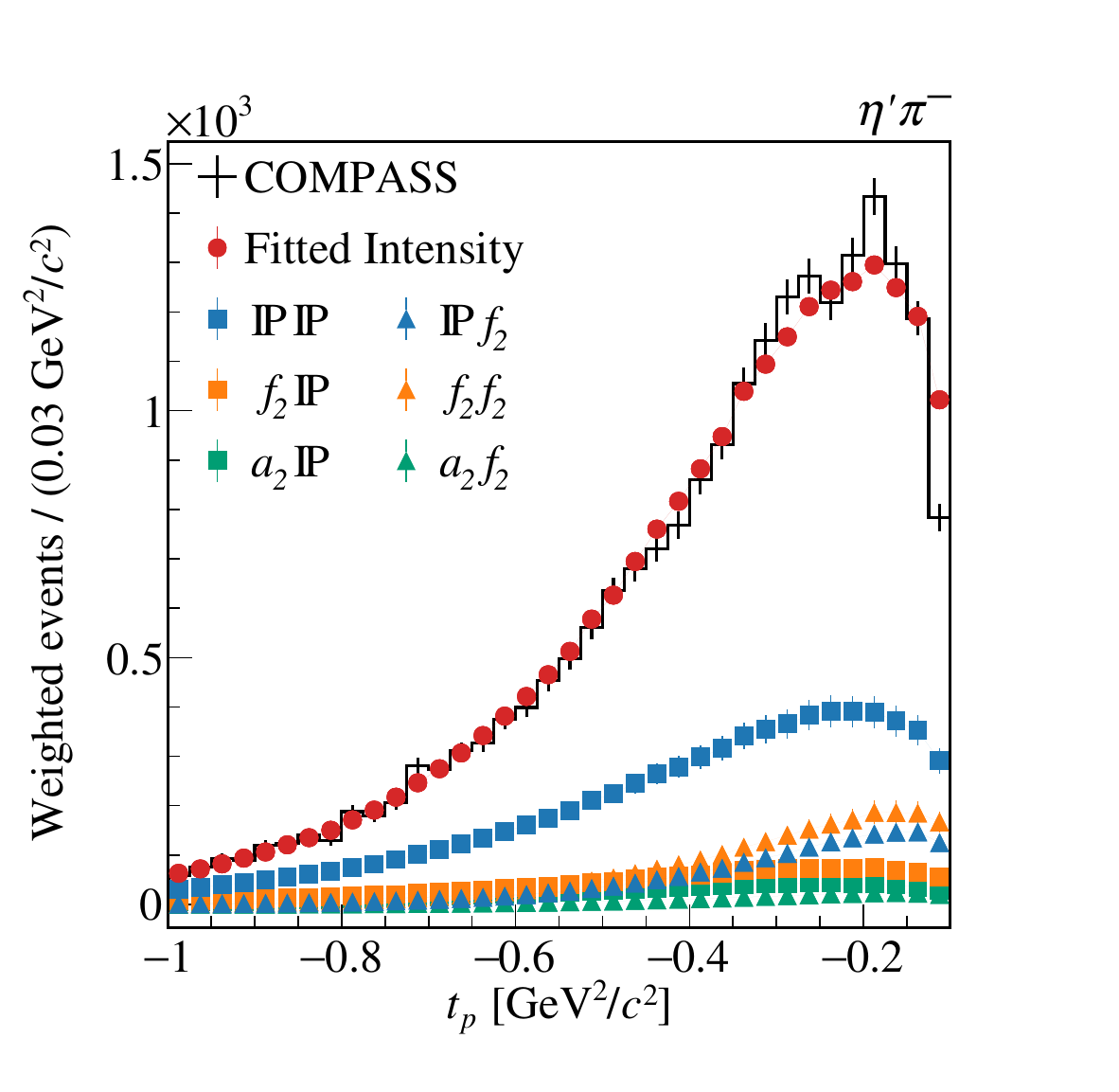} & 
\includegraphics[width=0.25\linewidth]{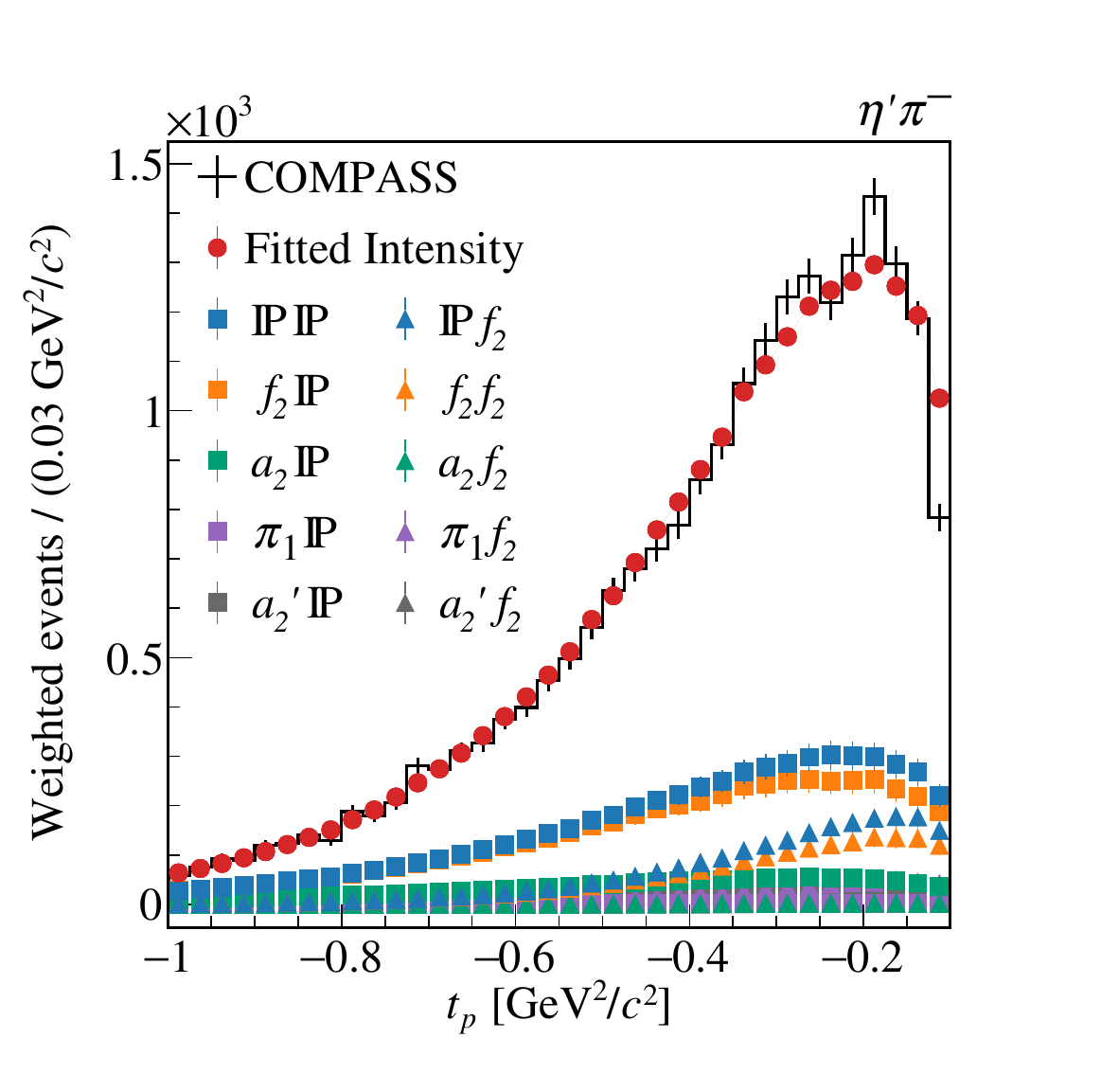} 
\end{tabular}
\caption{
$\eta^{\prime} \pi^-$ weighted intensity distributions dependence on
$t_p$. Individual contributions. Left: \mbox{KGR} fit;
Center-left: \mbox{$\text{KGR}+\pi_1+a_2^\prime$} fit; Center-right: \mbox{JPAC}
fit; Right: \mbox{$\text{JPAC}+\pi_1+a_2^\prime$} fit. Conventions as
in~\cref{figsup:etaprimeweight1}.
}
\label{figsup:etaprimeweight4_1}
\end{figure*}

\begin{figure*}[!h]
\begin{tabular}{cccc}
\includegraphics[width=0.25\linewidth]{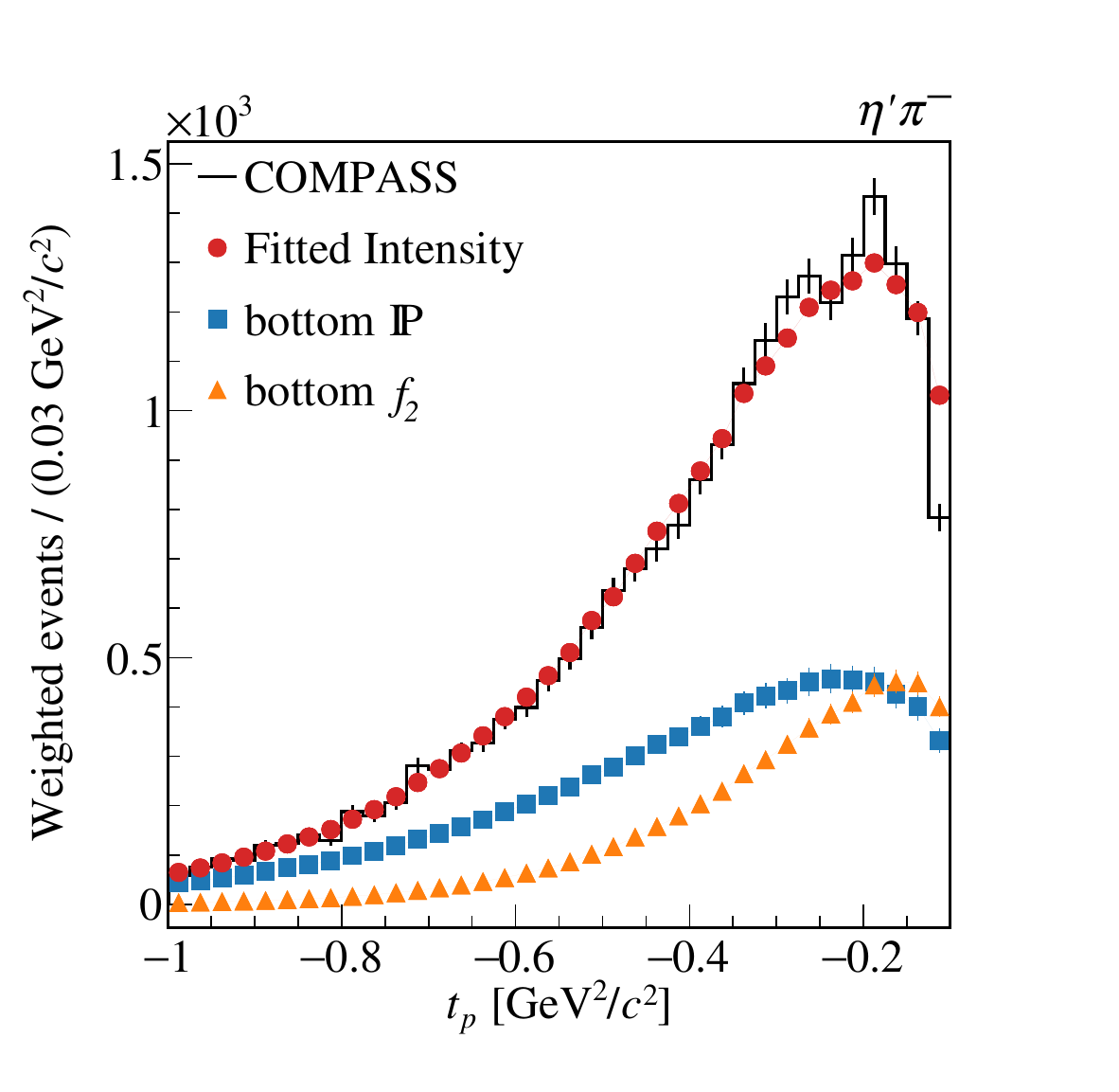} & 
\includegraphics[width=0.25\linewidth]{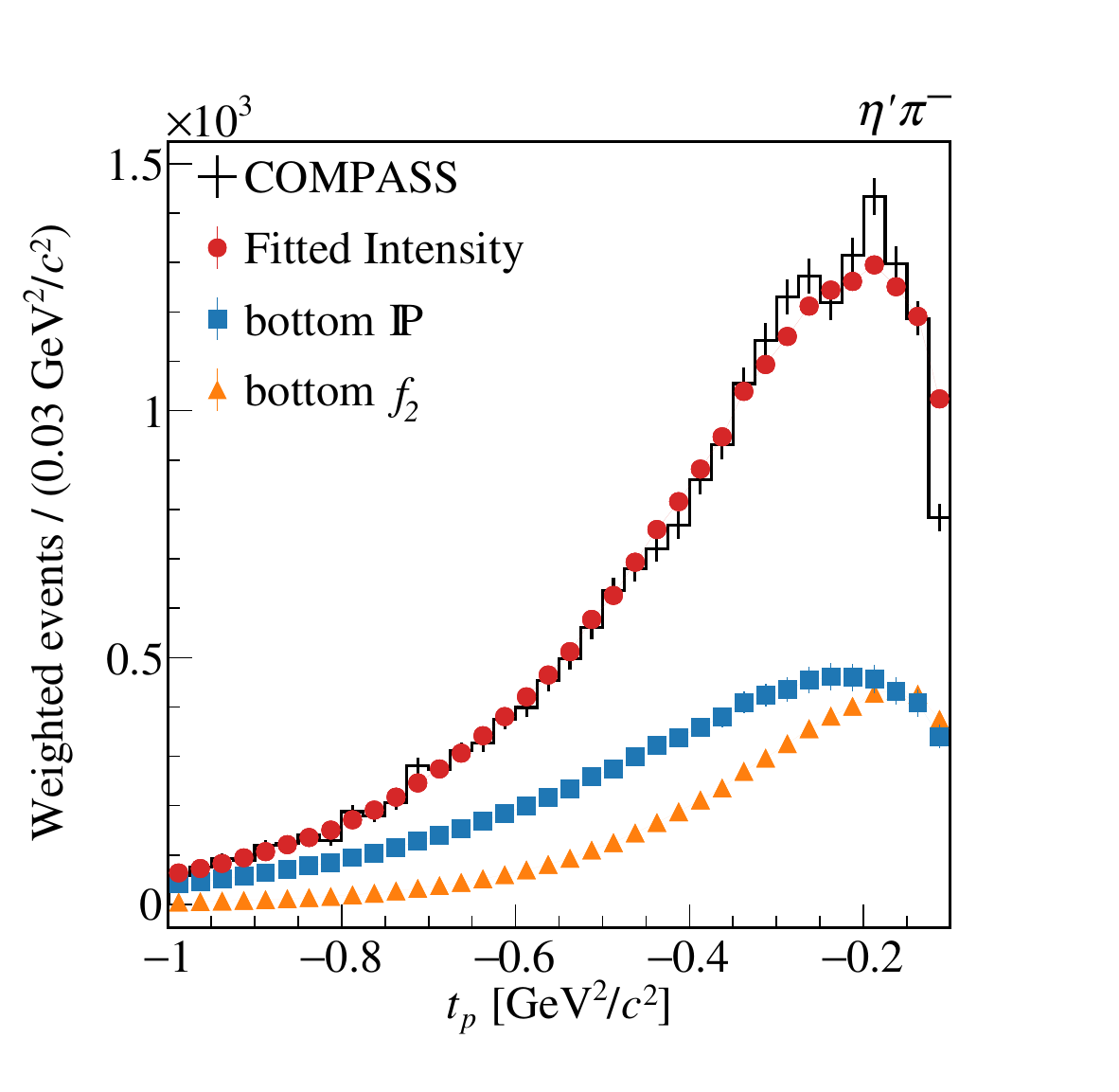} & 
\includegraphics[width=0.25\linewidth]{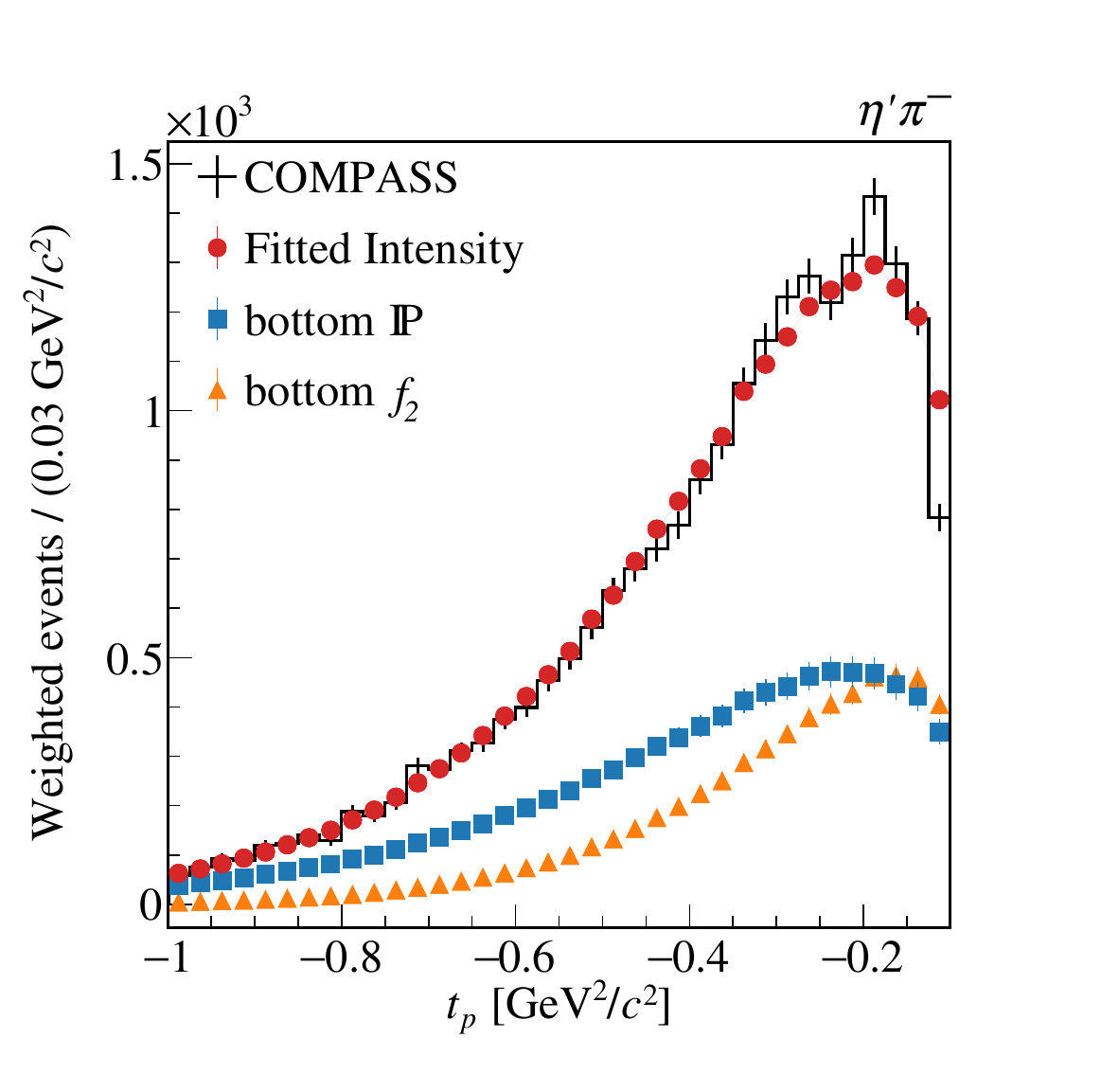} & 
\includegraphics[width=0.25\linewidth]{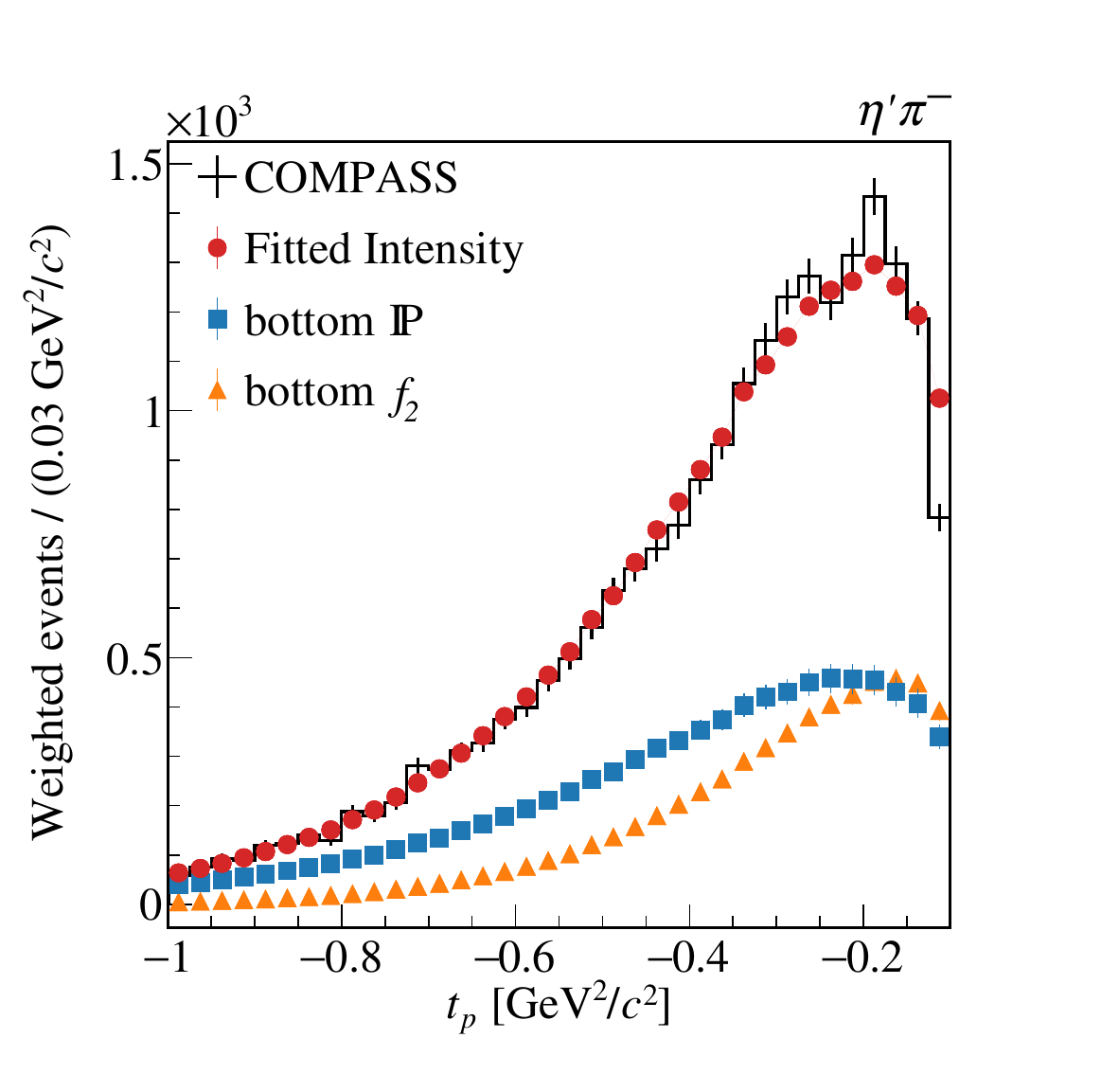} 
\end{tabular}
\caption{
$\eta^{\prime} \pi^-$ weighted intensity distributions dependence on
$t_p$. Bottom Reggeon exchange contributions coherently summed. Left: \mbox{KGR}
fit; Center-left: \mbox{$\text{KGR}+\pi_1+a_2^\prime$} fit;
Center-right: \mbox{JPAC} fit; Right: \mbox{$\text{JPAC}+\pi_1+a_2^\prime$}
fit. Conventions as in~\cref{figsup:etaweight4_2}.
}
\label{figsup:etaprimeweight4_2}
\end{figure*}

\begin{figure*}[!h]
\begin{tabular}{cccc}
\includegraphics[width=0.25\linewidth]{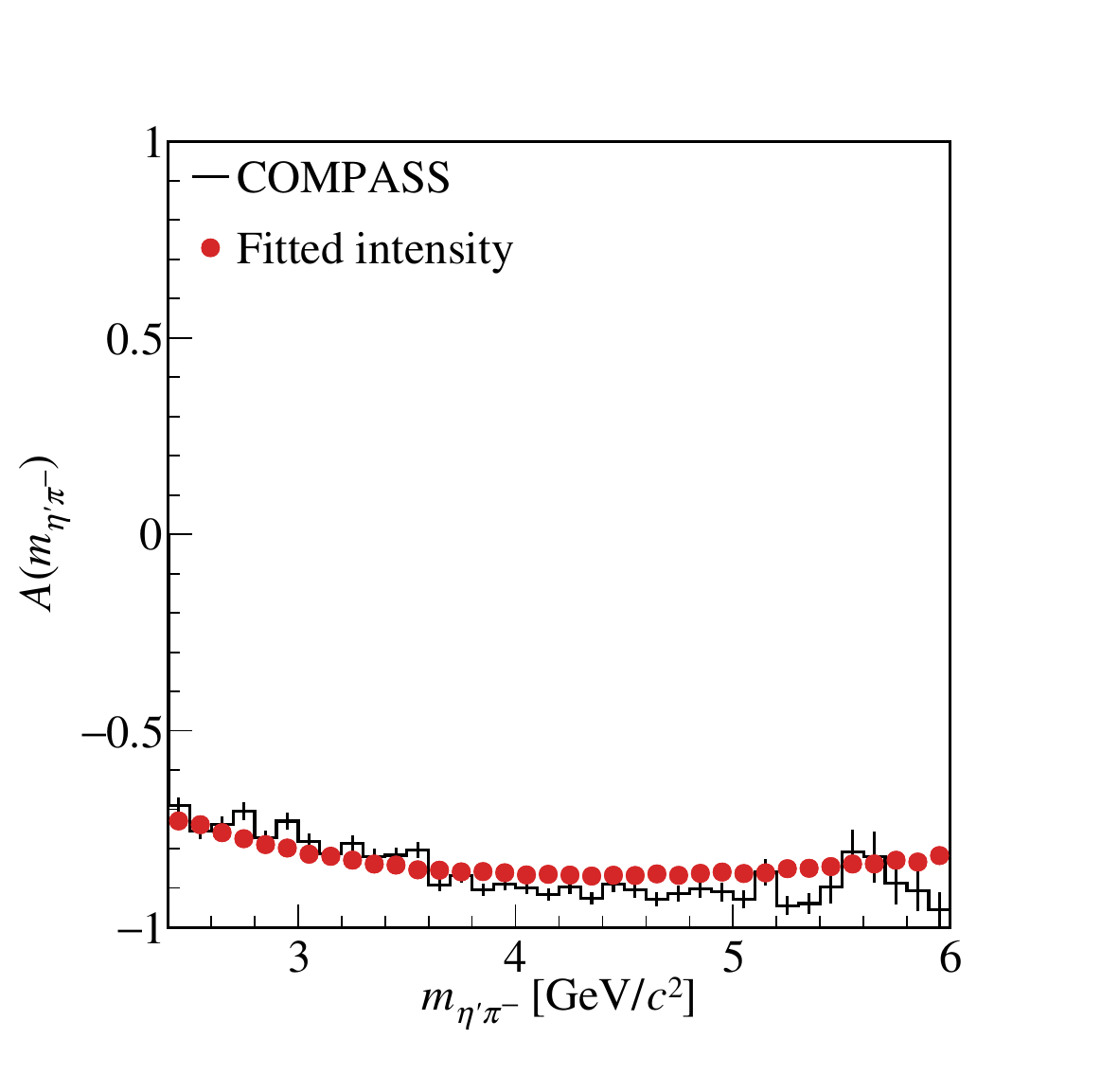} & 
\includegraphics[width=0.25\linewidth]{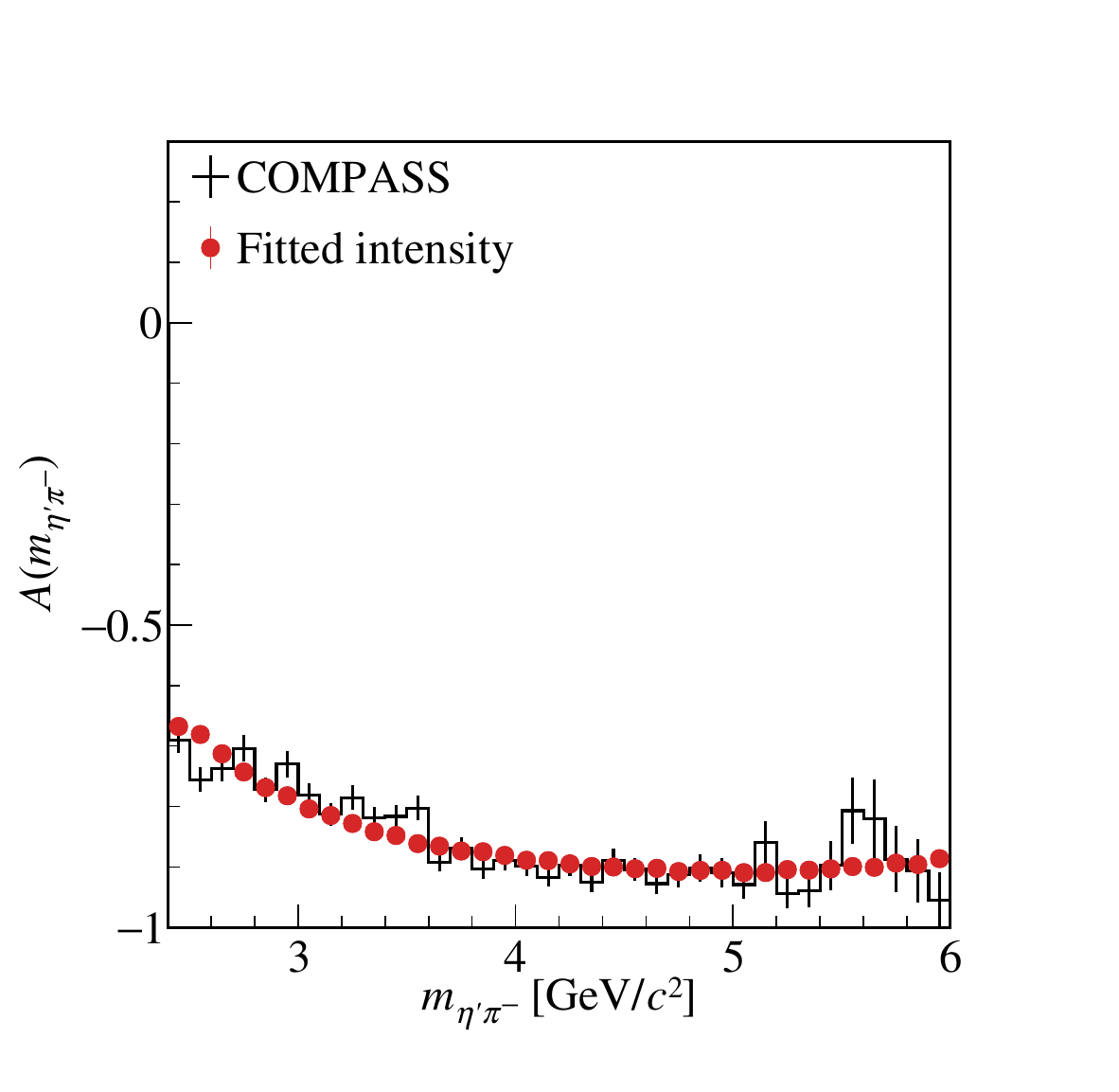} & 
\includegraphics[width=0.25\linewidth]{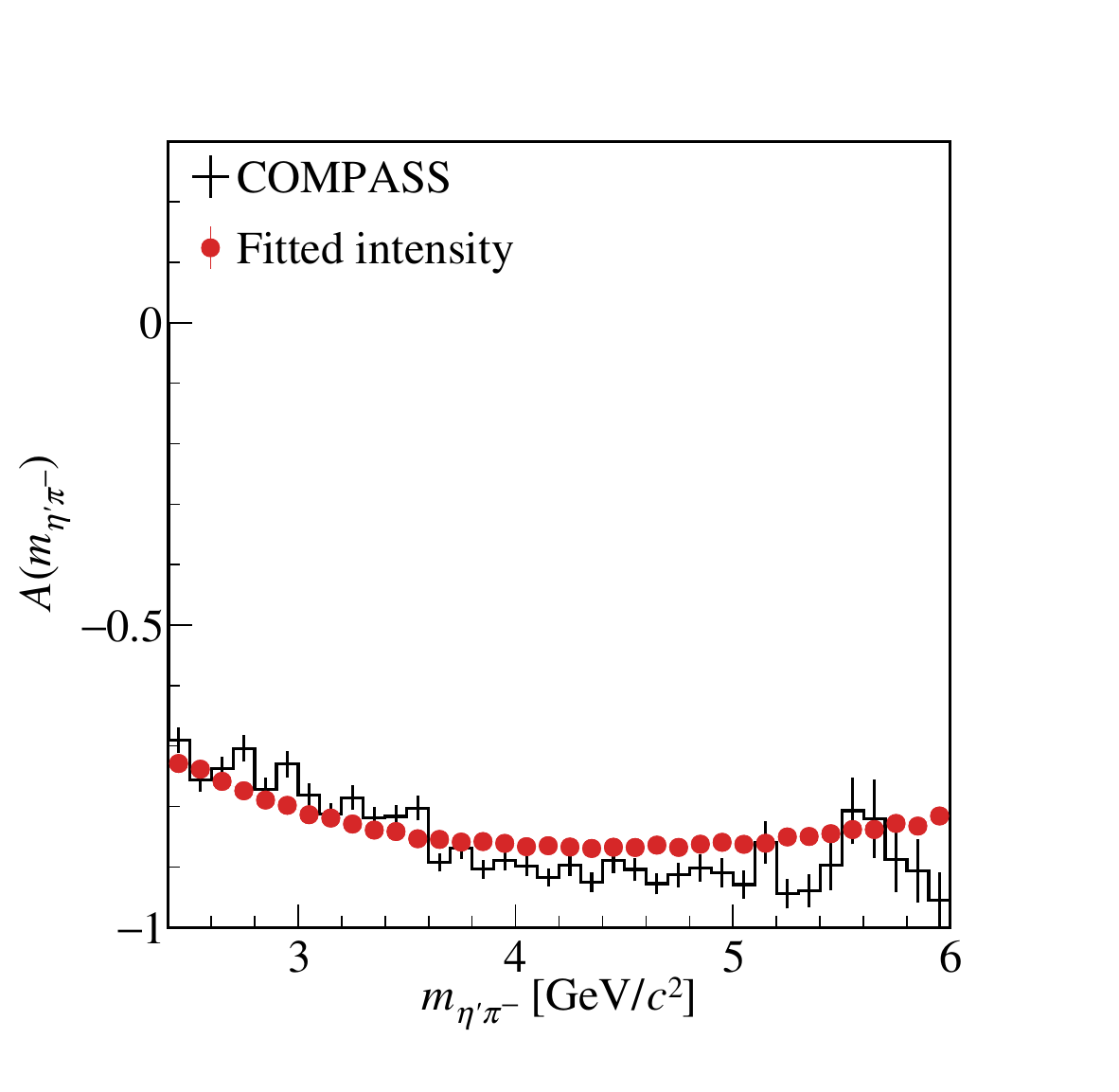} & 
\includegraphics[width=0.25\linewidth]{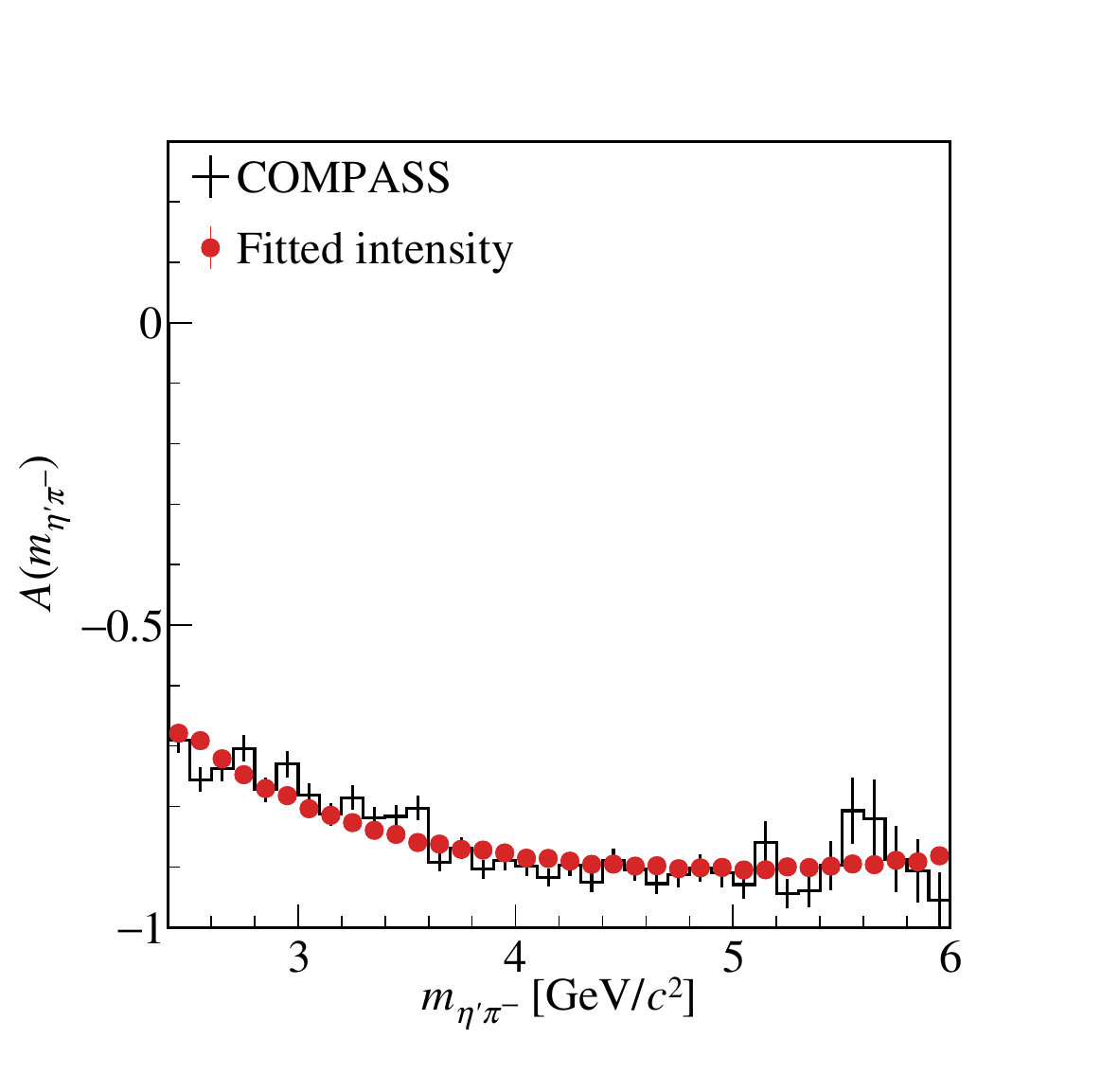} 
\end{tabular}
\caption{
$\eta^\prime \pi^-$ weighted forward-backward asymmetry distribution dependence
on $m_{\eta^\prime \pi}$. Left: \mbox{KGR} fit;
Center-left: \mbox{$\text{KGR}+\pi_1+a_2^\prime$} fit; Center-right: \mbox{JPAC}
fit; Right: \mbox{$\text{JPAC}+\pi_1+a_2^\prime$} fit. Conventions as
in~\cref{figsup:etaweight5}.
}
\label{figsup:etaprimeweight5}
\end{figure*}

In this section we compare the $\eta^\prime \pi^-$ weighted intensity
distributions of the fits using the \mbox{KGR}
(left), \mbox{$\text{KGR}+\pi_1+a_2^\prime$} (center-left), \mbox{JPAC}
(center-right) and \mbox{$\text{JPAC}+\pi_1+a_2^\prime$} (right) models to the
binned \compass data. The difference between \mbox{KGR} and \mbox{JPAC} fits
provide an estimation of the systematic uncertainty associated to the leading
ordinary exchanges. We present results depending on the invariant mass
distribution $m_{\eta^\prime \pi}$ (\cref{figsup:etaprimeweight1}),
$t_{\eta^\prime}$ (\cref{figsup:etaprimeweight2_1,figsup:etaprimeweight2_2}),
$t_\pi$ (\cref{figsup:etaprimeweight3_1,figsup:etaprimeweight3_2}.) and $t_p$
(\cref{figsup:etaprimeweight4_1,figsup:etaprimeweight4_2}.) We provide the
individual contributions of the exchanges as well as top $a_2$, top $f_2$, top
$\Pom$, top $a_2^\prime$ and top-$\pi_1$aggregates for $t_{\eta^\prime}$ and
$t_\pi$ distributions
(see~\cref{figsup:etaprimeweight2_2,figsup:etaprimeweight3_2}, and bottom $f_2$
and bottom $\Pom$ aggregates for $t_p$ distribution
(see \cref{figsup:etaprimeweight4_2}.) Finally, we show the results for the
forward-backward asymmetry in~\cref{figsup:etaprimeweight5}.

\clearpage
\newpage
\section*{Fits in the $m_{\eta \pi}=[4,6]\gev$ range}
\begin{figure*}[!h]
\begin{tabular}{cccc}
\includegraphics[width=0.25\linewidth]{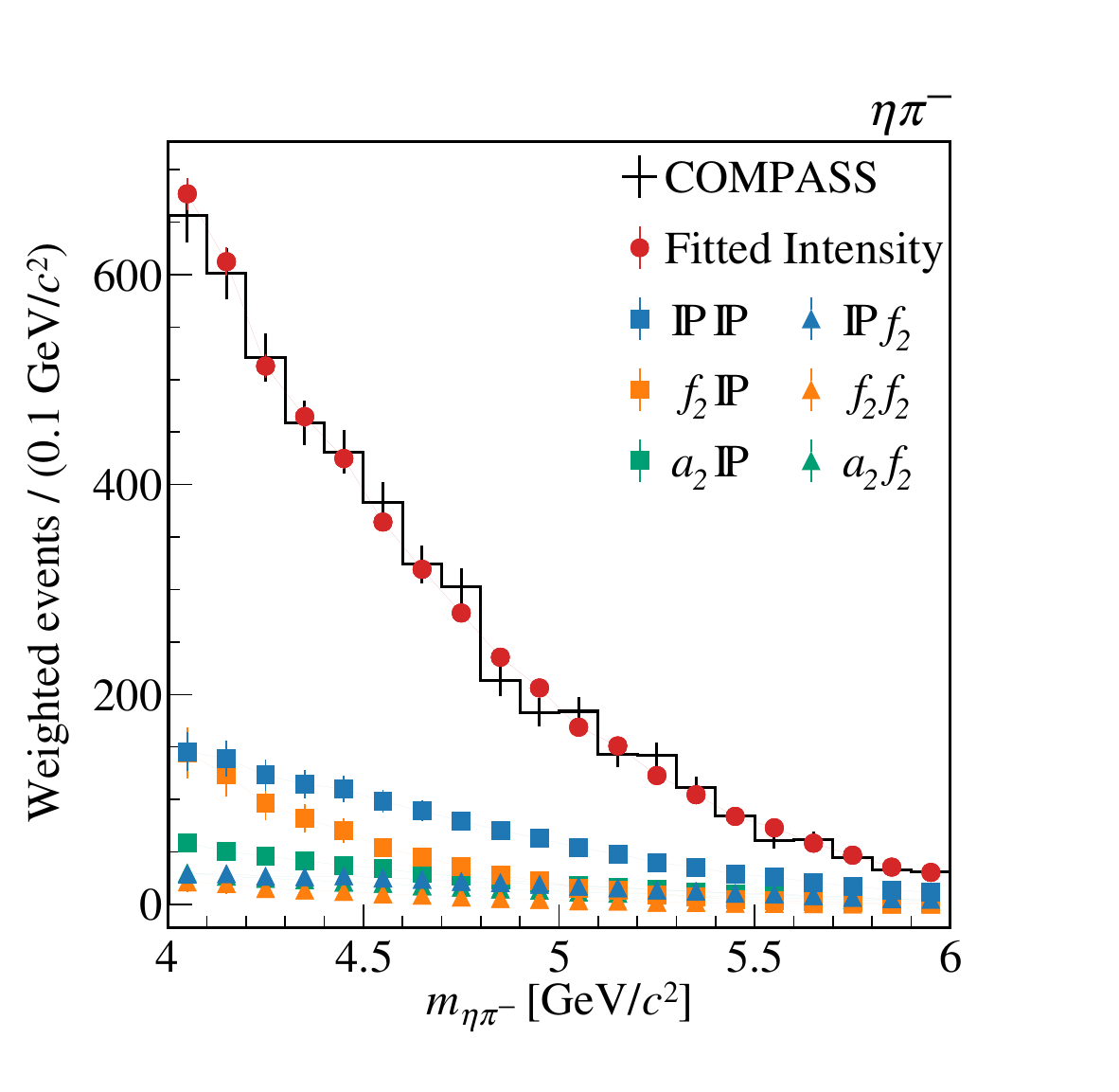} & 
\includegraphics[width=0.25\linewidth]{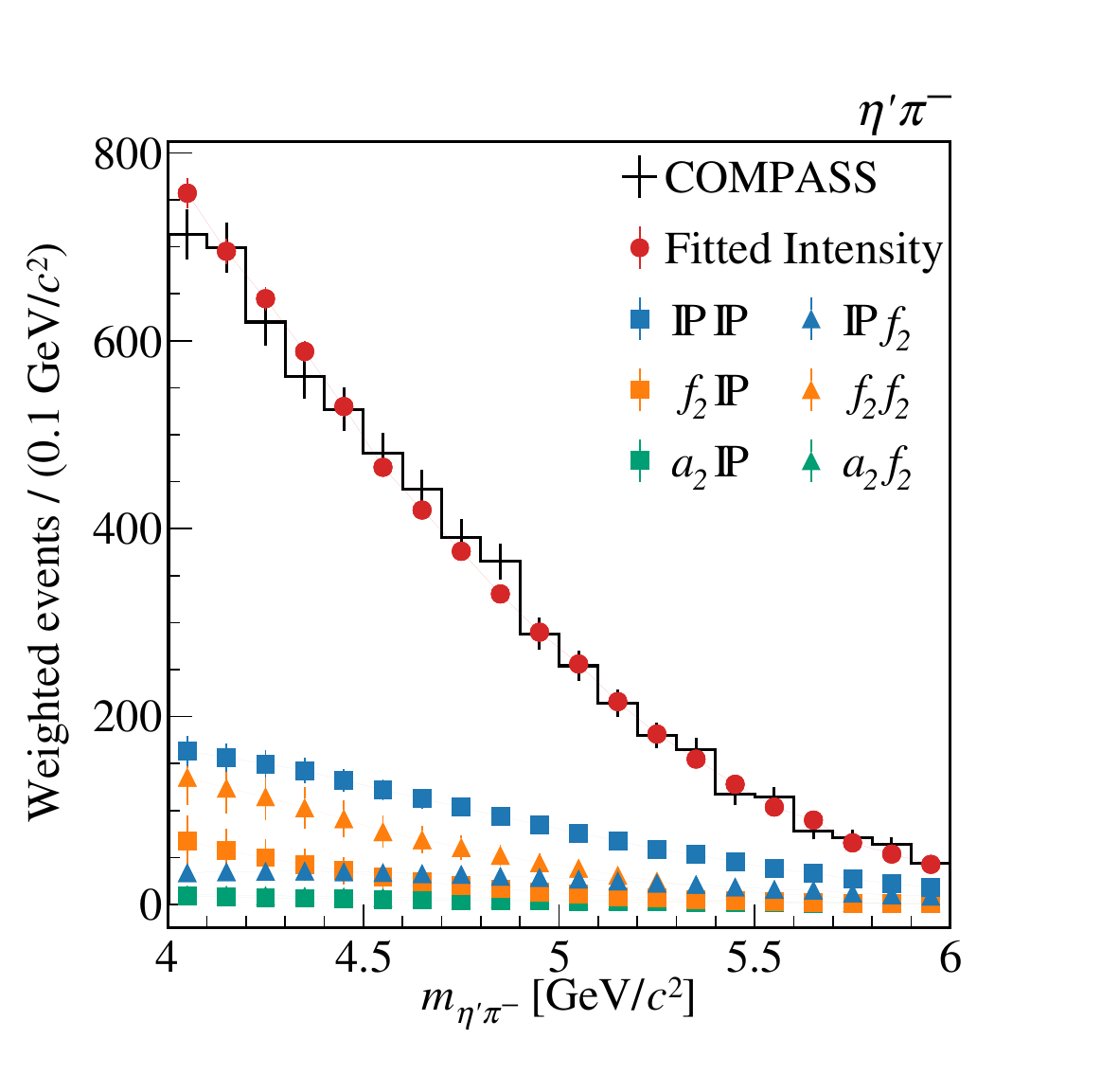} &
\includegraphics[width=0.25\linewidth]{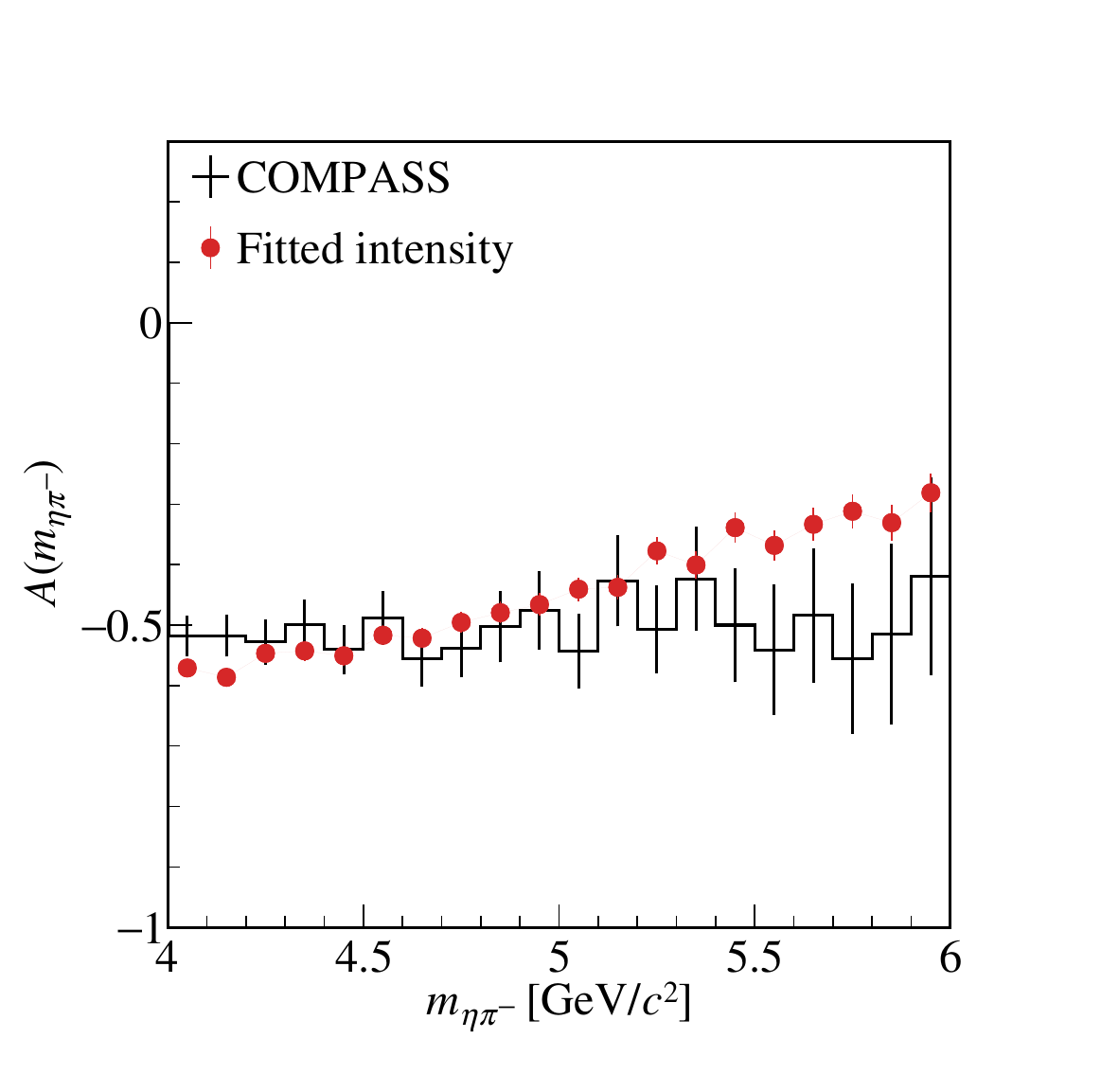} &
\includegraphics[width=0.25\linewidth]{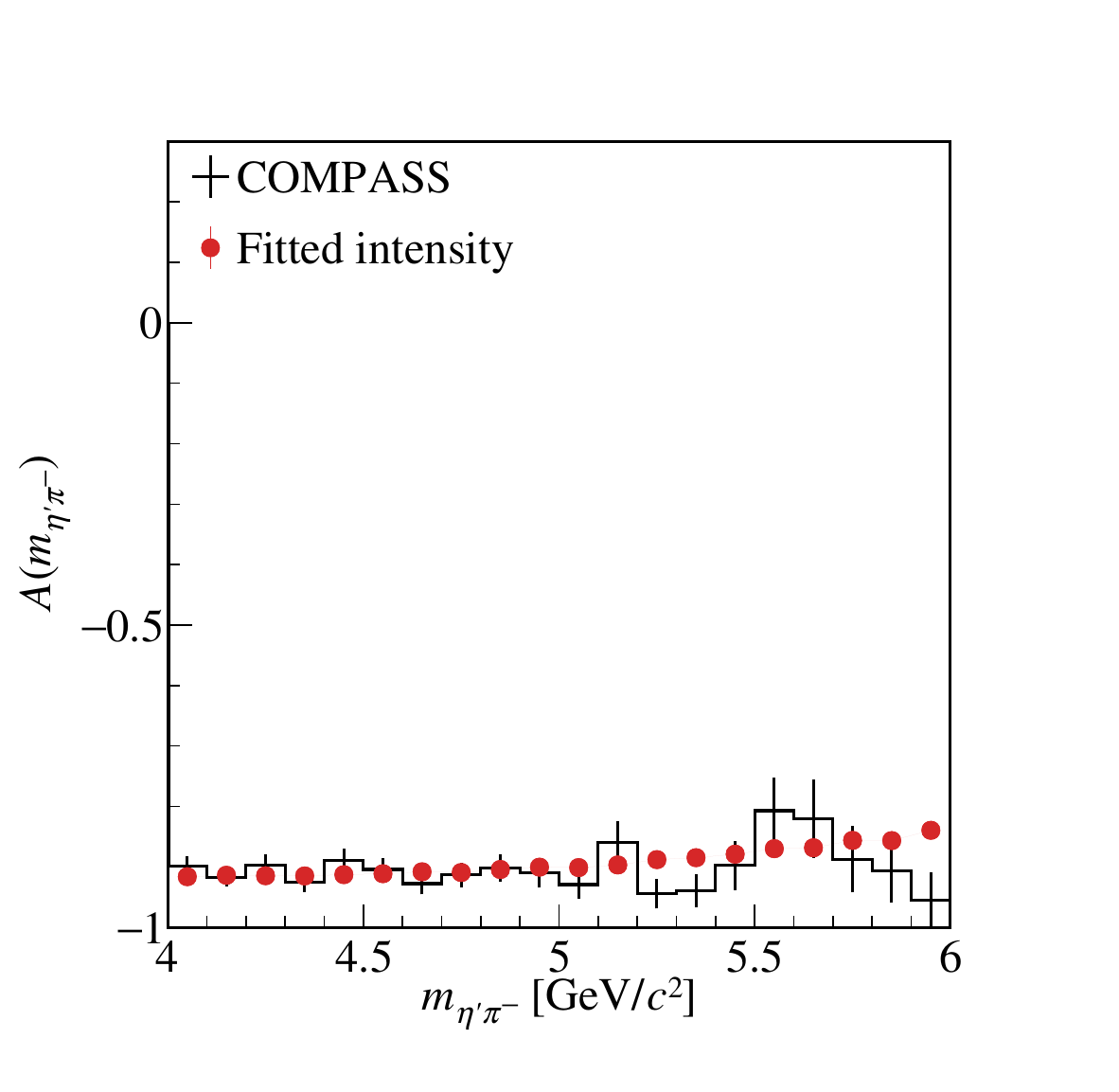}
\end{tabular}
\caption{
Left ($\eta \pi^-$) and center left ($\eta^\prime \pi^-$): Weighted intensity
distributions dependence on $m_{\eta \pi}$. Center right ($\eta \pi^-$) and
right ($\eta^\prime \pi^-$): Weighted asymmetry distributions dependence on
$m_{\eta \pi}$.
}
\label{figsup:he1}
\end{figure*}

For completeness, we show the results for the base model in the high invariant
mass region $m_{\eta \pi}=[4,6]\gev$ for both $\eta \pi^-$ and
$\eta^\prime \pi^-$ channels using the \mbox{KGR} Regge trajectories. Results
are similar for the \mbox{JPAC} trajectories. Overall results are good in this
energy range with some deviation in the description of the $\eta \pi^-$
asymmetry for the highest invariant masses, center-right plot
in~\cref{figsup:he1}.

\begin{figure*}[!h]
\begin{tabular}{cccc}
\includegraphics[width=0.25\linewidth]{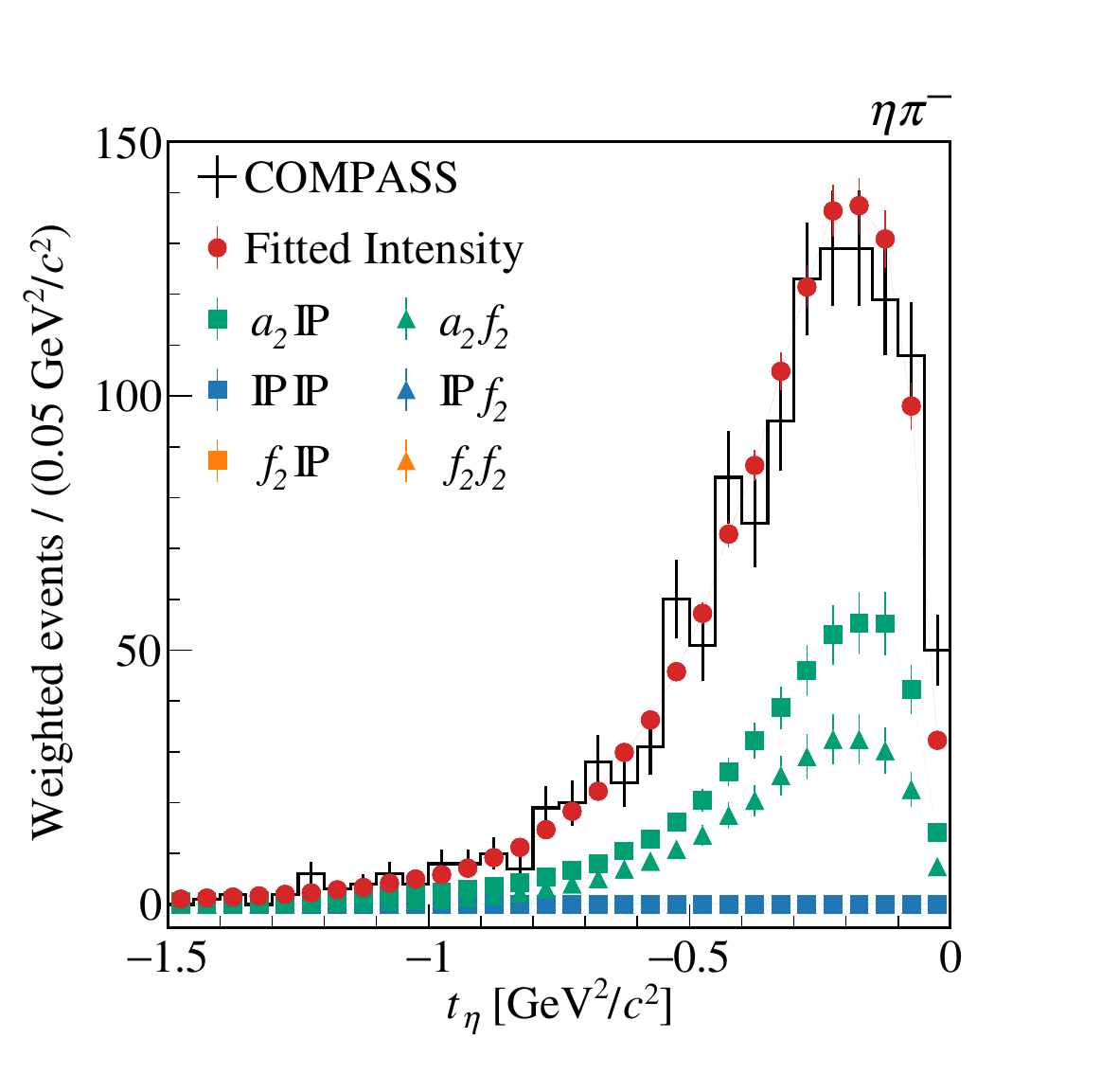} & 
\includegraphics[width=0.25\linewidth]{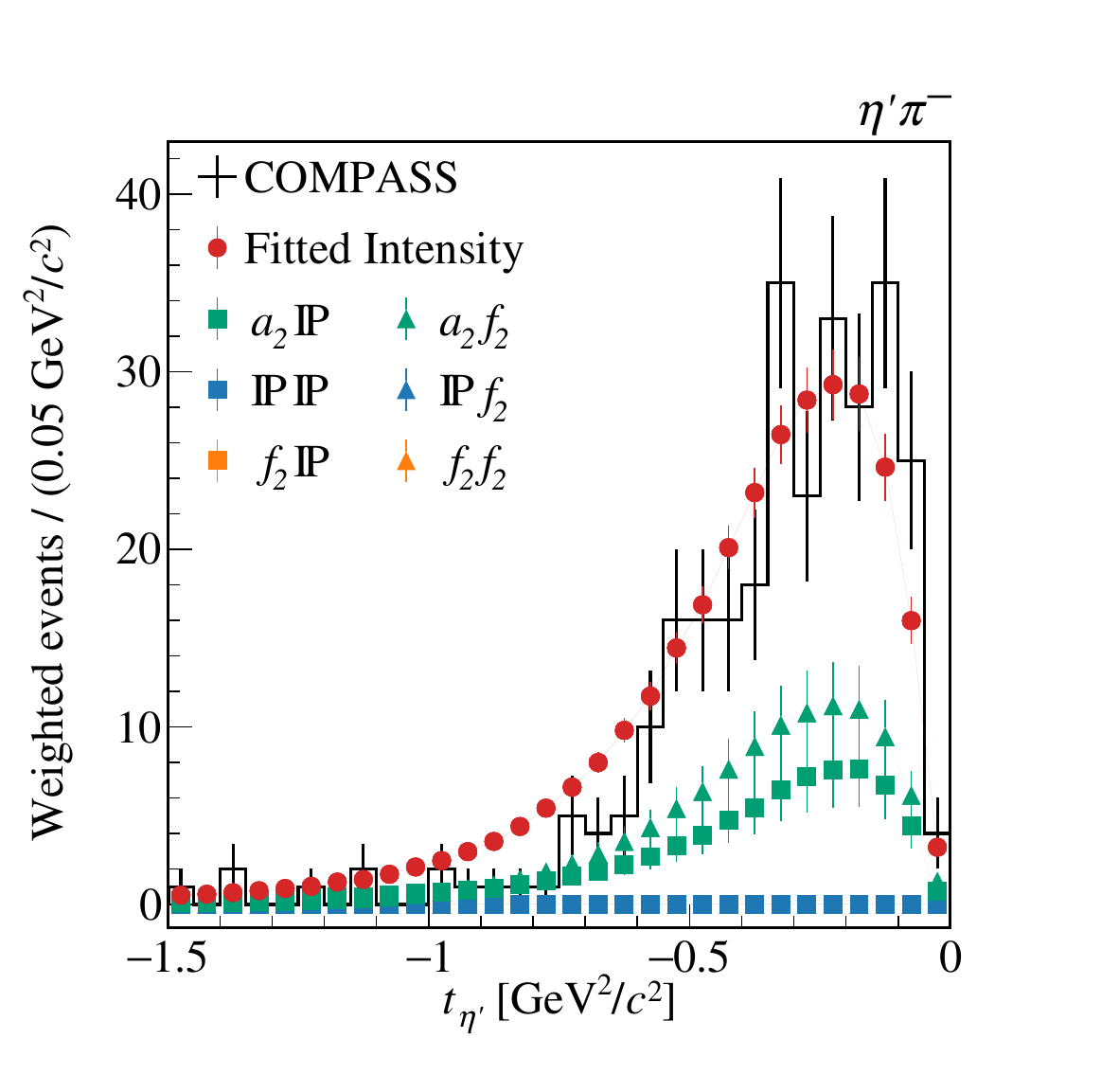} &
\includegraphics[width=0.25\linewidth]{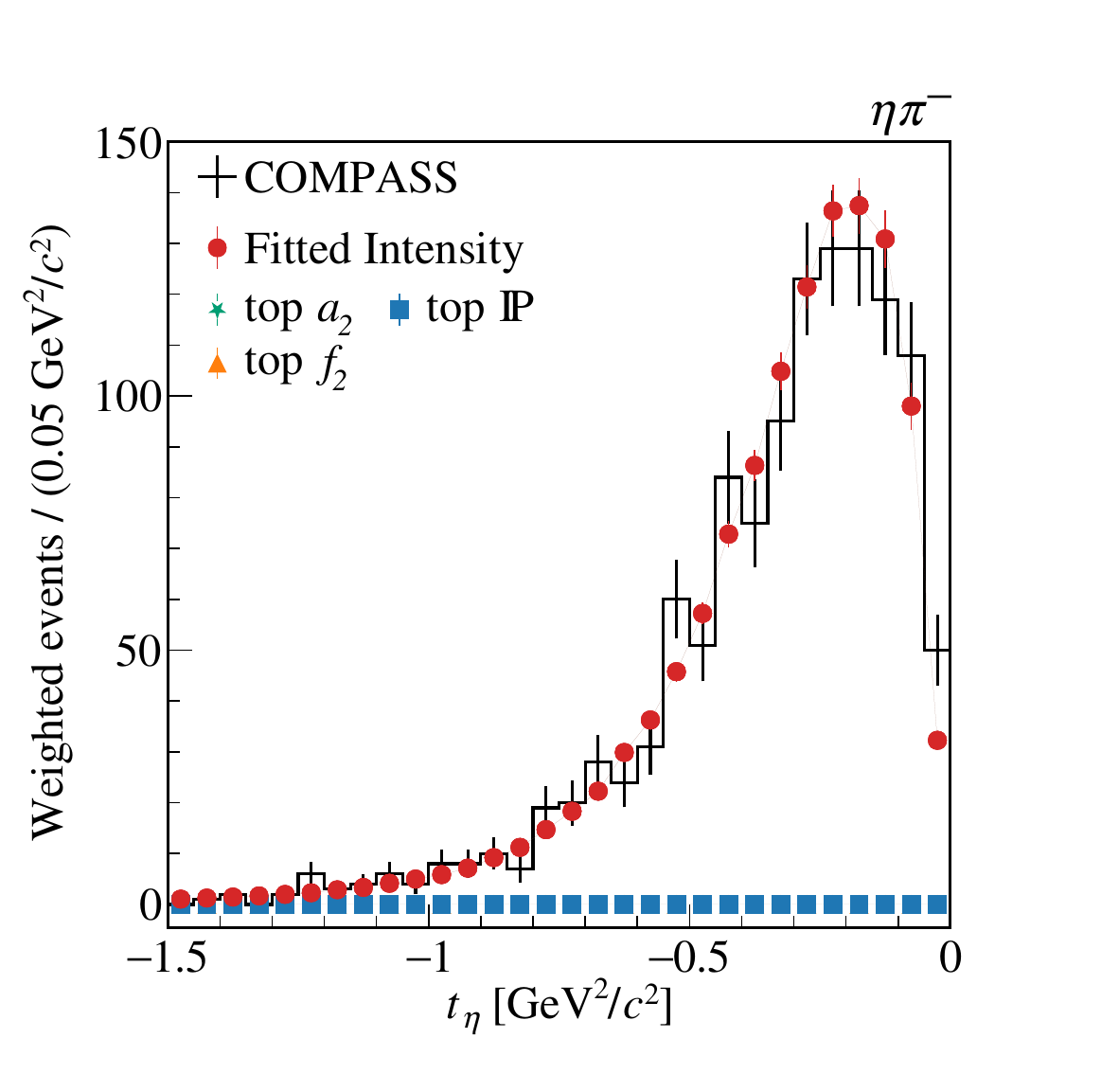} &
\includegraphics[width=0.25\linewidth]{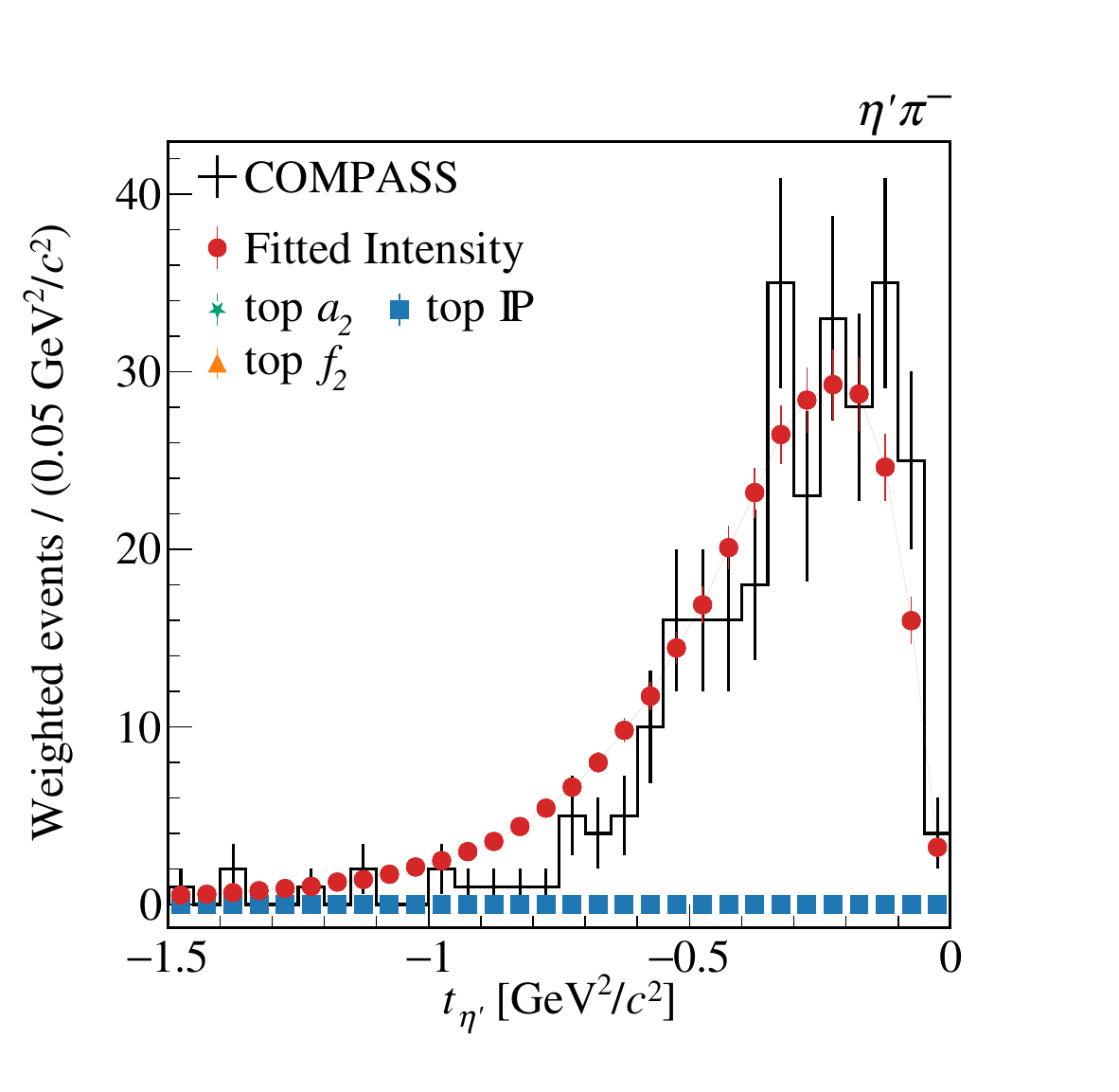}
\end{tabular}
\caption{
Weighted intensity distributions dependence on $t_{\eta}$. Left ($\eta \pi^-$)
and center left ($\eta^\prime \pi^-$): individual contributions; center right
($\eta \pi^-$) and right ($\eta^\prime \pi^-$): coherently summed top
contributions.
}
\label{figsup:he2}
\end{figure*}

\begin{figure*}[!h]
\begin{tabular}{cccc}
\includegraphics[width=0.25\linewidth]{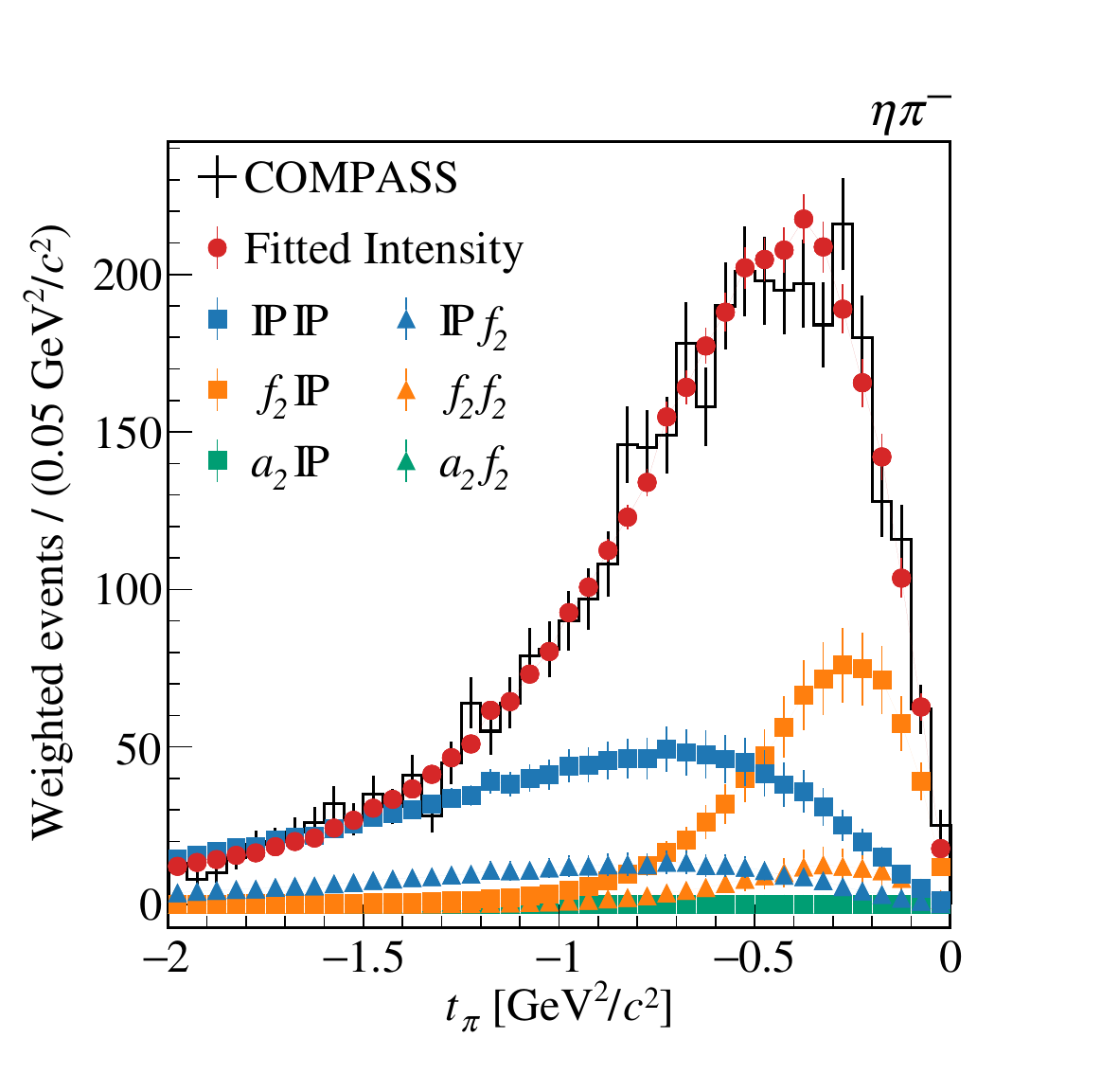} & 
\includegraphics[width=0.25\linewidth]{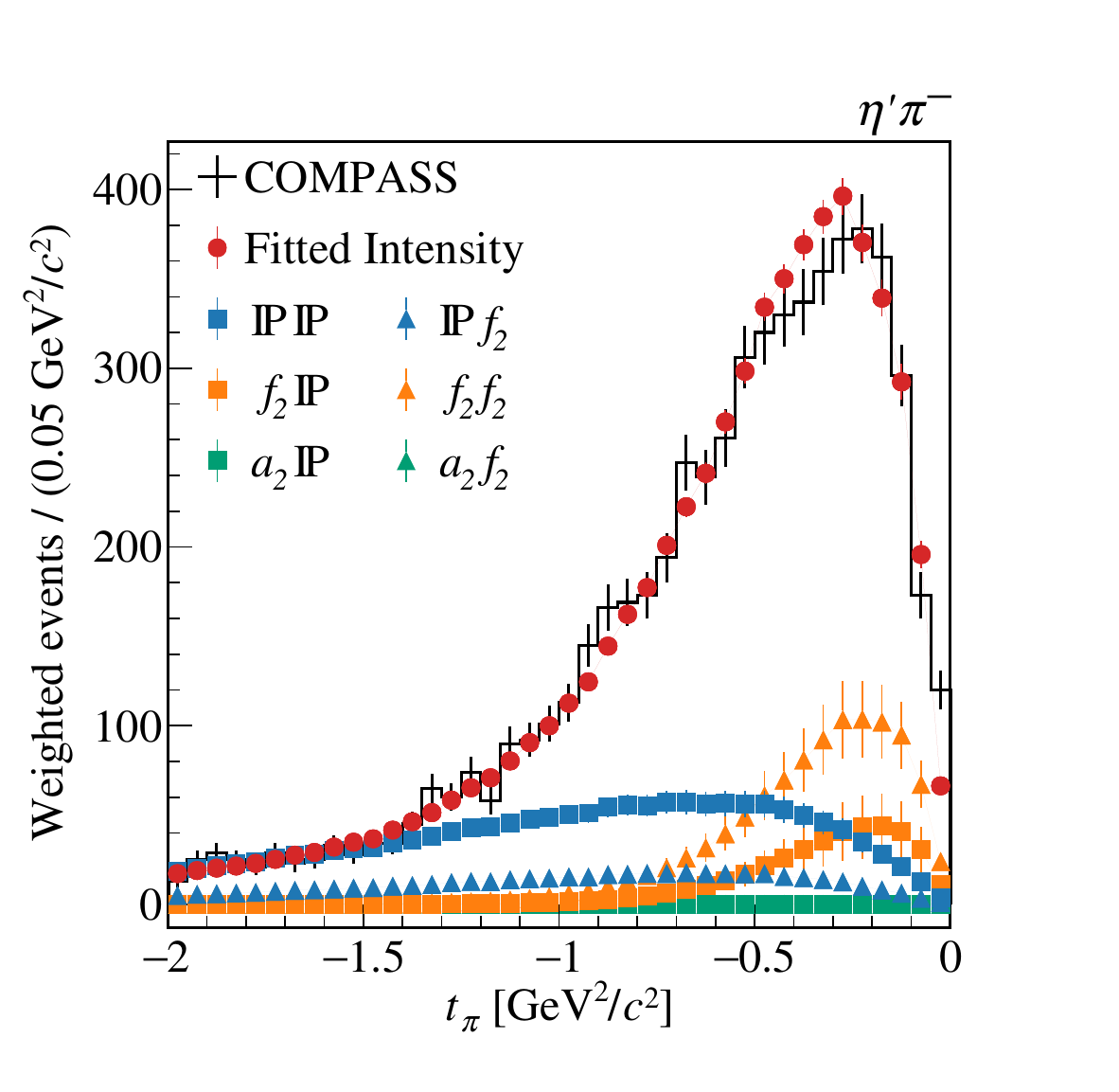} &
\includegraphics[width=0.25\linewidth]{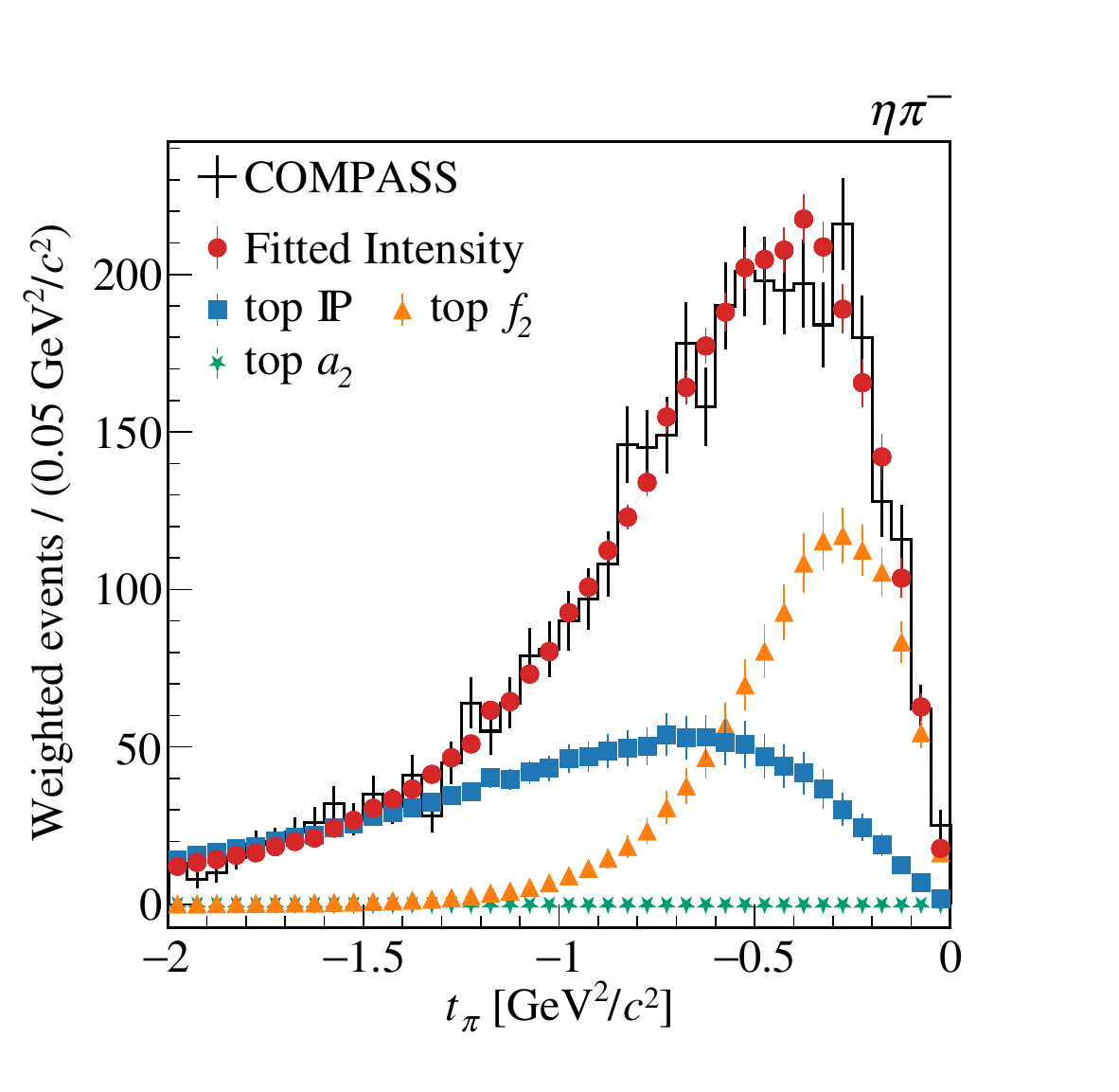} &
\includegraphics[width=0.25\linewidth]{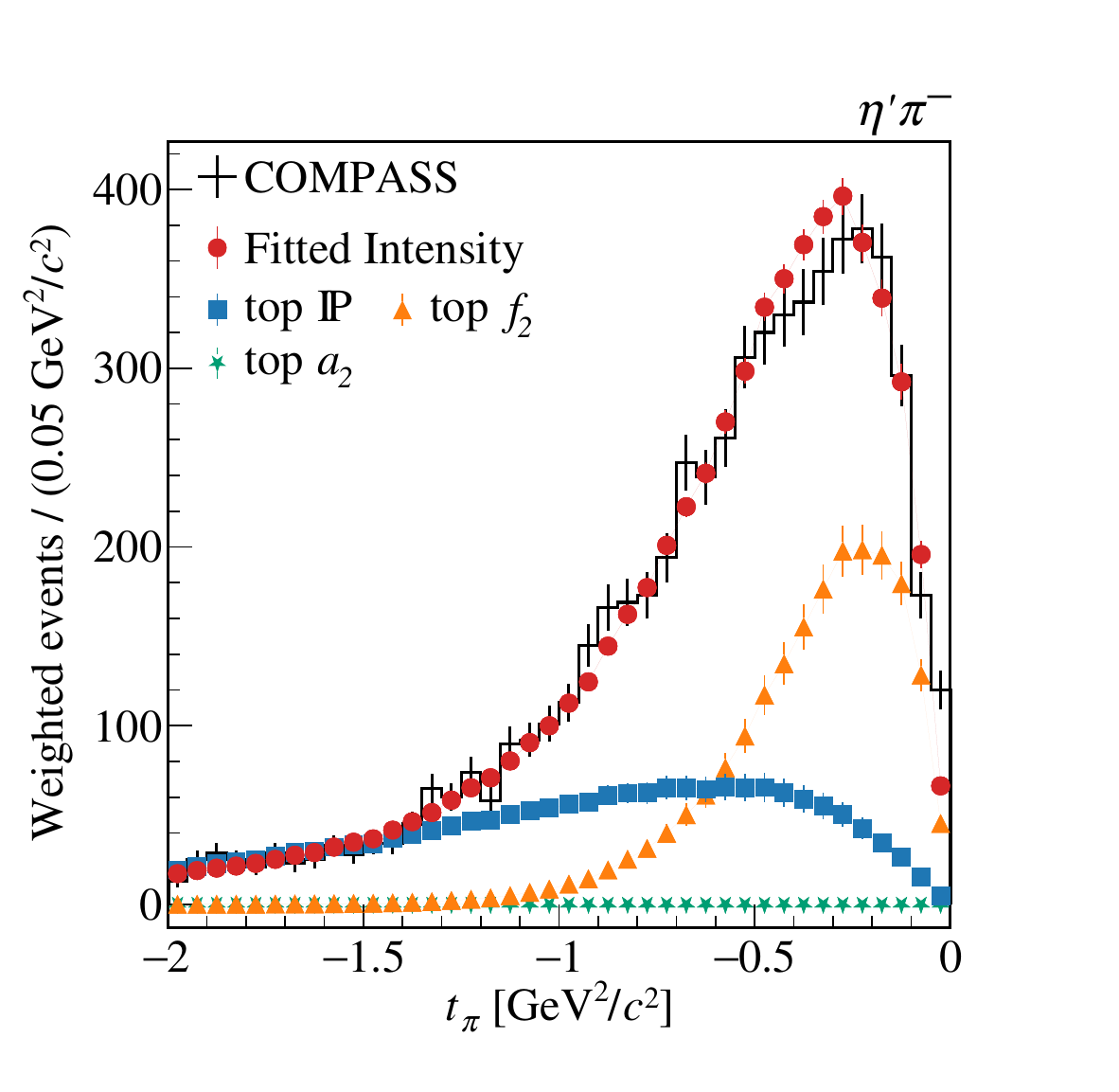}
\end{tabular}
\caption{
Weighted intensity distributions dependence on $t_{\pi}$. Left ($\eta \pi^-$)
and center left ($\eta^\prime \pi^-$): individual contributions; center right
($\eta \pi^-$) and right ($\eta^\prime \pi^-$): coherently summed top
contributions.
}
\label{figsup:he3}
\end{figure*}

\begin{figure*}[!h]
\begin{tabular}{cccc}
\includegraphics[width=0.25\linewidth]{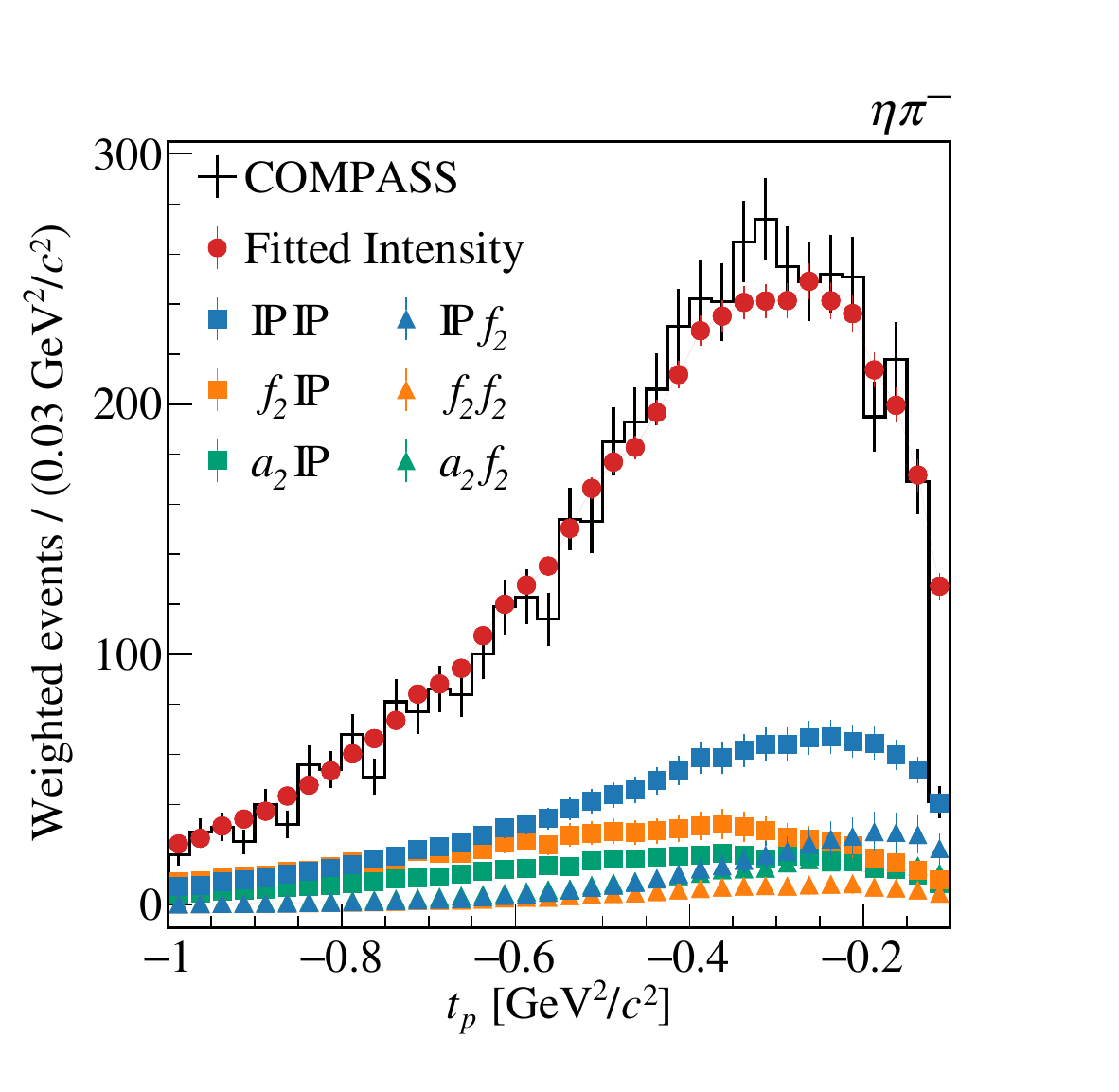} & 
\includegraphics[width=0.25\linewidth]{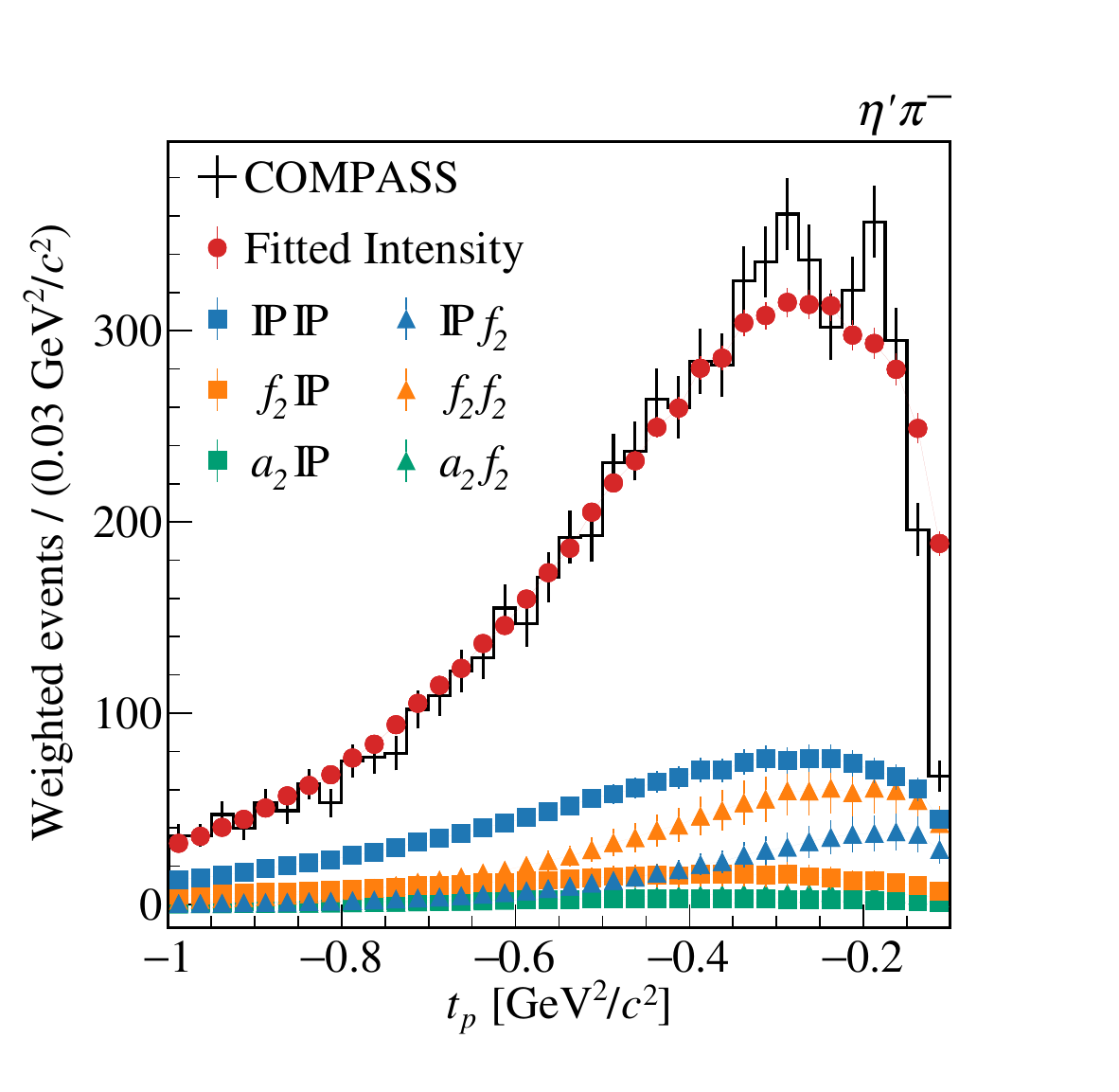} &
\includegraphics[width=0.25\linewidth]{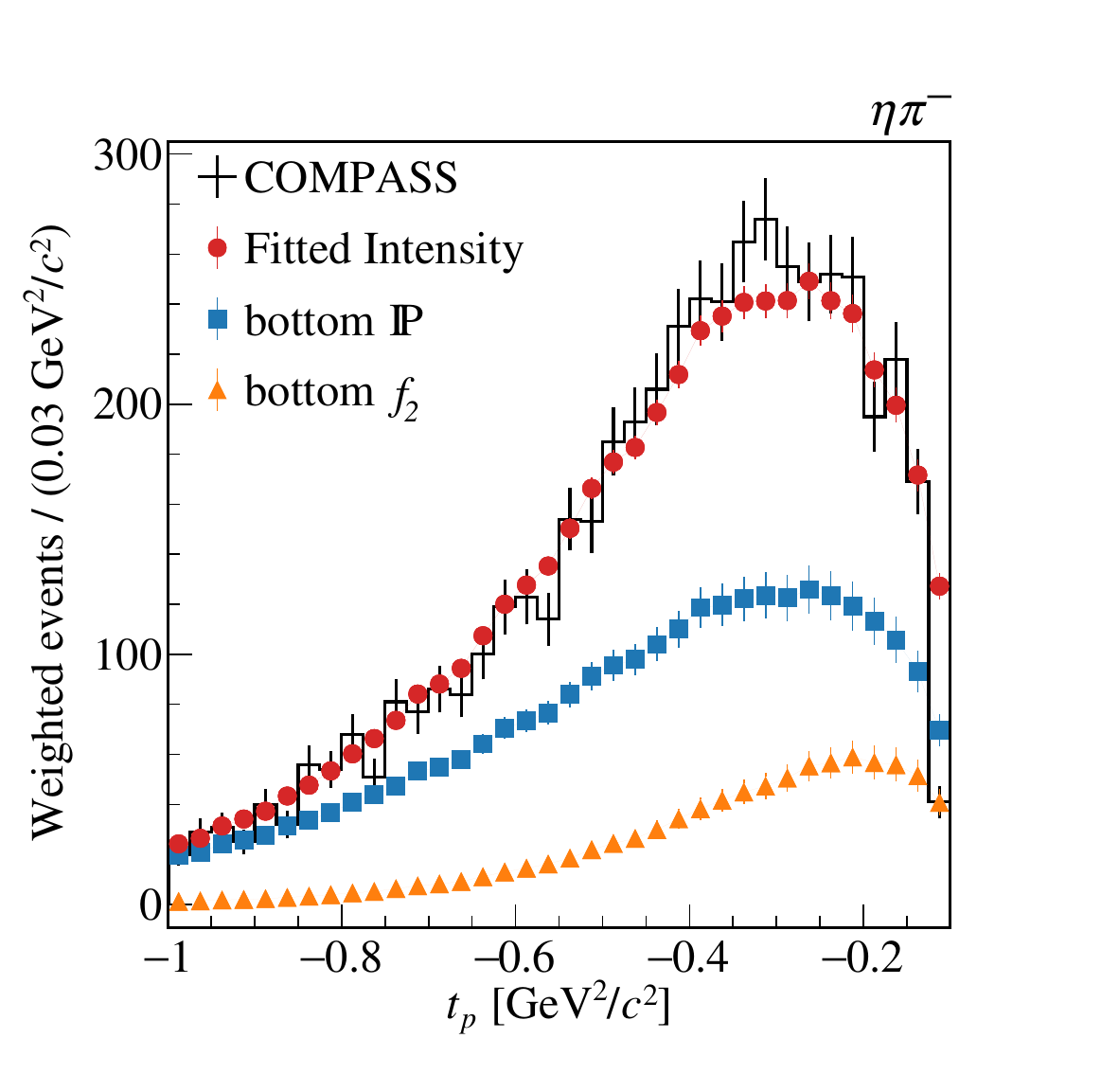} &
\includegraphics[width=0.25\linewidth]{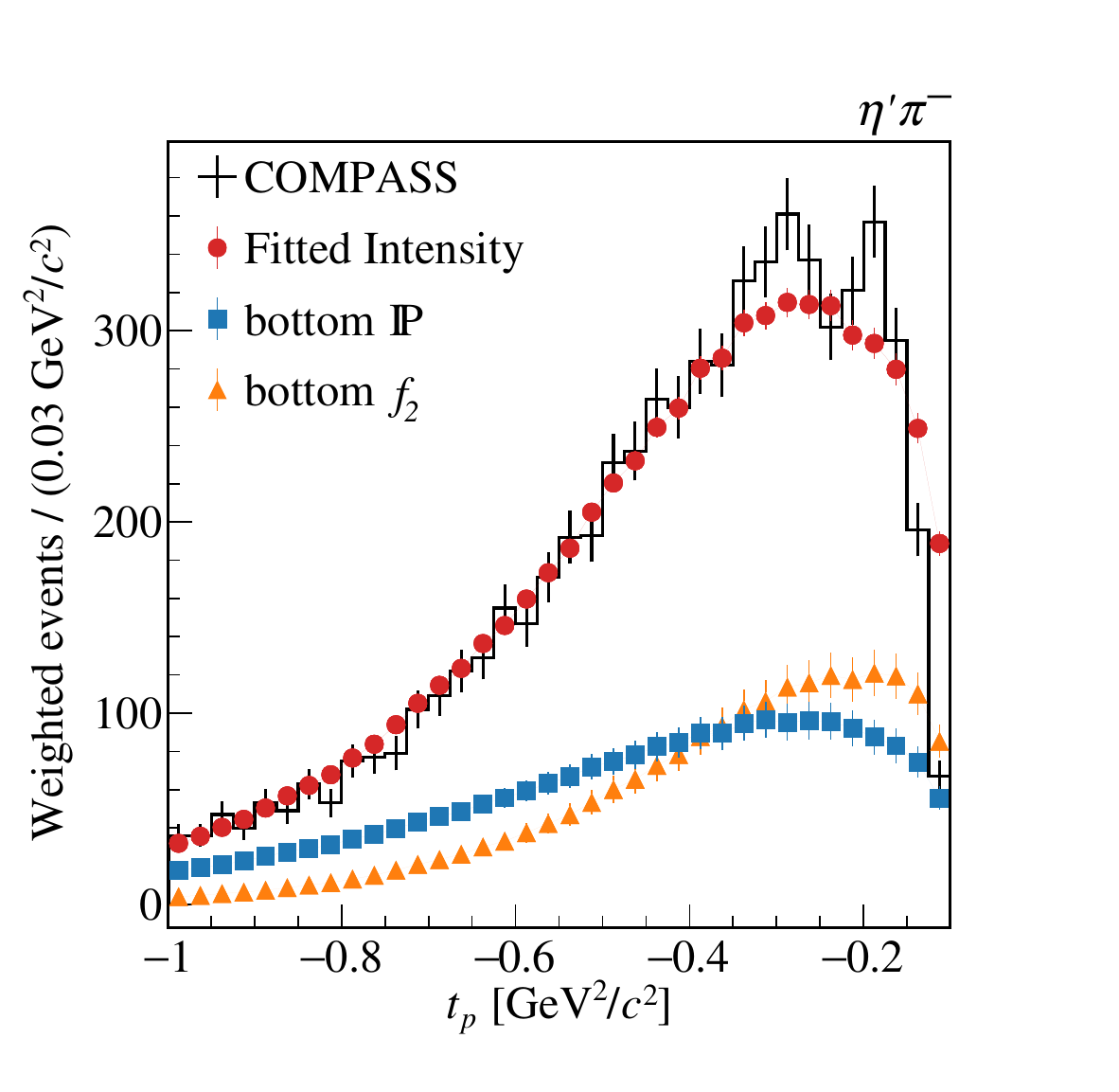}
\end{tabular}
\caption{
Weighted intensity distributions dependence on $t_{p}$. Left ($\eta \pi^-$) and
center left ($\eta^\prime \pi^-$): individual contributions; center right
($\eta \pi^-$) and right ($\eta^\prime \pi^-$): coherently summed top
contributions.
}
\label{figsup:he4}
\end{figure*}

\section*{Pomeron studies}
\begin{figure*}[!h]
\begin{tabular}{cccc}
\includegraphics[width=0.3\linewidth]{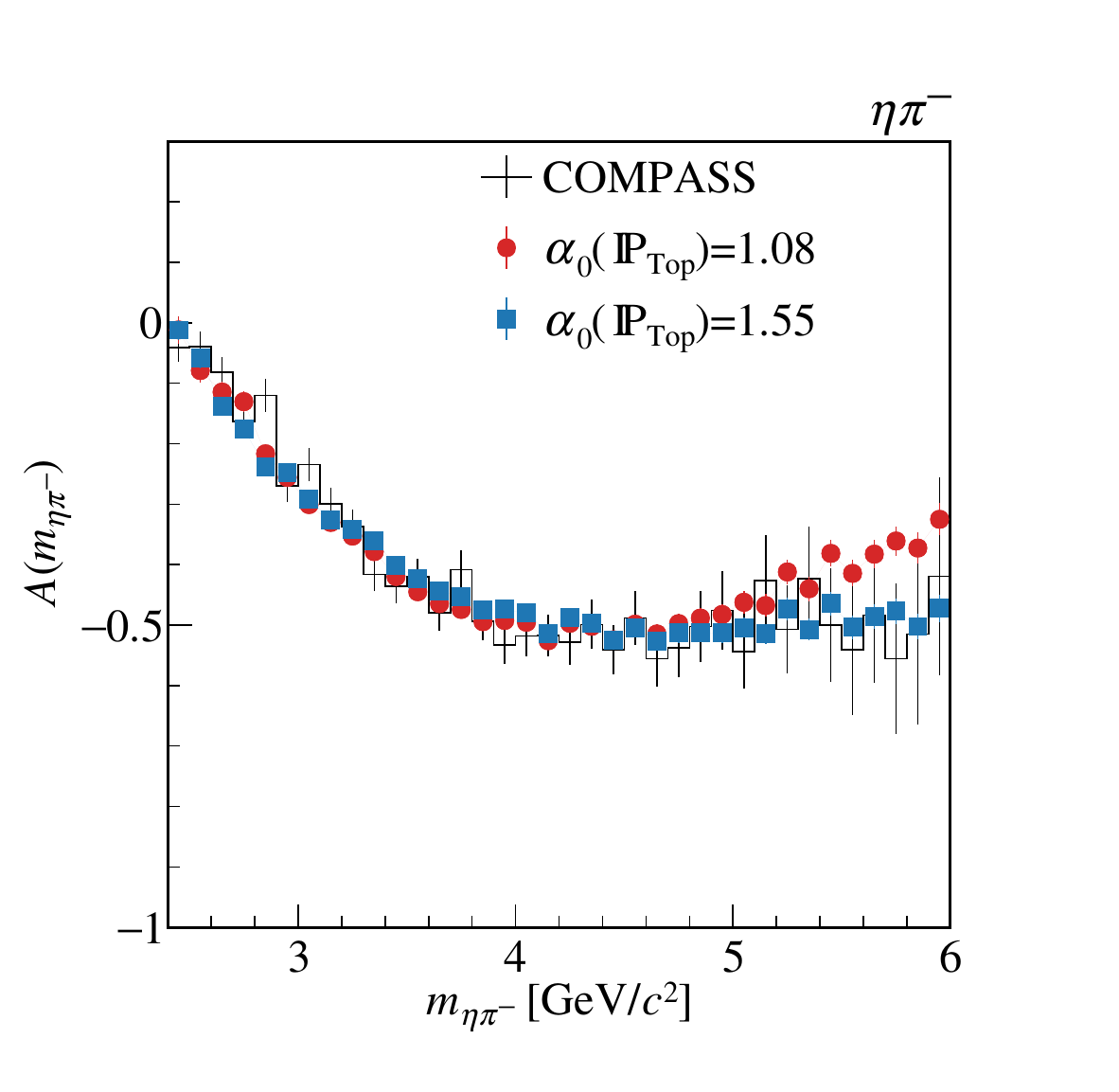} & 
\includegraphics[width=0.3\linewidth]{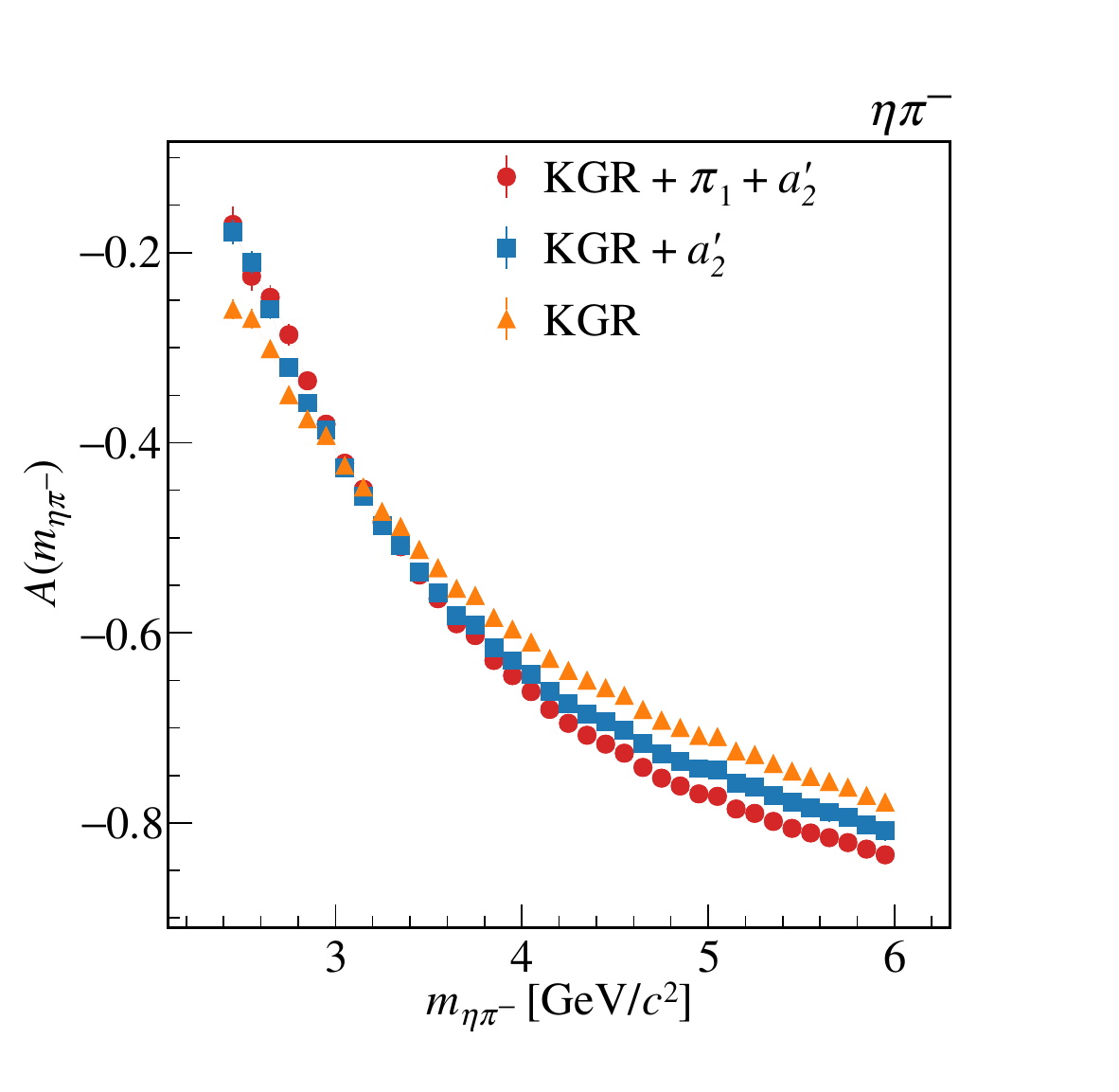}& 
\includegraphics[width=0.3\linewidth]{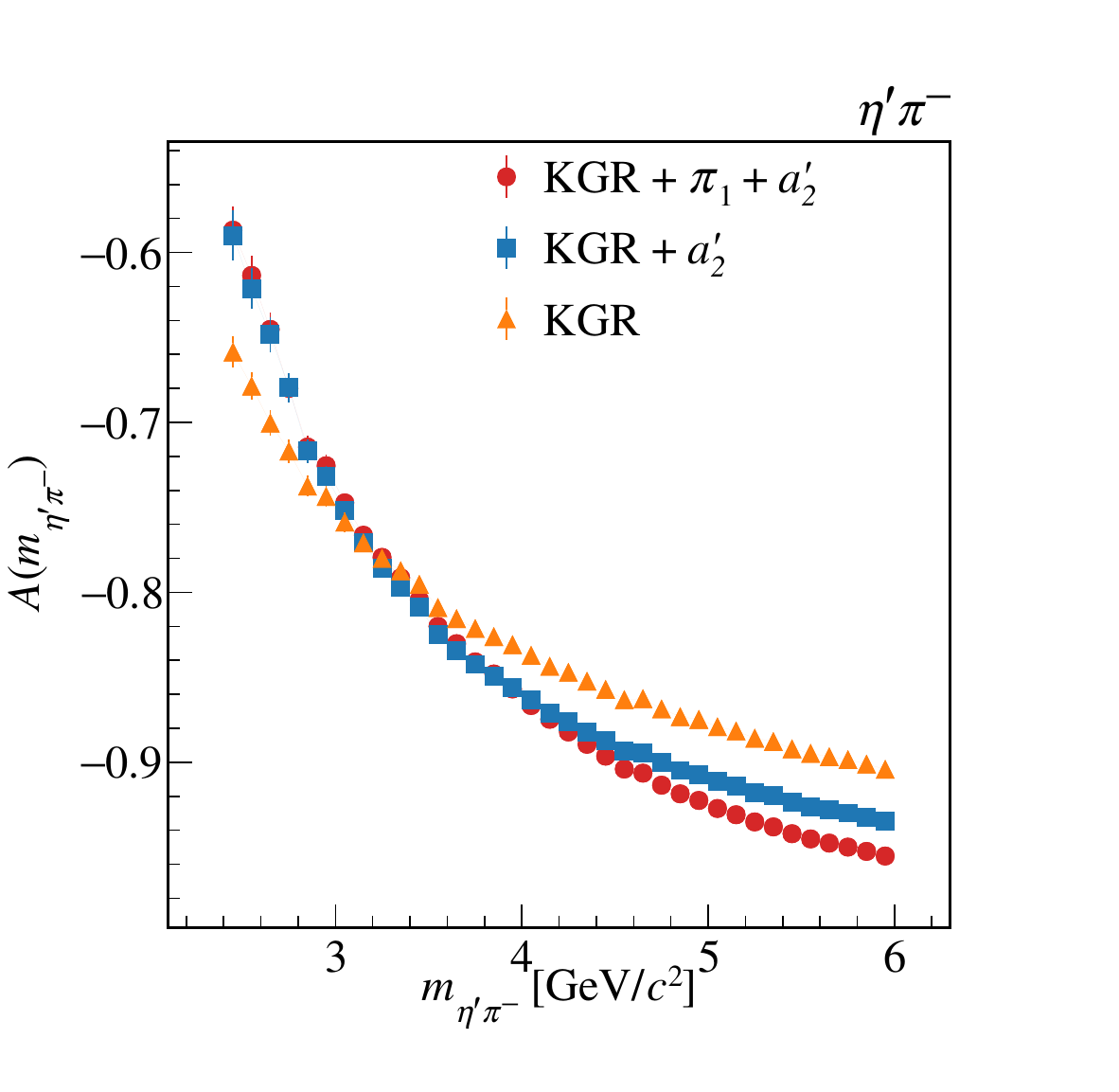} 
\end{tabular}
\caption{
Pomeron studies. Left: Impact of the variation of the Pomeron top exchange
intercept. Center: Theory without acceptance corrections for the $\eta \pi^-$
forward-backward asymmetry. Right: Theory without acceptance corrections for the
$\eta^\prime \pi^-$ forward-backward asymmetry.
}
\label{figsup:pomvar}
\end{figure*}

Finally, we study the impact of the Pomeron in the high energy behavior of the
asymmetry. Results are shown in \cref{figsup:pomvar}. First, on the left plot we
show the impact of changing the Pomeron intercept of the top Pomeron exchange
trajectory in the $\eta \pi^-$ asymmetry. This is an effective way to take into
account corrections to the leading pole approximation which can be relevant for
the $\eta^{\left(\prime \right)}\pi^-$ subsystem. We find that the agreement
with the data can be improved, although the uncorrected Pomeron exchange already
provides results within the experimental uncertainties. The center
($\eta \pi^-$) and right ($\eta^\prime \pi^-$) plots show the theoretical
forward-backward asymmetry for the \mbox{KGR}, \mbox{$\text{KGR}+a_2^\prime$}
and \mbox{$\text{KGR}+a_2^\prime+\pi_1$} fits, when it is not corrected by the
detector effects. It is apparent that the asymmetry tends to $-1$ as expected
from Pomeron dominance.

\end{document}